\newcommand{\lastrevised}{2013 May 26}
\newcommand{\vers}{7.3}
\newcommand{\subvers}{7.3.0}
\newcommand{\arxivnum}{xxxx.xxxx}
\newcommand{\ddscatv}{{\bf DDSCAT \vers}}
\newcommand{\ddscatsixzero}{{\bf DDSCAT 6.0}}

\newcommand{\ddscatsevenone}{{\bf DDSCAT 7.1}}
\newcommand{\ddscatseventwo}{{\bf DDSCAT 7.2}}
\newcommand{\ddscatseventhree}{{\bf DDSCAT 7.3}}
\newcommand{\ddscat}{{\bf DDSCAT}}
\newcommand{\beq}{\begin{equation}}
\newcommand{\eeq}{\end{equation}}
\newcommand{\beqa}{\begin{eqnarray}}
\newcommand{\eeqa}{\end{eqnarray}}

\newcommand{\abs}{{\rm abs}}
\newcommand{\aeff}{{a}_{\rm eff}}
\newcommand{\aone}{\hat{\bf a}_1}
\newcommand{\atwo}{\hat{\bf a}_2}

\newcommand{\balpha}{{\boldsymbol{\alpha}}}
\newcommand{\bA}{{\bf A}}
\newcommand{\bB}{{\bf B}}
\newcommand{\bE}{{\bf E}}
\newcommand{\bH}{{\bf H}}
\newcommand{\bk}{{\bf k}}
\newcommand{\bP}{{\bf P}}

\newcommand{\bx}{{\bf x}}
\newcommand{\DF}{{\rm DF}}
\newcommand{\ext}{{\rm ext}}
\newcommand{\Hz}{\,{\rm Hz}}
\newcommand{\Intel} {Intel\textsuperscript{\textregistered}}
\newcommand{\LF}{{\rm LF}}
\newcommand{\Matlab} {MATLAB\textsuperscript{\textregistered}}
\newcommand{\Microsoft} {Microsoft\textsuperscript{\textregistered}}
\newcommand{\micron}{{\mu{\rm m}}}
\newcommand{\NAG}{NAG\textsuperscript{\textregistered}}
\newcommand{\nm}{\,{\rm nm}}
\newcommand{\TF}{{\rm TF}}
\newcommand{\inc}{{\rm inc}}
\newcommand{\sca}{{\rm sca}}

\newcommand{\xlf}{\hat{\bf x}_\LF}
\newcommand{\ylf}{\hat{\bf y}_\LF}
\newcommand{\zlf}{\hat{\bf z}_\LF}
\newcommand{\xtf}{\hat{\bf x}_\TF}
\newcommand{\ytf}{\hat{\bf y}_\TF}
\newcommand{\ztf}{\hat{\bf z}_\TF}

\newcommand{\gtsim}{\lower.5ex\hbox{$\; \buildrel > \over \sim \;$}} 
\newcommand{\ltsim}{\lower.5ex\hbox{$\; \buildrel < \over \sim \;$}}

\documentclass[english,11pt]{article}
\usepackage[pdftex]{graphicx}
\usepackage{times}
\usepackage{color}
\usepackage{babel}
\usepackage[T1]{fontenc}       
\usepackage[latin9]{inputenc}

\newenvironment{lyxlist}[1]
{\begin{list}{}
{\settowidth{\labelwidth}{#1}
 \setlength{\leftmargin}{\labelwidth}
 \addtolength{\leftmargin}{\labelsep}
 }}
{\end{list}}

\makeatother

\usepackage{amsmath}
\usepackage{natbib}
\makeindex
\begin{document}
\title{
  \vspace*{-3em} 
	{\bf User Guide for the Discrete Dipole} \\
	{\bf Approximation Code \ddscatv}}
                                
\author{Bruce T. Draine \\
	Princeton University Observatory \\
	Princeton NJ 08544-1001 \\
	({\tt draine@astro.princeton.edu})
	\\
	and \\
	\\
	Piotr J. Flatau \\
        University of California San Diego \\
	Scripps Institution of Oceanography \\
	La Jolla CA 92093-0221 \\
	({\tt pflatau@ucsd.edu})
	}
\date{last revised: \lastrevised}
\maketitle
\vspace*{-2em}
\abstract{
	\ddscatv\ is a freely available open-source Fortran-90
	software package applying the
	``discrete dipole approximation'' (DDA) to calculate scattering
	and absorption of electromagnetic waves by targets with arbitrary
	geometries and complex refractive index.  The targets may be isolated
	entities (e.g., dust particles), but may also be 1-d or 2-d periodic
	arrays of ``target unit cells'', which can be used to study
	absorption, scattering, and electric fields around arrays of
	nanostructures.

	The DDA approximates
	the target by an array of polarizable points.
	The theory of the DDA and its implementation in \ddscat\ is
	presented in \citet{Draine_1988} and
        \citet{Draine+Flatau_1994}, and
	its extension to periodic structures
	in \citet{Draine+Flatau_2008a}.
        Efficient near-field calculations are carried out as described in
        \citet{Flatau+Draine_2012}.
	\ddscatv\ allows
	accurate calculations of electromagnetic scattering from targets
	with ``size parameters'' $2\pi \aeff/\lambda \ltsim 25$ provided the
	refractive index $m$ is not large compared to unity ($|m-1| \ltsim 2$).
	\ddscatv\ includes support for
	{\tt MPI}, {\tt OpenMP}, and the \Intel\ Math Kernel Library (MKL).

	\ddscat\ supports calculations for a variety of target geometries
	(e.g., ellipsoids, regular tetrahedra, rectangular solids, 
	finite cylinders, hexagonal prisms, etc.).
	Target materials may be both inhomogeneous and anisotropic.
	It is straightforward for the user to ``import'' arbitrary
	target geometries into the code.
	\ddscat\ automatically calculates total cross sections
	for absorption and scattering and selected elements of the
	Mueller scattering intensity 
	matrix for
	specified orientation of the target relative to the incident wave,
	and for specified scattering directions.
	\ddscatv\ can calculate scattering and absorption by targets
	that are periodic in one or two dimensions.
        \ddscatv\ can calculate and store $\bE$ and
        $\bB$ throughout a user-specified
        rectangular volume containing the target.
	A Fortran-90 code {\bf ddpostprocess} to support postprocessing of
        $\bP$, and nearfield $\bE$ and $\bB$, 
	is included in the distribution.

        \ddscatv\ differs from {\bf DDSCAT\,7.2} by offering two new
        options: 
        (1) The ``Filtered Coupled Dipole'' method 
        \citep{Piller+Martin_1998, Gay-Balmaz+Martin_2002}
        for
        DDA calculations.
        (2) Fast near-field calculations of $\bB$.
        In addition, a new postprocessing code 
        {\bf DDPOSTPROCESS.f90} is provided that is well-documented, and
        much more easily modifiable by the user.  As distributed,
        {\bf ddpostprocess} calculates the Poynting vector.
        
	This User Guide explains how to use \ddscatv\ (release \subvers)
        to carry out electromagnetic scattering calculations.
        If you publish results calculated using \ddscatv, 
        please cite relevant publications describing the
        methods, e.g., \citet{Draine+Flatau_1994},
        \citet{Draine+Flatau_2008a}, and \citet{Flatau+Draine_2012}.
	}

\newpage
\tableofcontents
\newpage
\section{Introduction\label{intro}}

DDSCAT is a software package to calculate scattering and
absorption of electromagnetic waves by targets with arbitrary geometries
using the ``discrete dipole approximation'' (DDA).
In this approximation the target is replaced by an array of point dipoles
(or, more precisely, polarizable points); the electromagnetic scattering
problem for an incident periodic wave interacting with this array of
point dipoles is then solved essentially exactly.
The DDA (sometimes referred to as the ``coupled dipole approximation'') 
was apparently first proposed by \citet{Purcell+Pennypacker_1973}.
DDA theory was reviewed and developed further by \citet{Draine_1988}, 
\citet{Draine+Goodman_1993}, reviewed by \citet{Draine+Flatau_1994},
and recently extended to periodic structures by
\citet{Draine+Flatau_2008a}.

\ddscatv, the current release of DDSCAT, is an open-source Fortran 90 
implementation of the DDA 
developed by the authors.\footnote{
	The release history of {\bf DDSCAT} is as follows:
	\begin{itemize}
	\vspace*{-0.4em}
	\item{\bf DDSCAT 4b}: Released 1993 March 12
	\vspace*{-0.4em} 
	\item{\bf DDSCAT 4b1}: Released 1993 July 9
	\vspace*{-0.4em} 
	\item{\bf DDSCAT 4c}: Although never announced, {\tt DDSCAT.4c} 
		was made available to a number of interested users
		beginning 1994 December 18
	\vspace*{-0.4em}
	\item{\bf DDSCAT 5a7}: Released 1996
	\vspace*{-0.4em}
	\item{\bf DDSCAT 5a8}: Released 1997 April 24
	\vspace*{-0.4em}
	\item{\bf DDSCAT 5a9}: Released 1998 December 15
	\vspace*{-0.4em}
	\item{\bf DDSCAT 5a10}: Released 2000 June 15
	\vspace*{-0.4em}
	\item{\bf DDSCAT 6.0}: Released 2003 September 2
	\vspace*{-0.4em}
	\item{\bf DDSCAT 6.1}: Released 2004 September 10
	\vspace*{-0.4em}
	\item{\bf DDSCAT 7.0}: Released 2008 September 1
        \vspace*{-0.4em}
        \item{\bf DDSCAT 7.1}: Released 2010 February 7
        \vspace*{-0.4em}
        \item{\bf DDSCAT 7.2}: Released 2012 February 15
        \vspace*{-0.4em}
        \item{\bf DDSCAT 7.2.1}: Released 2012 May 14
        \vspace*{-0.4em}
        \item{\bf DDSCAT 7.2.2}: Released 2012 June 3
        \vspace*{-0.4em}
        \item{\bf DDSCAT 7.3.0}: Released 2013 May 26
	\end{itemize}
	}
\ddscatv\ calculates absorption and scattering by isolated targets, or
targets that are periodic in one or two
dimensions, using methods described by \citet{Draine+Flatau_2008a}.

DDSCAT is intended to be a versatile tool, suitable for a wide variety
of applications including studies of interstellar dust, atmospheric aerosols,
blood cells, marine microorganisms, 
and nanostructure arrays.
As provided, \ddscatv\ should be usable 
for many applications without
modification, but the program is written in a modular form, so that
modifications, if required, should be fairly straightforward.

The authors make this code openly available to others, in the hope that it
will prove a useful tool.  We ask only that:
\begin{itemize}
\item If you publish results obtained using {{\bf DDSCAT}}, please 
	acknowledge the source of the code, and cite relevant papers,
	such as \citet{Draine_1988}, 
        \citet{Goodman+Draine+Flatau_1990},
        \citet{Draine+Flatau_1994}, 
        \citet{Draine+Flatau_2008a},
	and \citet{Flatau+Draine_2012}. 

\item If you discover any errors in the code or documentation, 
	please promptly communicate them to the authors.

\item You comply with the ``copyleft" agreement (more formally, the 
	GNU General Public License) of the Free Software Foundation: you may 
	copy, distribute, and/or modify the software identified as coming 
	under this agreement. 
	If you distribute copies of this software, you must give the 
	recipients all the rights which you have. 
	See the file {\tt doc/copyleft.txt} 
        distributed with the DDSCAT software.
\end{itemize}
We also strongly encourage you to send email to
{\tt draine@astro.princeton.edu} identifying  
yourself as a user of DDSCAT;  this will enable the authors to notify you of
any bugs, corrections, or improvements in DDSCAT.
Up-to-date information on DDSCAT 
and the latest version of \ddscatv\ can be found at\\
\hspace*{2em}{\tt http://code.google.com/p/ddscat/}\\

The current version, \ddscatv,
offers the option of using the DDA formulae from \citet{Draine_1988}, with
dipole polarizabilities determined from the Lattice Dispersion
Relation \citep{Draine+Goodman_1993,Gutkowicz-Krusin+Draine_2004}.
Alternatively, \ddscatv\ also allows the user to specify the
``filtered coupled dipole'' method of 
\citet{Piller+Martin_1998} and \citet{Gay-Balmaz+Martin_2002}, which may
give better results for targets with ``large'' refractive indices
$|m-1|\gtsim 2$.
 
The code incorporates Fast Fourier Transform (FFT) methods
\citep{Goodman+Draine+Flatau_1990}.
\ddscatv\ includes capability to calculate scattering and absorption
by targets that are periodic in one or two dimensions -- arrays of
nanostructures, for example.
The theoretical basis for application of the DDA to periodic structures
is developed in \citet{Draine+Flatau_2008a}.
\ddscatseventhree\ includes capability to efficiently perform 
``nearfield'' calculations of $\bE$ and $\bB$ in and around the target using
FFT methods, as described by \citet{Flatau+Draine_2012}.
A new postprocessing code, {\bf DDPOSTPROCESS.f90}, is included in the
\ddscatv\ distribution.

We refer you to the list of references at the end of this document for 
discussions
of the theory and accuracy of the DDA [in particular,
reviews by \citet{Draine+Flatau_1994} and \citet{Draine_2000a},
recent extension to 1-d and 2-d arrays by \citet{Draine+Flatau_2008a},
and comparison of the coupled dipole method with other DDA methods
(including the filtered coupled dipole method)
by \citet{Yurkin+Min+Hoekstra_2010}].

In \S\ref{sec:applicability} we summarize the applicability of the DDA, and 
in \S\ref{sec:DDSCATvers} we describe what the current release can calculate.

In \S\ref{sec:whats_new} we 
describe the principal changes between \ddscatv\ and the previous 
releases.  The succeeding sections contain instructions for:
\begin{itemize}
\item obtaining the source code (\S\ref{sec:downloading});
\item compiling and linking the code (\S\ref{sec:compiling});
\item information for \Microsoft\ Windows users (\S\ref{sec:windows});
\item running a sample calculation (\S\ref{sec:sample calculation});
\item modifying the parameter file to do your desired calculations
(\S\ref{sec:parameter_file});
\item specifying target orientation(s) (\S\ref{sec:target_orientation});
\item understanding the output from the sample calculation;
\item using {\bf DDPOSTPROCESS.f90} for postprocessing of solutions found
      by \ddscatv\
      (\S\ref{sec:ddpostprocess}).
\end{itemize}
The instructions for compiling, linking, and running will be appropriate for a
Linux system; slight changes will be necessary for non-Linux sites, 
but they are quite minor and should present no difficulty.

Finally, the current version of this 
User Guide can be obtained from\newline
{\tt http://arxiv.org/abs/\arxivnum}.

\medskip

{\bf Important Note:} \ddscatseventhree\ differs in a number of respects
from previous versions of \ddscat.  \ddscatseventhree\
includes support for both MPI and OpenMP, but -- as of this writing --
\ddscatseventhree\ has not yet been tested with MPI, and there has
been only limited testing with OpenMP.
\ddscatseventhree\ has been tested extensively on
single-processor systems, but if you are intending to use \ddscatseventhree\
with OpenMP or MPI, please proceed with caution -- 
do at least a few comparison calculations in
single-cpu mode to verify that the results obtained with OpenMP or MPI appear to
be correct.  If you do encounter problems with OpenMP or MPI, please document
them and communicate them to the authors.  And if you find that everything
appears to work properly, we'd like to know that too!

\section{Applicability of the DDA\label{sec:applicability}}
The principal advantage of the DDA is that it is completely flexible 
regarding the geometry of the target, being limited only by the need to 
use an interdipole separation $d$ small compared to 
(1) any structural lengths in the target, and
(2) the wavelength $\lambda$.
Numerical studies \citep{Draine+Goodman_1993,Draine+Flatau_1994,Draine_2000a}
indicate that the second criterion is adequately satisfied if
\beq
|m|kd< 1~~~,
\label{eq:mkd_max}
\eeq
where $m$ is the complex refractive index of the target
material, and $k\equiv2\pi/\lambda$, where $\lambda$ is the wavelength
{\it in vacuo}.
This criterion is valid provided that $|m-1| \ltsim 3$ or so.
When Im$(m)$ becomes large, the DDA solution tends to overestimate
the absorption cross section $C_\abs$, and it may be necessary to use
interdipole separations $d$ smaller than indicated by eq.\ (\ref{eq:mkd_max})
to reduce the errors in $C_\abs$ to acceptable values.

If accurate calculations of the scattering phase function
(e.g., radar or lidar cross sections)
are desired,
a more conservative criterion 
\beq
|m|kd < 0.5
\eeq
will usually ensure that differential scattering cross sections
$dC_\sca/d\Omega$ are accurate to within a few percent of the
average differential scattering cross section $C_\sca/4\pi$
\citep[see][]{Draine_2000a}.

Let $V$ be the actual volume of solid material in the target.\footnote{
   In the case of an infinite periodic target, $V$ is the volume of
   solid material in one ``Target Unit Cell''.}
If the target is represented by an array of $N$ dipoles, located on
a cubic lattice with lattice spacing $d$,
then 
\beq
V=Nd^3 ~~~.
\eeq
We characterize the size of the target by the ``effective radius''
\index{effective radius $\aeff$}
\index{$a_{\rm eff}$}
\beq
a_{\rm eff}\equiv(3V/4\pi)^{1/3} ~~~,
\eeq
the radius of an equal volume sphere.
A given scattering problem is then characterized by the
dimensionless ``size parameter''
\index{size parameter $x=ka_{\rm eff}$}
\beq
x\equiv ka_{\rm eff} = \frac{2\pi a_{\rm eff}}{\lambda} ~~~.
\eeq
The size parameter can be related to $N$ and $|m|kd$:
\beq
x\equiv{2\pi a_{\rm eff}\over\lambda} =
{62.04\over|m|}\left({N\over10^6}\right)^{1/3} \cdot |m|kd ~~~.
\eeq
Equivalently, the target size can be written
\beq
a_{\rm eff} = 9.873 {\lambda\over|m|}\left({N\over10^6}\right)^{1/3}
\cdot |m|kd~~~.
\eeq
Practical considerations of CPU speed and computer memory currently 
available on scientific workstations typically
limit the number
of dipoles employed to $N < 10^6$ (see \S\ref{sec:memory_requirements}
for limitations on $N$ due to available RAM); 
for a given $N$, the limitations on $|m|kd$ 
translate into limitations on the ratio of target size to wavelength.

\noindent
For calculations of total cross sections $C_\abs$ and $C_\sca$,
we require $|m|kd < 1$:
\beq
a_{\rm eff} < 9.88 {\lambda\over |m|}\left({N\over10^6}\right)^{1/3}
{\rm ~~or~~} x < {62.04\over|m|}\left({N\over10^6}\right)^{1/3} ~~~.
\eeq
For scattering phase function calculations, we require $|m|kd < 0.5$:
\beq
a_{\rm eff} < 4.94 {\lambda\over |m|}\left({N\over10^6}\right)^{1/3}
{\rm ~~~~or~~~~} x < {31.02\over|m|}\left({N\over10^6}\right)^{1/3} ~~~.
\eeq

It is therefore clear that the DDA is not suitable for very large values of
the size parameter 
$x$, or very large values of the refractive index $m$.
The primary utility of the DDA is for scattering by dielectric 
targets with sizes comparable to the wavelength.
As discussed by \citet{Draine+Goodman_1993}, \citet{Draine+Flatau_1994}, and
\citet{Draine_2000a},
total cross sections calculated with the DDA are 
accurate to a few percent provided
$N>10^4$ dipoles are used, criterion (\ref{eq:mkd_max}) is satisfied,
and the refractive index is not too large. 

For fixed $|m|kd$, the
accuracy of the approximation degrades with increasing $|m-1|$,
for reasons having to do with the surface polarization of the target,
as discussed by \citet{Collinge+Draine_2004}.
With the present code, good accuracy can be achieved for $|m-1| < 2$.

\begin{figure}[t]
\begin{center}
\vspace*{-0.9cm}
\includegraphics[width=8.3cm]{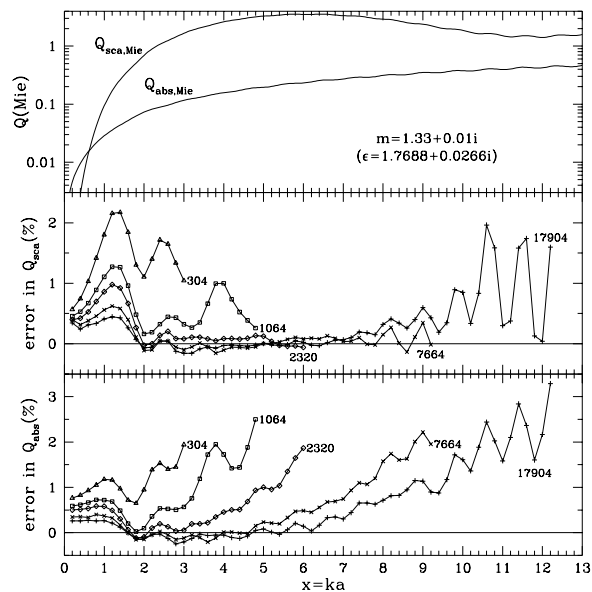}
\vspace*{-2.2cm}
\caption{\footnotesize
        Scattering and absorption for a sphere with
	$m=1.33+0.01i$.  The upper panel shows the exact values of $Q_\sca$
	and $Q_\abs$, obtained with Mie theory, as functions of $x=ka$.
	The middle and lower panels show fractional errors in $Q_\sca$ and
	$Q_\abs$, obtained using {{\bf DDSCAT}}\ with polarizabilities obtained
	from the Lattice Dispersion Relation, and labelled by the number $N$
	of dipoles in each pseudosphere.
	After Fig.\ 1 of \citet{Draine+Flatau_1994}.}
	\label{fig:Qm=1.33+0.01i}
\end{center}
\end{figure}
\begin{figure}[h]
\begin{center}
\vspace*{-0.9cm}
\includegraphics[width=8.3cm]{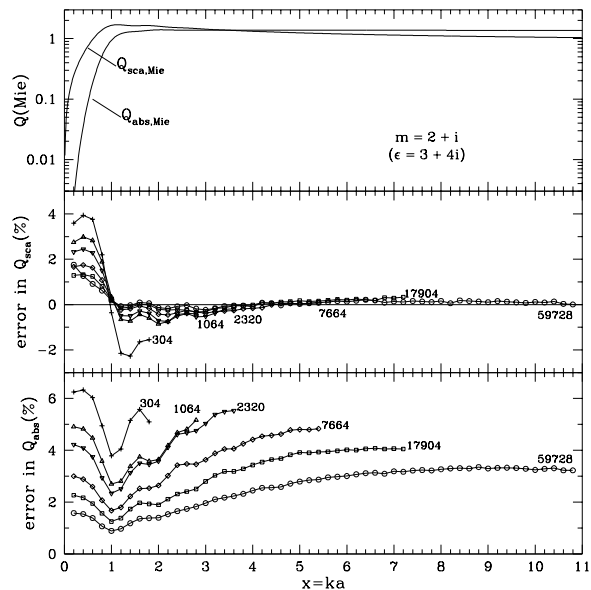}
\vspace*{-2.2cm}
\caption{\footnotesize
        Same as Fig.\ \protect{\ref{fig:Qm=1.33+0.01i}},
	but for $m=2+i$. After Fig.\ 2 of \citet{Draine+Flatau_1994}.
        Note: in the upper panel, the labels for $Q_{\rm sca,Mie}$ and
        $Q_{\rm abs,Mie}$ should be interchanged.}
	\label{fig:Qm=2+i}
\end{center}
\end{figure}
\begin{figure}[t]
\begin{center}
\vspace*{-0.9cm}
\includegraphics[width=8.3cm]{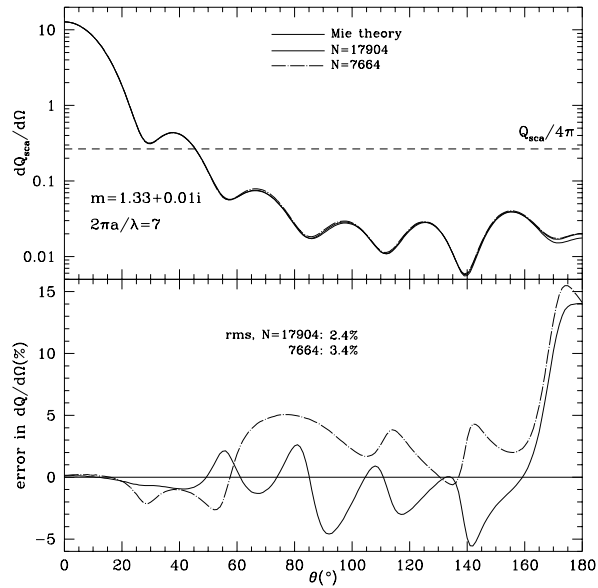}
\vspace*{-2.2cm}
\caption{\footnotesize
        Differential scattering cross section for 
	$m=1.33+0.01i$ pseudosphere and $ka=7$.
	Lower panel shows fractional error compared to exact Mie theory
	result.
	The $N=17904$ pseudosphere has $|m|kd=0.57$, and an rms fractional
	error in $d\sigma/d\Omega$ of 2.4\%.
	After Fig.\ 5 of \citet{Draine+Flatau_1994}.}
	\label{fig:dQdom=1.33+0.01i}
\end{center}
\end{figure}
\begin{figure}[h]
\begin{center}
\vspace*{-0.9cm}
\includegraphics[width=8.3cm]{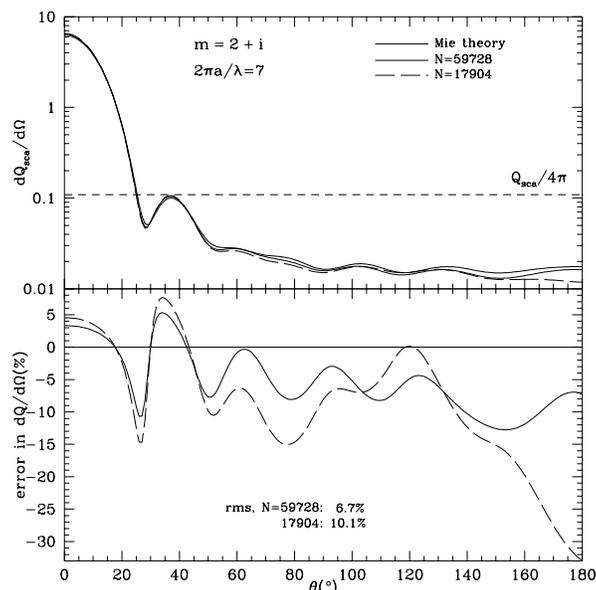}
\vspace*{-2.5cm}
\caption{\footnotesize
        Same as Fig.\ \protect{\ref{fig:dQdom=1.33+0.01i}}
	but for $m=2+i$.
	The $N=59728$ pseudosphere has $|m|kd=0.65$, and an rms fractional
	error in $d\sigma/d\Omega$ of 6.7\%.
	After Fig.\ 8 of \citet{Draine+Flatau_1994}.}
	\label{fig:dQdom=2+i}
\end{center}
\end{figure}

Examples illustrating the accuracy of the DDA are shown in 
Figs.\ \ref{fig:Qm=1.33+0.01i}--\ref{fig:Qm=2+i}, which show overall
scattering and absorption efficiencies as a function of wavelength for
different discrete dipole approximations to a sphere, with $N$ ranging
from 304 to 59728.
The DDA calculations assumed radiation incident along the (1,1,1)
direction in the ``target frame''.
Figs.\ {\ref{fig:dQdom=1.33+0.01i}--\ref{fig:dQdom=2+i} show the scattering
properties calculated with the DDA for $x=ka=7$.
Additional examples can be found in \citet{Draine+Flatau_1994} and 
\citet{Draine_2000a}.

As discussed below,
\ddscatv\ can also calculate scattering and absorption by targets that
are periodic in one or two directions -- for examples, see
\citet{Draine+Flatau_2008a}.

\section{DDSCAT \vers\label{sec:DDSCATvers}}
\subsection{ What Does It Calculate?}

\subsubsection{Absorption and Scattering by Finite Targets}
\ddscatv, like previous versions of \ddscat,
solves the problem of scattering and 
absorption by a finite target, represented by an array of
polarizable point dipoles, interacting with a monochromatic plane wave incident
from infinity.  
\ddscatv\ has the capability of automatically generating dipole
array representations for a variety of target geometries 
(see \S\ref{sec:target_generation}) and can also accept dipole
array representations of targets supplied by the user (although
the dipoles must be located on a cubic lattice).
The incident plane wave can have arbitrary elliptical
polarization (see \S\ref{sec:incident_polarization}), 
and the target can be arbitrarily oriented relative to the
incident radiation (see \S\ref{sec:target_orientation}).
The following quantities are calculated by \ddscatv\ :
\begin{itemize}
\item Absorption efficiency factor $Q_\abs\equiv C_\abs/\pi a_{\rm eff}^2$,
where $C_\abs$ is the absorption cross section;
\index{absorption efficiency factor $Q_\abs$}
\index{$Q_\abs$}
\item Scattering efficiency factor $Q_\sca\equiv C_\sca/\pi a_{\rm eff}^2$,
where $C_\sca$ is the scattering cross section;
\index{scattering efficiency factor $Q_\sca$}
\index{$Q_\sca$}
\item Extinction efficiency factor $Q_\ext\equiv Q_\sca+Q_\abs$;
\index{extinction efficiency factor $Q_\ext$}
\index{$Q_\ext$}
\item Phase lag efficiency factor $Q_{\rm pha}$, 
\index{phase lag efficiency factor $Q_{\rm pha}$}
\index{$Q_{\rm pha}$}
defined so that the phase-lag
(in radians) of a plane wave after propagating a distance $L$ is just
$n_{t}Q_{\rm pha}\pi a_{\rm eff}^2 L$, 
where $n_t$ is the number density of targets.
\item The 4$\times$4 Mueller scattering intensity matrix $S_{ij}$ 
describing the complete
scattering properties of the target for scattering directions specified
by the user (see \S\ref{sec:mueller_matrix}).
\item Radiation force efficiency vector ${\bf Q}_{\rm rad}$ 
(see \S\ref{sec:force and torque calculation}).
\item Radiation torque efficiency vector ${\bf Q}_\Gamma$
(see \S\ref{sec:force and torque calculation}).
\end{itemize}
In addition, the user can choose to have \ddscatv\ store the solution for
post-processing.  


\subsubsection{ Absorption and Scattering by Periodic Arrays of Finite 
               Structures}

\ddscatv\ includes the capability to solve the problem of scattering and 
absorption by an infinite target consisting of a 1-d or 2-d periodic array of 
finite structures, illuminated by an incident plane wave.  
The finite structures are themselves represented by arrays of point dipoles.

The electromagnetic scattering problem for periodic arrays is formulated by 
\citet{Draine+Flatau_2008a}, 
who show how the problem can be reduced to a finite system of linear 
equations, and solved by the same methods used for scattering by finite 
targets.

The far-field scattering properties of the 1-d and 2-d periodic arrays can be
conveniently represented by a generalization of the Mueller scattering matrix 
to the 1-d or 2-d periodic geometry -- see \citet{Draine+Flatau_2008a} for 
definition of $S_{ij}^{(1d)}(M,\zeta)$ and $S_{ij}^{(2d)}(M,N)$.
For targets with 1-d periodicity, \ddscatv\ calculates 
$S_{ij}^{(1d)}(M,\zeta)$ for user-specified $M$ and $\zeta$.
For targets with 2-d periodicity, \ddscatv\ calculates
$S_{ij}^{(2d)}(M,N)$ for both transmission and reflection, for user-specified
$(M,N)$.

As for finite targets, the user can choose to have \ddscatv\ store the 
calculated polarization field for post-processing.  

\subsection{ Application to Targets in Dielectric Media}
	\label{sec:target_in_medium}}

Let $\omega$ be the angular frequency of the incident radiation.
Beginning with
\ddscatseventwo, \ddscat\ facilitates calculation of absorption and scattering
by targets immersed in dielectric media (e.g., liquid water).
In the parameter file {\tt ddscat.par}, the user simply specifies
the refractive index $m_{\rm medium}$ of the ambient medium.
If the target is {\it in vacuo}, set $m_{\rm medium}=1$.
Otherwise, set $m_{\rm medium}$ to be the refractive index in the
ambient medium at frequency $\omega$.
For example, H$_2$O has $m_{\rm medium}=1.335$ near $\omega/2\pi=6\times10^{14}\Hz$,
the frequency corresponding to $\lambda_{\rm vac}=500\nm$ (green light).
At this frequency, air at STP has $m_{\rm medium}=1.00028$,
which can be taken to be 1 for most applications.

The wavelength provided in the file {\tt ddscat.par} should be the
{\it vacuum} wavelength $\lambda_{\rm vac}=2\pi c/\omega$ 
corresponding to the frequency $\omega$.

The dielectric function or refractive index for the target material
is provided
via a file, with the filename provided via {\tt ddscat.par}.
The file should give either the 
actual complex dielectric function $\epsilon_{\rm target}$ 
or actual complex refractive index 
$m_{\rm target}=\sqrt{\epsilon_{\rm target}}$
of the target material, as a function of
wavelength {\it in vacuo}.
\index{dielectric medium}

\underline{Internal} to \ddscatseventhree, the scattering calculation is 
carried out using
the {\it relative} dielectric function
\index{relative dielectric function}
\beq
\epsilon_{\rm rel}(\omega) = 
\frac{\epsilon_{\rm target}(\omega)}{\epsilon_{\rm medium}(\omega)} ~~~,
\eeq
{\it relative} refractive index:
\index{relative refractive index}
\beq
m_{\rm rel}(\omega) =
\frac{m_{\rm target}(\omega)}{m_{\rm medium}(\omega)}~~~,
\eeq
and wavelength in the ambient medium 
\beq
\lambda_{\rm medium}=\frac{\lambda_{\rm vac}}{m_{\rm medium}}~~~.
\label{eq:lambda_medium}
\eeq

The absorption, scattering, extinction, and phase lag 
efficiency factors $Q_\abs$,
$Q_\sca$, $Q_\ext$, and $Q_{\rm pha}$ calculated by {{\bf DDSCAT}}
will then be equal to the physical
cross sections for absorption, scattering, and extinction divided by
$\pi a_{\rm eff}^2$.
For example, the attenuation coefficient $\alpha$ for radiation
propagating through a diffuse medium with a number density $n_{t}$ of
scatterers will be just
\beq
\alpha = n_{t}Q_\ext\pi a_{\rm eff}^2
~~~.
\eeq
Similarly, the phase lag (in radians) after propagating a distance $L$ will be
$n_{t}Q_{\rm pha}\pi a_{\rm eff}^2 L$.

\medskip

The elements $S_{ij}$ of the 4$\times$4 Mueller scattering matrix ${\bf S}$
calculated by {{\bf DDSCAT}} for finite targets
will be correct for scattering in the medium:
\beq
{\bf I}_\sca = 
\left(\frac{\lambda_{\rm medium}}{2\pi r}\right)^2 
{\bf S}\cdot {\bf I}_{\rm in} ,
\eeq
where ${\bf I}_{\rm in}$ and ${\bf I}_\sca$ are the Stokes vectors for
the incident and scattered light (in the medium), 
$r$ is the distance from the target, and
$\lambda_{\rm medium}$ is the wavelength in the medium (eq.\ \ref{eq:lambda_medium}).
See \S\ref{sec:mueller_matrix} for a detailed discussion of the
Mueller scattering matrix.

The time-averaged 
radiative force and torque (see \S\ref{sec:force and torque calculation}) on a
finite target in a dielectric medium are
\beq
{\bf F}_{\rm rad} = {\bf Q}_{\rm pr}\pi a_{\rm eff}^2 u_{\rm rad} ~~~,
\eeq
\beq
{\bf \Gamma}_{\rm rad} = 
{\bf Q}_\Gamma \pi a_{\rm eff}^2 u_{\rm rad} \frac{\lambda_{\rm medium}}{2\pi} ~~~,
\eeq
where the time-averaged energy density is
\beq 
u_{\rm rad}=\epsilon_{\rm medium} \frac{|E_0|^2}{8\pi} ~~~ ,
\eeq
where $E_0\cos(\omega t+\phi)$
is the electric field of the incident plane wave in the medium.

The relationship between the microscopic and macroscopic fields 
$\bE_{\rm micro}$ and $\bE_{\rm macro}$ is discussed
in \S\ref{sec:micro_vs_macro} (see eq.\ \ref{eq:macro_vs_micro}).

\bigskip
\section{What's New?\label{sec:whats_new}}
\ddscatv\ differs from
\ddscatsevenone\ 
in several ways, including:

\begin{enumerate}
\item N.B.: The structure of the parameter file {\tt ddscat.par} 
        has again been changed: {\tt ddscat.par} files that were used with 
	\ddscatsevenone\ or \ddscatseventwo\ will need to be modified.
	See \S\ref{sec:parameter_file} and Appendix \ref{app:ddscat.par}.
\item As in \ddscatseventwo, \ddscatseventhree\ requires the user to 
specify the (real) refractive
index $m_{\rm ambient}$ of the ambient medium.  See \S\ref{sec:parameter_file}.
\item As with \ddscatseventwo, \ddscatseventhree\ 
includes support for two additional conjugate
gradient solvers (see \S\ref{sec:choice_of_algorithm}):
\begin{itemize}
\item GPBICG -- Implementation by Chamuet and Rahmani of the Complex
                Conjugate Gradient solver presented by 
                \citet{Tang+Shen+Zheng+Qiu_2004}.
\item QMRCCG -- Quasi-Minimum-Residual Complex Conjugate Gradient solver,
                adapted from fortran-90 implementation kindly made available 
                by P.C.\ Chaumet and A.\ Rahmani
                \citep{Chaumet+Rahmani_2009}.
\end{itemize}
\item As with \ddscatseventwo, \ddscatseventhree\ requires the user to 
    specify the maximum
    allowed number of iterations (this was previously hard-wired).
\item As with \ddscatseventwo, \ddscatseventhree\ supports
    fast calculations of the electric field within and near
    the target
    using FFT methods, as described by
    \citet{Flatau+Draine_2012}; see \S\ref{sec:nearfield}).
    The program DDFIELD that came with \ddscatsevenone\ 
    is no longer needed (and no longer supported).
\item \ddscatseventhree\ now includes the option of fast calculations
    of $\bB$ in and near the target using FFT methods
    (see \S\ref{sec:nearfield}).
\item As with \ddscatseventwo, 
    the \ddscatseventhree\ distribution includes a set of ``example''
    calculations, but a new example has been added:
\begin{itemize}
    \item {\bf FROM\_FILE}: target whose
          geometry is supplied via an ascii file
    \item {\bf ELLIPSOID}: sphere, with
          nearfield calculation of $\bE$ in and around the sphere.
    \item {\bf RCTGLPRSM}: rectangular brick
    \item {\bf SPH\_ANI\_N}: cluster of
          spheres, each characterized by an anisotropic dielectric function
    \item {\bf CYLNDRPBC}: infinite cylinder,
          calculated as a periodic array of disks.
    \item {\bf DSKRCTPBC}: doubly-periodic array of
          Au disks supported by a Si$_3$N$_4$ slab.
    \item {\bf RCTGL\_PBC}: doubly-periodic array of
          rectangular blocks.
    \item {\bf RCTGL\_PBC}: doubly-periodic array of
          rectangular blocks, with nearfield calculation of $\bE$
          in and around the sphere.
    \item {\bf SPHRN\_PBC}: doubly-periodic array of clusters of spheres.
    \item {\bf ELLIPSOID\_NEARFIELD}: same calculation as in the
          {\bf ELLIPSOID} example, followed by a nearfield calculation
          of $\bE$ in and around the sphere, followed by evaluation of
          $\bE$ along a track passing through the sphere, plus
          creation of ``VTK'' files for visualization.
    \item {\bf RCTGLPRSM\_NEARFIELD}: same calculation as in the
          {\bf RCTGLPRSM} example, followed by a nearfield calculation
          of $\bE$ in and around the target, followed by evaluation of
          $\bE$ along a track passing through the prism, plus
          creation of ``VTK'' files for visualization.
    \item {\bf RCTGL\_PBC\_NEARFIELD}: same calculation as in the
          {\bf RCTGL\_PBC} example, followed by a nearfield calculation
          of $\bE$ in and around the target, followed by evaluation of
          $\bE$ along a track passing through the prism, plus
          creation of ``VTK'' files for visualization.
    \item {\bf RCTGL\_PBC\_NEARFLD\_B}: same calculation as in the
          {\bf RCTGL\_PBC\_NARFIELD} example, 
          followed by evaluation of
          both $\bE$ and $\bB$ along a track passing through the prism, plus
          creation of ``VTK'' files for visualization.
    \end{itemize}
\item \ddscatseventhree\ is distributed with a program {\bf VTRCONVERT.f90}
    that supports visualization of target geometries using the
    Visualization Toolkit (VTK), an open-source, 
    freely-available software system
    for 3D computer graphics (http://www.vtk.org) and, specifically,
    ParaView (http://paraview.org).
\item \ddscatseventhree\ is distributed with a program {\bf DDPOSTPROCESS.f90}.
    \begin{itemize}
    \item {\bf DDPOSTPROCESS.f90} 
          allows the user to easily extract $\bE$ (and $\bB$ if it was
          also precalculated) at points along a line.
    \item If $\bB$ was precalculated, {\bf DDPOSTPROCESS.f90} calculates
      the time-averaged Poynting vector $c\langle\bE\times\bH\rangle/4\pi$
      at each point in the computational volume.
    \item {\bf DDPOSTPROCESS.f90} 
          creates ``VTK'' files for visualization of $|\bE|$,
          $|\bE|^2$, or $(c/4\pi)\langle\bE\times\bH\rangle$ 
          using the VTK tools.
    \end{itemize}
    
\end{enumerate}

\section{Downloading the Source Code and Example Calculations
         \label{sec:downloading}}

\ddscatv\ is written in standard Fortran-90 with a single extension: it uses
the Fortran-95 standard library call {\tt CPU\_TIME} to obtain timing
information.
\ddscatv\ is therefore portable to
any system having a f90 or f95 compiler.
It has been successfully
compiled with many different compilers, including {\tt gfortran},
{\tt g95}, {\tt ifort}, {\tt pgf77}, and \NAG {\tt f95}.

\index{Microsoft\textsuperscript{\textregistered}}
\index{Vista}
\index{Windows 7}
It is possible to use \ddscatv\ on PCs running \Microsoft\ Windows
operating systems, including Vista and Windows 7.
Section \ref{sec:windows} provides instructions for creating a unix-like
environment in which you can compile and run \ddscatv.

Alternatively, you may be able to obtain a precompiled native executable
-- see \S\ref{sec:native executable}.
More information on how to do this can be found at\\
\hspace*{4em}{\tt http://code.google.com/p/ddscat}

The remainder of this section will assume that the installation is taking
place on a Unix, Linux or Mac OSX system 
with the standard developer tools (e.g., {\tt tar}, {\tt make}, 
and a f90 or f95 compiler) installed.

\index{source code (downloading of)}
The complete source code for
\ddscatv\ is
provided in a single gzipped tarfile.
To obtain the source code, simply
point your browser to
{\tt http://code.google.com/p/ddscat/}
and download the latest release.

After downloading {\tt ddscat\subvers.tgz} 
into the directory where you would
like \ddscatv\ to reside (you should have at least 10 Mbytes of disk space
available), the source code can be installed as follows:


If you are on a Linux system, you should be able to type\\
\indent\indent{\tt tar xvzf ddscat\subvers.tgz}\\
which will``extract'' the files from the gzipped
tarfile.  If your version of ``tar'' doesn't support the ``z'' option,
then try\\
\indent\indent{\tt zcat ddscat\subvers.tgz | tar xvf -}\\
If neither of the above work on your system, try the two-stage procedure\\
\indent\indent{\tt gunzip ddscat\subvers.tgz} \\
\indent\indent{\tt tar xvf ddscat\subvers.tar} \\
The only disadvantage of the two-stage procedure is that it uses more disk
space, since after the second step you will have the uncompressed
tarfile {\tt ddscat\subvers.tar} -- about 3.8 Mbytes --
in addition to all the files you have extracted
from the tarfile -- another 4.6 Mbytes.

Any of the above approaches should 
create subdirectories {\tt src}, {\tt doc}, {\tt diel}, and {\tt examples\_exp}.
\begin{itemize}
\item {\tt src} contains the source code.
\item {\tt doc} contains documentation (including this UserGuide).
\item {\tt diel} contains a few sample files specifying refractive
indices or dielectric functions as functions of vacuum wavelength.
\end{itemize}

It is also recommended that you download {\tt ddscat\subvers\_examples.tgz},
followed by\\
\indent\indent{\tt tar xvzf ddscat\subvers\_examples.tgz}\\
Subdirectory {\tt examples\_exp} contains sample {\tt ddscat.par} files
as well as output files for various example problems, including
both isolated targets and infinite periodic targets.
It also includes sample {\tt ddpostprocess.par} files for running 
{\bf ddpostprocess}
to support visualization following nearfield calculations.




\section{Compiling and Linking on Unix/Linux Systems\label{sec:compiling}}
\index{compiling and linking}
\index{compiling and linking}
In the discussion below, it is assumed that the
\ddscatv\ source code has been installed in a directory
{\tt DDA/src}.
The instructions below assume that you are on a Unix, Linux, or Mac OSX 
system.

\subsection{\label{sec:Makefile}
            The default Makefile}
It is assumed that the system has the following already installed:
\begin{itemize}
\item a Fortran-90 compiler (e.g., {\tt gfortran}, {\tt g95},
\Intel\ {\tt ifort}, or \NAG {\tt f95}) .
\item {\tt cpp} -- the ``C preprocessor''.
\end{itemize}
There are a number of different ways to create an executable, depending
on what options the user wants:
\begin{itemize}
\item what compiler and compiler flags?
\item single-  or double-precision?
\item enable OpenMP?
\item enable MKL?
\item enable MPI?
\end{itemize}
Each of the above choices needs requires setting of appropriate
``flags'' in the Makefile.

The default Makefile has the following ``vanilla'' settings:
\begin{itemize}
\item {\tt gfortran -O2}
\item single-precision arithmetic
\item OpenMP not used
\item MKL not used
\item MPI not used.
\end{itemize}
To compile the code with the default settings, simply
position yourself in the directory
{\tt DDA/src}, and type\\
\hspace*{2em}{\tt make ddscat}\\
If you have {\tt gfortran} and {\tt cpp} installed on your system,
the above should work. You will get some warnings from the compiler,
but the code will compile and create an executable {\tt ddscat}.

If you wish to use a different compiler
(or compiler options) you will need to edit the file {\tt Makefile}
to change the choice of compiler (variable {\tt FC}),
compilation options (variable {\tt FFLAGS}), and possibly
and loader options (variable {\tt LDFLAGS}).
The file {\tt Makefile} as provided includes examples for compilers other than
{\tt gfortran}; you may be able to simply ``comment out'' the section of
{\tt Makefile} that was designed for {\tt gfortran}, and ``uncomment''
a section designed for another compiler (e.g., \Intel\ {\tt ifort}).

\subsection{ Optimization}
\index{Fortran compiler!optimization}
The performance of \ddscatv\ will benefit from optimization during compilation
and the user should enable the
appropriate compiler flags.

\subsection{ Single vs.\ Double Precision}
\index{precision: single vs. double}
\ddscatv\ is written to allow the user to easily generate either single- or
double-precision versions of the executable.  
For most purposes, the single-precision precision version of \ddscatv\
should work fine, but if you encounter a scattering 
problem where the single-precision version
of \ddscatv\ seems to be having trouble converging to a solution, you
may wish to try using the double-precision version -- this
can be beneficial in the event that round-off error is compromising
the performance of the conjugate-gradient solver.
Of course, the double precision version will demand about twice as much
memory as the single-precision version, 
and will take somewhat longer per iteration.

The only change required
is in the Makefile: for single-precision, set\\
\hspace*{2em}{\tt PRECISION = sp}\\
or for double-precision, set\\
\hspace*{2em}{\tt PRECISION = dp}\\
After changing the {\tt PRECISION} variable in the Makefile
(either sp $\rightarrow$ dp,
or dp $\rightarrow$ sp), it is necessary to recompile the
entire code.  Simply type\\
\hspace*{2em}{\tt make clean}\\
\hspace*{2em}{\tt make ddscat}\\
to create {\tt ddscat} with the appropriate precision.

\subsection{ OpenMP}
\index{OpenMP}

OpenMP is a standard for support of shared-memory parallel programming,
and can provide a performance advantage when using \ddscatv\ on
platforms with multiple cpus or multiple cores.
OpenMP is supported by many common compilers, such as {\tt gfortran}
and \Intel\ {\tt ifort}.

If you are using a multi-cpu (or multi-core) system with OpenMP 
({\tt www.openmp.org}) installed, you can compile \ddscatv\ with OpenMP
directives to parallelize some of the calculations.
To do so, simply change\\
\hspace*{2em}{\tt DOMP       =}\\
\hspace*{2em}{\tt OPENMP     =}\\
to\\
\hspace*{2em}{\tt DOMP       = -Dopenmp}\\
\hspace*{2em}{\tt OPENMP     = -openmp}\\
Note: {\tt OPENMP} is compiler-dependent: {\tt gfortran}, for instance,
requires\\
\hspace*{2em}{\tt OPENMP     = -fopenmp}\\

\index{{\tt OMP\_NUM\_THREADS}}
After compiling \ddscatv\ to use OpenMP,
it is necessary to specify the number of threads to be used.
To specify the number of threads, you need to set the environmental
variable {\tt OMP\_NUM\_THREADS}. 
This is done by a command in the shell (e.g., bash or
ksh).  For example, the number of threads would be set to two by
the command\\
\hspace*{2em}{\tt export OMP\_NUM\_THREADS=2}\\
The number of threads should not exceed the number of available ``cores''.
If you are executing \ddscatv\ on a multi-core system, 
you may wish to experiment to see how the execution ``wall-clock time'' 
varies depending on the number of threads.

\subsection{ MKL: the \Intel\ Math Kernel Library
            \label{sec:MKL}
            \index{MKL!\Intel\ MKL!DFTI}
            }

\Intel\ offers the Math Kernel Library (MKL) with the ifort compiler.
This library includes {\tt DFTI} for computing FFTs.
At least on some systems, {\tt DFTI} offers better performance than
the GPFA package.

To use the MKL library routine {\tt DFTI}:
\begin{itemize}
\item You must have MKL installed on your system.
\item You must obtain the routine {\tt mkl\_dfti.f90} and place a copy
      in the directory where you are compiling \ddscatv.  {\tt mkl\_dfti.f90}
      is \Intel\ proprietary software, so we cannot distribute it
      with the \ddscatv\ source code, but it
      should be available on any system with a license for the
      \Intel\ MKL library.
      If you cannot find it, ask your system administrator.
\item Edit the Makefile: define variables {\tt CXFFTMKL.f} and 
      {\tt CXFFTMKL.o} to:\\
      \hspace*{2em}{\tt CXFFTMKL.f = \$(MKL\_f)}\\
      \hspace*{2em}{\tt CXFFTMKL.f = \$(MKL\_o)}
\item Successful linking will require that the appropriate MKL libraries
be available, and that the string {\tt LFLAGS} in the Makefile be
defined so as to include these libraries.
Unfortunately, there is a lot of variation in how the MKL
libraries are installed on different systems.\\
{\footnotesize\tt -lmkl\_intel\_thread -lmkl\_core -lguide 
-lpthread -lmkl\_intel\_lp64}\\ 
appears to work on at least one installation.
On some installations,\\ 
{\footnotesize\tt -lmkl\_em64t -lmkl\_intel\_thread -lmkl\_core -lguide 
-lpthread -lmkl\_intel\_lp64}\\ 
seems to work.  The Makefile contains examples.
You may want to consult a guru who is familiar with the libraries
on your local system.
\Intel\ provides a website\\
\hspace*{2em}http://software.intel.com/sites/products/mkl/\\
that can assist you in figuring out 
the appropriate compiler and linker options
for your system.

\item type\\
      \hspace*{2em}{\tt make clean}\\
      \hspace*{2em}{\tt make ddscat}
\item The parameter file {\tt ddscat.par} should have {\tt FFTMKL} 
as the value of {\tt CMETHD}.
\end{itemize}

\subsection{ MPI: Message Passing Interface
            \label{sec:MPI}
            \index{MPI -- Message Passing Interface}
            }

\ddscatv\ includes support for parallelization under MPI.
MPI (Message Passing Interface) is a standard for communication between
processes.
More than one implementation of {\tt MPI} exists (e.g., {\tt mpich}
and {\tt openmpi}).  
{\tt MPI} support within \ddscatv\ is compliant with the MPI-1.2 and MPI-2
standards\footnote{\tt http://www.mpi-forum.org/}, 
and should be usable under any implementation of {\tt MPI}
that is compatible with those standards.

Many scattering calculations will require multiple orientations of the
target relative to the incident radiation.  For \ddscatv, such
calculations are  ``embarassingly parallel'', because they are carried
out essentially independently.  \ddscatv\ uses MPI so that scattering
calculations at a single wavelength but for multiple orientations can
be carried out in parallel, with the information for different orientations
gathered together for averaging etc.\ by the master process.  

If you intend
to use \ddscatv\ for only a single orientation, MPI offers no advantage
for \ddscatv\, so you should compile with MPI disabled.
However, if you intend to carry out calculations for multiple orientations,
{\it and} would like to do so in parallel over more than one cpu, {\it and}
you have MPI installed on your platform, then you will want to compile
\ddscatv\ with MPI enabled.
  
To compile with MPI disabled: in the Makefile, set\\
      \hspace*{2em}{\tt DMPI  =}\\
      \hspace*{2em}{\tt MIP.f = mpi\_fake.f90}\\
      \hspace*{2em}{\tt MPI.o = mpi\_fake.o}\\
To compile with MPI enabled: in the Makefile, set\\
      \hspace*{2em}{\tt DMPI  = -Dmpi}\\
      \hspace*{2em}{\tt MIP.f = \$(MPI\_f)}\\
      \hspace*{2em}{\tt MPI.o = \$(MPI\_o)}\\
and edit {\tt LFLAGS} as needed to make sure that you link to the
appropriate MPI library (if in doubt, consult your systems administrator).
The {\tt Makefile} in the distribution includes some examples, but 
library names and locations are often system-dependent.
Please do {\it not} direct questions regarding {\tt LFLAGS} to the authors --
ask your sys-admin or other experts familiar with your installation.

Note that the MPI-capable executable can also be used for ordinary serial 
calculations using a single cpu.

\subsection{ Device Numbers {\tt IDVOUT} and {\tt IDVERR}
\label{subsec:IDVOUT}}
\index{IDVOUT}
\index{IDVERR}
So far as we know, there are only one operating-system-dependent aspect of
\ddscatv: the device number to use for ``standard output".

The variables {\tt IDVOUT} and {\tt IDVERR} specify device numbers 
for ``running output" and ``error messages", respectively.
Normally these would both be set to the device number
for ``standard output" (e.g., writing to the screen if running interactively).
Variables {\tt IDVERR} are 
set by {\tt DATA} statements in the ``main" program
{\tt DDSCAT.f90} and in the output routine {\tt WRIMSG} (file {\tt wrimsg.f90}).  
The executable statement {\tt IDVOUT=0} initializes {\tt IDVOUT} to
0.  In the as-distributed version of {\tt DDSCAT.f90}, the statement

{\tt OPEN(UNIT=IDVOUT,FILE=CFLLOG)}

\noindent causes the output to {\tt UNIT=IDVOUT} to be buffered
and written to the file {\tt ddscat.log\_nnn}, where {\tt nnn=000} for
the first cpu, 001 for the second cpu, etc.
If it is desirable to have this output unbuffered for debugging purposes,
(so that the output will contain up-to-the-moment information)
simply comment out this {\tt OPEN} statement.

\section{\label{sec:windows}
         Information for Windows Users}

There are several options to run DDSCAT on \Microsoft\ Windows. 
One can purchase a
Fortran compiler such as the \Intel\ ifort compiler\footnote{%
   \Intel\ for some reason now refers to this as
   ``\Intel\ Visual Fortran Composer XE 2013''.} 
(http://software.intel.com/en-us/fortran-compilers/), or
the Portland Group PGF compiler (http://www.pgroup.com/).
However, these are not free. Below we discuss four methods which
give access to DDSCAT on \Microsoft\ Windows 
using Open Source applications. We have
tested each of these four methods.
We also remark on using the commercial \Intel\ ifort compiler under Windows.

\begin{table}[h]
\begin{center}
\caption{\label{tab:DDSCAT_on_windows}
         DDSCAT on \Microsoft\ Windows Platforms}
{\footnotesize
\begin{tabular}{|c|c|c|c|c|}
\hline 
Method & Compiler & Advantage & Problems & Comments\tabularnewline
\hline 
\hline 
Native & gfortran & pre-compiled & limiting options & simplest\tabularnewline
\hline 
MINGW & gfortran & compile with user options & compilation step & simple\tabularnewline
\hline 
Cygwin & gfortran,G95 & compile with user options & compilation step & simple\tabularnewline
\hline 
Virtualbox/UBUNTU & gfortran, G95, intel & full LINUX access & learning curve & difficult\tabularnewline
\hline 
\end{tabular}
}
\end{center}
\end{table}

\subsection{\label{sec:native executable}
            Native executable}

We provide (on {\tt http://code.google.com/p/ddscat/}) 
a self-extracting executable which includes a pre-compiled, 
ready-to-run,
DDSCAT executable for \Microsoft\ Windows. 
The advantage is that the user avoids the
possibly difficult compilation
step, and has immediate access to DDSCAT. 

Beginning with release 7.2 we package the Windows distribution using
{}``Inno Setup5'' (http://www.jrsoftware.org) which is a free
installer for Windows programs.  This will install a Windows native
executable {\tt ddscat.exe} as well as source code, documentation, and
relevant test examples. This is by far the the simplest way to get
DDSCAT running on a Windows system. However, we provide only single
precision version without optimization.

Thus, serious DDSCAT
users may, at some stage, need to recompile the code as outlined below.
The executable should be executed using the windows ``{\tt cmd}'' command
which opens a separate shell window. One then has to change directory to
where DDSCAT is placed, and execute DDSCAT invoking

\medskip

\indent\indent ddscat.exe

\subsection{Compilation using MINGW}

The executable file discussed in {}``native executable'' section
was compiled using {}``gfortran'' in the MINGW environment 
(http://www.mingw.org/).
\Microsoft\ Windows users may choose to recompile code using MINGW. MINGW and
MSYS provide several tools crucial for compilation of DDSCAT to a native
windows executable. MinGW (\textquotedbl{}Minimalist GNU for
Windows\textquotedbl{}) is a minimalist development environment for
native \Microsoft\ Windows applications. It provides a complete Open
Source programming tool set which is suitable for the development
of native MS-Windows applications, and which do not depend on any
3rd-party libraries. 

MSYS (\textquotedbl{}Minimal SYStem\textquotedbl{}),
is a Bourne Shell command line interpreter system. Offered as an alternative
to Window's cmd.exe, this provides a general purpose command line
environment, which is particularly suited to use with MinGW, for porting
applications to the MS-Windows platform. We suggest using the installer
{\tt mingw-get-inst} (for example {\tt mingw-get-inst-20111118.exe}) which is
a simple Graphical User Interface installer that installs MinGW and
MSYS. During the GUI phase of the installation select the  default options, 
but from
the following list you need to select Fortran Compiler, MSYS Basic
System and MinGW Developer ToolKit:
\begin{enumerate}
\item MinGW Compiler Suite C Compiler optional
\item C++ Compiler 
\item {\bf optional Fortran Compiler} 
\item optional ObjC Compiler 
\item {\bf optional MSYS Basic System} 
\item {\bf optional MinGW Developer Toolkit}
\end{enumerate}
You will have to add PATH to bin directories as described in the MINGW.

Once you have MINGW and MSYS installed, in {}``all programs'' there is now the
MinGW program {}``MinGW shell''. Once you open this shell one can
now compile Fortran code using {}``gfortran''. The ``make'' and
``tar'' utilities are available. 
The command {}``pwd'' shows which directory corresponds to the initial
{}``MinGW'' shell. For example it can be {}``/home/Piotr'' which
corresponds to windows directory {}``c:\textbackslash{}mingw\textbackslash{}msys\textbackslash{}1.0\textbackslash{}home\textbackslash{}piotr''.

You now need to copy ddscat.tar files and untar them. To make a native
executable, edit ``Makefile'' so that the ``LFLAGS'' string is defined to be
\begin{lyxlist}{00.00.0000}
\item [{LFLAGS=-static-libgcc}] -static-libgfortran
\end{lyxlist}
and then execute {}``make all''. The resulting {\tt ddscat.exe} doesn't require
any non-windows libraries. This can be checked with the

\indent\indent {objdump} -x ddscat.exe | grep {}``DLL''\\
command.

\subsection{Compilation using CYGWIN}

Another option is provided by ``CYGWIN'',
an easily installable UNIX-like emulation package.
It is available from http://www.cygwin.com/ . It installs authomatically.
However, during installation one has to specify installation of several
packages incuding the f95 compiler {\bf gfortran}, the {\bf make} utility, 
the {\bf tar} utility, and {\bf nano}. 
Once installed, you will be able
to open the CYGWIN shell and make DDSCAT using the standard
Linux commands as discussed in this manual (see \S\ref{sec:compiling}).

\subsection{UBUNTU and Virtualbox}

By far the most comprehensive solution to running DDSCAT on windows
is to install UBUNTU Linux under Oracle Virtualbox. First install
Oracle Virtualbox (from https://www.virtualbox.org/) and then
install UBUNTU Linux
(from http://www.ubuntu.com/). You will be able to run a full LINUX
environment
on your Windows computer 
You can add (using {}``synaptic file manager'')
gfortran and many other packages including graphics. 
Once a f90 or f95 compiler has been installed, you can compile \ddscat\ as
described in \S\ref{sec:compiling}.

\subsection{Compilation in Windows}

In addition to public domain options (gfortran) for compilation of \ddscatv\ of
under Linux environments running under Windows, 
we have also tested direct Windows compilation of DDSCAT with
\Intel\ Fortran
Composer XE2013 (in the USA Educational price was \$399). 
We used the
``command prompt'' option (IFORT) to compile the code. We were able to
compile the code with the standard Makefile provided in the DDSCAT
distribution.
We were able to compile the code with OpenMP and MKL options as well.

For the MKL option using IFORT one has to specify links to libraries
using interface defined at
http://software.intel.com/sites/products/mkl/
For example, the appropriate line in the Makefile may need to look like\\
\hspace*{2em}LFLAGS\hspace*{1em}= mkl\_intel\_c\_dll.lib mkl\_sequential\_dll.lib mkl\_core\_dll.lib\\
Subroutine mkl\_dfti.f90 is available with INTEL fortran in the subdirectory
program files (x86)/intel/composer xe/mkl/include/mkl\_dfti.f90

\section{A Sample Calculation: {\tt RCTGLPRSM}
         \label{sec:sample calculation}}

When the tarfile is unpacked, it will create four directories: {\tt src},
{\tt doc}, {\tt diel}, and {\tt examples\_exp}.
The {\tt examples\_exp} directory has a number of subdirectories,
each with
files for a sample calculation.
Here we focus on the files in the subdirectory {\tt RCTGLPRSM}. 
To follow this, go to the {\tt DDA} directory (the directory in which
you unpacked the tarfile) and

\indent\indent{\tt cd examples\_exp}

\noindent to enter the {\tt examples\_exp} directory, and

\indent\indent{\tt cd RCTGLPRSM}

\noindent to enter the {\tt RCTGLPRSM} directory.

\section{The Parameter File {\tt ddscat.par}
        \label{sec:parameter_file}}
\index{ddscat.par parameter file}

It is assumed that you are positioned in the {\tt examples\_exp/RCTGLPRSM}.
directory, as per \S\ref{sec:sample calculation}.
The file
{\tt ddscat.par} 
(see also Appendix \ref{app:ddscat.par})
provides parameters to the program {\tt ddscat}:
for example, {\tt examples\_exp/RCTGLPRSM/ddscat.par} :
{\scriptsize
\begin{verbatim}
' ========= Parameter file for v7.3 ===================' 
'**** Preliminaries ****'
'NOTORQ' = CMDTRQ*6 (NOTORQ, DOTORQ) -- either do or skip torque calculations
'PBCGS2' = CMDSOL*6 (PBCGS2, PBCGST, GPBICG, PETRKP, QMRCCG) -- CCG method
'GPFAFT' = CMDFFT*6 (GPFAFT, FFTMKL) -- FFT method
'GKDLDR' = CALPHA*6 (GKDLDR, LATTDR, FLTRCD) -- DDA method
'NOTBIN' = CBINFLAG (NOTBIN, ORIBIN, ALLBIN) -- specify binary output
'**** Initial Memory Allocation ****'
100 100 100 = dimensioning allowance for target generation
'**** Target Geometry and Composition ****'
'RCTGLPRSM' = CSHAPE*9 shape directive
16 32 32  = shape parameters 1 - 3
1         = NCOMP = number of dielectric materials
'../diel/Au_evap' = file with refractive index 1
'**** Additional Nearfield calculation? ****'
0 = NRFLD (=0 to skip nearfield calc., =1 to calculate nearfield E)
0.0 0.0 0.0 0.0 0.0 0.0 (fract. extens. of calc. vol. in -x,+x,-y,+y,-z,+z)
'**** Error Tolerance ****'
1.00e-5 = TOL = MAX ALLOWED (NORM OF |G>=AC|E>-ACA|X>)/(NORM OF AC|E>)
'**** maximum number of iterations allowed ****'
300     = MXITER
'**** Interaction cutoff parameter for PBC calculations ****'
1.00e-2 = GAMMA (1e-2 is normal, 3e-3 for greater accuracy)
'**** Angular resolution for calculation of <cos>, etc. ****'
0.5	= ETASCA (number of angles is proportional to [(3+x)/ETASCA]^2 )
'**** Vacuum wavelengths (micron) ****'
0.5000 0.5000 1 'LIN' = wavelengths (first,last,how many,how=LIN,INV,LOG)
'**** Refractive index of ambient medium'
1.000 = NAMBIENT
'**** Effective Radii (micron) **** '
0.246186 0.246186 1 'LIN' = aeff (first,last,how many,how=LIN,INV,LOG)
'**** Define Incident Polarizations ****'
(0,0) (1.,0.) (0.,0.) = Polarization state e01 (k along x axis)
2 = IORTH  (=1 to do only pol. state e01; =2 to also do orth. pol. state)
'**** Specify which output files to write ****'
1 = IWRKSC (=0 to suppress, =1 to write ".sca" file for each target orient.
'**** Prescribe Target Rotations ****'
0.    0.   1  = BETAMI, BETAMX, NBETA  (beta=rotation around a1)
0.    0.   1  = THETMI, THETMX, NTHETA (theta=angle between a1 and k)
0.    0.   1  = PHIMIN, PHIMAX, NPHI (phi=rotation angle of a1 around k)
'**** Specify first IWAV, IRAD, IORI (normally 0 0 0) ****'
0   0   0    = first IWAV, first IRAD, first IORI (0 0 0 to begin fresh)
'**** Select Elements of S_ij Matrix to Print ****'
6	= NSMELTS = number of elements of S_ij to print (not more than 9)
11 12 21 22 31 41	= indices ij of elements to print
'**** Specify Scattered Directions ****'
'LFRAME' = CMDFRM (LFRAME, TFRAME for Lab Frame or Target Frame)
2 = NPLANES = number of scattering planes
0.   0. 180.  5 = phi, thetan_min, thetan_max, dtheta (in deg) for plane 1
90.  0. 180.  5 = phi, thetan_min, thetan_max, dtheta (in deg) for plane 2
\end{verbatim}}

Here we discuss the general structure of {\tt ddscat.par}
(see also Appendix \ref{app:ddscat.par}).
\subsection{ Preliminaries}
{\tt ddscat.par} starts by setting the values of five strings:
\begin{itemize}
\item CMDTRQ specifying whether or not radiative torques
      are to be calculated (e.g., NOTORQ)
\item CMDSOL specifying the CCG method (e.g., PBCGS2)
      (see section \S\ref{sec:choice_of_algorithm}). 
\item CMDFFT specifying the FFT method (e.g., GPFAFT)
      (see section \S\ref{sec:choice_of_fft}).
\index{GPFAFT -- see ddscat.par}
\index{ddscat.par!GPFAFT}
\item CALPHA specifying the DDA method (e.g., GKDLDR or FLTRCD)
      (see \S\ref{sec:DDA_method}).
\index{GKDLDR -- see ddscat.par}
\index{LATTDR -- see ddscat.par}
\index{ddscat.par!GKDLDR}
\index{ddscat.par!LATTDR}
\item CBINFLAG specifying whether to write out binary files (e.g., NOTBIN)
\index{NOTBIN -- see ddscat.par}
\index{ddscat.par!NOTBIN}
\end{itemize}
\subsection{ Initial Memory Allocation}
Three integers\\
\hspace*{3em}{\tt MXNX  MXNY  MXNZ}\\
are given that need to be equal to or larger than
the anticipated target size (in lattice units) in the $\xtf$, $\ytf$, and
$\ztf$ directions.  This is required only for initial memory allocation for
the target-generation stage of the calculation.
For example,\\
\hspace*{3em}100 100 100\\
would require initial memory allocation of only $\sim100~$MBytes.
Note that after the target geometry has been determined, \ddscatv\ will proceed
to reallocate as much memory as is actually required for the calculation.

\subsection{ Target Geometry and Composition}  
\begin{itemize}
\item CSHAPE specifies the target geometry (e.g., RCTGLPRSM)
\end{itemize}
As provided,
the file {\tt ddscat.par} 
is set up to calculate scattering by a 32$\times$64$\times$64 
rectangular array of 131072
dipoles.

The user must specify {\tt NCOMP}, the number of
different compositions (i.e., dielectric functions) that will be used.
This is then followed by {\tt NCOMP} lines, with each line giving the name 
(in quotes) of a dielectric function file.  In our example, 
{\tt NCOMP} is set to 1.
The dielectric function of the target material is provided in the file
{\tt ../diel/Au\_evap}, 
which gives the refractive index of evaporated Au
over a range of wavelengths.  The file Au\_evap is located in subdirectory
examples\_exp/diel .

\index{NCOMP!ddscat.par}
\index{diel.tab -- see ddscat.par}
\index{ddscat.par!diel.tab}

\subsection{ \label{sec:nearfield calc}
            Additional Nearfield Calculation?}
The user set NRFLD = 0 or 1 to indicate whether 
the first DDSCAT calculation (solving for the
target polarization, and absorption and scattering cross sections) should
automatically be followed by a second ``nearfield''
calculation to evaluate the electric field $\bE$
throughout a rectangular volume containing the original target.
\begin{itemize}
\item If {\tt NRFLD} = 0 , the nearfield calculation will not be done.\\
The next line in {\tt ddscat.par} will be read but not used.
\item If {\tt NRFLD} = 1 , nearfield $\bE$ will be calculated and stored.
\item If {\tt NRFLD} = 2 , nearfield $\bE$ and $\bB$ will be calculated and
stored.
\end{itemize}
The next line in {\tt ddscat.par} then specifies 6 non-negative 
numbers, $r_{1}$, $r_{2}$, $r_{3}$, $r_{4}$, $r_{5}$, $r_{6}$
specifying the increase
in size of the nearfield computational volume relative to the circumscribing
rectangular volume used for the original solution.
If the original volume is $X_1\leq x \leq X_2$, $Y_1\leq y\leq Y_2$,
$Z_1\leq z \leq Z_2$, then the nearfield   
calculation will be done in a volume
$(X_1-r_{1}L_x) \leq x \leq (X_2+r_{2}L_x)$,
$(Y_1-r_{3}L_y) \leq y \leq (Y_2+r_{4}L_y)$,
$(Z_1-r_{5}L_z) \leq z \leq (Z_2+r_{6}L_z)$,
where 
$L_x\equiv (X_2-X_1)$,
$L_y\equiv (Y_2-Y_1)$,
$L_z\equiv (Z_2-Z_1)$.
\index{nearfield calculation}
\index{ddscat.par!NRFLD}

\subsection{ Error Tolerance}
{\tt TOL} = the error tolerance.  Conjugate gradient iteration will proceed
until the linear equations are solved to a fractional error {\tt TOL}.
The sample calculation has {\tt TOL}=$10^{-5}$.

\subsection{ Maximum Number of Iterations}
\index{MXITER -- see ddscat.par}
\index{ddscat.par!MXITER}

{\tt MXITER} = maximum number of complex-conjugate-gradient iterations allowed.
As there are $3N$ equations to solve, {\tt MXITER} should never be larger
than $3N$, but in practice should be much smaller.  As a default we suggest
setting {\tt MXITER}=100, but for some problems that converge very slowly
you may need to use a larger value of {\tt MXITER}.

\subsection{ Interaction Cutoff Parameter for PBC Calculations}
{\tt GAMMA} = parameter limiting certain summations that are required for
periodic targets \citep[see]{Draine+Flatau_2008a}.  
{\bf The value of {\tt GAMMA} has no effect on 
computations for finite targets} --
it can be set to any value, including $0$).

For targets that are periodic in 1 or 2 dimensions,
{\tt GAMMA} needs to be small for high accuracy, but
the required cpu time increases as {\tt GAMMA} becomes smaller.
{\tt GAMMA} = $10^{-2}$ is reasonable for initial calculations,
but you may want to experiment to see if the results you are interested
in are sensitive to the value of {\tt GAMMA}.
\citet{Draine+Flatau_2008a} show examples of how computed results for scattering
can depend on the value of $\gamma$.
\index{GAMMA -- see ddscat.par}
\index{ddscat.par!GAMMA}

\subsection{ Angular Resolution for Computation of 
            $\langle\cos\theta\rangle$, etc.}
The parameter {\tt ETASCA} determines the selection of scattering angles used
for computation of certain angular averages, such as 
$\langle\cos\theta\rangle$ and $\langle\cos^2\theta\rangle$, 
and the radiation pressure force 
(see \S\ref{sec:force and torque calculation}) and radiative torque 
(if {\tt CMDTRQ=DOTORQ}).
Small values of {\tt ETASCA} result in increased accuracy but also cost
additional computation time.
{\tt ETASCA}=0.5 generally gives accurate results.

If accurate computation of $\langle\cos\theta\rangle$
or the radiation pressure force is not required, the user can set
{\tt ETASCA} to some large number, e.g. 10, to minimize unnecessary
computation. 

\subsection{ Vacuum Wavelengths}
Wavelengths $\lambda$ (in vacuo) are specified in one line in {\tt ddscat.par}
consisting of values for 4 variables:\\
\hspace*{2em}
{\tt WAVINI}~ {\tt WAVEND}~ {\tt NWAV}~ {\tt CDIVID}\\
where {\tt WAVINI} and {\tt WAVEND} are real numbers, {\tt NWAV} is an
integer, 
and {\tt CDIVID} is a character variable.
\begin{itemize}
\item If {\tt CDIVID = 'LIN'}, the $\lambda$ will be uniformly spaced
beween {\tt WAVINI} and {\tt WAVEND}.
\item If {\tt CDIVID = 'INV'}, the $\lambda$ will be uniformly spaced in
$1/\lambda$ between {\tt WAVINI} and {\tt WAVEND}
\item If {\tt CDIVID = 'LOG'}, the $\lambda$ will be uniformly spaced
in $\log(\lambda)$ between {\tt WAVINI} and {\tt WAVEND}
\item If {\tt CDIVID = 'TAB'}, the $\lambda$ will be read from a
user-supplied file {\tt wave.tab}, with one wavelength per line.
For this case, the values of {\tt WAVINI}, {\tt WAVEND}, 
and {\tt NWAV} will be disregarded.
\end{itemize}
The sample {\tt ddscat.par} file specifies that the calculations be done for a
single wavelength ($\lambda=0.50$).  
{\bf The units must be the same as the wavelength
units used in the file specifying the refractive index.}  In this case,
we are using $\micron$.

\subsection{\label{sec:nambient} 
            Refractive Index of Ambient Medium, $m_{\rm medium}$}

\index{ambient medium!refractive index}
\index{NAMBIENT}

In some cases the target of interest will be immersed in a transparent
medium (e.g., water) with a refractive index different from vacuum.
The refractive index $m_{\rm medium}$ should be specified here.  As the
medium is assumed to be transparent, $m_{\rm medium}$ is a real number.
\ddscat\ will calculate the scattering properties for the target
immersed in the ambient medium.
If the target is located in a vaccum, set $m_{\rm medium}=1$.

\subsection{ Target Size $\aeff$}

Note that in \ddscat\ the ``effective radius'' 
$a_{\rm eff}$ 
\index{effective radius $\aeff$}
\index{$a_{\rm eff}$}
is the radius of a sphere of
equal volume -- i.e., a sphere of volume $Nd^3$ , where $d$ 
is the lattice spacing
and $N$ is the number of occupied (i.e., non-vacuum) 
lattice sites in the target.
Thus the effective radius $\aeff = (3N/4\pi)^{1/3}d$ .
Our target should have a thickness $a=0.5\micron$ in the $\xtf$ direction.
If the rectangular solid is $a\times b\times c$, with $a:b:c::32:64:64$,
then $V=abc=4a^3$.  Thus
$\aeff=(3V/4\pi)^{1/3}=(3/\pi)^{1/3}a = 0.49237\micron$.

The target sizes $\aeff$ to be studied are specified on one line of
{\tt ddscat.par} consisting of 4 variables:
\hspace*{2em}{\tt AEFFINI AEFFEND NRAD CDIVID}\\
where {\tt AEFFINI} and {\tt AEFFEND} are real numbers, {\tt NRAD} is
an integer, and {\tt CDIVID} is a character variable.
\begin{itemize}
\item If {\tt CDIVID = 'LIN'}, the $\aeff$ will be uniformly spaced
beween {\tt AEFFINI} and {\tt AEFFEND}.
\item If {\tt CDIVID = 'INV'}, the $\aeff$ will be uniformly spaced in
$1/\aeff$ between {\tt AEFFINI} and {\tt AEFFEND}
\item If {\tt CDIVID = 'LOG'}, the $\aeff$ will be uniformly spaced
in $\log(\aeff)$ between {\tt AEFFINI} and {\tt AEFFEND}
\item If {\tt CDIVID = 'TAB'}, the $\aeff$ will be read from a
user-supplied file {\tt aeff.tab}, with one value of $\aeff$ per line,
beginning on line 2.
For this case, the values of {\tt AEFFINI}, {\tt AEFFEND}, 
and {\tt NRAD} will be disregarded (\ddscat\ will read and use all the
values of $\aeff$ in the table (beginning on line 2).
\end{itemize}
The sample {\tt ddscat.par} file specifies that the calculations be done
for a single $\aeff=0.49237\micron$.
The target is a rectangular solid with aspect ratio 1:2:2.  The
thickness in the x direction is $(\pi/3)^{1/3}\aeff=0.500\micron$.

\subsection{ Incident Polarization}
The incident radiation is always assumed to propagate along the $\xlf$ axis --
the $x$-axis in
the ``Lab Frame''.  
The sample {\tt ddscat.par} file specifies incident polarization
state ${\hat{\bf e}}_{01}$ to be along the $\ylf$ axis 
(and consequently polarization state ${\hat{\bf e}}_{02}$
will automatically be taken to be along the $\zlf$ axis).  
{\tt IORTH=2} in {\tt ddscat.par}
calls for calculations to be carried out for both incident polarization
states (${\hat{\bf e}}_{01}$ and ${\hat{\bf e}}_{02}$
-- see \S\ref{sec:incident_polarization}).



\subsection{\label{sec:target_orientation}
            Target Orientation}
The target is assumed to have two vectors ${\hat{\bf a}}_1$ and
${\hat{\bf a}}_2$ embedded in it; ${\hat{\bf a}}_2$ is perpendicular
to ${\hat{\bf a}}_1$.  
\index{$\hat{\bf a}_1$, $\hat{\bf a}_2$ target vectors}
For the present target shape {\tt RCTGLPRSM},
the vector ${\hat{\bf
a}}_1$ is along the $\xtf$ axis of the target, and the vector
${\hat{\bf a}}_2$ is along the $\ytf$ axis (see \S\ref{sec:RCTGLPRSM}).
The target orientation in the Lab Frame is set by three angles: $\beta$,
$\Theta$, and $\Phi$, defined and discussed below in
\S\ref{sec:target_orientation}.  
\index{$\Theta$ -- target orientation angle}
\index{$\Phi$ -- target orientation angle}
\index{$\beta$ -- target orientation angle}
Briefly, the polar angles $\Theta$
and $\Phi$ specify the direction of ${\hat{\bf a}}_1$ in the Lab
Frame.  The target is assumed to be rotated around ${\hat{\bf a}}_1$
by an angle $\beta$.  The sample {\tt ddscat.par} file specifies
$\beta=0$ and $\Phi=0$ (see lines in {\tt ddscat.par} specifying
variables {\tt BETA} and {\tt PHI}), and calls for three values of the
angle $\Theta$ (see line in {\tt ddscat.par} specifying variable {\tt
THETA}).  \ddscat\ chooses $\Theta$ values uniformly spaced
in $\cos\Theta$.  In this case we specify 0 0 1 : we obtain
only one value: $\Theta=0$.
\footnote{Had we specified 0 90 3 we would have obtained 
three values of $\Theta$ between 0
and $90^\circ$, unformly-spaced in $\cos\theta$: $\Theta=0$, $60^\circ$, and $90^\circ$.}

\subsection{ Starting Values of {\tt IWAV}, {\tt IRAD}, {\tt IORI}}
Normally we begin the calculation with {\tt IWAVE}=0, {\tt IRAD}=0, and
{\tt IORI}=0.  However, under some circumstances, a prior
calculation may have completed some of the cases.  If so, the user
can specify starting values of {\tt IWAV}, {\tt IRAD}, {\tt IORI};
the computations will begin with this case, and continue.

\subsection{ Which Mueller Matrix Elements?}
The sample parameter file specifies that \ddscat\ should calculate 6 distinct
Mueller scattering matrix elements $S_{ij}$, with 
11, 12, 21, 22, 31, 41 being the chosen values of $ij$.

\subsection{ What Scattering Directions?}
\subsubsection{Isolated Finite Targets}
For finite targets, the user may specify the scattering directions in 
either the Lab Frame ({\tt 'LFRAME'})
or the Target Frame ({\tt 'TFRAME'}).

For finite targets, such as specified in this sample {\tt ddscat.par},
the $S_{ij}$ are to be calculated for scattering directions specified
by angles $(\theta,\phi)$.

The sample {\tt ddscat.par} specifies that 2 scattering planes
are to be specified:
the first has $\phi=0$ and the second has $\phi=90^\circ$; for
each scattering plane
$\theta$ values run from 0 to $180^\circ$
in increments of $10^\circ$.

\subsubsection{1-D Periodic Targets}

For periodic targets, the scattering directions {\tt must} be specified in the
Target Frame: {\tt 'TFRAME' = CMDFRM} .

Scattering from 1-d periodic targets is discussed in detail by
\citet{Draine+Flatau_2008a}.
For periodic targets, the user does not specify scattering planes.
For 1-dimensional targets, the user specifies scattering cones, corresponding
to different scattering orders $M$. 
For each scattering cone $M$, the user specifies
$\zeta_{\rm min}$, $\zeta_{\rm max}$, $\Delta\zeta$
(in degrees);
the azimuthal angle $\zeta$ will run from
$\zeta_{\rm min}$ to $\zeta_{max}$, in increments of $\Delta\zeta$.
For example:
{\footnotesize
\begin{verbatim}
'TFRAME' = CMDFRM (LFRAME, TFRAME for Lab Frame or Target Frame)
1 = number of scattering cones
0.  0. 180. 0.05  = OrderM zetamin zetamax dzeta for scattering cone 1
\end{verbatim}
}
\subsubsection{2-D Periodic Targets}

For targets that are periodic in 2 dimensions, 
the scattering directions {\tt must} be specified in the
Target Frame: {\tt 'TFRAME' = CMDFRM} .

Scattering from targets that are periodic in 2 dimensions
is discussed in detail by
\citet{Draine+Flatau_2008a}.
For 2-D periodic targets, the user specifies the diffraction orders 
$(M,N)$ for
transmitted radiation: the code will automatically calculate
the scattering matrix elements $S_{ij}^{(2d)}(M,N)$ for both
transmitted and reflected radiation for each $(M,N)$ specified by the user.
For example:
{\footnotesize
\begin{verbatim}
'TFRAME' = CMDFRM (LFRAME, TFRAME for Lab Frame or Target Frame)
1 = number of scattering orders
0.  0. = OrderM OrderN for scattered radiation
\end{verbatim}
}

\section{Running \ddscatv\ Using the Sample {\tt ddscat.par} File}

It is again assumed that you are in directory
{\tt ../DDA/examples\_exp/RCTGLPRSM}, as per \S\ref{sec:sample calculation}.
The {\tt ddscat} executable (created as per the instructions in
\S\ref{sec:compiling}) is assumed to be {\tt ../DDA/src/ddscat}

\subsection{ Single-Process Execution}

To execute the program on a UNIX system (running either {\tt sh} 
or {\tt csh}), simply create a symbolic link by typing

\indent\indent {\tt ln -s ../../src/ddscat ddscat}

\noindent or you could simply move the previously-created executable into the 
current directory (assumed to be {../DDA/examples\_exp/RCTGLPRSM/) by
typing

\indent\indent {\tt mv ../../src/ddscat ddscat}

\noindent Then, to perform the calculation, type

\indent\indent {\tt ddscat >\& ddscat.out \&}

\noindent which will redirect the ``standard output'' to the file {\tt
ddscat.out}, and run the calculation in the background.  

The sample calculation 
[32x64x64=131072 dipole target, 
3 target orientations, 
two incident polarizations for each orientation, 
with scattering (Mueller matrix elements $S_{ij}$)
calculated for 37 distinct scattering directions], 
requires 672 cpu~sec to complete on a 2.53 GHz cpu.
for the assumed Au composition, between 28 and 32
iterations were required for the complex conjugate
gradient solver to converge to the specified error 
tolerance of {\tt TOL = 1.e-5} for each orientation and incident polarization.

\subsection{ Code Execution Under MPI}
\index{MPI -- Message Passing Interface!code execution}

Local installations of {\tt MPI} will vary -- you should consult with
someone familiar with way {\tt MPI} is installed and used on your system.

At Princeton University Dept.\ of Astrophysical Sciences we use {\tt PBS}
(Portable Batch System)\footnote{\tt http://www.openpbs.org}
to schedule jobs.\footnote{%
   As of this writing (2013.05), 
   this information on batch scheduling is several years old.
   It may have been superseded by new practices since we last checked.}
{\tt MPI} jobs are submitted using {\tt PBS} by first creating a 
shell script such as the following example file {\tt pbs.submit}:\\

\noindent
\hspace*{3em}\#!/bin/bash\\
\hspace*{3em}\#PBS -l nodes=2:ppn=1\\
\hspace*{3em}\#PBS -l mem=1200MB,pmem=300MB\\
\hspace*{3em}\#PBS -m bea\\
\hspace*{3em}\#PBS -j oe\\
\hspace*{3em}cd ~\$PBS\_O\_WORKDIR\\
\hspace*{3em}/usr/local/bin/mpiexec ~ddscat\\

\noindent
The lines beginning with {\tt\#PBS -l} specify the required resources:\\
{\tt\#PBS -l nodes=2:ppn=1} specifies that 2 nodes are to be used,
with 1 processor per node.\\
{\tt\#PBS -l mem=1200MB,pmem=300MB} specifies that the total memory
required ({\tt mem}) is 1200MB, 
and the maximum physical memory used by any single
process ({\tt pmem}) is 300MB.  The actual definition of {\tt mem} is
not clear, but in practice it seems that it should be set equal to
2$\times$({\tt nodes})$\times$({\tt ppn})$\times$({\tt pmem}) .\\
{\tt\#PBS -m bea} specifies that PBS should send email when the job begins
({\tt b}), and when it ends ({\tt e}) or aborts ({\tt a}).\\
{\tt\#PBS -j oe} specifies that the output from stdout and stderr will be
merged, intermixed, as stdout.

\noindent This example assumes that the executable {\tt dscat} is located
in the same directory where the code is to execute and write its output.
If {\tt ddscat} is located in another directory, simply give the full pathname.
to it.
The {\tt qsub} command is used to submit the {\tt PBS} job:\\
\\
\hspace*{2em}{\tt qsub pbs.submit}\\

As the calculation proceeds, the usual 
output files will be written to this directory: for each wavelength, 
target size, and target orientation,
there will be a file
{\tt w}{\it aaa}{\tt r}{\it bbb}{\tt k}{\it ccc}{\tt .sca}, where
{\it aaa=000, 001, 002, ...} specifies the wavelength,
{\it bbb=000, 001, 002, ...} specifies the target size,
and
{\it cc=000, 001, 002, ...} specifies the orientation.
For each wavelength and target size there will also be a file
{\tt w}{\it aaa}{\tt r}{\it bbb}{\tt ori.avg} with orientationally-averaged
quantities.
Finally, there will also be tables {\tt qtable},
and {\tt qtable2} with orientationally-averaged cross sections for each
wavelength and target size.

In addition, each processor employed will write to its own log file
{\tt ddscat.log\_}{\it nnn}, where {\it nnn=000, 001, 002, ...}.
These files contain information
concerning cpu time consumed by different parts of the calculation,
convergence to the specified error tolerance, etc.  If you are uncertain
about how the calculation proceeded, examination of these log files
is recommended.

\section{Output Files}

\subsection{ ASCII files\label{subsec:ascii}}

If you run DDSCAT using the command\hfill\break \indent\indent{\tt
ddscat >\& ddscat.log \&} \hfill\break you will have various types of
ASCII files when the computation is complete:
\begin{itemize}
\item a file {\tt ddscat.log};
\index{ddscat.out}
\item a file {\tt mtable};
\index{mtable -- output file}
\item a file {\tt qtable};
\index{qtable -- output file}
\item a file {\tt qtable2};
\index{qtable2 file}
\item files {\tt w}{\it xxx}{\tt r}{\it yyy}{\tt ori.avg} 
	(one, {\tt w000r000ori.avg}, for the sample calculation);
\index{w000r000ori.avg file -- output file}
\item if {\tt ddscat.par} specified {\tt IWRKSC}=1, there will also be
	files {\tt w}{\it xxx}{\tt r}{\it yyy}{\tt k}{\it zzz}{\tt .sca} (1 for
	the sample calculation: {\tt w000r000k000.sca}lcl, {\tt w000r000k001.sca},
	{\tt w000r000k002.sca}).
\index{w000r000k000.sca -- output file}
\index{w000r000k000.pol{\it n} -- output file}
\end{itemize}

The file {\tt ddscat.out} will contain minimal information (it may in fact
be empty).

The file {\tt ddscat.log\_000} will contain any error messages generated as
well as a running report on the progress of the calculation, including
creation of the target dipole array.  During the iterative
calculations, $Q_\ext$, $Q_\abs$, and $Q_{\rm pha}$ are printed after
each iteration; you will be able to judge the degree to which
convergence has been achieved.  Unless {\tt TIMEIT} has been disabled,
there will also be timing information.
If the {\tt MPI} option is used to run the code on multiple cpus, there
will be one file of the form {\tt ddscat.log\_nnn} for each of the
cpus, with {\tt nnn=000,001,002,...}.
\index{ddscat.log\_000 -- output file}

The file {\tt mtable} contains a summary of the dielectric constant
used in the calculations.
\index{mtable -- output file}

The file {\tt qtable} contains a summary of the
orientationally-averaged values of $Q_\ext$, $Q_\abs$, $Q_\sca$,
$g(1)=\langle\cos(\theta_s)\rangle$,
$\langle\cos^2(\theta_s)\rangle$, $Q_{\rm bk}$, and $N_\sca$.  
\index{qtable -- output file}
Here $Q_\ext$,
$Q_\abs$, and $Q_\sca$ are the extinction, absorption, and
scattering cross sections divided by $\pi a_{\rm eff}^2$.  $Q_{\rm bk}$ is
the differential cross section for backscattering (area per sr)
divided by $\pi a_{\rm eff}^2$.
$N_\sca$ is the number of scattering directions used for averaging
over scattering directions (to obtain $\langle\cos\theta\rangle$, etc.)
(see \S\ref{sec:averaging_scattering}).

The file {\tt qtable2} contains a summary of the
orientationally-averaged values of $Q_{\rm pha}$, $Q_{\rm pol}$, and
$Q_{\rm cpol}$.  
\index{qtable2 -- output file}
Here $Q_{\rm pha}$ is the ``phase shift'' cross section
divided by $\pi a_{\rm eff}^2$ \citep[see definition in]{Draine_1988}.
$Q_{\rm pol}$ is the ``polarization efficiency factor'', equal to the
difference between $Q_\ext$ for the two orthogonal polarization
states.  We define a ``circular polarization efficiency factor''
$Q_{\rm cpol}\equiv Q_{\rm pol}Q_{\rm pha}$, 
since an optically-thin medium with a
small twist in the alignment direction will produce circular
polarization in initially unpolarized light in proportion to
$Q_{\rm cpol}$.

For each wavelength and size, \ddscatv\ produces a file with a
name of the form\break{\tt w{\it xxx}r{\it yyy}ori.avg}, where index
{\it xxx} (=000, 001, 002....)  designates the wavelength and index {\it
yyy} (=000, 001, 002...) designates the ``radius''; this file contains $Q$
values and scattering information averaged over however many target
orientations have been specified (see \S\ref{sec:target_orientation}.
\index{w000r000ori.avg -- output file}
The file {\tt w000r000ori.avg} produced by the sample calculation is
provided below in Appendix \ref{app:w000r000ori.avg}.

In addition, if {\tt ddscat.par} has specified {\tt IWRKSC}=1 (as for
the sample calculation), \ddscatv\ will generate files with
names of the form {\tt w{\it xxx}r{\it yyy}k{\it zzz}.avg}, where {\it
xxx} and {\it yyy} are as before, and index {\it zzz}
enumerates the
target orientations.\footnote{%
   The number of digits in {\it zzz} is only as many as are needed to
   specify the total number of different orientations considered.  
   If the total number of orientations is $\leq10$,
   then {\it zzz} will have only a single digit, if the number of orientations
   is between 11 and 100, {\it zzz} wll have two digits, etc.}
These files contain $Q$ values and
scattering information for {\it each} of the target orientations.
\index{IWRKSC -- see ddscat.par}
\index{ddscat.par!IWRKSC}  
\index{w000r000k0.sca -- output file}
The structure of each of these files is very similar to that of the {\tt
w{\it xxx}r{\it yyy}ori.avg} files.  Because these files may not be of
particular interest, and take up disk space, you may choose to set
{\tt IWRKSC}=0 in future work.  However, it is suggested that you run
the sample calculation with {\tt IWRKSC}=1.

The sample {\tt ddscat.par} file specifies {\tt IWRKSC}=1 and calls
for use of 1 wavelength, 1 target size, and averaging over 3 target
orientations.  Running \ddscatv\ with the sample {\tt
ddscat.par} file will therefore generate files {\tt w000r000k000.sca},
{\tt w000r000k001.sca}, and {\tt w000r000k002.sca} .  To understand the
information contained in one of these files, please consult Appendix
\ref{app:w000r000k000.sca}, which contains an example of the file {\tt
w000r000k000.sca} produced in the sample calculation.

\subsection{ Binary Option\label{subsec:binary}}

It is possible to output an ``unformatted'' or ``binary'' file ({\tt
dd.bin}) with fairly complete information, including header and data
sections.  This is accomplished by specifying either {\tt ALLBIN} or
{\tt ORIBIN} in {\tt ddscat.par} .
\index{ddscat.par!ORIBIN}
\index{ddscat.par!ALLBIN}

Subroutine {\tt writebin.f90} provides an example of how this can be
done.  The ``header'' section contains dimensioning and other
variables which do not change with wavelength, particle geometry, and
target orientation.  The header section contains data defining the
particle shape, wavelengths, particle sizes, and target orientations.
If {\tt ALLBIN} has been specified, the ``data'' section contains, for
each orientation, Mueller matrix results for each scattering
direction.  The data output is limited to actual dimensions of arrays;
e.g. {\tt nscat,4,4} elements of Muller matrix are written rather than
{\tt mxscat,4,4}.  This is an important consideration when writing
postprocessing codes.

\section{Choice of Iterative Algorithm\label{sec:choice_of_algorithm}}
\index{iterative algorithm}
\index{conjugate gradient algorithm}
As discussed elsewhere \citep[e.g.,][]{Draine_1988,Draine+Flatau_1994}, 
the problem of
electromagnetic scattering of an incident wave ${\bf E}_{\rm inc}$ by an
array of $N$ point dipoles can be cast in the form
\beq
{\bf A} {\bf P} = {\bf E}
\label{eq:AP=E}
\eeq
where ${\bf E}$ is a $3N$-dimensional (complex) vector of the incident
electric field ${\bf E}_{\rm inc}$ at the $N$ lattice sites, ${\bf P}$ is
a $3N$-dimensional (complex) vector of the (unknown) dipole
polarizations, and ${\bf A}$ is a $3N\times3N$ complex matrix.

Because $3N$ is a large number, direct methods for solving this system
of equations for the unknown vector ${\bf P}$ are impractical, but
iterative methods are useful: we begin with a guess (typically, ${\bf
P}=0$) for the unknown polarization vector, and then iteratively
improve the estimate for ${\bf P}$ until equation (\ref{eq:AP=E}) is
solved to some error criterion.  The error tolerance may be specified
as
\beq
{|{\bf A}^\dagger {\bf A} {\bf P} - {\bf A}^\dagger {\bf E}| \over
| {\bf A}^\dagger {\bf E} |}
 < h 
\label{eq:err_tol}~~~,
\eeq
where ${\bf A}^\dagger$ is the Hermitian conjugate of ${\bf A}$
[$(A^\dagger)_{ij} \equiv (A_{ji})^*$], and $h$ is the error
tolerance.  We typically use $h=10^{-5}$ in order to satisfy
eq.(\ref{eq:AP=E}) to high accuracy.  The error tolerance $h$ can be
specified by the user through the parameter {\tt TOL} in
the parameter file {\tt ddscat.par} (see Appendix \ref{app:ddscat.par}).
\index{error tolerance}
\index{ddscat.par!TOL}

A major change in going from {{\bf DDSCAT}}{\bf .4b} to {\bf 5a} (and
subsequent versions) was the implementation of several different
algorithms for iterative solution of the system of complex linear
equations.  \ddscatv\ is now
structured to permit solution algorithms to be treated in a fairly
``modular'' fashion, facilitating the testing of different algorithms.
A number of algorithms were compared by \citet{Flatau_1997}\footnote{ A postscript
copy of this report -- file {\tt cg.ps} -- is distributed with the
\ddscatv\ documentation.}; two of them ({\tt PBCGST} and {\tt PETRKP}) 
performed well and were
made available to the user in the \ddscatsixzero\ release.
\ddscatsevenone\ introduced a third option, {\tt PBCGS2}.
\ddscatseventwo\ and \ddscatseventhree\ include 
two more CCG options: {\tt GPBICG} and {\tt QMRCCG}.
The choice of algorithm is made by specifying one of the options (here in
alphabetical order):
\begin{itemize}
\item {\tt GPBICG} -- Generalized Product-type methods based on Bi-CG
      \citet{Zhang_1997}.  We use an implementation suggested by
      \citet{Tang+Shen+Zheng+Qiu_2004}, and coded by P.C. Chaumet and
      A. Rahmani \citep{Chaumet+Rahmani_2009}.
      We are grateful to P.C. Chaumet and A. Rahmani for making their
      code available.
      
\item {\tt PBCGS2} --  BiConjugate Gradient with Stabilization as implemented 
      in the routine ZBCG2 by M.A. Botchev, University of Twente.
      This is based on the PhD thesis of D.R. Fokkema, and on work
      by \citet{Sleijpen+vanderVorst_1995,Sleijpen+vanderVorst_1996}.
\item {\tt PBCGST} -- Preconditioned BiConjugate Gradient with 
	STabilitization method from the Parallel Iterative Methods 
	(PIM) package created by R. Dias da Cunha and T. Hopkins.
\index{PBCGST algorithm}
\index{ddscat.par!PBCGST}
\index{PIM package}
\item {\tt PETRKP} -- the complex conjugate gradient algorithm of 
	\citet{Petravic+Kuo-Petravic_1979}, as coded in the Complex Conjugate 
	Gradient package (CCGPACK) created by P.J. Flatau.
	This is the algorithm discussed by \citet{Draine_1988} and used in 
	the earliest versions of {{\bf DDSCAT}}.
\index{PETRKP algorithm}
\index{ddscat.par!PETRKP}
\item {\tt QMRCCG} -- the quasi-minimum-residual complex conjugate
       gradient algorithm, based on f77 code written by
       P.C. Chaumet and A. Rahmani, converted here to f90 and adapted
       to single/double precision.
\end{itemize}
All five methods work fairly well.
Our experience suggests that {\tt PBCGS2} is generally fastest and 
best-behaved, and we recommend that the user try it first.
There have been claims that {\tt QMRCCG} and/or {\tt GPBICG}
are faster, but this has not been our experience.
{\tt PETRKP} is slow but may prove stable on some problems where
more aggressive algorithms become unstable.  
We have not carried out systematic studies of the relative performance
of the different algorithms -- the user is encouraged to experiment.

If a fast algorithm the case it runs into numerical difficulties, {\tt PBCGST} and
{\tt PETRKP} are available as alternatives.\footnote{
The Parallel Iterative Methods (PIM) by Rudnei Dias da Cunha ({\tt
rdd@ukc.ac.uk}) and Tim Hopkins ({\tt trh@ukc.ac.uk}) is a collection
of Fortran 77 routines designed to solve systems of linear equations
on parallel and scalar computers using a variety of iterative methods
(available at \hfill\break {\tt http://chasqueweb.ufrgs.br/~rudnei.cunha/pim.html}).
PIM offers a number of iterative methods, including
\begin{itemize}
\item the stabilised version of Bi-Conjugate-Gradients, BICGSTAB 
\citep{van_der_Vorst_1992},
\item the restarted version of BICGSTAB, RBICGSTAB 
\citet{Sleijpen+Fokkema_1993} 
\end{itemize}
The source code for these methods is distributed with {\tt DDSCAT} but
only {\tt PBCGST} and {\tt PETRKP} can be called directly via {\tt
ddscat.par}. It is possible to add other options by
changing the code in {\tt getfml.f90}~. \citet{Flatau_1997} has compared
the convergence rates of a number of different methods.
A helpful introduction to
conjugate gradient methods is provided by the report ``Conjugate
Gradient Method Without Agonizing Pain" by Jonathan R. Shewchuk,
available as a postscript file: {\tt
   ftp://REPORTS.ADM.CS.CMU.EDU/usr0/anon/1994/CMU-CS-94-125.ps}.
}

\section{Choice of FFT Algorithm\label{sec:choice_of_fft}}
\index{FFT algorithm}
\index{FFTW}
\index{GPFA}
\ddscatv\ offers two FFT options:
(1) the GPFA FFT algorithm developed by Dr. Clive Temperton 
\citep{Temperton_1992},
\footnote{The GPFA code contains a parameter {\tt LVR} which is set in 
          {\tt data} statements in the routines {\tt gpfa2f}, {\tt gpfa3f}, 
          and {\tt gpfa5f}.  {\tt LVR} is supposed to be optimized to 
          correspond to the ``length of a vector register'' on vector 
          machines.  As delivered, this parameter is set to 64, which is 
          supposed to be appropriate for Crays other than the C90.  
          For the C90, 128 is supposed to be preferable (and perhaps 
	  ``preferable'' should be read as
	  ``necessary'' -- there is some basis for fearing that results
	  computed on a C90 with {\tt LVR} other than 128 run the risk of being
	  incorrect!)
          The value of {\tt LVR} is not critical for scalar machines, as 
          long as it is fairly large.  We found little difference between 
          {\tt LVR}=64 and 128 on a Sparc 10/51, on an Ultrasparc 170, and 
          on an \Intel\ Xeon cpu.  You may wish to
          experiment with different {\tt LVR} values on your computer
          architecture.  To change {\tt LVR}, you need to edit {\tt gpfa.f90} 
          and change the three {\tt data} statements where {\tt LVR} is set.}
and (2) the \Intel\ MKL routine {\tt DFTI}. 

\begin{figure}[h]
\begin{center}
\vspace*{-0.9cm}
\includegraphics[width=9.0cm]{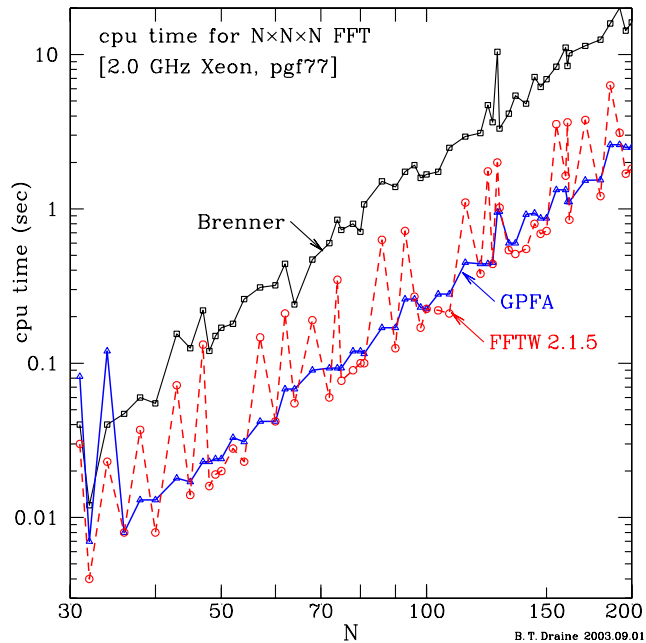}
\vspace*{-2.2cm}
\caption{\label{fig:fft_timings}
        \footnotesize
	Comparison of cpu time required by 3 different FFT
	implementations.
	It is seen that the GPFA and FFTW implementations have comparable
	speeds, much faster than Brenner's FFT implementation.
	}
\end{center}
\end{figure}

The GPFA routine is portable and quite fast:
Figure 5 compares the speed of three FFT implementations: Brenner's, GPFA,
and FFTW ({\tt http://www.fftw.org}).  
We see that while for some cases {\tt FFTW 2.1.5} is
faster than the GPFA algorithm, the difference is only marginal.
The FFTW code and GPFA code are quite comparable in
performance -- for some cases the GPFA code is faster, for other cases
the FFTW code is faster.
For target dimensions which are factorizable as $2^i3^j5^k$ (for integer
$i$, $j$, $k$), the GPFA
and FFTW 
codes have the same memory requirements.
For targets with extents $N_x$, $N_y$, $N_z$ 
which are not factorizable as $2^i3^j5^k$, the GPFA code
needs to ``extend'' the computational volume to have values of $N_x$,
$N_y$, and $N_z$ which are factorizable by 2, 3, and 5.
For these cases, GPFA requires somewhat more memory than
FFTW.
However, the fractional difference in required memory 
is not large, since integers factorizable
as $2^i3^j5^k$ occur fairly 
frequently.\footnote{\label{fn:235}2, 3, 4, 5, 6, 8, 9, 10, 12,
	15, 16, 18, 20, 24, 25, 27, 30, 32, 36, 40,
	45, 48, 50, 54, 60, 64, 72, 75, 80, 81, 90, 
	96, 100, 108, 120, 125, 128, 135,
	144, 150, 160, 162, 180, 192, 200  216, 225,
        240, 243, 250, 256, 270, 288, 300, 320, 324, 360, 375, 384, 400, 405,
        432, 450, 480, 486, 500, 512, 540, 576, 600, 625, 640, 648, 675, 720,
        729, 750, 768, 800, 810, 864, 900, 960, 972, 1000, 1024, 1080, 1125,
        1152, 1200, 1215, 1250, 1280, 1296, 1350, 1440, 1458, 1500, 1536,
        1600, 1620, 1728, 1800, 1875, 1920, 1944, 2000, 2025, 2048, 2160,
        2187, 2250, 2304, 2400, 2430, 2500, 2560, 2592, 2700, 2880, 2916,
        3000, 3072, 3125, 3200, 3240, 3375, 3456, 3600, 3645, 3750, 3840,
        3888, 4000, 4050, 4096 are the integers $\leq 4096$
	which are of the form $2^i3^j5^k$.}
[Note: This ``extension'' of the target volume occurs automatically and
is transparent to the user.]

\ddscatv\ offers a new FFT option: the \Intel\ Math Kernel
Library {\tt DFTI}.  This is tuned for optimum performance,
and appears to offer real performance advantages on modern multi-core cpus.
With this now available, the FFTW option, which had been included in
\ddscat\ {\bf 6.1}, has been removed from
\ddscatv. 

The choice of FFT implementation is obtained by specifying one of:
\begin{itemize}
\item{\tt FFTMKL} to use the \Intel\ MKL routine {\tt DFTI} 
        (see \S\ref{sec:MKL}).
	{\bf This is recommended, but requires that the \Intel\ Math Kernel
	Library be
	installed on your system}.
\item{\tt GPFAFT} to use the GPFA algorithm \citep{Temperton_1992}.
        {\bf This
	is not quite as fast as FFTMKL, but is written in plain Fortran-90.
        It is a perfectly good alternative if the \Intel\ Math Kernel
        Library is not available on your system.}
\end{itemize}

\section{Choice of DDA Method\label{sec:DDA_method}}
\index{DDA method}
\index{Filtered Coupled Dipole method}
\index{FLTRCD}
\index{LATTDR}
\index{GKDLDR}
\index{ddscat.par!LATTDR}
\index{ddscat.par!GKDLDR}

\subsection{Point Dipoles: Options LATTDR and GKDLDR}

Earlier versions of {\bf DDSCAT} (up to and including {\bf DDSCAT\,7.2}
treated the well-defined problem
of absorption and scattering by an array of polarizable points
\citep{Purcell+Pennypacker_1973,Draine_1988,Draine+Flatau_1994}, where
the target is divided up into finite elements, each represented by
a polarizable point.
The problem is then fully characterized by the geometric distribution of
the polarizable points, the polarizability $\alpha$ of each point,
and the incident electromagnetic wave.
The polarizability $\alpha$ is chosen according to some prescription.
Earlier versions of {\bf DDSCAT} offered as options the ``Lattice Dispersion
Relation'' prescription of
\citet{Draine+Goodman_1993}, and the modified Lattice Dispersion Relation
prescription of \citet{Gutkowicz-Krusin+Draine_2004}.
\index{GKDLDR}
\index{polarizabilities!GKDLDR}
\index{ddscat.par!GKDLDR}
Option {\tt GKDLDR} specifies that the polarizability be prescribed
by the ``Lattice Dispersion Relation'', with the polarizability
found by \citet{Gutkowicz-Krusin+Draine_2004}, 
who corrected a subtle error in
the analysis of \citet{Draine+Goodman_1993}.
For $|m|kd\ltsim 1$, 
the {\tt GKDLDR}  polarizability differs slightly from the {\tt LATTDR} 
polarizability, but the differences in calculated scattering cross sections
are relatively small, as can be seen from Figure \ref{fig:GKDLDR_vs_LATTDR}.
We recommend option {\tt GKDLDR}.

\index{LATTDR}
\index{polarizabilities!LATTDR}
\index{ddscat.par!LATTDR}
Users wishing to compare can invoke option {\tt LATTDR} to specify that the 
``Lattice Dispersion Relation'' of \citet{Draine+Goodman_1993}
be employed to determine the dipole
polarizabilities.
This polarizability also works well.

This approach works well provided the refractive index $m$ of the
target material is not too large.  However, when $|m|$ is large,
both of these methods perform poorly.

\subsection{Filtered Coupled Dipole: option FLTRCD}
\index{ddscat.par!FLTRCD}
\index{FLTRCD -- filtered coupled dipole method}

\citet{Piller+Martin_1998} proposed 
the ``filtered coupled dipole'' (FCD) method as an approach that would work
better for targets with large refractive indices.  This method continues
to represent a finite target by an array of polarizable points, but 
with the electric field generated by each point differing from the
field of a true point dipole by virtue of having component of high
spatial frequency ``filtered out''.
\citet{Gay-Balmaz+Martin_2002} revisited the FCD method, correcting some
typographical errors in \citet{Piller+Martin_1998}.
\citet{Yurkin+Min+Hoekstra_2010} carried out a comparison of the FCD method
with the point dipole method and showed that the FCD method could be used
for targets with large refractive indices where the the point dipole method
failed.

\ddscatv\ offers the filtered coupled dipole method as an option
({\tt FLTRDD}).  When this option is selected, the dipole polarizabilities
are assigned by
\beq
\alpha_j = \frac{\alpha^{\rm (CM)}}{1 + D}
~~~,
\eeq
where $\alpha^{\rm (CM)}$ is the Clausius-Mossotti polarizability
\beq
\alpha^{\rm (CM)} \equiv \frac{3d^3}{4\pi}\frac{(m_j^2-1)}{(m_j^2+2)}
~~~,
\eeq
where $m_j$ is the complex refractive index at lattice site $j$.
The correction term $D$ is given by
\beq
D = \frac{\alpha^{\rm (CM)}}{d^3}
              \left[
              \frac{4}{3}(kd)^2+
              \frac{2}{3\pi}\ln\left[\frac{\pi-kd}{\pi+kd}\right]+
              \frac{2}{3}i(kd)^3
              \right]
~~~,
\eeq
\citep[see][eq.\ 9]{Yurkin+Min+Hoekstra_2010}.
Here $d$ is the lattice spacing, and $k\equiv\omega/c$.

\begin{figure}[t]
\begin{center}
\vspace*{-0.9cm}
\includegraphics[width=8.5cm]{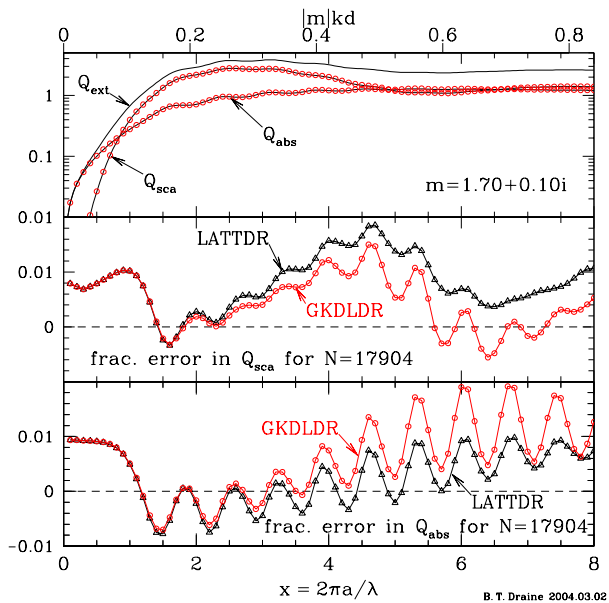}
\vspace*{-2.2cm}
\caption{\footnotesize
         Scattering and absorption for a $m=1.7+0.1i$ sphere,
         calculated using two prescriptions for the polarizability:
	 LATTDR is the lattice dispersion relation result of 
         \citet{Draine+Goodman_1993}.
	 GKDLDR is the lattice dispersion relation result
	 of \citet{Gutkowicz-Krusin+Draine_2004}.
	 Results are shown as a function of scattering parameter
	 $x=2\pi a/\lambda$;
	 the upper scale gives values of $|m|kd$.
	 We see that the cross sections calculated with these two
	 prescriptions are quite similar for $|m|kd\ltsim 0.5$.
	 For other examples see \citet{Gutkowicz-Krusin+Draine_2004}.
	\label{fig:GKDLDR_vs_LATTDR}}
\end{center}
\end{figure}

\section{Dielectric Functions\label{sec:dielectric_func}}
\index{dielectric function of target material}
\index{refractive index of target material}
In order to assign the appropriate dipole polarizabilities, \ddscatv\
must be given the refractive index $m$ or dielectric constant $\epsilon$
of the material (or
materials) of which the target of interest is composed.
This information is supplied to \ddscatv\
through a table (or tables), read by subroutine {\tt DIELEC} in file
{\tt dielec.f90}, and providing either the complex refractive index
$m=n+ik$ or complex dielectric function
$\epsilon=\epsilon_1+i\epsilon_2$ as a function of wavelength
$\lambda$.  Since $m=\epsilon^{1/2}$, or $\epsilon=m^2$, the user must
supply either $m$ or $\epsilon$.

\ddscatv\ can calculate scattering and absorption by targets with anisotropic
dielectric functions, with arbitrary orientation of the optical axes relative
to the target shape. See \S\ref{sec:composite anisotropic targets}.

The table containing the dielectric function
information should give $m$ or $\epsilon$
as a function of the {\it wavelength in vacuo}.

The table formatting is intended to be quite flexible.  The first line
of the table consists of text, up to 80 characters of which will be
read and included in the output to identify the choice of dielectric
function.  (For the sample problem, it consists of simply the
statement {\tt m = 1.33 + 0.01i}.)  The second line consists of 5
integers; either the second and third {\it or} the fourth and fifth
should be zero.
\begin{itemize}
\item The first integer specifies which column the wavelength is stored in.
\item The second integer specifies which column Re$(m)$ is stored in.
\item The third integer specifies which column Im$(m)$ is stored in.
\item The fourth integer specifies which column Re$(\epsilon)$ is stored in.
\item The fifth integer specifies which column Im$(\epsilon)$ is stored in.
\end{itemize}
If the second and third integers are zeros, then {\tt DIELEC} will
read Re$(\epsilon)$ and Im$(\epsilon)$ from the file; if the fourth
and fifth integers are zeros, then Re$(m)$ and Im$(m)$ will be read
from the file.

The third line of the file is used for column headers, and the data
begins in line 4.  {\it There must be at least 3 lines of data:} even
if $m$ or $\epsilon$ is required at only one wavelength, please supply two
additional ``dummy'' wavelength entries in the table so that the
interpolation apparatus will not be confused.

As discussed in \S\ref{sec:target_in_medium}, \ddscat\ can scattering
for targets embedded in dielectric media.  The refractive index of the
ambient medium is specified by the value of {\tt NAMBIENT} in the
parameter file {\tt ddscat.par} (see \S\ref{sec:nambient}).

Here is an example of a refractive index file for Au:
{\footnotesize
\begin{verbatim}
Gold, evaporated (Johnson & Christy 1972, PRB 6, 4370)
1 2 3 0 0 = columns for wave, Re(n), Im(n), eps1, eps2
wave(um) Re(n)  Im(n)   eps1   eps2
0.5486   0.43   2.455   -5.84  2.11
0.5209   0.62   2.081   -3.95  2.58
0.4959   1.04   1.833   -2.28  3.81
0.4714   1.31   1.849   -1.70  4.84
0.4509   1.38   1.914   -1.76  5.28
0.4305   1.45   1.948   -1.69  5.65
0.4133   1.46   1.958   -1.70  5.72
0.3974   1.47   1.952   -1.65  5.74
0.3815   1.46   1.933   -1.60  5.64
0.3679   1.48   1.895   -1.40  5.61
0.3542   1.50   1.866   -1.23  5.60
0.3425   1.48   1.871   -1.31  5.54
0.3315   1.48   1.883   -1.36  5.57
0.3204   1.54   1.898   -1.23  5.85
0.3107   1.53   1.893   -1.24  5.79
0.3009   1.53   1.889   -1.23  5.78
\end{verbatim}
}

\section{Calculation of $\langle\cos\theta\rangle$, Radiative Force, and 
         Radiation Torque
	\label{sec:force and torque calculation}}

In addition to solving the scattering problem for a dipole array,
\ddscat\ can compute the three-dimensional force ${\bf F}_{\rm rad}$ 
and torque ${\bf \Gamma}_{\rm rad}$ exerted on this array by the
incident and scattered radiation fields.  
The radiation torque calculation is carried
out, after solving the scattering problem, only if {\tt DOTORQ} has
been specified in {\tt ddscat.par}.  
\index{DOTORQ}
\index{ddscat.par!DOTORQ}
For each incident polarization
mode, the results are given in terms of dimensionless efficiency
vectors ${\bf Q}_{\rm pr}$ and ${\bf Q}_{\Gamma}$, defined by
\index{radiative force efficiency vector ${\bf Q}_{\rm pr}$}
\index{radiative torque efficiency vector ${\bf Q}_\Gamma$}
\index{${\bf Q}_{\rm pr}$ -- radiative force efficiency vector}
\index{${\bf Q}_\Gamma$ -- radiative torque efficiency vector}
\beq
{\bf Q}_{\rm pr} \equiv {{\bf F}_{\rm rad} \over 
\pi \aeff^2 u_{\rm rad}} ~~~,
\eeq
\beq
{\bf Q}_\Gamma \equiv {k{\bf \Gamma}_{\rm rad} \over
\pi \aeff^2 u_{\rm rad}} ~~~,
\eeq
where ${\bf F}_{\rm rad}$ and ${\bf \Gamma}_{\rm rad}$ are the time-averaged
force and torque on the dipole array, $k=2\pi/\lambda$ is the
wavenumber {\it in vacuo}, and $u_{\rm rad} = E_0^2/8\pi$ is the
time-averaged energy density for an incident plane wave with amplitude
$E_0 \cos(\omega t + \phi)$.  The radiation pressure efficiency vector
can be written
\beq
{\bf Q}_{\rm pr} = Q_\ext{\hat{\bf k}} - Q_\sca{\bf g} ~~~,
\eeq
where ${\hat{\bf k}}$ is the direction of propagation of the incident
radiation, and the vector {\bf g} is the mean direction of propagation
of the scattered radiation:
\beq\label{eq:gvec}
{\bf g} = {1\over C_\sca}\int d\Omega
{dC_\sca({\hat{\bf n}},{\hat{\bf k}})\over d\Omega} {\hat{\bf n}} ~~~,
\eeq
where $d\Omega$ is the element of solid angle in scattering direction
${\hat{\bf n}}$, and $dC_\sca/d\Omega$ is the differential scattering
cross section.
The components of ${\bf Q}_{\rm pr}$ are reported in the Target Frame:
$Q_{{\rm pr},1}\equiv {\bf F}_{\rm rad}\cdot \xtf$,
$Q_{{\rm pr},2}\equiv {\bf F}_{\rm rad}\cdot \ytf$,
$Q_{{\rm pr},3}\equiv {\bf F}_{\rm rad}\cdot \ztf$.

Equations for the evaluation of the radiative force and torque are
derived by \citet{Draine+Weingartner_1996}.  It is important to note that
evaluation of ${\bf Q}_{\rm pr}$ and ${\bf Q}_\Gamma$ involves averaging
over scattering directions to evaluate the linear and angular momentum
transport by the scattered wave.  This evaluation requires appropriate
choices of the parameter {\tt ETASCA} -- see
\S\ref{sec:averaging_scattering}.
\index{ETASCA}
\index{ddscat.par!ETASCA}

In addition, \ddscat\ calculates $\langle\cos\theta\rangle$ [the first
component of the vector $g$ in eq.\ (\ref{eq:gvec})] and the
second moment $\langle\cos^2\theta\rangle$.
These two moments are useful measures of the anisotropy of the scattering.
For example, \citet{Draine_2003b} gives an analytic approximation to the
scattering phase function of dust mixtures that is parameterized by
the two moments
$\langle\cos\theta\rangle$ and $\langle\cos^2\theta\rangle$.

\section{Memory Requirements \label{sec:memory_requirements}}
\index{memory requirements}
\index{MXNX,MXNY,MXNZ}

The memory requirements are determined by the size of the ``computational
volume'' -- this is a rectangular region, of size 
{\tt NX}$\times${\tt NY}$\times${\tt NZ} that is large enough to contain
the target.  If using the {\tt GPFAFT} option, then {\tt NX},
{\tt NY}, {\tt NZ} are also required to have only 2,3, and 5 as prime factors
(see footnote \ref{fn:235}).

In single precision, the memory requirement for \ddscatv\ is approximately 
\beq
(35.+0.0010\times{\tt NX}\times{\tt NY}\times{\tt NZ}) {\rm ~Mbytes}  
~~~~{\rm for~single~precision}
\eeq
\beq
(42+0.0020\times{\tt NX}\times{\tt NY}\times{\tt NZ}) {\rm ~Mbytes}
~~~~{\rm for~double~precision}
\eeq
Thus, in single precision, a
48$\times$48$\times$48 calculation requires $\sim$146 MBytes.

The memory is allocated dynamically -- once the target has been created,
\ddscatv\ will determine just how much overall memory is needed, and
will allocate it.  However, the user must provide information (via 
{\tt ddscat.par}) to allow \ddscatv\ to allocate sufficiently large arrays
to carry out the initial target creation.  Initially, the only arrays that
will be allocated are those related to the target geometry, so it is
OK to be quite generous in this initial allowance, as the memory
required for the target generation step is small compared to the
memory required to carry out the full scattering calculation.
\index{memory requirements}

\section{Target Geometry: The Target Frame
         \label{sec:target geometry}}
\index{Target Frame}

The geometry of the target is specified by the locations of the
lattice sites where polarizable points (``dipoles'') are located.
The list of occupied sites will be generated internally by \ddscat\ if
the user selects one of the ``built-in'' target geometries, but the
user can also use the target option {\tt FROM\_FILE} to read in the
list of occupied site locations and composition information.

Every target is defined by a list of ``occupied'' lattice sites
$(i,j,k)_n$, $n=1,...,N$.
In the ``Target Frame'' (TF), these sites have physical locations
$(x,y,z)_n = [(i,j,k)_n + (x_0,y_0,z_0)]\times d$,
where $d$ is the lattice constant (in physical units)
and $(x_0,y_0,z_0)$ is a vector that gives the physical
location corresponding to $(i,j,k)=(0,0,0)$.
Thus, the vector $\bx_0$ specifies the physical location of the TF 
``lattice coordinate'' origin 
$(0,0,0)_\TF$.
The vector $\bx_0$ 
is specified for 
for each of the ``built-in'' target geometries.
For targets provided externally through the {\tt FROM\_FILE}
option (see \S\ref{sec:FROM_FILE}), the ``target file'' must
include the three components of the vector $\bx_0$.

\section{Target Orientation \label{sec:target_orientation}}
\index{target orientation}
\index{Lab Frame (LF)}

Recall that we define a ``Lab Frame'' (LF) in which the incident
radiation propagates in the $+x$ direction.  
For purposes of discussion we will always
let unit vectors $\xlf$, $\ylf$, $\zlf=\xlf\times\ylf$ 
be the three coordinate axes of the LF.
 
In {\tt ddscat.par} one
specifies the first polarization state ${\hat{\bf e}}_{01}$ (which
obviously must lie in the $y,z$ plane in the LF); {{\bf DDSCAT}}\
automatically constructs a second polarization state ${\hat{\bf
e}}_{02} = \xlf \times {\hat{\bf e}}_{01}^*$ orthogonal to
${\hat{\bf e}}_{01}$.
Users will often find it convenient to let polarization
vectors ${\hat{\bf e}}_{01}={\hat{\bf y}}$, ${\hat{\bf
e}}_{02}={\hat{\bf z}}$ (although this is not mandatory -- see
\S\ref{sec:incident_polarization}).

\begin{figure}[h]
\begin{center}
\vspace*{-0.9cm}
\includegraphics[width=12.cm]{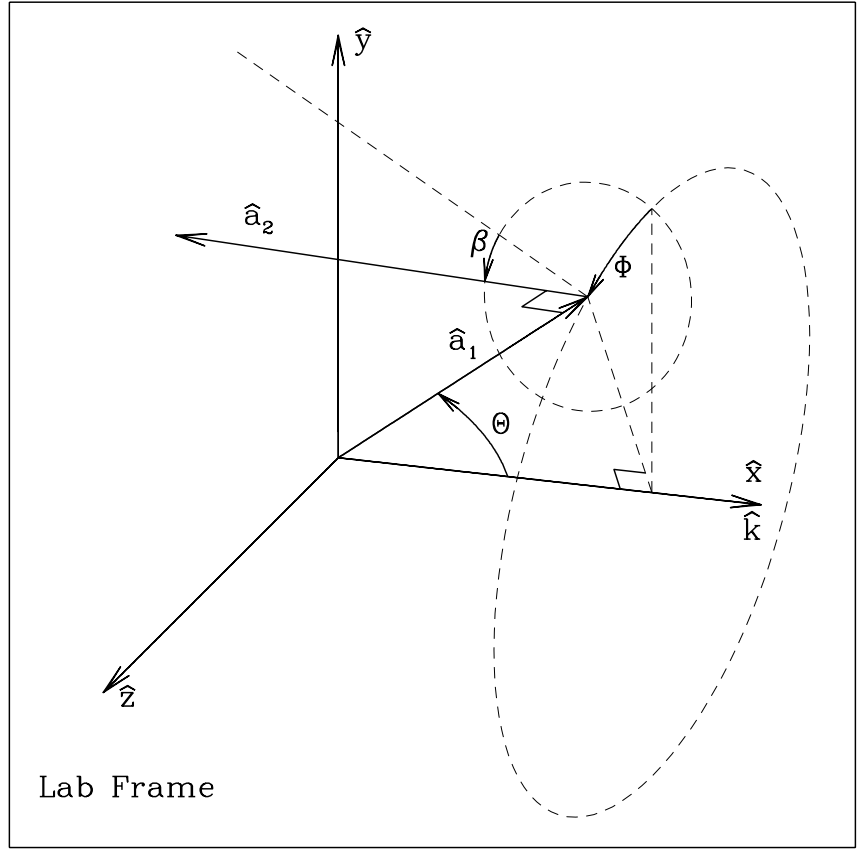}
\vspace*{-2.2cm}
\caption{\label{fig:target_orientation}
         \footnotesize
        Target orientation in the Lab Frame.  ${\hat{\bf x}}=\xlf$ is the
	direction of propagation of the incident radiation, and ${\hat{\bf
	y}}=\ylf$ is the direction of the real component (at $x_\LF=0$, $t=0$)
	of the first incident
	polarization mode.  In this coordinate system, the orientation of
	target axis ${\hat{\bf a}}_1$ is specified by angles $\Theta$ and
	$\Phi$.  With target axis ${\hat{\bf a}}_1$ fixed, the orientation of
	target axis ${\hat{\bf a}}_2$ is then determined by angle $\beta$
	specifying rotation of the target around ${\hat{\bf a}}_1$.  When
	$\beta=0$, ${\hat{\bf a}}_2$ lies in the ${\hat{\bf a}}_1$,$\xlf$
	plane.
	}
\end{center}
\end{figure}

Recall that definition of a target involves specifying two unit
vectors, ${\hat{\bf a}}_1$ and ${\hat{\bf a}}_2$, which are imagined
to be ``frozen'' into the target.  We require ${\hat{\bf a}}_2$ to be
orthogonal to ${\hat{\bf a}}_1$.  Therefore we may define a ``Target
Frame" (TF) defined by the three unit vectors ${\hat{\bf a}}_1$,
${\hat{\bf a}}_2$, and ${\hat{\bf a}}_3 = {\hat{\bf a}}_1 \times
{\hat{\bf a}}_2$ .
\index{Target Frame (TF)}

For example, when {{\bf DDSCAT}}\ creates a 32$\times$24$\times$16
rectangular solid, it fixes ${\hat{\bf a}}_1$ to be along the longest
dimension of the solid, and ${\hat{\bf a}}_2$ to be along the
next-longest dimension.

{\bf Important Note:} for periodic targets, \ddscatv\ requires that the
periodic target have ${\hat{\bf a}}_1=\xtf$ and ${\hat{\bf a}}_2=\ytf$.

Orientation of the target relative to the incident radiation can in
principle be determined two ways:
\begin{enumerate}
\item specifying the direction of ${\hat{\bf a}}_1$ and 
      ${\hat{\bf a}}_2$ in the LF, or
\item specifying the directions of $\xlf$ (incidence
	direction) and $\ylf$ in the TF.
\end{enumerate}
\ddscat\ uses method 1.: the angles $\Theta$, $\Phi$, and
$\beta$ are specified in the file {\tt ddscat.par}.  The target is
oriented such that the polar angles $\Theta$ and $\Phi$ specify the
direction of ${\hat{\bf a}}_1$ relative to the incident direction
$\xlf$, where the $\xlf$,$\ylf$ plane has
$\Phi=0$.  Once the direction of ${\hat{\bf a}}_1$ is specified, the
angle $\beta$ then specifies how the target is to rotated around the
axis ${\hat{\bf a}}_1$ to fully specify its orientation.  A more
extended and precise explanation follows:

\subsection{ Orientation of the Target in the Lab Frame}
\index{target orientation}
\index{Lab Frame}
\index{$\Theta$ -- target orientation angle}
\index{$\Phi$ -- target orientation angle}
\index{$\beta$ -- target orientation angle}
DDSCAT uses three angles, $\Theta$, $\Phi$, and $\beta$, to specify
the directions of unit vectors ${\hat{\bf a}}_1$ and ${\hat{\bf a}}_2$
in the LF (see Fig.\ \ref{fig:target_orientation}).

$\Theta$ is the angle between ${\hat{\bf a}}_1$ and $\xlf$.

When $\Phi=0$, ${\hat{\bf a}}_1$ will lie in the $\xlf,\ylf$
plane.  When $\Phi$ is nonzero, it will refer to
the rotation of ${\hat{\bf a}}_1$ around $\xlf$: e.g.,
$\Phi=90^\circ$ puts ${\hat{\bf a}}_1$ in the $\xlf,\zlf$ plane.

When $\beta=0$, ${\hat{\bf a}}_2$ will lie in the 
$\xlf,{\hat{\bf a}}_1$ plane, in such a way that when $\Theta=0$ and
$\Phi=0$, ${\hat{\bf a}}_2$ is in the $\ylf$ direction: e.g,
$\Theta=90^\circ$, $\Phi=0$, $\beta=0$ has 
${\hat{\bf a}}_1=\ylf$ 
and ${\hat{\bf a}}_2=-\xlf$.  Nonzero $\beta$ introduces
an additional rotation of ${\hat{\bf a}}_2$ around ${\hat{\bf a}}_1$:
e.g., $\Theta=90^\circ$, $\Phi=0$, $\beta=90^\circ$ has 
${\hat{\bf a}}_1=\ylf$ and ${\hat{\bf a}}_2=\zlf$.

Mathematically:
\begin{eqnarray}
{\hat{\bf a}}_1 &=&   \xlf \cos\Theta 
+ \ylf \sin\Theta \cos\Phi 
+ \zlf \sin\Theta \sin\Phi
	\\
{\hat{\bf a}}_2 &=& - \xlf \sin\Theta \cos\beta 
+ \ylf [\cos\Theta \cos\beta \cos\Phi-\sin\beta \sin\Phi] \nonumber\\
&&+ \zlf [\cos\Theta \cos\beta \sin\Phi+\sin\beta \cos\Phi]
	\\
{\hat{\bf a}}_3 &=&   \xlf \sin\Theta \sin\beta 
- \ylf [\cos\Theta \sin\beta \cos\Phi+\cos\beta \sin\Phi] \nonumber\\
  &&         - \zlf [\cos\Theta \sin\beta \sin\Phi-\cos\beta \cos\Phi]
\end{eqnarray}
or, equivalently:
\begin{eqnarray}
\xlf &=&   {\hat{\bf a}}_1 \cos\Theta
           - {\hat{\bf a}}_2 \sin\Theta \cos\beta
           + {\hat{\bf a}}_3 \sin\Theta \sin\beta \\
\ylf &=&   {\hat{\bf a}}_1 \sin\Theta \cos\Phi
           + {\hat{\bf a}}_2 [\cos\Theta \cos\beta \cos\Phi-\sin\beta \sin\Phi]
\nonumber\\
&&         - {\hat{\bf a}}_3 [\cos\Theta \sin\beta \cos\Phi+\cos\beta \sin\Phi]
\\
\zlf &=&   {\hat{\bf a}}_1 \sin\Theta \sin\Phi
           + {\hat{\bf a}}_2 [\cos\Theta \cos\beta \sin\Phi+\sin\beta \cos\Phi]
\nonumber\\
&&         - {\hat{\bf a}}_3 [\cos\Theta \sin\beta \sin\Phi-\cos\beta \cos\Phi]
\end{eqnarray}

\subsection{ Orientation of the Incident Beam in the Target Frame}
\index{Target Frame}

Under some circumstances, one may wish to specify the target
orientation such that $\xlf$ (the direction of propagation of
the radiation) and $\ylf$ (usually the first polarization
direction) and $\zlf$ (= $\xlf \times \ylf$) 
refer to certain directions in the TF.  Given the definitions of
the LF and TF above, this is simply an exercise in coordinate
transformation.  For example, one might wish to have the incident
radiation propagating along the (1,1,1) direction in the TF (example
14 below).  Here we provide some selected examples:
\begin{enumerate}
\item$\xlf= {\hat{\bf a}}_1$, 
     $\ylf= {\hat{\bf a}}_2$, 
     $\zlf= {\hat{\bf a}}_3$ : 
	$\Theta=  0$, $\Phi+\beta= 0$
\item$\xlf= {\hat{\bf a}}_1$, 
     $\ylf= {\hat{\bf a}}_3$, 
     $\zlf=-{\hat{\bf a}}_2$ : 
	$\Theta=  0$, $\Phi+\beta= -90^\circ$
\item$\xlf= {\hat{\bf a}}_2$, 
     $\ylf= {\hat{\bf a}}_1$, 
     $\zlf=-{\hat{\bf a}}_3$ : 
	$\Theta= 90^\circ$, $\beta=180^\circ$, $\Phi=  0$
\item$\xlf= {\hat{\bf a}}_2$, 
     $\ylf= {\hat{\bf a}}_3$, 
     $\zlf= {\hat{\bf a}}_1$ : 
	$\Theta= 90^\circ$, $\beta=180^\circ$, $\Phi= 90^\circ$
\item$\xlf= {\hat{\bf a}}_3$, 
     $\ylf= {\hat{\bf a}}_1$, 
     $\zlf= {\hat{\bf a}}_2$ : 
	$\Theta= 90^\circ$, $\beta=90^\circ$, $\Phi=  0$
\item$\xlf= {\hat{\bf a}}_3$, 
     $\ylf= {\hat{\bf a}}_2$, 
     $\zlf=-{\hat{\bf a}}_1$ : 
	$\Theta= 90^\circ$, $\beta=90^\circ$, $\Phi=-90^\circ$
\item$\xlf=-{\hat{\bf a}}_1$, 
     $\ylf= {\hat{\bf a}}_2$, 
     $\zlf=-{\hat{\bf a}}_3$ : 
	$\Theta=180^\circ$, $\beta-\Phi=180^\circ$
\item$\xlf=-{\hat{\bf a}}_1$, 
     $\ylf= {\hat{\bf a}}_3$, 
     $\zlf= {\hat{\bf a}}_2$ : 
	$\Theta=180^\circ$, $\beta-\Phi= 90^\circ$
\item$\xlf=-{\hat{\bf a}}_2$, 
     $\ylf= {\hat{\bf a}}_1$, 
     $\zlf= {\hat{\bf a}}_3$ : 
	$\Theta= 90^\circ$, $\beta=  0$, $\Phi=  0$
\item$\xlf=-{\hat{\bf a}}_2$, 
     $\ylf= {\hat{\bf a}}_3$, 
     $\zlf=-{\hat{\bf a}}_1$ : 
	$\Theta= 90^\circ$, $\beta=  0$, $\Phi=-90^\circ$
\item$\xlf=-{\hat{\bf a}}_3$, 
     $\ylf= {\hat{\bf a}}_1$, 
     $\zlf=-{\hat{\bf a}}_2$ : 
	$\Theta= 90^\circ$, $\beta=-90^\circ$, $\Phi=  0$
\item$\xlf=-{\hat{\bf a}}_3$, 
     $\ylf= {\hat{\bf a}}_2$, 
     $\zlf= {\hat{\bf a}}_1$ : 
	$\Theta= 90^\circ$, $\beta=-90^\circ$, $\Phi= 90^\circ$
\item$\xlf=({\hat{\bf a}}_1+{\hat{\bf a}}_2)/\surd2$, 
     $\ylf={\hat{\bf a}}_3$, 
     $\zlf=({\hat{\bf a}}_1-{\hat{\bf a}}_2)/\surd2$ : 
	$\Theta=45^\circ$, $\beta=180^\circ$, $\Phi=90^\circ$
\item$\xlf=({\hat{\bf a}}_1+{\hat{\bf a}}_2+{\hat{\bf a}}_3)/\surd3$, 
	$\ylf=({\hat{\bf a}}_1-{\hat{\bf a}}_2)/\surd2$, 
	$\zlf=({\hat{\bf a}}_1+{\hat{\bf a}}_2-2{\hat{\bf a}}_3)/\surd6$ :\\
	$\Theta=54.7356^\circ$, $\beta=135^\circ$, $\Phi=30^\circ$.
\end{enumerate}

\subsection{ Sampling in $\Theta$, $\Phi$, and $\beta$\label{subsec:sampling}}
\index{orientational sampling in $\beta$, $\Theta$, and $\Phi$}
\index{ddscat.par -- BETAMI,BETAMX,NBETA}
\index{ddscat.par -- THETMI,THETMX,NTHETA}
\index{ddscat.par -- PHIMIN,PHIMAX,NPHI}
The present version, \ddscatv, chooses the angles $\beta$,
$\Theta$, and $\Phi$ to sample the intervals ({\tt BETAMI,BETAMX}),
({\tt THETMI,THETMX)}, ({\tt PHIMIN,PHIMAX}), where {\tt BETAMI}, {\tt
BETAMX}, {\tt THETMI}, {\tt THETMX}, {\tt PHIMIN}, {\tt PHIMAX} are
specified in {\tt ddscat.par} .  The prescription for choosing the
angles is to:
\begin{itemize}
\item uniformly sample in $\beta$;
\item uniformly sample in $\Phi$;
\item uniformly sample in $\cos\Theta$.
\end{itemize}
This prescription is appropriate for random orientation of the target,
within the specified limits of $\beta$, $\Phi$, and $\Theta$.

Note that when \ddscatv\ chooses angles it handles $\beta$ and
$\Phi$ differently from $\Theta$.
The range for $\beta$ is divided into {\tt NBETA} intervals, and the
midpoint of each interval is taken.  Thus, if you take {\tt
BETAMI}=0, {\tt BETAMX}=90, {\tt NBETA}=2 you will get
$\beta=22.5^\circ$ and $67.5^\circ$.  Similarly, if you take
{\tt PHIMIN}=0, {\tt PHIMAX}=180, {\tt NPHI}=2 you will get
$\Phi=45^\circ$ and $135^\circ$.

Sampling in $\Theta$ is done quite differently from sampling in
$\beta$ and $\Phi$.  First, as already mentioned above, \ddscatv\ 
samples uniformly in $\cos\Theta$, not $\Theta$.
Secondly, the sampling depends on whether {\tt NTHETA} is even or odd.
\begin{itemize}
\item If {\tt NTHETA} is odd, then the values of $\Theta$ selected
	include the extreme values {\tt THETMI} and {\tt THETMX}; thus, 
        {\tt THETMI}=0, {\tt THETMX}=90, {\tt NTHETA}=3 will give you
	$\Theta=0,60^\circ,90^\circ$.
\item If {\tt NTHETA} is even, then the range of $\cos\Theta$ will be
	divided into {\tt NTHETA} intervals, and the midpoint of each interval
	will be taken; thus, {\tt THETMI}=0, {\tt THETMX}=90, {\tt NTHETA}=2
	will give you $\Theta=41.41^\circ$ and $75.52^\circ$
	[$\cos\Theta=0.25$ and $0.75$].
\end{itemize}
The reason for this is that if odd {\tt NTHETA} is specified, then the
``integration'' over $\cos\Theta$ is performed using Simpson's rule
for greater accuracy.  If even {\tt NTHETA} is specified, then the
integration over $\cos\Theta$ is performed by simply taking the
average of the results for the different $\Theta$ values.

If averaging over orientations is desired, it is recommended that the
user specify an {\it odd} value of {\tt NTHETA} so that Simpson's rule
will be employed.

\section{Orientational Averaging\label{sec:orientational_averaging}}
\index{orientational averaging}

{{\bf DDSCAT}} has been constructed to facilitate the computation of
orientational averages.  How to go about this depends on the
distribution of orientations which is applicable.

\subsection{ Randomly-Oriented Targets}
\index{orientational averaging!randomly-oriented targets}
For randomly-oriented targets, we wish to compute the orientational
average of a quantity $Q(\beta,\Theta,\Phi)$:
\beq
\langle Q \rangle = {1\over 8\pi^2}\int_0^{2\pi}d\beta
\int_{-1}^1 d\cos\Theta
\int_0^{2\pi}d\Phi ~ Q(\beta,\Theta,\Phi) ~~~.
\eeq
To compute such averages, all you need to do is edit the file {\tt
ddscat.par} so that DDSCAT knows what ranges of the angles $\beta$,
$\Theta$, and $\Phi$ are of interest.  For a randomly-oriented target
with no symmetry, you would need to let $\beta$ run from 0 to
$360^\circ$, $\Theta$ from 0 to $180^\circ$, and $\Phi$ from 0 to
$360^\circ$.

For targets with symmetry, on the other hand, the ranges of $\beta$,
$\Theta$, and $\Phi$ may be reduced.  First of all, remember that
averaging over $\Phi$ is relatively ``inexpensive", so when in doubt
average over 0 to $360^\circ$; most of the computational ``cost" is
associated with the number of different values of ($\beta$,$\Theta$)
which are used.  Consider a cube, for example, with axis ${\hat{\bf
a}}_1$ normal to one of the cube faces; for this cube $\beta$ need run
only from 0 to $90^\circ$, since the cube has fourfold symmetry for
rotations around the axis ${\hat{\bf a}}_1$.  Furthermore, the angle
$\Theta$ need run only from 0 to $90^\circ$, since the orientation
($\beta$,$\Theta$,$\Phi$) is indistinguishable from ($\beta$,
$180^\circ-\Theta$, $360^\circ-\Phi$).

For targets with symmetry, the user is encouraged to test the
significance of $\beta$,$\Theta$,$\Phi$ on targets with small numbers
of dipoles (say, of the order of 100 or so) but having the desired
symmetry.

\subsection{ Nonrandomly-Oriented Targets}
\index{orientational averaging!nonrandomly-oriented targets}

Some special cases (where the target orientation distribution is
uniform for rotations around the $x$ axis = direction of propagation
of the incident radiation), one may be able to use \ddscatv\ 
with appropriate choices of input parameters.  More generally,
however, you will need to modify subroutine {\tt ORIENT} to generate a
list of {\tt NBETA} values of $\beta$, {\tt NTHETA} values of
$\Theta$, and {\tt NPHI} values of $\Phi$, plus two weighting arrays
{\tt WGTA(1-NTHETA,1-NPHI)} and {\tt WGTB(1-NBETA)}.  Here {\tt WGTA}
gives the weights which should be attached to each ($\Theta$,$\Phi$)
orientation, and {\tt WGTB} gives the weight to be attached to each
$\beta$ orientation.  Thus each orientation of the target is to be
weighted by the factor {\tt WGTA}$\times${\tt WGTB}.  For the case of
random orientations, \ddscat\ chooses $\Theta$ values which
are uniformly spaced in $\cos\Theta$, and $\beta$ and $\Phi$ values
which are uniformly spaced, and therefore uses uniform
weights\hfill\break \indent\indent {\tt WGTB}=1./{\tt
NBETA}\hfill\break 

\noindent When {\tt NTHETA} is even, {{\bf DDSCAT}}\ sets\hfill\break
\indent\indent {\tt WGTA}=1./({\tt NTHETA}$\times${\tt
NPHI})\hfill\break 

\noindent but when {\tt NTHETA} is odd, {{\bf DDSCAT}}\ uses Simpson's
rule when integrating over $\Theta$ and \hfill\break \indent\indent
{\tt WGTA}= (1/3 or 4/3 or 2/3)/({\tt NTHETA}$\times${\tt NPHI})

Note that the program structure of {{\bf DDSCAT}}\ may not be ideally
suited for certain highly oriented cases.  If, for example, the
orientation is such that for a given $\Phi$ value only one $\Theta$
value is possible (this situation might describe ice needles oriented
with the long axis perpendicular to the vertical in the Earth's
atmosphere, illuminated by the Sun at other than the zenith) then it
is foolish to consider all the combinations of $\Theta$ and $\Phi$
which the present version of {{\bf DDSCAT}}\ is set up to do.  We hope
to improve this in a future version of {{\bf DDSCAT}}.

\section{Target Generation: Isolated Finite Targets
         \label{sec:target_generation}}
\index{target generation}
DDSCAT contains routines to generate dipole arrays representing finite
targets of various geometries, including spheres, ellipsoids,
rectangular solids, cylinders, hexagonal prisms, tetrahedra, two
touching ellipsoids, and three touching ellipsoids.  The target type
is specified by variable {\tt CSHAPE} on line 9 of {\tt ddscat.par},
up to 12 target shape parameters ({\tt SHPAR$_1$}, {\tt SHPAR$_2$}, {\tt
SHPAR}$_3$, ...) on line 10.
\index{ddscat.par!CSHAPE}
\index{ddscat.par!SHPAR$_1$,SHPAR$_2$,...}
The target geometry is most conveniently described in a
coordinate system attached to the target which we refer to as the
``Target Frame'' (TF), with orthonormal 
unit vectors $\xtf$, $\ytf$, $\ztf\equiv\xtf\times\ytf$.
Once the target is generated, the
orientation of the target in the Lab Frame is accomplished as
described in \S\ref{sec:target_orientation}.

Every target generation routine will specify 
\begin{itemize}
\item The ``occupied'' lattice sites;
\item The composition associated with each occupied lattice site;
\item Two ``target
axes'' $\aone$ and $\atwo$ that are used as references when specifying
the target orientation; and
\item The location of the Target Frame origin of coordinates.
\end{itemize}
Target geometries currently supported include:
\begin{itemize}
\item {\bf FROM\_FILE} : isotropic target material(s), geometry read from file
                         (\S\ref{sec:FROM_FILE})
\item {\bf ANIFRMFIL} : anisotropic target material(s), geometry read from file
                        (\S\ref{sec:ANIFRMFIL})
\item {\bf ANIELLIPS} : anisotropic ellipsoid (\S\ref{sec:ANIELLIPS})
\item {\bf ANI\_ELL\_2} : two touching anisotropic ellipsoids (single 
                          composition) (\S\ref{sec:ANI_ELL_2})
\item {\bf ANI\_ELL\_3} : three touching anisotropic ellipsoids (single 
                          composition) (\S\ref{sec:ANI_ELL_3})
\item {\bf ANIRCTNGL} : anisotropic brick (\S\ref{sec:ANIRCTNGL})
\item {\bf CONELLIPS} : two concentric ellipsoids) (\S\ref{sec:CONELLIPS})
\item {\bf CYLINDER1} : finite cylinder (\S\ref{sec:CYLINDER1})
\item {\bf CYLNDRCAP} : cylinder with hemispherical end-caps 
                        (\S\ref{sec:CYLNDRCAP})
\item {\bf DSKRCTNGL} : disk resting on a brick (\S\ref{sec:DSKRCTNGL})
\item {\bf DW1996TAR} : 13-block target used by \citet{Draine+Weingartner_1996}
                        (\S\ref{sec:DW1996TAR})
\item {\bf ELLIPSOID} : ellipsoid (including spheroid and sphere)
                        (\S\ref{sec:ELLIPSOID})
\item {\bf ELLIPSO\_2} : two touching ellipsoids, different compositions allowed
                         (\S\ref{sec:ELLIPSO_2})
\item {\bf ELLIPSO\_3} : three touching ellipsoids, different compositions
                         allowed (\S\ref{sec:ELLIPSO_3})
\item {\bf HEX\_PRISM} : finite hexagonal prism (\S\ref{sec:HEX_PRISM})
\item {\bf LAYRDSLAB} : multilayer rectangular slab (\S\ref{sec:LAYRDSLAB})
\item {\bf MLTBLOCKS} : collection of cubic blocks (\S\ref{sec:MLTBLOCKS})
\item {\bf RCTGLPRSM} : rectangular prism (i.e., brick) (\S\ref{sec:RCTGLPRSM})
\item {\bf RCTGLBLK3} : stack of 3 rectangular blocks (\S\ref{sec:RCTGLBLK3})
\item {\bf SLAB\_HOLE} : rectangular slab with cylindrical hole
                         (\S\ref{sec:SLAB_HOLE})
\item {\bf SPHERES\_N} : collection of N spheres (\S\ref{sec:SPHERES_N})
\item {\bf SPHROID\_2} : two touching spheroids, different compositions allowed
                         (\S\ref{sec:SPHROID_2})
\item {\bf SPH\_ANI\_N} : collection of N anisotropic spheres 
                          (\S\ref{sec:SPH_ANI_N})
\item {\bf TETRAHDRN} : tetrahedron (\S\ref{sec:TETRAHDRN})
\item {\bf TRNGLPRSM} : triangular prism (\S\ref{sec:TRNGLPRSM})
\item {\bf UNIAXICYL} : finite cylinder of uniaxial material 
                        (\S\ref{sec:UNIAXICYL})
\end{itemize}
Each is described below.

\index{target shape options!FROM\_FILE}
\subsection{ FROM\_FILE = Target composed of possibly anisotropic material,
            defined by list of dipole locations and ``compositions'' 
	    obtained from a file
            \label{sec:FROM_FILE}}
        If anisotropic, the ``microcrystals'' in the target are assumed
	to be aligned with the principal axes of the dielectric tensor
	parallel to $\xtf$, $\ytf$, and $\ztf$.
        This option causes {{\bf DDSCAT}} to read the target geometry
	and composition information from a file {\tt shape.dat}
	instead of automatically generating one of the geometries for
	which DDSCAT has built-in target generation capability.
	The {\tt shape.dat} file is read by routine {\tt REASHP}
	(file {\tt reashp.f90}).
	The file {\tt shape.dat} gives the number $N$ of dipoles in the
	target, the components of the 
	``target axes'' $\hat{\bf a}_1$ and $\hat{\bf a}_2$ in
	the Target Frame (TF), the vector $x_0(1-3)$ determining the
	correspondence between the integers {\tt IXYZ} and
	actual coordinates in the TF, and specifications for the
	location and ``composition'' of each dipole.
	The user can customize {\tt REASHP} as needed to conform to the
	manner in which the target description is stored in file
	{\tt shape.dat}.
	However, as supplied, {\tt REASHP} expects the file {\tt shape.dat}
	to have the following structure:
	\begin{itemize}
	  \item one line containing a description; the first 67 characters will
	        be read and printed in various output statements
	  \item {\tt N} = number of dipoles in target
	  \item $a_{1x}$ $a_{1y}$ $a_{1z}$ = x,y,z components (in TF)
	    of $\bf{a}_1$
	\item $a_{2x}$ $a_{2y}$ $a_{2z}$ = x,y,z components (in TF)
	    of $\bf{a}_2$
        \item $d_x/d$ ~$d_y/d$ ~$d_z/d$ = 1. 1. 1. = relative spacing of
              dipoles in $\xtf$, $\ytf$, $\ztf$ directions
	\item $x_{0x}$ ~$x_{0y}$ ~$x_{0z}$ = TF coordinates
              $x_{\rm TF}/d$ ~$y_{\rm TF}/d$ ~$z_{\rm TF}/d$ corresponding to
	      lattice site {\tt IXYZ}= 0 0 0
	\item (line containing comments)
	\item {\footnotesize
	      $dummy$ {\tt IXYZ(1,1) IXYZ(1,2) IXYZ(1,3) 
		ICOMP(1,1) ICOMP(1,2) ICOMP(1,3)}
              }
	\item {\footnotesize
	      $dummy$ {\tt IXYZ(2,1) IXYZ(2,2) IXYZ(2,3) 
		ICOMP(2,1) ICOMP(2,2) ICOMP(2,3)}
              }
	\item {\footnotesize
              $dummy$ {\tt IXYZ(3,1) IXYZ(3,2) IXYZ(3,3)
		ICOMP(3,1) ICOMP(3,2) ICOMP(3,3)}
              }
	\item ...
	\item {\footnotesize
              $dummy$ {\tt IXYZ(J,1) IXYZ(J,2) IXYZ(J,3)
		ICOMP(J,1) ICOMP(J,2) ICOMP(J,3)}
              }
	\item ...
	\item {\footnotesize
              $dummy$ {\tt IXYZ(N,1) IXYZ(N,2) IXYZ(N,3) 
		ICOMP(N,1) ICOMP(N,2) ICOMP(N,3)}
              }
	\end{itemize}
	where $dummy$ is a number (integer or floating point)that might,
        for example, identify the dipole.  This number will {\it not}
        be used in any calculations.\\
	If the target material at location {\tt J} is isotropic, 
	{\tt ICOMP(J,1)}, {\tt ICOMP(J,2)}, and
	{\tt ICOMP(J,3)} have the same value.

{\footnotesize
\begin{verbatim}
--- demo file for target option FROM_FILE (homogeneous,isotropic target) ---
8       = NAT
1.000   0.000   0.000   = target vector a1 (in TF)
0.000   1.000   0.000   = target vector a2 (in TF)
1.      1.      1.      = d_x/d  d_y/d  d_z/d  (normally 1 1 1)
0.5     0.5     0.5      = X0(1-3) = location in lattice of "target origin"
J    JX   JY   JZ  ICOMPX,ICOMPY,ICOMPZ
1     0    0    0   1  1  1
2     0    0    1   1  1  1
3     0    1    0   1  1  1
4     0    1    1   1  1  1
5     1    0    0   1  1  1
6     1    0    1   1  1  1
7     1    1    0   1  1  1
8     1    1    1   1  1  1
\end{verbatim}}

\noindent The above sample target consists of 8 dipoles arranged to represent a cube.\\
This example is homogeneous: All sites have composition 1\\
The target origin {\tt X0} is set to be at the center of the target\\
Note that ICOMPX,ICOMPY,ICOMPZ could differ, allowing treatment
of anisotropic targets, provided the dielectric tensor at each location
is diagonal in the TF.

{\footnotesize
\begin{verbatim}
--- demo file for target option FROM_FILE (inhomogeneous, isotropic target) ---
8       = NAT
1.000   0.000   0.000   = target vector a1 (in TF)
0.000   1.000   0.000   = target vector a2 (in TF)
1.      1.      1.      = d_x/d  d_y/d  d_z/d  (normally 1 1 1)
0.5     0.5     0.5      = X0(1-3) = location in lattice of "target origin"
J    JX   JY   JZ  ICOMPX,ICOMPY,ICOMPZ
1     0    0    0   1  1  1
2     0    0    1   1  1  1
3     0    1    0   1  1  1
4     0    1    1   1  1  1
5     1    0    0   2  2  2
6     1    0    1   2  2  2
7     1    1    0   2  2  2
8     1    1    1   2  2  2
\end{verbatim}}

\noindent This sample target consists of 8 dipoles 
arranged to represent a cube.\\
This example is inhomogeneous: The lower half of the cube (JX=0) has isotropic 
composition 1\\
The upper half of the cube (JX=1) has isotropic composition 2\\
The target origin {\tt X0} is set to be at the center of the target.\\
Note that ICOMPX, ICOMPY, ICOMPZ can be different, allowing treatment
of anisotropic targets, provided the dielectric tensor at each location
is diagonal in the TF.

\subsubsection{\bf Sample calculation in directory examples\_exp/FROM\_FILE}

Subdirectory {\tt examples\_exp/FROM\_FILE} contains {\tt ddscat.par} 
for calculation 
of scattering by a $0.5\micron\times1\micron\times1\micron$ Au block,
represented by a $32\times64\times64 = 131072$ dipole array, as well as
the output files from the calculation.
The target geometry is input via the file {\tt shape.dat}.

This target has $V=0.5\micron^3$, and $\aeff=(3V/4\pi)^{1/3}=0.49237\micron$.
The calculation is for an incident wavelength $\lambda=0.50\micron$;
the Au has refractive index $m=0.9656+1.8628i$.  The CCG method used is
{\tt PBCGS2}; the two orthogonal polarization require 29 and 30 iterations,
respectively, to converge to the specified tolerance {\tt TOL = 1e-5}.
The computation used 165 MB of RAM, 
and required 208 cpu~sec on a 2.53 GHz cpu.
  
N.B.: This is the same physical problem as the example in
{\tt examples\_exp/RCTGLPRSM} (see \S\ref{sec:example RCTGLPRSM}), differing
only in that in the present calculation 
the target geometry is input through the file {\tt shape.dat}
rather than generated by {\tt ddscat}.

\index{target shape options!ANIFRMFIL}
\subsection{ ANIFRMFIL = General anistropic target defined by list of dipole 
            locations,``compositions'', and material orientations obtained 
            from a file
            \label{sec:ANIFRMFIL}}
	This option causes {{\bf DDSCAT}}\ to read the target geometry
	information from a file {\tt shape.dat} instead of automatically
	generating one of the geometries listed below.  
	The file {\tt shape.dat} gives the number $N$ of dipoles in the
	target, the components of the 
	``target axes'' $\hat{\bf a}_1$ and $\hat{\bf a}_2$ in
	the Target Frame (TF), the vector $x_0(1-3)$ determining the
	correspondence between the integers {\tt IXYZ} and
	actual coordinates in the TF, and specifications for the
	location and ``composition'' of each dipole.
	For each dipole $J$, the file {\tt shape.dat} provides  the location
	{\tt IXYZ($J$,1-3)}, the composition identifier integer 
	{\tt ICOMP($J$,1-3)} specifying the
	dielectric function corresponding to the three principal axes of
	the dielectric tensor, and angles $\Theta_\DF$,
	$\Phi_\DF$, and $\beta_\DF$ specifying the orientation of the
	local 
	\index{Dielectric Frame (DF)}
	``Dielectric Frame'' (DF) relative to the ``Target Frame'' (TF)
	(see \S\ref{sec:composite anisotropic targets}).
	The DF is the reference frame in which the dielectric tensor is
	diagonalized.
	The Target Frame is the reference frame in which we specify the
	dipole locations.

	The {\tt shape.dat}
	file is read by routine {\tt REASHP} (file {\tt reashp.f90}).
	The user can customize {\tt REASHP} as needed to conform to the
	manner in which the target geometry is stored in file {\tt shape.dat}.
	However, as supplied, {\tt REASHP} expects the file {\tt shape.dat}
	to have the following structure:
	\begin{itemize}
	\item one line containing a description; the first 67 characters will
		be read and printed in various output statements.
	\item {\tt N} = number of dipoles in target
	\item $a_{1x}$ $a_{1y}$ $a_{1z}$ = x,y,z components (in Target Frame) 
	of $\bf{a}_1$
	\item $a_{2x}$ $a_{2y}$ $a_{2z}$ = x,y,z components (in Target Frame) 
	of $\bf{a}_2$
        \item $d_x/d$ ~$d_y/d$ ~$d_z/d$ = 1. 1. 1. = relative spacing of
              dipoles in $\xtf$, $\ytf$, $\ztf$ directions
	\item $x_{0x}$~ $x_{0y}$~ $x_{0z}$ = TF coordinates
              $x_{\rm TF}/d$~ $y_{\rm TF}/d$~ $z_{\rm TF}/d$ corresponding to
	      lattice site {\tt IXYZ}= 0 0 0
	\item (line containing comments)
	\item {\footnotesize
	      $dummy$ {\tt IXYZ(1,1-3) 
		ICOMP(1,1-3)
	        THETADF(1) PHIDF(1) BETADF(1)}}
	\item {\footnotesize
	      $dummy$ {\tt IXYZ(2,1-3) 
		ICOMP(2,1-3)
	        THETADF(2) PHIDF(2) BETADF(2)}}
	\item {\footnotesize
              $dummy$ {\tt IXYZ(3,1-3)
		ICOMP(3,1-3)
                THETADF(3) PHIDF(3) BETADF(3)}}
	\item ...
	\item {\footnotesize
              $dummy$ {\tt IXYZ(J,1-3)
		ICOMP(J,1-3)
                THETADF(J) PHIDF(J) BETADF(J)}}
	\item ...
	\item {\footnotesize
              $dummy$ {\tt IXYZ(N,1-3) 
		ICOMP(N,1-3)
                THETADF(N) PHIDF(N) BETADF(N)}}
	\end{itemize}
	Where $dummy$ is a number (either integer or floating point) that
        might, for example, give the number identifying the dipole.  This
        number will {\it not} be used in any calculations.\\
	{\tt THETADF PHIDF BETADF} should be given in {\bf radians}.\\

        Here is an example of the first few lines of a target description
        file suitable for target option {\tt ANIFRMFIL}:

{\footnotesize
\begin{verbatim}
--- demo file for target option ANIFRMFIL (this line is for comments) ---
8       = NAT
1.000   0.000   0.000   = target vector a1 (in TF)
0.000   1.000   0.000   = target vector a2 (in TF)
1.      1.      1.      = d_x/d  d_y/d  d_z/d  (normally 1 1 1)
0.      0.      0.      = X0(1-3) = location in lattice of "target origin"
J    JX   JY   JZ  ICOMP(J,1-3) THETADF PHIDF BETADF
1     0    0    0   1  1  1    0.     0.      0.
2     0    0    1   1  1  1    0.     0.      0.
3     0    1    0   1  1  1    0.     0.      0.
4     0    1    1   1  1  1    0.     0.      0.
5     1    0    0   2  3  3    0.5236 1.5708  0.
6     1    0    1   2  3  3    0.5236 1.5708  0.
7     1    1    0   2  3  3    0.5236 1.5708  0.
8     1    1    1   2  3  3    0.5236 1.5708  0.
\end{verbatim}}

This sample target consists of 8 dipoles arranged to represent a cube.\\
Half of the cube (dipoles with JX=0) has isotropic composition 1.  
For this case,
the angles THETADF, PHIDF, BETADF do not matter, and it convenient to
set them all to zero.\\
The other half of the cube (dipoles with JX=1) 
consists of a uniaxial material, with dielectric
function 2 for E fields parallel to one axis (the ``c-axis''), 
and dielectric function 3 for E fields perpendicular
to the c-axis. The c-axis is
$30^o$ (0.5236 radians) away from $\xtf$, 
and lies in the $\xtf$-$\ztf$ plane (having been
rotated by 1.5708 radians around $\xtf$).\\
Note that ICOMP(J,K) can be different for K=1,3, allowing treatment
of anisotropic targets, provided the dielectric tensor at each location
is diagonal in the TF.

\index{target shape options!ANIELLIPS}
\subsection{ ANIELLIPS =  Homogeneous, anisotropic ellipsoid.
         \label{sec:ANIELLIPS}}
	{\tt SHPAR$_1$}, {\tt SHPAR$_2$},
	{\tt SHPAR}$_3$ define the ellipsoidal boundary:
\beq
	\left(\frac{x_\TF/d}{{\tt SHPAR}_1}\right)^2+
        \left(\frac{y_\TF/d}{{\tt SHPAR}_2}\right)^2+
        \left(\frac{z_\TF/d}{{\tt SHPAR}_3}\right)^2 = \frac{1}{4}
        ~~~,
\eeq
The TF origin is located at the centroid of the ellipsoid.
\index{target shape options!ANI\_ELL\_2}
\subsection{ ANI\_ELL\_2 = Two touching, homogeneous, anisotropic 
	    ellipsoids, with distinct compositions
            \label{sec:ANI_ELL_2}}
	Geometry as for {\tt ELLIPSO\_2};
	{\tt SHPAR$_1$}, {\tt SHPAR$_2$}, {\tt SHPAR}$_3$ 
        have same meanings as for {\tt ELLIPSO\_2}.  
	Target axes ${\hat{\bf a}}_1=(1,0,0)_\TF$ 
	and ${\hat{\bf a}}_2=(0,1,0)_\TF$.\\
	Line connecting ellipsoid centers is 
	$\parallel \hat{\bf a}_1 = \xtf$.\\
	TF origin is located between ellipsoids, at point of contact.\\
	It is assumed that (for both ellipsoids) the dielectric tensor
	is diagonal in the TF.
	User must set {\tt NCOMP}=6 and provide 
	$xx$, $yy$, $zz$ components of dielectric tensor for 
	first ellipsoid, and $xx$, $yy$, $zz$ components of 
	dielectric tensor for second 
	ellipsoid (ellipsoids are in order of increasing $x_\TF$).
\index{target shape options!ANI\_ELL\_3}
\subsection{ ANI\_ELL\_3 = Three touching homogeneous, anisotropic ellipsoids 
	    with same size and orientation but distinct dielectric tensors
            \label{sec:ANI_ELL_3}}
	{\tt SHPAR$_1$}, {\tt SHPAR$_2$}, {\tt SHPAR}$_3$ 
        have same meanings as for {\tt ELLIPSO\_3}.\\  
	Target axis ${\hat{\bf a}}_1=(1,0,0)_\TF$ 
	(along line of ellipsoid centers), and 
	${\hat{\bf a}}_2=(0,1,0)_\TF$.\\
	TF origin is located at center of middle ellipsoid.\\
	It is assumed that dielectric tensors 
	are all diagonal in the TF.
	User must set {\tt NCOMP}=9 and provide $xx$, $yy$, $zz$ 
	elements of dielectric tensor
	for first ellipsoid, $xx$, $yy$, $zz$ elements for second ellipsoid, 
	and $xx$, $yy$, $zz$ elements for third ellipsoid 
	(ellipsoids are in order of increasing $x_\TF$).
\index{target shape options!ANIRCTNGL}
\subsection{ ANIRCTNGL = Homogeneous, anisotropic, rectangular solid
            \label{sec:ANIRCTNGL}}
	x, y, z lengths/$d$ = {\tt SHPAR$_1$}, {\tt SHPAR$_2$}, 
        {\tt SHPAR}$_3$.\\
	Target axes ${\hat{\bf a}}_1=(1,0,0)_\TF$ 
	and ${\hat{\bf a}}_2=(0,1,0)_\TF$ 
	in the TF. \\
        $(x_\TF,y_\TF,z_\TF)=(0,0,0)$ at middle of upper target surface,
	(where ``up'' = $\xtf$).  (The target surface is taken to be $d/2$
	about the upper dipole layer.)\\
	Dielectric tensor is assumed to be diagonal in the target frame.\\
	User must set {\tt NCOMP}=3 and
	supply names of three files for $\epsilon$ as a function
	of wavelength or energy: first for $\epsilon_{xx}$,
	second for $\epsilon_{yy}$, and third for $\epsilon_{zz}$,

\subsubsection{\bf Sample calculation in directory examples\_exp/ANIRCTNGL}

Subdirectory {\tt examples\_exp/ANIRCTNGL} contains {\tt ddscat.par}
for calculation
of scattering by a $0.1\micron\times0.2\micron\times0.2\micron$ rectangular
brick ($\aeff=0.098475\micron$)
with an anisotropic dielectric tensor: $m=1.33+0.01i$ for
$\bE\parallel\xtf$ and $\bE\parallel\ytf$, and
$m=1.50+0.01i$ for $\bE \parallel \ztf$.
Radiation is incident with $\bk_0\parallel\xtf$, with
$\lambda=0.5\micron$.
The {\tt ddscat.par} file is as follows:
{\scriptsize
\begin{verbatim}
' ============ Parameter file for v7.3 ==================='
'**** Preliminaries ****'
'NOTORQ' = CMDTRQ*6 (DOTORQ, NOTORQ) -- either do or skip torque calculations
'PBCGS2' = CMDSOL*6 (PBCGS2, PBCGST, PETRKP) -- CCG method
'GPFAFT' = CMETHD*6 (GPFAFT, FFTMKL) -- FFT method
'GKDLDR' = CALPHA*6 (GKDLDR, LATTDR, FLTRCD) -- DDA method
'NOTBIN' = CBINFLAG (ALLBIN, ORIBIN, NOTBIN)
'**** Initial Memory Allocation ****'
10 20 20 = upper bound on target extent
'**** Target Geometry and Composition ****'
'ANIRCTNGL' = CSHAPE*9 shape directive
10 20 20 = shape parameters SHPAR1, SHPAR2, SHPAR3
3         = NCOMP = number of dielectric materials
'../diel/m1.33_0.01' = name of file containing dielectric function
'../diel/m1.33_0.01'
'../diel/m1.50_0.01'
'**** Additional Nearfield calculation? ****'
0 = NRFLD (=0 to skip nearfield calc., =1 to calculate nearfield E)
0.0 0.0 0.0 0.0 0.0 0.0 (fract. extens. of calc. vol. in -x,+x,-y,+y,-z,+z)
'**** Error Tolerance ****'
1.00e-5 = TOL = MAX ALLOWED (NORM OF |G>=AC|E>-ACA|X>)/(NORM OF AC|E>)
'**** maximum number of iterations allowed ****'
300     = MXITER
'**** Interaction cutoff parameter for PBC calculations ****'
5.00e-3 = GAMMA (1e-2 is normal, 3e-3 for greater accuracy)
'**** Angular resolution for calculation of <cos>, etc. ****'
2.0     = ETASCA (number of angles is proportional to [(2+x)/ETASCA]^2 )
'**** Vacuum wavelengths (micron) ****'
0.5 0.5 1 'LIN' = wavelengths (first,last,how many,how=LIN,INV,LOG)
'**** Refractive index of ambient medium'
1.000 = NAMBIENT
'**** Effective Radii (micron) **** '
0.098475 0.098457 1 'LIN' = eff. radii (first, last, how many, how=LIN,INV,LOG)
'**** Define Incident Polarizations ****'
(0,0) (1.,0.) (0.,0.) = Polarization state e01 (k along x axis)
2 = IORTH  (=1 to do only pol. state e01; =2 to also do orth. pol. state)
'**** Specify which output files to write ****'
1 = IWRKSC (=0 to suppress, =1 to write ".sca" file for each target orient.
'**** Prescribe Target Rotations ****'
 0.   0.  1  = BETAMI, BETAMX, NBETA (beta=rotation around a1)
 0.   0.  1  = THETMI, THETMX, NTHETA (theta=angle between a1 and k)
 0.   0.  1  = PHIMIN, PHIMAX, NPHI (phi=rotation angle of a1 around k)
'**** Specify first IWAV, IRAD, IORI (normally 0 0 0) ****'
0   0   0    = first IWAV, first IRAD, first IORI (0 0 0 to begin fresh)
'**** Select Elements of S_ij Matrix to Print ****'
6       = NSMELTS = number of elements of S_ij to print (not more than 9)
11 12 21 22 31 41       = indices ij of elements to print
'**** Specify Scattered Directions ****'
'LFRAME' = CMDFRM*6 ('LFRAME' or 'TFRAME' for Lab Frame or Target Frame)
2 = number of scattering planes
0.  0. 180. 30 = phi, thetan_min, thetan_max, dtheta (in degrees) for plane A
90. 0. 180. 30 = phi, ... for plane B
\end{verbatim}
}
This calculation required 0.22 cpu~sec on a 2.53 GHz cpu.
\index{target shape options!CONELLIPS}
\subsection{ CONELLIPS = Two concentric ellipsoids
            \label{sec:CONELLIPS}}
	{\tt SHPAR$_1$}, {\tt SHPAR$_2$},
	{\tt SHPAR}$_3$ = lengths/$d$ of the {\it outer} ellipsoid
        along the $\xtf$, $\ytf$, $\ztf$ axes;\\
	{\tt SHPAR}$_4$, {\tt SHPAR}$_5$, {\tt SHPAR}$_6$ 
	= lengths/$d$ of the
	{\it inner} ellipsoid along the $\xtf$, $\ytf$, $\ztf$ axes.\\
	Target axes ${\hat{\bf a}}_1=(1,0,0)_\TF$, 
                    ${\hat{\bf a}}_2=(0,1,0)_\TF$.\\
	TF origin is located at centroids of ellipsoids.\\
	The ``core" within the inner ellipsoid is composed of isotropic 
	material 1; 
	the ``mantle" between inner and outer
	ellipsoids is composed of isotropic material 2.\\
	User must set {\tt NCOMP}=2 and provide dielectric functions for 
	``core'' and ``mantle'' materials.
\index{target shape options!CYLINDER1}
\subsection{ CYLINDER1 = Homogeneous, isotropic finite cylinder
            \label{sec:CYLINDER1}}
	{\tt SHPAR$_1$} = length/$d$,
	{\tt SHPAR$_2$} = diameter/$d$, with\\
	{\tt SHPAR}$_3$ = 1 for cylinder axis 
        $\hat{\bf a}_1\parallel \xtf$:
	$\hat{\bf a}_1=(1,0,0)_\TF$
	            and $\hat{\bf a}_2=(0,1,0)_\TF$;\\
	{\tt SHPAR}$_3$ = 2 for cylinder axis 
        $\hat{\bf a}_1\parallel \ytf$:
        $\hat{\bf a}_1=(0,1,0)_\TF$
	            and $\hat{\bf a}_2=(0,0,1)_\TF$;\\
	{\tt SHPAR}$_3$ = 3 for cylinder axis 
        $\hat{\bf a}_1\parallel \ztf$:
	$\hat{\bf a}_1=(0,0,1)_\TF$
	            and $\hat{\bf a}_2=(1,0,0)_\TF$ in the TF.\\
	TF origin is located at centroid of cylinder.\\
	User must set {\tt NCOMP}=1.
\index{target shape options!CYLNDRCAP}
\subsection{ CYLNDRCAP = Homogeneous, isotropic finite cylinder with 
            hemispherical endcaps.
            \label{sec:CYLNDRCAP}}
	{\tt SHPAR$_1$} = cylinder length/$d$ 
	({\it not} including end-caps!) and
	{\tt SHPAR$_2$} = cylinder diameter/$d$,
	with cylinder axis 
	$={\hat{\bf a}}_1=(1,0,0)_\TF$
	and ${\hat{\bf a}}_2=(0,1,0)_\TF$.
	The total length along the target axis (including the endcaps) is
	({\tt SHPAR$_1$}+{\tt SHPAR$_2$})$d$. \\
	TF origin is located at centroid of cylinder.\\
	User must set {\tt NCOMP}=1.
\index{target shape options!DSKRCTNGL}
\subsection{ DSKRCTNGL = Disk on top of a homogeneous rectangular slab
            \label{sec:DSKRCTNGL}}
            This option causes {{\bf DDSCAT}}\ to create a target consisting
	    of a disk of composition 1
	    resting on top of a rectangular block of composition 2.
	    Materials 1 and 2 are assumed to be homogeneous and isotropic.\\
	    {\tt ddscat.par} should set {\tt NCOMP} to 2 .
	    \ \\
	    The cylindrical disk has thickness {\tt SHPAR$_1$}$\times d$ in the
	    x-direction, and diameter {\tt SHPAR$_2$}$\times d$.
	    The rectangular block is assumed to have thickness
	    {\tt SHPAR}$_3$$\times$d in the x-direction, length
	    {\tt SHPAR}$_4$$\times d$ in the y-direction,
	    and length {\tt SHPAR}$_5$$\times d$ in the z-direction.
	    The lower surface of the cylindrical disk is in the $x=0$ plane.
	    The upper surface of the slab is also in the $x=0$ plane.

	    The Target Frame origin (0,0,0) is located where the symmetry
	    axis of the disk intersects the $x=0$ plane (the upper surface
	    of the slab, and the lower surface of the disk).
	    In the Target Frame, 
	    dipoles representing the rectangular block are located at
	    $(x/d,y/d,z/d)=(j_x+0.5,j_y+\Delta_y,j_z+\Delta_z)$, where
	    $j_x$, $j_y$, and $j_z$ are integers. $\Delta_y=0$ or 0.5
	    depending on whether {\tt SHPAR}$_4$ is even or odd.
	    $\Delta_z=0$ or 0.5 depending on whether {\tt SHPAR}$_5$ is
	    even or odd.

            Dipoles representing the disk are located at\\
	    \hspace*{1.0cm}$x/d = 0.5 , 1.5 ,..., 
	    [{\rm int}({\tt SHPAR}_4+0.5)-0.5]$

	    As always, the physical size of the target is fixed by
	    specifying the value of the effective radius
	    $\aeff\equiv(3V_{\rm T}/4\pi)^{1/3}$,
	    where $V_{\rm T}$ 
	    is the total volume of solid material in the target.
	    For this geometry, the number of dipoles in the target will
	    be approximately
	    $N=[{\tt SHPAR}_1\times{\tt SHPAR}_2\times{\tt SHPAR}_3 +
	    (\pi/4)(({\tt SHPAR}_4)^2\times{\tt SHPAR}_5)]$,
	    although the exact number may differ because of the
	    dipoles are required to be located on a rectangular lattice.
	    The dipole spacing $d$ in physical units
	    is determined from the specified value of $\aeff$ and
	    the number $N$ of dipoles in the target:
	    $d=(4\pi/3N)^{1/3}\aeff$.
	    This option requires 5 shape parameters:\newline
	    The pertinent line in {\tt ddscat.par} should read\\
	    \ \\
	{\tt SHPAR$_1$ SHPAR$_2$ SHPAR$_3$ SHPAR$_4$ SHPAR}$_5$\\
	    \ \\
	where\\
	{\tt SHPAR$_1$} = [disk thickness (in $\xtf$ direction)]/$d$ 
        [material 1]\\
	{\tt SHPAR$_2$} = (disk diameter)/$d$\\
	{\tt SHPAR}$_3$ = (brick thickness in $\xtf$ direction)/$d$ 
	[material 2]\\
	{\tt SHPAR}$_4$ = (brick thickness in $\ytf$ direction)/$d$\\
	{\tt SHPAR}$_5$ = (brick thickness in $\ztf$ direction)/$d$\\
	\ \\
	The overall size of the target (in terms of numbers of
	dipoles) is determined by parameters 
	{\tt (SHPAR1+SHPAR4), SHPAR$_2$}, and {\tt SHPAR}$_3$.
	The periodicity in the TF $y$ and $z$ directions is determined
	by parameters {\tt SHPAR}$_4$ and {\tt SHPAR}$_5$.\\
	The physical size of the TUC is specified by the value of
	$\aeff$ (in physical units, e.g. cm), 
	specified in the file {\tt ddscat.par}.

	The ``computational volume'' is determined by
	({\tt SHPAR$_1$+SHPAR}$_4$)$\times${\tt SHPAR$_2$}$\times${\tt SHPAR}$_3$.

	The target axes (in the TF) 
	are set to ${\hat{\bf a}}_1 = \xtf = (1,0,0)_\TF$ --
	i.e., normal to the ``slab'' -- and
	${\hat{\bf a}}_2 = \ytf= (0,1,0)_\TF$.
	The orientation of the incident radiation relative to the target
	is, as for all other targets, set by the usual orientation
	angles  $\Theta$, $\Phi$, and $\beta$ 
	(see \S\ref{sec:target_orientation} above); for example,
	$\Theta=0$ would be for radiation incident normal to the slab.

\index{target shape options!DW1996TAR}
\subsection{ DW1996TAR = 13 block target used by 
             \citet{Draine+Weingartner_1996}.
            \label{sec:DW1996TAR}}
	Single, isotropic material.
	Target geometry was used in study by \citet{Draine+Weingartner_1996} 
	of radiative torques on irregular grains.
	${\hat{\bf a}}_1$ and ${\hat{\bf a}}_2$ 
	are principal axes with largest and
	second-largest moments of inertia.
	User must set {\tt NCOMP=1}.
	Target size is controlled by shape parameter {\tt SHPAR(1)} = 
	width of one block in lattice units.\\
	TF origin is located at centroid of target.
\index{ELLIPSOID}
\index{target shape options!ELLIPSOID}
\subsection{ ELLIPSOID = Homogeneous, isotropic ellipsoid.
            \label{sec:ELLIPSOID}}
	``Lengths'' 
	{\tt SHPAR$_1$}, {\tt SHPAR$_2$},
	{\tt SHPAR}$_3$ in the $x$, $y$, $z$ directions in the TF:
\beq
	\left(\frac{x_\TF}{{\tt SHPAR}_1 d}\right)^2+
        \left(\frac{y_\TF}{{\tt SHPAR}_2 d}\right)^2+
        \left(\frac{z_\TF}{{\tt SHPAR}_3 d}\right)^2 = \frac{1}{4}
        ~~~,
\eeq
	where $d$ is the interdipole spacing.\\
	The target axes are set to ${\hat{\bf a}}_1=(1,0,0)_\TF$ and 
	${\hat{\bf a}}_2=(0,1,0)_\TF$.\\
	Target Frame origin = centroid of ellipsoid.\\
	User must set {\tt NCOMP}=1 on line 9 of {\tt ddscat.par}.\\
	A {\bf homogeneous, isotropic sphere} is obtained by setting 
	{\tt SHPAR$_1$} = {\tt SHPAR$_2$} = {\tt SHPAR}$_3$ = diameter/$d$.

\subsubsection{\bf Sample calculation in directory examples\_exp/ELLIPSOID}

The directory {\tt examples\_exp/ELLIPSOID} contains {\tt ddscat.par} for
calculation of scattering by a sphere with refractive index $m=1.5+0.01i$
and
$2\pi a/\lambda=5$, represented by a $N=59728$ dipole
pseudosphere just fitting within a
$48\times48\times48$ computational volume, as well as the
output files from the calculation.
The calculation with $2\pi a/\lambda=5$ has $|m|kd=0.309$.
The computation used 144 MB of RAM and required 63 cpu~sec on a 2.53 GHz cpu.

\subsubsection{\label{sec:ELLIPSOID_NEARFIELD}
               \bf Sample calculation in directory 
               examples\_exp/ELLIPSOID\_NEARFIELD}

The directory {\tt examples\_exp/ELLIPSOID\_NEARFIELD} contains 
{\tt ddscat.par} for calculation of 
(1) far-field scattering and (2) $\bE$ in and near the target
for a sphere with refractive index $m=0.96+1.01i$
(refractive index of Au at $\lambda=0.5\micron$)
and
$2\pi a/\lambda=5$.
The spherical target is represented by a $N=59728$ dipole
pseudosphere just fitting within a
$48\times48\times48$ computational volume, as well as the
output files from the calculation.

In physical units with $\lambda=0.5\micron$, 
$\aeff=5\times\lambda/2\pi=0.39789\micron$.
The calculation with $2\pi a/\lambda=5$ has $|m|kd=0.309$.
The computation used 144 MB of RAM and required 60 cpu~sec on a 2.53 GHz cpu.

The nearfield calculation is specified to extend throughout a computational
volume extending the original $48d\times48d\times48d$ computational volume
by 50\% in all directions, to become a $96d\times96d\times96d$ volume
centered on the sphere.
$\bE$ is evaluated at all points in this volume.
The nearfield calculation used 62 MB of RAM and required just 9.6 cpu~sec.
The nearfield calculation creates the binary files
{\tt w000r000k000.E1} and {\tt w000r000k000.E2}, one for each of
the two incident polarizations.

After the nearfield calculation is complete, the program {\tt ddpostprocess}
is used to read the file {\tt w000r000k000.E1} 
(specified in {\tt ddpostprocess.par}) and extract $\bE$ at
501 points along a line specified in {\tt ddpostprocess.par} -- the line
runs along the $\xtf$ axis through the center of the sphere.  The results
are shown in Figure \ref{fig:ellipsoid_E^2}.

\begin{figure}
\begin{center}
\vspace*{-0.1cm}
\includegraphics[width=8.cm,angle=270]{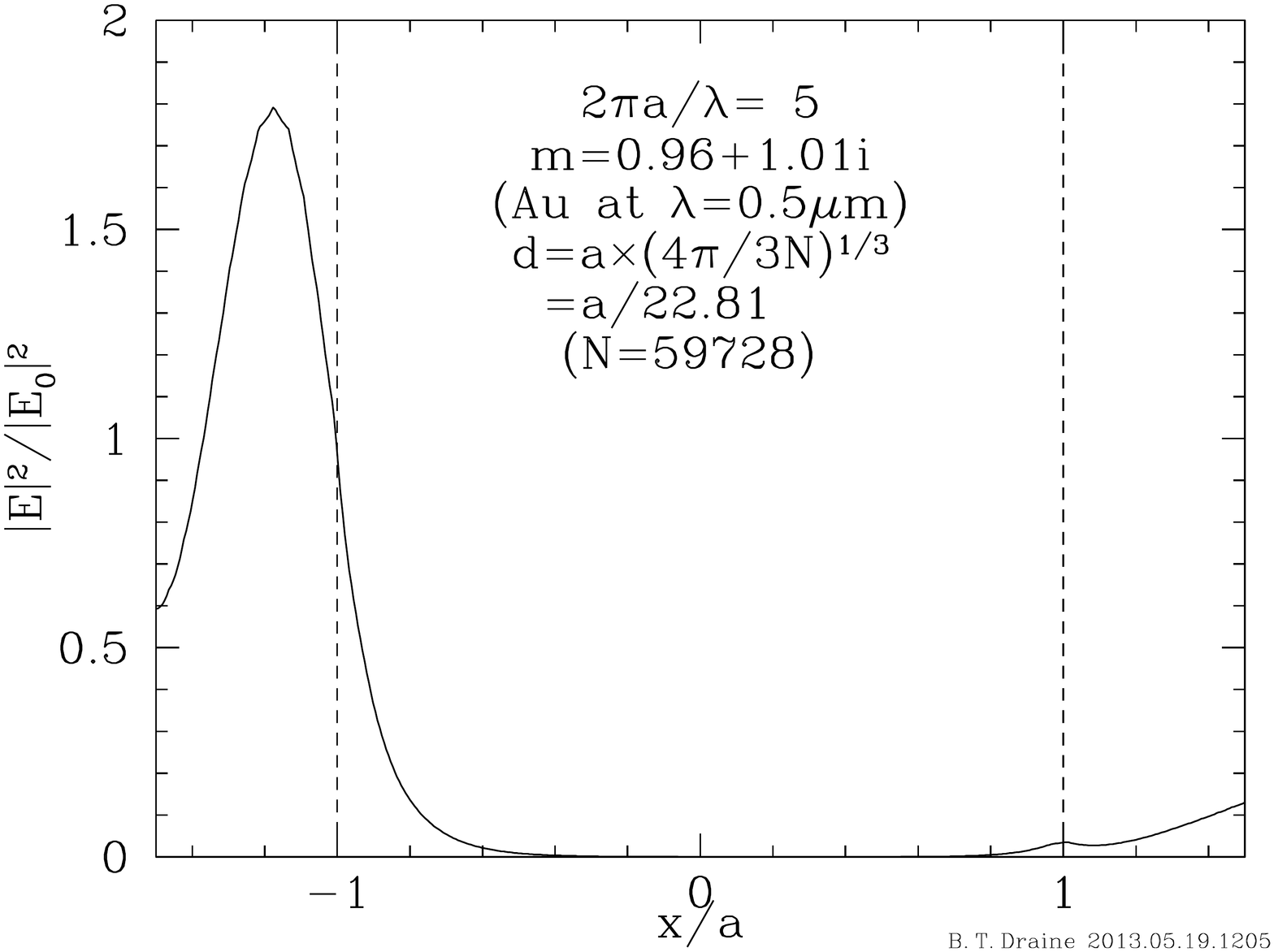}
\vspace*{-0.4cm}
\caption{\label{fig:ellipsoid_E^2}
         \footnotesize
        Normalized electric field intensity $|\bE|^2/|\bE_0|^2$ along a line
        parallel to the direction of propagation, and passing through
        the center of an Au sphere of radius $a=0.3979\micron$, for
        light with wavelength $\lambda=0.5\micron$.
        The calculation in examples\_exp/ELLIPSOID\_NEARFIELD was
        done with dipole spacing $d=a/48.49=0.00821\micron$.
        }
\end{center}
\end{figure}

\index{target shape options!ELLIPSO\_2}
\subsection{ ELLIPSO\_2 = Two touching, homogeneous, isotropic ellipsoids,
	    with distinct compositions
            \label{sec:ELLIPSO_2}}
	{\tt SHPAR$_1$}, {\tt SHPAR$_2$},
        {\tt SHPAR}$_3$=x-length/$d$, y-length/$d$,
	$z$-length/$d$ of one ellipsoid.
	The two ellipsoids have identical shape, size, and orientation,
	but distinct dielectric functions.
	The line connecting ellipsoid centers is along the $\xtf$-axis.
	Target axes ${\hat{\bf a}}_1=(1,0,0)_\TF$ 
	[along line connecting ellipsoids]
	and ${\hat{\bf a}}_2=(0,1,0)_\TF$.\\
	Target Frame origin = midpoint between ellipsoids
        (where ellipsoids touch).\\
	User must set {\tt NCOMP}=2 and provide dielectric function file names
	for both ellipsoids. 
	Ellipsoids are in order of increasing $x_\TF$:
	first dielectric function is for ellipsoid with 
	center at negative $x_\TF$, second dielectric function for 
	ellipsoid with center at positive $x_\TF$.
\index{target shape options!ELLIPSO\_3}
\subsection{ ELLIPSO\_3 = Three touching homogeneous, isotropic ellipsoids 
	    of equal size and orientation, but distinct compositions
            \label{sec:ELLIPSO_3}}
	{\tt SHPAR$_1$}, {\tt SHPAR$_2$}, {\tt SHPAR}$_3$ have same meaning
        as for {\tt ELLIPSO\_2}.
	Line connecting ellipsoid centers is parallel to $\xtf$ axis.  
	Target axis ${\hat{\bf a}}_1=(1,0,0)_\TF$ 
	(along line of ellipsoid centers), and
	${\hat{\bf a}}_2=(0,1,0)_\TF$.\\
	Target Frame origin = centroid of middle ellipsoid.\\
	User must set {\tt NCOMP}=3 and provide (isotropic) dielectric 
	functions for first, second, and third ellipsoid.

\index{target shape options!HEX\_PRISM}
\subsection{ HEX\_PRISM = Homogeneous, isotropic hexagonal prism
            \label{sec:HEX_PRISM}}
	{\tt SHPAR$_1$} = (Length of prism)/$d$ = (distance between hexagonal
	faces)/$d$,\\
	{\tt SHPAR$_2$} = (distance between opposite vertices of one hexagonal
	face)/$d$ = 2$\times$hexagon side/$d$.\\
	{\tt SHPAR}$_3$ selects one of 6 orientations of the prism in the 
        Target Frame (TF).\\
	Target axis ${\hat{\bf a}}_1$ is along the prism axis (i.e., normal to
	the hexagonal faces), and
	target axis ${\hat{\bf a}}_2$ is normal to one of the rectangular
	faces.  There are 6 options for {\tt SHPAR}$_3$:\\
	{\tt SHPAR}$_3$ = 1 for ${\hat{\bf a}}_1\parallel\hat{\bf x}_{\rm TF}$
         and ${\hat{\bf a}}_2\parallel\hat{\bf y}_{\rm TF}$\quad;\quad
	{\tt SHPAR}$_3$ = 2 for ${\hat{\bf a}}_1\parallel\hat{\bf x}_{\rm TF}$ 
        and ${\hat{\bf a}}_2\parallel\hat{\bf z}_{\rm TF}$\quad;\\
	{\tt SHPAR}$_3$ = 3 for ${\hat{\bf a}}_1\parallel\hat{\bf y}_{\rm TF}$ 
        and ${\hat{\bf a}}_2\parallel\hat{\bf x}_{\rm TF}$\quad;\quad
	{\tt SHPAR}$_3$ = 4 for ${\hat{\bf a}}_1\parallel\hat{\bf y}_{\rm TF}$
        and ${\hat{\bf a}}_2\parallel\hat{\bf z}_{\rm TF}$\quad;\\
	{\tt SHPAR}$_3$ = 5 for ${\hat{\bf a}}_1\parallel\hat{\bf z}_{\rm TF}$
        and ${\hat{\bf a}}_2\parallel\hat{\bf x}_{\rm TF}$\quad;\quad
	{\tt SHPAR}$_3$ = 6 for ${\hat{\bf a}}_1\parallel\hat{\bf z}_{\rm TF}$
        and ${\hat{\bf a}}_2\parallel\hat{\bf y}_{\rm TF}$\\
	TF origin is located at the centroid of the target.\\
	User must set {\tt NCOMP}=1.
\index{target shape options!LAYRDSLAB}
\subsection{ LAYRDSLAB = Multilayer rectangular slab
            \label{sec:LAYRDSLAB}}
        Multilayer rectangular slab with overall x, y, z
	lengths 
	$a_x = {\tt SHPAR}_1\times d$ \\
	$a_y = {\tt SHPAR}_2\times d$,\\
	$a_z = {\tt SHPAR}_3\times d$.\\
	Upper surface is at $x_\TF=0$, lower surface at $x_\TF=-{\tt SHPAR}_1\times d$\\
	${\tt SHPAR}_4$ = fraction which is composition 1 (top layer).\\
	${\tt SHPAR}_5$ = fraction which is composition 2 (layer below top)\\
	${\tt SHPAR}_6$ = fraction which is composition 3 (layer below comp 2)\\
	$1-({\tt SHPAR}_4+{\tt SHPAR}_5+{\tt SHPAR}_6)$ = fraction which is
	composition 4 (bottom layer).\\
	To create a bilayer slab, just set 
	${\tt SHPAR}_5={\tt SHPAR}_6=0$\\
	To create a trilayer slab, just set ${\tt SHPAR}_6=0$\\
	User must set {\tt NCOMP}=2,3, or 4 and provide dielectric function
	files for each of the two layers.
	Top dipole layer is at $x_\TF=-d/2$.
	Origin of TF is at center of top surface.
\index{target shape options!MLTBLOCKS}
\subsection{ MLTBLOCKS = Homogeneous target constructed from cubic ``blocks''
            \label{sec:MLTBLOCKS}}
	Number and location of blocks are specified in separate file 
	{\tt blocks.par} with following structure:\\
	\hspace*{2em}one line of comments (may be blank)\\
	\hspace*{2em}{\tt PRIN} (= 0 or 1 -- see below)\\
	\hspace*{2em}{\tt N} (= number of blocks) \\
	\hspace*{2em}{\tt B}  (= width/$d$ of one block) \\
	\hspace*{2em}$x_\TF$ $y_\TF$ $z_\TF$ 
		(= position of 1st block in units of {\tt B}$d$)\\
	\hspace*{2em}$x_\TF$ $y_\TF$ $z_\TF$ 
	(= position of 2nd block in units of {\tt B}$d$) )\\
	\hspace*{2em}... \\
	\hspace*{2em}$x_\TF$ $y_\TF$ $z_\TF$ 
	        (= position of {\tt N}th block in units of {\tt B}$d$)\\
	If {\tt PRIN}=0, then ${\hat{\bf a}}_1=(1,0,0)_\TF$, 
	${\hat{\bf a}}_2=(0,1,0)_\TF$.
	If {\tt PRIN}=1, then ${\hat{\bf a}}_1$ and ${\hat{\bf a}}_2$ are set 
	to principal 
	axes with largest and second largest moments of inertia,
	assuming target to be of uniform density.
	User must set {\tt NCOMP=1}.
\index{target shape options!RCTGLPRSM}
\subsection{ RCTGLPRSM = Homogeneous, isotropic, rectangular solid
            \label{sec:RCTGLPRSM}}
	x, y, z lengths/$d$ = {\tt SHPAR$_1$}, {\tt SHPAR$_2$}, 
        {\tt SHPAR}$_3$.\\
	Target axes ${\hat{\bf a}}_1=(1,0,0)_\TF$ 
	and ${\hat{\bf a}}_2=(0,1,0)_\TF$.\\
	TF origin at center of upper surface of solid:
	target extends from $x_\TF/d=-{\tt SHPAR}_1$ to 0,\\
	$y_\TF/d$ from $-0.5\times{\tt SHPAR_2}$ to $+0.5\times{\tt SHPAR_2}$\\
	$z_\TF/d$ from $-0.5\times{\tt SHPAR_3}$ to $+0.5\times{\tt SHPAR_3}$\\
	User must set {\tt NCOMP}=1.

\subsubsection{\label{sec:example RCTGLPRSM}
               \bf Sample calculation in directory examples\_exp/RCTGLPRSM}
The directory {\tt examples\_exp/RCTGLPRSM} contains {\tt ddscat.par} for
 calculation
of scattering by a $0.25\micron\times0.5\micron\times0.5\micron$ Au block,
represented by a $16\times32\times32$ dipole array, together with output
files from the calculation.
The Au has refractive index $m=0.9656+1.8628i$.
The DDA calculation has $|m|kd=0.4120$.
The calculation used 52 MB of RAM, and required 10.0 cpu~sec on a 2.53 GHz
cpu.

\subsubsection{\label{sec:example RCTGLPRSM_NEARFIELD}
               \bf Sample calculation in directory
               examples\_exp/RCTGLPRSM\_NEARFIELD} 

The directory {\tt examples\_exp/RCTGLPRSM\_NEARFIELD} contains 
{\tt ddscat.par} for the same scattering problem as in 
{\tt examples\_exp/RCTGLPRSM}, but also calling for nearfield calculation
of $\bE$ throughout a $0.5\micron\times1.0\micron\times1.0\micron$
volume centered on the target (i.e., fractional extension of 50\% in
$+x_\TF,-x_\TF,+y_\TF,-y_\TF,+z_\TF,-z_\TF$ directions).  The $\bE$
field is evaluated on a grid with spacing $d$ (which includes all the
dipole locations); the results for the two orthogonal polarizations
are written into the binary files {\tt w000r000k000.E1} and {\tt
  w000r000k000.E2}.  The complete calculation used 11.4 cpu~sec on a
2.53 GHz cpu.

\index{target shape options!RCTGLBLK3}
\subsection{ RCTGLBLK3 = Stack of 3 rectangular blocks, with centers on
            the $\xtf$ axis.
            \label{sec:RCTGLBLK3}}
        Each block consists of a distinct material.  There are 9 shape
	parameters:\\
        {\tt SHPAR$_1$} = (upper solid thickness in $\xtf$ direction)/$d$ 
        [material 1]\\
        {\tt SHPAR$_2$} = (upper solid width in $\ytf$ direction)/$d$\\
        {\tt SHPAR}$_3$ = (upper solid width in $\ztf$ direction)/$d$\\
        {\tt SHPAR}$_4$ = (middle solid thickness in $\xtf$ direction)/$d$ 
        [material 2]\\
        {\tt SHPAR}$_5$ = (middle solid width in $\ytf$ direction)/$d$\\
        {\tt SHPAR}$_6$ = (middle solid width in $\ztf$ direction)/$d$\\
        {\tt SHPAR}$_7$ = (lower solid thickness in $\xtf$ direction)/$d$ 
        [material 3]\\
        {\tt SHPAR}$_8$ = (lower solid width in $\ytf$ direction)/$d$\\
        {\tt SHPAR}$_9$ = (lower solid width in $\ztf$ direction)/$d$\\
	TF origin is at center of top surface of material 1.
\index{target shape options!SLAB\_HOLE}
\subsection{ SLAB\_HOLE = Rectangular slab with a cylindrical hole.
            \label{sec:SLAB_HOLE}}
        The target consists of a rectangular block with a cylindrical
        hole with the axis passing through the centroid and aligned
        with the $\xtf$ axis.
        The block dimensions are 
        $a\times b\times c$.
        The cylindrical hole has radius $r$. 
        The pertinent line in {\tt ddscat.par} should read\\
        {\tt SHPAR$_1$ SHPAR$_2$ SHPAR$_3$ SHPAR$_4$}\\
        where\\
        {\tt SHPAR$_1$} = $a/d$ ($d$ is the interdipole spacing) \\
        {\tt SHPAR$_2$} = $b/a$ \\
        {\tt SHPAR$_3$} = $c/a$ \\
        {\tt SHPAR$_4$} = $r/a$ \\
        Ideally, {\tt SHPAR}$_1$,
        {\tt SHPAR}$_2\times${\tt SHPAR}$_1$,
        {\tt SHPAR}$_3\times${\tt SHPAR}$_1$ will be integers (so that the
        cubic lattice can accurately approximate the desired target
        geometry), and
        {\tt SHPAR}$_4\times${\tt SHPAR}$_1$ will be large enough for
        the circular cross section to be well-approximated.\\
        The TF origin is at the center of the top surface (the top surface
        lies in the $\ytf-\ztf$ plane, and extends from 
        $y_{\TF}=-b/2$ to $+b/2$,
        and $z_\TF=-c/2$ to $+c/2$).
        The cylindrical hole axis runs from $(x_\TF=0,y_\TF=0,z_\TF=0)$ to
        $(x_\TF=-a,y_\TF=0,z_\TF=0)$.
\index{target shape options!SPHERES\_N}
\subsection{ SPHERES\_N = Multisphere target = union of $N$
	    spheres of single isotropic material
            \label{sec:SPHERES_N}}
	Spheres may overlap if desired.
	The relative locations and sizes of these spheres are
	defined in an external file, whose name (enclosed in single quotes) 
	is passed through {\tt ddscat.par}.  The length of the file name
	should not exceed 80 characters.  
	The pertinent line in {\tt ddscat.par} should read\\
	{\tt SHPAR$_1$ SHPAR$_2$} {\it 'filename'} (quotes must be used)\\
	where {\tt SHPAR$_1$} = target diameter in $x$ direction 
	(in Target Frame) in units of $d$\\
	{\tt SHPAR$_2$}= 0 to have $a_1=(1,0,0)_\TF$, $a_2=(0,1,0)_\TF$.\\ 
	{\tt SHPAR$_2$}= 1 to use principal axes of moment of inertia
	tensor for $a_1$ (largest $I$) and $a_2$ (intermediate $I$).\\
	{\it filename} is the name of the file specifying the locations and
	relative sizes of the spheres.\\
	The overall size of the multisphere target (in terms of numbers of
	dipoles) is determined by parameter {\tt SHPAR$_1$}, which is
	the extent of the multisphere target in the $x$-direction, in
	units of the lattice spacing $d$.\\
	The file {\it `filename'} should have the
	following structure:\\ \\
	\hspace*{2em}$N$ (= number of spheres)\\
	\hspace*{2em}line of comments (may be blank)\\
	\hspace*{2em}line of comments (may be blank) [N.B.: changed from v7.0.7]\\
	\hspace*{2em}line of comments (may be blank) [N.B.: changed from v7.0.7]\\
	\hspace*{2em}line of comments (may be blank) [N.B.: changed from v7.0.7]\\
	\hspace*{2em}$x_1$ $y_1$ $z_1$ $a_1$ (arb. units)\\
	\hspace*{2em}$x_2$ $y_2$ $z_2$ $a_2$ (arb. units)\\
	\hspace*{2em} ... \\
	\hspace*{2em}$x_N$ $y_N$ $z_N$ $a_N$ (arb. units)\\ \\
	where $x_j$, $y_j$, $z_j$ are the coordinates (in the TF)
	of the center of sphere $j$,
	and $a_j$ is the radius of sphere $j$.\\
	Note that $x_j$, $y_j$, $z_j$, $a_j$ ($j=1,...,N$) establish only
	the {\it shape} of the $N-$sphere target.  For instance,
	a target consisting of two touching spheres with the line between
	centers parallel to the $x$ axis could equally well be
	described by lines 6 and 7 being\\
	\hspace*{2em}0 ~0 ~0 ~0.5\\
	\hspace*{2em}1 ~0 ~0 ~0.5\\
	or\\
	\hspace*{2em}0 ~0 ~0 ~1\\
	\hspace*{2em}2 ~0 ~0 ~1\\
	The actual size (in physical units) is set by the value
	of $a_{\rm eff}$ specified in {\tt ddscat.par}, where, as
	always, $a_{\rm eff}\equiv (3 V/4\pi)^{1/3}$, where $V$ is the
	total volume of material in the target.\\
	Target axes $\hat{\bf a}_1$ and $\hat{\bf a}_2$ are set to
	be principal axes of moment of inertia tensor (for uniform density), 
	where $\hat{\bf a}_1$
	corresponds to the largest eigenvalue, and $\hat{\bf a}_2$ to the
	intermediate eigenvalue.\\
	The TF origin is taken to be located at the volume-weighted
	centroid.\\
	User must set {\tt NCOMP}=1.

\subsubsection{\label{sec:example SPHERES_N}
               \bf Sample calculation in directory
               examples\_exp/SPHERES\_N}

The directory {\tt examples\_exp/SPHERES\_N} contains {\tt ddscat.par}
for a sample scattering problem using target option {\tt SPHERES\_N}.
{\tt ddscat.par} specifies that the locations of the spheres is to be 
read in from the file {\tt BAM2.16.1.targ}, which contains locations and
radii of 16 spheres in a cluster formed by ``Ballistic Aggregation with 2
Migrations'' (see \citet{Shen+Draine+Johnson_2008} for a description of this
procedure for producting random aggregates).
The spheres are assumed to be composed of material with refractive
index $m=1.33+0.01i$.
The effective radius $\aeff=0.25198\micron$, so that each sphere has
a radius $\aeff/16^{1/3}=0.10\micron$.  The calculation is done
for wavelength $\lambda=0.6\micron$, so that each monomer has
$x=2\pi a/\lambda = 1.047$, and $|m|kd=0.3386$.
Mueller matrix elements are evaluated for
two scattering planes.

This calculation required 1.7 cpu~sec on a 2.53 GHz cpu.

\index{target shape options!SPHROID\_2}
\subsection{ SPHROID\_2 = Two touching homogeneous, isotropic spheroids,
	   with distinct compositions
           \label{sec:SPHROID_2}}
	First spheroid has length {\tt SHPAR$_1$} along symmetry axis, 
        diameter {\tt SHPAR$_2$} perpendicular to symmetry axis.
	Second spheroid has length {\tt SHPAR}$_3$ along symmetry axis, 
	diameter {\tt SHPAR}$_4$ perpendicular to symmetry axis.  
	Contact point is on line connecting centroids.  
	Line connecting centroids is in $\xtf$ direction.
	Symmetry axis of first spheroid is in $\ytf$ direction.  
	Symmetry axis of second spheroid is in direction 
	$\ytf\cos({\tt SHPAR}_5)+\ztf\sin({\tt SHPAR}_5)$,
	and {\tt SHPAR}$_5$ is in degrees.  
	If {\tt SHPAR}$_6=0.$, then target axes ${\hat{\bf a}}_1=(1,0,0)_\TF$,
	${\hat{\bf a}}_2=(0,1,0)_\TF$. 
	If {\tt SHPAR}$_6=1.$, then axes ${\hat{\bf a}}_1$ and 
        ${\hat{\bf a}}_2$ are set to 
	principal axes with largest and 2nd largest moments of inertia assuming
	spheroids to be of uniform density.\\
	Origin of TF is located between spheroids, at point of contact.\\
	User must set {\tt NCOMP}=2 and provide dielectric function files for 
	each spheroid.
\index{target shape options!SPH\_ANI\_N}
\subsection{ SPH\_ANI\_N = Multisphere target consisting of the union of $N$
	    spheres of various materials, possibly anisotropic
            \label{sec:SPH_ANI_N}}
	Spheres may NOT overlap.
	The relative locations and sizes of these spheres are
	defined in an external file, whose name (enclosed in single quotes)
	is passed through {\tt ddscat.par}.  The length of the file name
	should not exceed 80 characters.
	Target axes $\hat{\bf a}_1$ and $\hat{\bf a}_2$ are set to
	be principal axes of moment of inertia tensor (for uniform density), 
	where $\hat{\bf a}_1$
	corresponds to the largest eigenvalue, and $\hat{\bf a}_2$ to the
	intermediate eigenvalue.\\
	The TF origin is taken to be located at the volume-weighted
	centroid.\\
	The pertinent line in {\tt ddscat.par} should read\\
	{\tt SHPAR$_1$ SHPAR$_2$} {\it `filename'} (quotes must be used)\\
	where {\tt SHPAR$_1$} = target diameter in $x$ direction 
	(in Target Frame) in units of $d$\\
	{\tt SHPAR$_2$}= 0 to have $a_1=(1,0,0)_\TF$, $a_2=(0,1,0)_\TF$ 
	in Target Frame.\\
	{\tt SHPAR$_2$}= 1 to use principal axes of moment of inertia
	tensor for $a_1$ (largest $I$) and $a_2$ (intermediate $I$).\\
	{\it filename} is the name of the file specifying the locations and
	relative sizes of the spheres.\\
	The overall size of the multisphere target (in terms of numbers of
	dipoles) is determined by parameter {\tt SHPAR$_1$}, which is
	the extent of the multisphere target in the $x$-direction, in
	units of the lattice spacing $d$.
	The file {\it `filename'} should have the
	following structure:\\ \\
	\hspace*{2em}$N$ (= number of spheres)\\
	\hspace*{2em}line of comments (may be blank)\\
		\hspace*{2em}line of comments (may be blank) [N.B.: changed from v7.0.7]\\
	\hspace*{2em}line of comments (may be blank) [N.B.: changed from v7.0.7]\\
	\hspace*{2em}line of comments (may be blank) [N.B.: changed from v7.0.7]\\
        \hspace*{2em}$x_1$~ $y_1$~ $z_1$~ $r_1$~ $Cx_1$~ $Cy_1$~ $Cz_1$~ 
	$\Theta_{\DF,1}$~ $\Phi_{\DF,1}$~ $\beta_{\DF,1}$\\
	\hspace*{2em}$x_2$~ $y_2$~ $z_2$~ $r_2$~ $Cx_2$~ $Cy_2$~ $Cz_2$~ 
	$\Theta_{\DF,2}$~ $\Phi_{\DF,2}$~ $\beta_{\DF,2}$\\
	\hspace*{2em} ... \\
	\hspace*{2em}$x_N$ $y_N$ $z_N$ $r_N$ $Cx_N$ $Cy_N$ $Cz_N$ 
	$\Theta_{\DF,N}$ $\Phi_{\DF,N}$ $\beta_{\DF,N}$\\ \\
	where $x_j$, $y_j$, $z_j$ are the coordinates of the center,
	and $r_j$ is the radius of sphere $j$ (arbitrary units),
	$Cx_j$, $Cy_j$, $Cz_j$ are integers specifying the ``composition''
	of sphere $j$ in the $x,y,z$ directions in the ``Dielectric Frame''
	\index{Dielectric Frame (DF)}
	(see \S\ref{sec:composite anisotropic targets})
	of sphere $j$, and 
	\index{$\Theta_\DF$}
	$\Theta_{\DF,j}$ 
	\index{$\Phi_\DF$}
	$\Phi_{\DF,j}$ 
	\index{$\beta_\DF$}
	$\beta_{\DF,j}$
	are angles (in radians) specifying orientation of the 
	dielectric frame (DF) of
	sphere $j$ relative to the Target Frame.
	Note that $x_j$, $y_j$, $z_j$, $r_j$ ($j=1,...,N$) establish only
	the {\it shape} of the $N-$sphere target, just as for
	target option {\tt NSPHER}.
	The actual size (in physical units) is set by the value
	of $a_{\rm eff}$ specified in {\tt ddscat.par}, where, as
	always, $a_{\rm eff}\equiv (3 V/4\pi)^{1/3}$, where $V$ is the
	volume of material in the target.\\
	User must set {\tt NCOMP} to the number of different dielectric 
	functions being invoked (i.e., the range of $\{Cx_j,Cy_j,Cz_j\}$.

	Note that while the spheres can be anisotropic and of differing
	composition, they can of course also be isotropic and of a single
	composition, in which case the relevant lines in the file
	{\it 'filename'} would be simply\\ \\
	\hspace*{2em}$N$ (= number of spheres)\\
	\hspace*{2em}line of comments (may be blank)\\
	\hspace*{2em}line of comments (may be blank) [N.B.: changed from v7.0.7]\\
	\hspace*{2em}line of comments (may be blank) [N.B.: changed from v7.0.7]\\
	\hspace*{2em}line of comments (may be blank) [N.B.: changed from v7.0.7]\\
	\hspace*{2em}$x_1$~ $y_1$~ $z_1$~ $r_1$~ 1 ~1 ~1 ~0 ~0 ~0\\
	\hspace*{2em}$x_2$~ $y_2$~ $z_2$~ $r_2$~ 1 ~1 ~1 ~0 ~0 ~0\\
	\hspace*{2em} ... \\
	\hspace*{2em}$x_N$ $y_N$ $z_N$ $r_N$ ~1 ~1 ~1 ~0 ~0 ~0\\

\subsubsection{\bf Sample calculation in directory examples\_exp/SPH\_ANI\_N}

Subdirectory {\tt examples\_exp/SPH\_ANI\_N} contains {\tt ddscat.par} for
calculating scattering by a random aggregate of 64 spheres
[aggregated according following the ``BAM2'' aggregation process
described by \citet{Shen+Draine+Johnson_2008}].
32 of the spheres are assumed to consist of ``astrosilicate'', and
32 of crystalline graphite, with random orientations for each of the 32
graphite spheres.
Each sphere is assumed to have a radius $0.050\micron$.
The entire cluster is represented by $N=7947$ dipoles, or about 124 dipoles
per sphere.
Scattering and absorption are calculated for $\lambda=0.55\micron$.
\index{target shape options!TETRAHDRN}
\subsection{ TETRAHDRN = Homogeneous, isotropic tetrahedron
            \label{sec:TETRAHDRN}}
	{\tt SHPAR$_1$}=length/$d$ 
	of one edge. 
	Orientation: one face parallel to $\ytf,\ztf$ plane,
	opposite ``vertex" is in $+\xtf$ 
	direction, and one edge is parallel to $\ztf$.
	Target axes ${\hat{\bf a}}_1=(1,0,0)_\TF$ [emerging from one vertex] 
        and ${\hat{\bf a}}_2=(0,1,0)_\TF$ [emerging from an edge] in the TF.
	User must set {\tt NCOMP}=1.
\index{target shape options!TRNGLPRSM}
\subsection{ TRNGLPRSM = Triangular prism of homogeneous, isotropic material
            \label{sec:TRNGLPRSM}}
	{\tt SHPAR$_1$, SHPAR$_2$, SHPAR$_3$, SHPAR$_4$} $= a/d$, $b/a$, 
        $c/a$, $L/a$\\
	The triangular cross section has sides of width $a$, $b$, $c$.
	$L$ is the length of the prism.
	$d$ is the lattice spacing.
	The triangular cross-section 
	has interior angles $\alpha$, $\beta$, $\gamma$
	(opposite sides $a$, $b$, $c$) given by
	$\cos\alpha=(b^2+c^2-a^2)/2bc$, $\cos\beta=(a^2+c^2-b^2)/2ac$,
	$\cos\gamma=(a^2+b^2-c^2)/2ab$.
	In the Target Frame, the prism axis is in the $\hat{\bf x}$ direction,
	the normal to the rectangular face of width $a$ is (0,1,0), 
	the normal to the rectangular face of width $b$ is
	$(0,-\cos\gamma,\sin\gamma)$, and
	the normal to the rectangular face of width  $c$ is
	$(0,-\cos\beta,-\sin\beta)$.
\index{target shape options!UNIAXICYL}
\subsection{ UNIAXICYL = Homogeneous finite cylinder with uniaxial anisotropic 
	    dielectric tensor
            \label{sec:UNIAXICYL}}
	{\tt SHPAR$_1$}, {\tt SHPAR$_2$} have same meaning as for
	{\tt CYLINDER1}.
	Cylinder axis $={\hat{\bf a}}_1=(1,0,0)_\TF$, 
	${\hat{\bf a}}_2=(0,1,0)_\TF$. 
	It is assumed that the dielectric tensor $\epsilon$ 
	is diagonal in the TF,
	with $\epsilon_{yy}=\epsilon_{zz}$.
	User must set
	{\tt NCOMP}=2.
	Dielectric function 1 is for ${\bf E} \parallel {\bf\hat{a}}_1$
	(cylinder axis), dielectric function 2 is for 
	${\bf E} \perp {\bf\hat{a}}_1$.
\bigskip
\index{target routines: modifying}
\subsection{ Modifying Existing Routines or Writing New Ones}

The user should be able to easily modify these routines, or write new
routines, to generate targets with other geometries.  The user should
first examine the routine {\tt target.f90} and modify it to call any new
target generation routines desired.  Alternatively, targets may be
generated separately, and the target description (locations of dipoles
and ``composition" corresponding to x,y,z dielectric properties at
each dipole site) read in from a file by invoking the option {\tt
FROM\_FILE} in {\tt ddscat.f90}.

Note that it will also be necessary to modify the routine {\tt reapar.f90}
so that it will accept whatever new target option is added to the
target generation code
.
\bigskip
\subsection{ Testing Target Generation using CALLTARGET}
\index{CALLTARGET}
It is often desirable to be able to run the target generation routines
without running the entire {{\bf DDSCAT}}\ code.  We have therefore
provided a program {\tt CALLTARGET} which allows the user to generate
targets interactively; to create this executable just type\footnote{
	Non-Linux sites: The source code for {\tt CALLTARGET} is in the file
	{\tt CALLTARGET.f90}.  You must compile and link {\tt CALLTARGET.f90}, 
	{\tt ddcommon.f90}, {\tt dsyevj3.f90}, {\tt errmsg.f90}, 
	{\tt gasdev.f90}, {\tt p\_lm.f90},
        {\tt prinaxis.f90}, {\tt ran3.f90}, {\tt reashp.f90}, {\tt sizer.f90},
        {\tt tar2el.f90}, {\tt tar2sp.f90}, {\tt tar3el.f90}, 
        {\tt taranirec.f90},
        {\tt tarblocks.f90}, {\tt tarcel.f90}, {\tt tarcyl.f90}, 
        {\tt tarcylcap.f90}, 
        {\tt tarell.f90}, {\tt target.f90}, {\tt targspher.f90}, 
        {\tt tarhex.f90}, {\tt tarnas.f90}, {\tt tarnsp.f90}, 
	{\tt tarpbxn.f90}, {\tt tarprsm.f90}, {\tt tarrctblk3.f90}, 
        {\tt tarrecrec.f90}, {\tt tarslblin.f90}, 
        {\tt tartet.f90}, and {\tt wrimsg.f90}.
	}  
\hfill\break \indent\indent{\tt make calltarget} .\hfill\break 
The program {\tt calltarget} is to be run interactively; the prompts are
self-explanatory.  You may need to edit the code to change the device
number {\tt IDVOUT} as for {\tt DDSCAT} (see \S\ref{subsec:IDVOUT}
above).

\index{target.out -- output file}
After running, {\tt calltarget} will leave behind an ASCII file 
{\tt target.out}
which is a list of the occupied lattice sites in the last target generated.
The format of {\tt target.out} is the same as the format of the {\tt shape.dat}
files read if option {\tt FROM\_FILE} is used (see above).
Therefore you can simply \\
\indent\indent{\tt mv target.out shape.dat} \\
and then use {\tt DDSCAT}
with the option {\tt FROM\_FILE} (or option {\tt ANIFRMFIL} in the case
of anisotropic target materials with arbitrary orientation relative to
the Target Frame)
in order to input a target shape generated by {\tt CALLTARGET}.

\index{CALLTARGET and PBC}
Note that {\tt CALLTARGET} -- designed to generate finite targets -- 
can be used with some of the ``PBC'' target options 
(see \S\ref{sec:target_generation_PBC} below) to generate a list
of dipoles in the TUC.
At the moment, {\tt CALLTARGET} has support for target options
{\tt BISLINPBC}, {\tt DSKBLYPBC}, and {\tt DSKRCTPBC}.

\section{Target Generation: Periodic Targets
         \label{sec:target_generation_PBC}}
\index{target generation: infinite periodic targets}
A periodic target consists of a ``Target Unit Cell'' (TUC) which is then
repeated in either the $\ytf$ direction, the $\ztf$ direction, or
both.  Please see \citet{Draine+Flatau_2008a} for illustration of how
periodic targets are assembled out of TUCs, and how the scattering from these
targets in different diffraction orders $M$ or $(M,N)$
is constrained by the periodicity.

The following options for the TUC geometry are included in \ddscat:
\begin{itemize}
\item {\bf FRMFILPBC} : TUC geometry read from file (\S\ref{sec:FRMFILPBC})
\item {\bf ANIFILPBC} : TUC geometry read from file, anisotropic materials 
                        supported (\S\ref{sec:ANIFILPBC})
\item {\bf BISLINPBC} : TUC = bilayer slab (\S\ref{sec:BISLINPBC})
\item {\bf CYLNDRPBC} : TUC = finite cylinder (\S\ref{sec:CYLNDRPBC})
\item {\bf DSKBLYPBC} : TUC = disk plus bilayer slab (\S\ref{sec:DSKBLYPBC})
\item {\bf DSKRCTPBC} : TUC = disk plus brick (\S\ref{sec:DSKRCTPBC})
\item {\bf HEXGONPBC} : TUC = hexagonal prism (\S\ref{sec:HEXGONPBC})
\item {\bf LYRSLBPBC} : TUC = layered slab (up to 4 layers) 
                        (\S\ref{sec:LYRSLBPBC})
\item {\bf RCTGL\_PBC} : TUC = brick (\S\ref{sec:RCTGL_PBC})
\item {\bf RECRECPBC} : TUC = brick resting on brick (\S\ref{sec:RECRECPBC})
\item {\bf SLBHOLPBC} : TUC = brick with cylindrical hole 
                        (\S\ref{sec:SLBHOLPBC})
\item {\bf SPHRN\_PBC} : TUC = N spheres (\S\ref{sec:SPHRN_PBC})
\item {\bf TRILYRPBC} : TUC = three stacked bricks (\S\ref{sec:TRILYRPBC})
\end{itemize}
Each option is decribed in detail below.
\index{target shape options!FRMFILPBC}
\subsection{ FRMFILPBC = periodic target with TUC geometry and composition
            input from a file
            \label{sec:FRMFILPBC}}
        The TUC can have arbitrary geometry and inhomogeneous
        composition, and is assumed to repeat periodically in either 
        1-d (y or z) or 2-d (y and z).\\
        The pertinent line in {\tt ddscat.par} should read\\
        {\tt SHPAR$_1$ SHPAR$_2$} {\it 'filename'} (quotes must be used)\\
        {\tt SHPAR}$_1$ = $P_y/d$~~~ 
            ($P_y$ = periodicity in $\ytf$ direction)\\
        {\tt SHPAR}$_2$ = $P_z/d$~~~
            ($P_z$ = periodicity in $\ztf$ direction)\\
        {\it filename} is the name of the file specifying the locations
        of the dipoles, and the ``composition" at each dipole location.  The
        composition can be anisotropic, but the dielectric tensor must
        be diagonal in the TF.  The shape and composition of the TUC
        are provided {\it exactly} as for target option {\tt FROM$\_$FILE}
        -- see \S\ref{sec:FROM_FILE}\\
        If {\tt SHPAR}$_1=0$ then the target does {\it not} repeat in the
            $\ytf$ direction.\\
        If {\tt SHPAR}$_2=0$ then the target does {\it not} repeat in the
            $\ztf$ direction.\\

\subsubsection{\bf Sample calculation in directory examples\_exp/FRMFILPBC}

Subdirectory {\tt examples\_exp/FRMFILPBC} contains {\tt ddscat.par}
and {\tt shape.dat} for the same scattering calculation as in 
{\tt examples\_exp/DSKRCTPBC}: a slab of Si$_3$N$_4$ glass, thickness
$0.05\micron$, supporting a periodic array of Au disks, with center-to-center
distance $0.08\micron$.  Here the slab and Au disk
geometry is input via a file {\tt shape.dat}.  {\tt shape.dat} is
a copy of the file {\tt target.out} created after running the
calculation in {\tt examples\_exp/DSKRCTPBC}.

\index{target shape options!ANIFILPBC}
\subsection{ ANIFILPBC = general anisotropic periodic target with TUC 
            geometry and composition input from a file
            \label{sec:ANIFILPBC} }
        The TUC can have arbitrary geometry and inhomogeneous
        composition, and is assumed to repeat periodically in either 
        1-d (y or z) or 2-d (y and z).\\
        The pertinent line in {\tt ddscat.par} should read\\
        {\tt SHPAR$_1$ SHPAR$_2$} {\it 'filename'} (quotes must be used)\\
        {\tt SHPAR}$_1$ = $P_y/d$~~~ 
            ($P_y$ = periodicity in $\ytf$ direction)\\
        {\tt SHPAR}$_2$ = $P_z/d$~~~
            ($P_z$ = periodicity in $\ztf$ direction)\\
        {\it filename} is the name of the file specifying the locations
        of the dipoles, and the "composition" at each dipole location.  The
        composition can be anisotropic, and the dielectric tensor need not
        be diagonal in the TF.  The shape and composition of the TUC
        are provided {\it exactly} as for target option {\tt ANIFRMFIL} --
        see \S\ref{sec:ANIFRMFIL}\\
        If {\tt SHPAR}$_1=0$ then the target does {\it not} repeat in the
            $\ytf$ direction.\\
        If {\tt SHPAR}$_2=0$ then the target does {\it not} repeat in the
            $\ztf$ direction.\\
\index{target shape options!BISLINPBC}
\subsection{ BISLINPBC = Bi-Layer Slab with Parallel Lines}
        \label{sec:BISLINPBC}
        The target consists of a bi-layer slab, on top of which there is
        a ``line'' with rectangular cross-section.\\
	The ``line'' on top
	is composed of material 1, has height $X_1$ (in the $\xtf$ direction),
	width $Y_1$ (in the $\ytf$ direction), and
	is infinite in extent in the $\ztf$ direction.\\
	The bilayer slab has width $Y_2$ (in the $\ytf$ direction).
	It is consists of a layer of thickness $X_2$ of material 2, on top
	of a layer of material 3 with thickness $X_3$.\\
	{\tt SHPAR}$_1$ = $X_1/d$~~~~ ($X_1$ = thickness of line)\\
	{\tt SHPAR}$_2$ = $Y_1/d$~~~~ ($Y_1$ = width of line)\\
	{\tt SHPAR}$_3$ = $X_2/d$~~~~ ($X_2$ = thickness of upper layer of 
	slab)\\
	{\tt SHPAR}$_4$ = $X_3/d$~~~~ ($X_3$ = thickness of lower layer of 
	slab)\\
	{\tt SHPAR}$_5$ = $Y_2/d$~~~~ ($Y_2$ = width of slab)\\
	{\tt SHPAR}$_6$ = $P_y/d$~~~~ ($P_y$ = periodicity in $\ytf$ 
        direction).\\

	If {\tt SHPAR}$_6$ = 0, the target is NOT periodic in the $\ytf$
	direction, consisting of a single column, infinite in the $\ztf$
	direction.
\index{target shape options!CYLNDRPBC}
\index{periodic boundary conditions}
\subsection{ CYLNDRPBC = Target consisting of homogeneous cylinder
                     repeated in
                     target y and/or z directions using
                     periodic boundary conditions}
            \label{sec:CYLNDRPBC}
            This option causes {{\bf DDSCAT}}\ to create a target consisting
	    of an infinite array of cylinders.
	    The individual cylinders 
	    are assumed to be homogeneous and isotropic,
	    just as for option RCTNGL (see \S\ref{sec:RCTGLPRSM}).
	    \ \\
	    Let us 
	    refer to a single cylinder as the Target Unit Cell (TUC).
	    The TUC 
	    is then repeated in the target y- and/or z-directions, with 
	    periodicities {\tt PYD}$\times d$\index{PYD} and 
	    {\tt PZD}$\times d$,\index{PZD} where $d$ is the lattice spacing.
	    To repeat in only one direction, set either {\tt PYD} or
	    {\tt PZD} to zero.\\
	    This option requires 5 shape parameters:
	    The pertinent line in {\tt ddscat.par} should read\\
	    \ \\
	{\tt SHPAR$_1$ SHPAR$_2$ SHPAR$_3$ SHPAR$_4$ SHPAR}$_5$\\
	    \ \\
	where {\tt SHPAR$_1$,SHPAR$_2$,SHPAR$_3$,SHPAR$_4$,SHPAR$_5$} are 
        numbers:\\
	{\tt SHPAR}$_1$ = cylinder length along axis (in units of $d$)
	in units of $d$\\
	{\tt SHPAR$_2$} = cylinder diameter/$d$\\
	{\tt SHPAR}$_3$ = 1 for cylinder axis $\parallel \xtf$\\
        \hspace*{5em}= 2 for cylinder axis $\parallel \ytf$\\
        \hspace*{5em}= 3 for cylinder axis $\parallel \ztf$
	                 (see below)\\
	{\tt SHPAR}$_4$ = {\tt PYD} = periodicity/$d$ in $\ytf$ direction 
        ( = 0 to suppress repetition)\\
	{\tt SHPAR}$_5$ = {\tt PZD} = periodicity/$d$ in $\ztf$ direction 
        ( = 0 to suppress repetition)\\
	\ \\
	The overall size of the TUC (in terms of numbers of
	dipoles) is determined by parameters 
	{\tt SHPAR1} and {\tt SHPAR$_2$}.
	The orientation of a single cylinder is determined by {\tt SHPAR}$_3$.
	The periodicity in the TF $y$ and $z$ directions is determined
	by parameters {\tt SHPAR}$_4$ and {\tt SHPAR}$_5$.\\
	The physical size of the TUC is specified by the value of
	$\aeff$ (in physical units, e.g. cm), 
	specified in the file {\tt ddscat.par}, with the usual
	correspondence $d=(4\pi/3N)^{1/3}\aeff$, where $N$ is the number
	of dipoles in the TUC.

	With target option {\tt CYLNDRPBC}, the target
	becomes a periodic structure, of infinite extent.
	\begin{itemize}
	\item If ${\tt NPY}>0$ and ${\tt NPZ}=0$, 
              then the target cylindrical TUC
              repeats in the $\ytf$ direction,
	      with periodicity ${\tt NPY}\times d$.
        \item If ${\tt NPY}=0$ and ${\tt NPZ}>0$
	      then the target cylindrical TUC
	      repeats in the $\ztf$ direction,
	       with periodicity ${\tt NPZ}\times d$.
        \item If ${\tt NPY}>0$ and ${\tt NPZ}>0$
	      then the target cylindrical TUC
	      repeats in the $\ytf$ direction,
	      with periodicity ${\tt NPY}\times d$, and in
	      the $\ztf$ direction,
	       with periodicity ${\tt NPZ}\times d$.
        \end{itemize}

	\noindent
	{\bf Target Orientation:} The target axes (in the TF) 
	are set to ${\hat{\bf a}}_1 = (1,0,0)_\TF$
	and
	${\hat{\bf a}}_2 = (0,1,0)_\TF$.
        Note that ${\hat{\bf a}}_1$ does {\bf not} necessarily
	coincide with the cylinder axis: individual
	cylinders may have any of 3 different orientations in the TF.\\
	\ \\

	\noindent
	{\bf Example 1:} One could construct a single infinite cylinder
	\index{infinite cylinder}
	with the following two lines in {\tt ddscat.par}:\\
	\ \\
	100  1 100 \\
	1.0\ \ \  100.49\ \ \  2\ \ \  1.0\ \ \  0.\\
	\ \\
	The first line ensures that there will be enough memory allocated
	to generate the target.
	The TUC would be a thin circular ``slice'' containing just one layer
	of dipoles.  The diameter of the circular slice would be about 
	100.49$d$ in
	extent, so the TUC would have approximately
	$(\pi/4)\times(100.49)^2=7931$ dipoles 
	(7932 in the actual realization) within a $100\times1\times100$
	``extended target volume''.
	The TUC would be oriented with the cylinder axis in the $\ytf$ 
        direction
	({\tt SHPAR3=2}) and the structure would repeat in the $\ytf$
         direction
	with a period of $1.0\times d$.
	{\tt SHPAR}$_5$=0 means that there will be no repetition in the $z$
	direction.
	As noted above, the ``target axis'' vector 
	$\hat{\bf a}_1 = \xtf$.
	
	Note that {\tt SHPAR}$_1$, {\tt SHPAR}$_2$, {\tt SHPAR}$_4$, and 
        {\tt SHPAR}$_5$ need not be integers.  
        However, {\tt SHPAR}$_3$, determining the orientation of the cylinders 
        in the TF, can only take on the values 1,2,3.

	The orientation of the incident radiation relative to the target
	is, as for all other targets, set by the usual orientation
	angles $\beta$, $\Theta$, and $\Phi$ 
	(see \S\ref{sec:target_orientation} above); for example,
	$\Theta=0$ would be for radiation incident normal to the periodic
	structure.



\subsubsection{\bf Sample calculation in directory examples\_exp/CYLNDRPBC}

The subdirectory {\tt examples\_exp/CYLNDRPBC} contains {\tt ddscat.par}
for calculating scattering by an infinite cylinder with $m=1.33+0.01i$ for 
$x\equiv 2\pi R/\lambda=5$, where $R$ is the cylinder radius.
Thus $R=x\lambda/2\pi$.
We set $\lambda=1$.

The TUC is a disk of thickness $d$. 
The sample calculation calls for the cylinder diameter $2R$ to be $64.499d$,
where $d$ is the dipole spacing.

With this choice, the TUC (disk of thickness $d$)
turns out to be represented by $N=3260$ dipoles.
Thus $\pi R^2 = N d^2$, or $d=(\pi/N)^{1/2} R$.

The volume of the TUC is $V=Nd^3$.
The effective radius of the TUC is 
$\aeff=(3V/4\pi)^{1/3} =
(3N/4\pi)^{1/3} d = 
(3N/4\pi)^{1/3} (\pi/N)^{1/2} R = 
(3/4\pi)^{1/3} \pi^{1/2} R N^{-1/6}$.
With $R=x\lambda/2\pi$ we have
$\aeff=(3/4\pi)^{1/3} \pi^{1/2} (x\lambda/2\pi) N^{-1/6} = 0.22723$
for $x=5$ and $\lambda=1$.

With the standard error tolerance {\tt TOL=1.0e-5}, 
the calculation converges in {\tt IT=7} iterations for each incident
polarization.
The scattering properties are reported in {\tt w000r000k000.sca}.
The calculation used required 40 MB of RAM, and used 39 cpu~sec 
on a 2.53 GHz cpu.
\index{target shape options!DSKBLYPBC}
\index{periodic boundary conditions}
\subsection{ DSKBLYPBC = Target consisting of a periodic array of disks on top 
            of a two-layer rectangular slabs.}
            \label{sec:DSKBLYPBC}
            This option causes {{\bf DDSCAT}}\ to create a target consisting
	    of a periodic or biperiodic array of Target Unit Cells (TUCs),  
            each TUC consisting of a disk of composition 1
	    resting on top of a rectangular block consisting of
	    two layers: composition 2 on top and composition 3 below.
	    Materials 1, 2, and 3 are assumed to be homogeneous and isotropic.
	    \ \\
	    This option requires 8 shape parameters:\newline
	    The pertinent line in {\tt ddscat.par} should read\\
	    \ \\
	${\tt SHPAR}_1~~{\tt SHPAR}_2~~{\tt SHPAR}_3~~{\tt SHPAR}_4~~{\tt SHPAR}_5~~{\tt SHPAR}_6~~{\tt SHPAR}_7~~{\tt SHPAR}_8$\\
	    \ \\
	where\\
	{\tt SHPAR}$_1$ = disk thickness in $x$ direction 
	(in Target Frame) in units of $d$\\
	{\tt SHPAR}$_2$ = (disk diameter)/$d$\\
	{\tt SHPAR}$_3$ = (upper slab thickness)/$d$\\
	{\tt SHPAR}$_4$ = (lower slab thickness)/$d$\\
	{\tt SHPAR}$_5$ = (slab extent in $\ytf$ direction)/$d$\\
	{\tt SHPAR}$_6$ = (slab extent in $\ztf$ direction)/$d$\\
	{\tt SHPAR}$_7$ = period in $\ytf$ direction/$d$\\
	{\tt SHPAR}$_8$ = period in $\ztf$ direction/$d$\\
	\ \\
	The physical size of the TUC is specified by the value of
	$\aeff$ (in physical units, e.g. cm), 
	specified in the file {\tt ddscat.par}.\\
	\ \\
	The ``computational volume'' is determined by\\
	({\tt SHPAR$_1$+SHPAR$_3$+SHPAR$_4$})$\times${\tt SHPAR}$_5\times${\tt SHPAR}$_6$.\\
	\ \\
	    The lower surface of the cylindrical disk is in the $x=0$ plane.
	    The upper surface of the slab is also in the $x=0$ plane.
	    It is required that 
            ${\tt SHPAR}_2\leq{\rm min}({\tt SHPAR}_4,{\tt SHPAR}_5)$.\\
	    \ \\
	    The Target Frame origin (0,0,0) is located where the symmetry
	    axis of the disk intersects the $x=0$ plane (the upper surface
	    of the slab, and the lower surface of the disk).\\
	    In the Target Frame, 
            dipoles representing the disk are located at\\
	    \hspace*{1.0cm}$x/d = 0.5 , 1.5 ,..., 
	    [{\rm int}({\tt SHPAR}_1+0.5)-0.5]$\\
	    and at $(y,z)$ values \\
	    \hspace*{1.0cm}$y/d = \pm 0.5, \pm 1.5, ...$ and\\
	    \hspace*{1.0cm}$z/d = \pm 0.5, \pm 1.5, ...$ satisfying\\
	    \hspace*{1.0cm}$(y^2+z^2) \leq ({\tt SHPAR}_2/2)^2 d^2$.
	    
	    Dipoles representing the rectangular slab
	    are located at
	    $(x/d,y/d,z/d)=(j_x+0.5,j_y+\Delta y,j_z+\Delta z)$, where
	    $j_x$, $j_y$, and $j_z$ are integers. $\Delta y=0$ or 0.5
	    depending on whether {\tt SHPAR}$_5$ is even or odd.
	    $\Delta z=0$ or 0.5 depending on whether {\tt SHPAR}$_6$ is
	    even or odd.\\
	    \ \\
	    The TUC 
	    is repeated in the target y- and z-directions, with 
	    periodicities {\tt SHPAR$_7$}$\times d$ and 
            {\tt SHPAR}$_8$$\times d$.

	    As always, the physical size of the target is fixed by
	    specifying the value of the effective radius
	    $\aeff\equiv(3V_{\rm TUC}/4\pi)^{1/3}$,
	    where $V_{\rm TUC}$ 
	    is the total volume of solid material in one TUC.
	    For this geometry, the number of dipoles in the target will
	    be approximately
	    $$N=
	    (\pi/4)\times{\tt SHPAR}_1\times({\tt SHPAR}_2)^2+
	    [{\tt SHPAR}_3+{\tt SHPAR}_4]
	    \times{\tt SHPAR}_5\times{\tt SHPAR}_6 $$
	    although the exact number may differ because of the
	    dipoles are required to be located on a rectangular lattice.
	    The dipole spacing $d$ in physical units
	    is determined from the specified value of $\aeff$ and
	    the number $N$ of dipoles in the target:
	    $d=(4\pi/3N)^{1/3}\aeff$.\\
	The target axes (in the TF) 
	are set to ${\hat{\bf a}}_1 = \xtf = (1,0,0)_\TF$ --
	i.e., normal to the ``slab'' -- and
	${\hat{\bf a}}_2 = \ytf= (0,1,0)_\TF$.
	The orientation of the incident radiation relative to the target
	is, as for all other targets, set by the usual orientation
	angles $\beta$, $\Theta$, and $\Phi$ 
	(see \S\ref{sec:target_orientation} above); for example,
	$\Theta=0$ would be for radiation incident normal to the slab.

\index{target shape options!DSKRCTPBC}
\index{periodic boundary conditions}
\subsection{ DSKRCTPBC = Target consisting of homogeneous rectangular brick plus
                     a disk, 
                     extended in
                     target y and z directions using
                     periodic boundary conditions}
            \label{sec:DSKRCTPBC}
            This option causes {{\bf DDSCAT}}\ to create a target consisting
	    of a biperiodic array of Target Unit Cells.  Each Target
	    Unit Cell (TUC) consists of a disk of composition 1
	    resting on top of a rectangular block of composition 2.
	    Materials 1 and 2 are assumed to be homogeneous and isotropic.
	    \ \\
	    The cylindrical disk has thickness {\tt SHPAR$_1$}$\times d$ in the
	    x-direction, and diameter {\tt SHPAR$_2$}$\times d$.
	    The rectangular block is assumed to have thickness
	    {\tt SHPAR}$_3$$\times$d in the x-direction,
	    extent {\tt SHPAR}$_4$$\times d$ in the y-direction,
	    and extent {\tt SHPAR}$_5$$\times d$ in the z-direction.
	    The lower surface of the cylindrical disk is in the $x=0$ plane.
	    The upper surface of the slab is also in the $x=0$ plane.
	    It is required that 
            {\tt SHPAR$_2$}$\leq$min({\tt SHPAR4,SHPAR}$_5$).

	    The Target Frame origin (0,0,0) is located where the symmetry
	    axis of the disk intersects the $x=0$ plane (the upper surface
	    of the slab, and the lower surface of the disk).
	    In the Target Frame, 
	    dipoles representing the rectangular block are located at
	    $(x/d,y/d,z/d)=(j_x+0.5,j_y+\Delta y,j_z+\Delta z)$, where
	    $j_x$, $j_y$, and $j_z$ are integers. $\Delta y=0$ or 0.5
	    depending on whether {\tt SHPAR}$_4$ is even or odd.
	    $\Delta z=0$ or 0.5 depending on whether {\tt SHPAR}$_5$ is
	    even or odd.\\
	    \hspace*{1.0cm}
	    $j_x = -[{\rm int}({\tt SHPAR}_3+0.5)] , ... , -1$.
	     \\
	    \hspace*{1.0cm}
	    $j_y = -[{\rm int}(0.5\times{\tt SHPAR}_4-0.5) + 1]$ , ... ,
            $j_z = -[{\rm int}({\tt SHPAR}_4+0.5)-0.5]$\\
	    \hspace*{1.0cm}
	    $y/d = -[{\rm int}(0.5\times{\tt SHPAR}_4+0.5)-0.5] , ... , 
                    [{\rm int}(0.5\times{\tt SHPAR}_4+0.5)-0.5]$\\
            \hspace*{1.0cm}
	    $z/d =  -[{\rm int}(0.5\times{\tt SHPAR}_5+0.5)-0.5] , ... , 
                     [{\rm int}(0.5\times{\tt SHPAR}_5+0.5)-0.5]$\\
            where ${\rm int}(x)$ is the greatest integer less than or equal
	    to $x$.
            Dipoles representing the disk are located at\\
	    \hspace*{1.0cm}$x/d = 0.5 , 1.5 ,..., 
	    [{\rm int}({\tt SHPAR}_4+0.5)-0.5]$\\
	    and at $(y,z)$ values \\
	    \hspace*{1.0cm}$y/d = \pm 0.5, \pm 1.5, ...$ and\\
	    \hspace*{1.0cm}$z/d = \pm 0.5, \pm 1.5, ...$ satisfying\\
	    \hspace*{1.0cm}$(y^2+z^2) \leq ({\tt SHPAR}_5/2)^2 d^2$.
	    
	    The TUC 
	    is repeated in the target y- and z-directions, with 
	    periodicities {\tt SHPAR$_6$}$\times d$ and 
	    {\tt SHPAR}$_7$$\times d$.
	    As always, the physical size of the target is fixed by
	    specifying the value of the effective radius
	    $\aeff\equiv(3V_{\rm TUC}/4\pi)^{1/3}$,
	    where $V_{\rm TUC}$ 
	    is the total volume of solid material in one TUC.
	    For this geometry, the number of dipoles in the target will
	    be approximately
	    $N=[{\tt SHPAR}_1\times{\tt SHPAR}_2\times{\tt SHPAR}_3 +
	    (\pi/4)(({\tt SHPAR}_4)^2\times{\tt SHPAR}_5)]$,
	    although the exact number may differ because of the
	    dipoles are required to be located on a rectangular lattice.
	    The dipole spacing $d$ in physical units
	    is determined from the specified value of $\aeff$ and
	    the number $N$ of dipoles in the target:
	    $d=(4\pi/3N)^{1/3}\aeff$.
	    This option requires 7 shape parameters:\newline
	    The pertinent line in {\tt ddscat.par} should read\\
	    \ \\
	{\tt SHPAR$_1$ SHPAR$_2$ SHPAR$_3$ SHPAR$_4$ SHPAR$_5$ SHPAR$_6$
         SHPAR$_7$}\\
	    \ \\
	where\\
	{\tt SHPAR$_1$} = [disk thickness (in $\xtf$ direction)]/$d$ 
        [material 1]\\ 
	{\tt SHPAR$_2$} = (disk diameter)/$d$\\
	{\tt SHPAR}$_3$ = (brick thickness in $\xtf$ direction)/$d$ 
        [material 2]\\
	{\tt SHPAR}$_4$ = (brick length in $\ytf$ direction)/$d$\\
	{\tt SHPAR}$_5$ = (brick length in $\ztf$ direction)/$d$\\
	{\tt SHPAR}$_6$ = periodicity in $\ytf$ direction/$d$\\
	{\tt SHPAR}$_7$ = periodicity in $\ztf$ direction/$d$\\
	\ \\
	The overall extent of the TUC (the ``computational volume'')
	is determined by parameters 
	{\tt (SHPAR1 +SHPAR4), max(SHPAR$_2$,SHPAR$_4$}), and 
	{\tt max(SHPAR$_3$,SHPAR$_5$)}.
	The periodicity in the TF $y$ and $z$ directions is determined
	by parameters {\tt SHPAR}$_6$ and {\tt SHPAR}$_7$.\\
	The physical size of the TUC is specified by the value of
	$\aeff$ (in physical units, e.g. cm -- the same unit
        as used to specify the wavelength), 
	specified in the file {\tt ddscat.par}.

	The target
	is a periodic structure, of infinite extent in the target
	y- and z- directions.
	The target axes (in the TF) 
	are set to ${\hat{\bf a}}_1 = \xtf = (1,0,0)_\TF$ --
	i.e., normal to the ``slab'' -- and
	${\hat{\bf a}}_2 = \ytf= (0,1,0)_\TF$.
	The orientation of the incident radiation relative to the target
	is, as for all other targets, set by the usual orientation
	angles $\beta$, $\Theta$, and $\Phi$ 
	(see \S\ref{sec:target_orientation} above); for example,
	$\Theta=0$ would be for radiation incident normal to the slab.


\subsubsection{\bf Sample calculation in directory examples\_exp/DSKRCTPBC}

Subdirectory {\tt examples\_exp/DSKRCTPBC} contains {\tt ddscat.par} for
calculating scattering by a $0.0500\micron$ thick Si$_3$N$_4$ slab supporting
a doubly periodic array of Au disks, with periodicity $0.0800\micron$,
disk diameter $0.0400\micron$, and disk height $0.0200\micron$, for
light with wavelength $\lambda=0.5320\micron$.

The TUC consists of a rectangular block of Si$_3$N$_4$, of dimension
$15d\times24d\times24d$ (8640 dipoles), 
supporting an Au disk of thickness $6d$ and
a diameter $\sim$$12d$ (672 dipoles).  With a ``diameter'' of only $12d$,
the cross section of the ``disk'' is only roughly circular; each layer
of the disk contains 112 dipoles.

The volume of the ideal TUC is $V_{\rm TUC}=(0.08)^2\times0.05 + \pi (0.02)^2\times0.02=3.4513\times10^{-4}\micron^3$.  Thus we set $\aeff=(3V_{\rm TUC}/4\pi)^{1/3}=4.3514\times10^{-2}\micron$.
The DDA calculation has $kd=.039377$.

The radiation is incident at an angle of $60^\circ$ relative to the
surface normal.
The entire calculation required 2400 cpu~sec on a 2.53 GHz cpu.
Most of the cpu time was spent computing the effective $\bA$ matrix
for the calculation, which requires extensive summations; 
once this was obtained, the solution was found
in 31 and 33 iterations, respectively, for the two incident polarizations,
requiring $\sim$8 cpu~sec.

\index{target shape options!HEXGONPBC}
\index{periodic boundary conditions}
\subsection{ HEXGONPBC = Target consisting of homogeneous hexagonal prism
                     repeated in
                     target y and/or z directions using
                     periodic boundary conditions}
            \label{sec:HEXGONPBC}
            This option causes {{\bf DDSCAT}}\ to create a target consisting
	    of a periodic or biperiodic array of hexagonal prisms.
	    The individual prisms are assumed to be homogeneous and isotropic,
	    just as for option RCTNGL (see \S\ref{sec:RCTGLPRSM}).
	    \ \\
	    Let us 
	    refer to a single hexagonal prism as the Target Unit Cell (TUC).
	    The TUC 
	    is then repeated in the target y- and z-directions, with 
	    periodicities {\tt PYD}$\times d$\index{PYD} and 
	    {\tt PZD}$\times d$,\index{PZD} where $d$ is the lattice spacing.
	    To repeat in only one direction, set either {\tt PYD} or
	    {\tt PZD} to zero.\\
	    This option requires 5 shape parameters:
	    The pertinent line in {\tt ddscat.par} should read\\
	    \ \\
	{\tt SHPAR$_1$ SHPAR$_2$ SHPAR3 SHPAR4 SHPAR}$_5$\\
	    \ \\
	where{\tt SHPAR$_1$}, {\tt SHPAR$_2$}, {\tt SHPAR}$_3$, 
        {\tt SHPAR}$_4$, {\tt SHPAR}$_5$ are numbers: \\
	{\tt SHPAR$_1$} = prism length along prism axis (in units of $d$)
	in units of $d$\\
	{\tt SHPAR$_2$} = 2$\times$length of one hexagonal side/$d$\\
	{\tt SHPAR}$_3$ = 1,2,3,4,5 or 6 to specify prism orientation
	                 in the TF (see below)\\
	{\tt SHPAR}$_4$ = {\tt PYD} = periodicity in TF $y$ direction/$d$\\
	{\tt SHPAR}$_5$ = {\tt PZD} = periodicity in TF $z$ direction/$d$\\
	\ \\
	The overall size of the TUC (in terms of numbers of
	dipoles) is determined by parameters 
	{\tt SHPAR$_1$, SHPAR$_2$}, and {\tt SHPAR}$_3$.
	The periodicity in the TF $y$ and $z$ directions is determined
	by parameters {\tt SHPAR}$_4$ and {\tt SHPAR}$_5$.\\
	The physical size of the TUC is specified by the value of
	$\aeff$ (in physical units, e.g. cm), 
	specified in the file {\tt ddscat.par}, with the usual
	correspondence $d=(4\pi/3N)^{1/3}\aeff$, where $N$ is the number
	of dipoles in the TUC.

	With target option {\tt HEXGONPBC}, the target
	becomes a periodic structure, of infinite extent in the target
	y- and z- directions (assuming both {\tt NPY} and {\tt NPZ} are
	nonzero).

	The target axes (in the TF) 
	are set to ${\hat{\bf a}}_1 = \xtf = (1,0,0)_\TF$ --
	i.e., normal to the ``slab'' -- and
	${\hat{\bf a}}_2 = \ytf = (0,1,0)_\TF$.

	The individual
	hexagons may have any of 6 different orientations relative to the
	slab: 
	Let unit vectors $\hat{\bf h}$ be $\parallel$ to the axis of the 
	hexagonal prism, and 
	let unit vector $\hat{\bf f}$ be normal to one of the rectangular
	faces of the hexagonal prism.
	Then\\
	{\tt SHPAR3=1} for $\hat{\bf h}\parallel \hat{\bf x}_{\rm TF}$,
	                  $\hat{\bf f}\parallel \hat{\bf y}_{\rm TF}$\\
	{\tt SHPAR3=2} for $\hat{\bf h}\parallel \hat{\bf x}_{\rm TF}$,
	                  $\hat{\bf f}\parallel \hat{\bf z}_{\rm TF}$\\
	{\tt SHPAR3=3} for $\hat{\bf h}\parallel \hat{\bf y}_{\rm TF}$,
	                  $\hat{\bf f}\parallel \hat{\bf x}_{\rm TF}$\\
	{\tt SHPAR3=4} for $\hat{\bf h}\parallel \hat{\bf y}_{\rm TF}$,
	                  $\hat{\bf f}\parallel \hat{\bf z}_{\rm TF}$\\
	{\tt SHPAR3=5} for $\hat{\bf h}\parallel \hat{\bf z}_{\rm TF}$,
	                  $\hat{\bf f}\parallel \hat{\bf x}_{\rm TF}$\\
	{\tt SHPAR3=6} for $\hat{\bf h}\parallel \hat{\bf z}_{\rm TF}$,
	                  $\hat{\bf f}\parallel \hat{\bf y}_{\rm TF}$\\

	For example, one could construct a single infinite hexagonal column
	\index{infinite hexagonal column}
	with the following line in {\tt ddscat.par}:\\
	\ \\
	2.0\ \ \ 100.0\ \ \  3\ \ \  2.0\ \ \  0.\\
	\ \\
	The TUC would be a thin hexagonal ``slice'' containing two layers
	of dipoles.  The edges of the hexagon would be about 50$d$ in
	extent, so the TUC would have approximately 
	$(3\sqrt{3}/2)\times50^2\times2=
	12990$ dipoles (13024 in the actual realization)
	within a $90\times2\times100$ ``extended target volume''.
	The TUC would be oriented with the hexagonal axis in the 
	$\hat{\bf y}_{\rm TF}$ 
	direction, with $\hat{\bf z}_{\rm TF}$ normal to a rectangular
	faces of the prism
	({\tt SHPAR3=3}), and the structure would repeat in the 
	$\hat{\bf y}_{\rm TF}$ direction
	with a period of $2\times d$ ({\tt SHPAR4=2.0}).
	{\tt SHPAR5=0} means that there will be no repetition in the 
	$\hat{\bf z}_{\rm TF}$
	direction.
	
	Note that {\tt SHPAR$_1$}, {\tt SHPAR$_2$}, {\tt SHPAR}$_4$, and 
        {\tt SHPAR}$_5$ need not be integers.  
        However, {\tt SHPAR}$_5$, determining the
	orientation of the prisms in the TF, can only take on the values
	1,2,3,4,5,6.

	{\bf Important Note:} 
	For technical reasons, {\tt PYD} and {\tt PZD} must not be smaller than
	the ``extended'' target extent in the $\hat{\bf y}_{\rm TF}$ 
	and $\hat{\bf z}_{\rm TF}$ directions.
	When the GPFAFT\index{GPFAFT} option is used 
	for the 3-dimensional FFT calculations, 
	the extended target volume always
	has dimensions/$d = 2^a3^b5^c$, where $a$, $b$, and $c$ are
	nonnegative 
	integers, with (dimension/$d)\geq 1$).  

	The orientation of the incident radiation relative to the target
	is, as for all other targets, set by the usual orientation
	angles $\beta$, $\Theta$, and $\Phi$ 
	(see \S\ref{sec:target_orientation} above); for example,
	$\Theta=0$ would be for radiation incident normal to the periodic
	structure.

\index{target shape options!LYRSLBPBC}
\index{periodic boundary conditions}
\subsection{ LYRSLBPBC = Target consisting of layered slab,
                     extended in
                     target y and z directions using
                     periodic boundary conditions}
            \label{sec:LYRSLBPBC}
            This option causes {{\bf DDSCAT}}\ to create a target consisting
	    of an array of multilayer bricks, layered in the $\xtf$ direction.
	    The size of each brick 
	    in the $\ytf$ and $\ztf$ direction is specified.
	    Up to 4 layers are allowed.\\
	    The bricks are repeated in the $\ytf$ and $\ztf$ direction
	    with a specified periodicity.
	    If $L_y=P_y$ and $L_z=P_z$, then the target
	    consists of a continuous multilayer slab.  For this case,
	    it is most economical to set $L_y/d=L_z/d=P_y/d=P_z/d=1$.\\
            If $P_y=0$, then repetition in the $\ytf$ direction is
            suppressed -- the target repeats only in the $\ztf$ direction.\\
            If $P_z=0$, then repetition in the $\ztf$ direction is
            suppressed -- the target repeats only in the $\ytf$ direction.\\
	    The upper surface of the slab is asssume to be located at
	    $x_\TF=0$.  The lower surface of the slab is at
	    $x_\TF=-L_x=-$~{\tt SHPAR1}$\times d$.\\
	    The multilayer slab geometry is specified with 9 parameters.
	    The pertinent line in {\tt ddscat.par} should read \\
	    \ \\
	{\tt SHPAR$_1$ SHPAR$_2$ SHPAR$_3$ SHPAR$_4$ SHPAR$_5$ SHPAR$_6$
             SHPAR$_7$ SHPAR$_8$ SHPAR$_9$}\\
	    \ \\
	where\\
	{\tt SHPAR}$_1$ = $L_x/d$ = (brick thickness in $\xtf$ direction)$/d$\\
	{\tt SHPAR}$_2$ = $L_y/d$ = (brick extent in $\ytf$ direction)/$d$\\
	{\tt SHPAR}$_3$ = $L_z/d$ = (brick extent in $\ztf$ direction)/$d$\\
	{\tt SHPAR}$_4$ = fraction of the slab with composition 1\\
	{\tt SHPAR}$_5$ = fraction of the slab with composition 2\\
	{\tt SHPAR}$_6$ = fraction of the slab with composition 3\\
	{\tt SHPAR}$_7$ = fraction of the slab with composition 4\\
	{\tt SHPAR}$_8$ = $P_y/d$ = (periodicity in $\ytf$ direction)/$d$
             (0 to suppress repetition in $\ytf$ direction)\\
	{\tt SHPAR}$_9$ = $P_z/d$ = (periodicity in $\ztf$ direction)/$d$
             (0 to suppress repetition in $\ztf$ direction)\\
	\ \\
	For a slab with only one layer, set {\tt SHPAR}$_5=0$,
	{\tt SHPAR}$_6=0$, {\tt SHPAR}$_7=0$.\\
	For a slab with only two layers, set {\tt SHPAR}$_6=0$ and
	{\tt SHPAR}$_7=0$.\\
	For a slab with only three layers, set {\tt SHPAR}$_7=0$.\\
	The user must set {\tt NCOMP} equal to
	the number of nonzero thickness layers.\\
	The number $N$ of dipoles in one TUC is
	$N={\rm nint}({\tt SHPAR}_1)\times{\rm nint}({\tt SHPAR}_2)\times
	{\rm nint}({\tt SHPAR}_3)$.
	
	The fractions {\tt SHPAR$_4$}, {\tt SHPAR}$_5$, {\tt SHPAR}$_7$,
	and {\tt SHPAR}$_7$ {\it must} sum to 1.
	The number of dipoles in each of the layers will be integers
	that are close to ${\rm  nint}({\tt SHPAR}_1*{\tt SHPAR}_4)$, 
	${\rm  nint}({\tt SHPAR}_1*{\tt SHPAR}_5)$, 
	${\rm  nint}({\tt SHPAR}_1*{\tt SHPAR}_6)$, 
	${\rm  nint}({\tt SHPAR}_1*{\tt SHPAR}_7)$, 

	The physical size of the TUC is specified by the value of
	$\aeff$ (in physical units, e.g. cm), 
	specified in the file {\tt ddscat.par}.  Because of the way
	$\aeff$ is defined 
	($4\pi \aeff^3/3 \equiv Nd^3$),
	it should be set to
	$$ \aeff = (3/4\pi)^{1/3} N^{1/3} d = (3/4\pi)^{1/3}
	(L_x L_y L_z)^{1/3}$$.

	With target option {\tt LYRSLBPBC}, the target
	becomes a periodic structure, of infinite extent in the target
	y- and z- directions.
	The target axes (in the TF) 
	are set to ${\hat{\bf a}}_1 = (1,0,0)_\TF$ --
	i.e., normal to the ``slab'' -- and
	${\hat{\bf a}}_2 = (0,1,0)_\TF$.
	The orientation of the incident radiation relative to the target
	is, as for all other targets, set by the usual orientation
	angles $\beta$, $\Theta$, and $\Phi$ 
	(see \S\ref{sec:target_orientation} above); for example,
	$\Theta=0$ would be for radiation incident normal to the slab.

	For this option, there are only two allowed scattering directions,
	corresponding to transmission and specular reflection.
	\ddscat\ will calculate both the transmission and reflection
	coefficients.\\
	The last two lines in {\tt ddscat.par} should appear as in the
	following example {\tt ddscat.par} file.  This example is
	for a slab with two layers: the slab is 26 dipole layers thick;
	the first layer comprises 76.92\%
	of the thickness, the second layer 23.08\% of the thickness.
	The wavelength is $0.532\mu$m, the thickness is
	$L_x=(4\pi/3)^{1/3}N_x^{2/3}\aeff=(4\pi/3)^{1/3}(26)^{2/3}0.009189=
	0.1300\micron$.\\
	The upper layer thickness is $0.2308L_x=0.0300\micron$\\
	{\tt ddscat.par} below is set up to calculate a single orientation:
	in the Lab Frame, 
	the target is rotated through an angle $\Theta=120^\circ$, with
	$\Phi=0$.  In this orientation, the incident radiation is propagating
	in the $(-0.5,0.866,0)$ direction in the Target Frame, so that it
	is impinging on target layer 2 (Au).
	\\
{\footnotesize\begin{verbatim}
' =========== Parameter file for v7.3 ===================' 
'**** PRELIMINARIES ****'
'NOTORQ' = CMTORQ*6 (DOTORQ, NOTORQ) -- either do or skip torque calculations
'PBCGS2' = CMDSOL*6 (PBCGS2, PBCGST, PETRKP) -- CCG method
'GPFAFT' = CMETHD*6 (GPFAFT, FFTMKL) -- FFT method
'GKDLDR' = CALPHA*6 (GKDLDR, LATTDR, FLTRCD) -- DDA method
'NOTBIN' = CBINFLAG (ALLBIN, ORIBIN, NOTBIN)
'**** Initial Memory Allocation ****'
26  1  1  = upper bounds on size of TUC 
'**** Target Geometry and Composition ****'
'LYRSLBPBC' = CSHAPE*9 shape directive
26 1 1 0.7692 0.2308 0 0 1 1 = shape parameters SHPAR1 - SHPAR9
2         = NCOMP = number of dielectric materials
'/u/draine/work/DDA/diel/Eagle_2000' = refractive index 1
'/u/draine/work/DDA/diel/Au_evap'    = refractive index 2
'**** Additional Nearfield calculation? ****'
0 = NRFLD (=0 to skip nearfield calc., =1 to calculate nearfield E)
0 0 0 0 0 0 (fract. extens. of calc. vol. in -x,+x,-y,+y,-z,+z)
'**** Error Tolerance ****'
1.00e-5 = TOL = MAX ALLOWED (NORM OF |G>=AC|E>-ACA|X>)/(NORM OF AC|E>)
'**** Maximum number of iterations ****'
100     = MXITER
'**** Interaction cutoff parameter for PBC calculations ****'
5.00e-3 = GAMMA (1e-2 is normal, 3e-3 for greater accuracy)
'**** Angular resolution for calculation of <cos>, etc. ****'
0.5	= ETASCA (number of angles is proportional to [(3+x)/ETASCA]^2 )
'**** Wavelengths (micron) ****'
0.5320 0.5320 1 'INV' = wavelengths (first,last,how many,how=LIN,INV,LOG)
'**** Effective Radii (micron) **** '
0.009189 0.009189 1 'LIN' = eff. radii (first,last,how many,how=LIN,INV,LOG)
'**** Define Incident Polarizations ****'
(0,0) (1.,0.) (0.,0.) = Polarization state e01 (k along x axis)
2 = IORTH  (=1 to do only pol. state e01; =2 to also do orth. pol. state)
'**** Specify which output files to write ****'
1 = IWRKSC (=0 to suppress, =1 to write ".sca" file for each target orient.
'**** Prescribe Target Rotations ****'
0.   0.   1  = BETAMI, BETAMX, NBETA (beta=rotation around a1)
120. 120. 1  = THETMI, THETMX, NTHETA (theta=angle between a1 and k)
0.   0.   1  = PHIMIN, PHIMAX, NPHI (phi=rotation angle of a1 around k)
'**** Specify first IWAV, IRAD, IORI (normally 0 0 0) ****'
0   0   0    = first IWAV, first IRAD, first IORI (0 0 0 to begin fresh)
'**** Select Elements of S_ij Matrix to Print ****'
6	= NSMELTS = number of elements of S_ij to print (not more than 9)
11 12 21 22 31 41	= indices ij of elements to print
'**** Specify Scattered Directions ****'
'TFRAME' = CMDFRM (LFRAME, TFRAME for Lab Frame or Target Frame)
1 = number of scattering orders
0.  0. = (M,N) for scattering
\end{verbatim}}
\index{target shape options!RCTGL\_PBC}
\index{periodic boundary conditions}
\subsection{ RCTGL\_PBC = Target consisting of homogeneous rectangular brick, 
                     extended in
                     target y and z directions using
                     periodic boundary conditions}
            \label{sec:RCTGL_PBC}
            This option causes {{\bf DDSCAT}}\ to create a target consisting
	    of a biperiodic array of rectangular bricks.  
	    The bricks are assumed to be homogeneous and isotropic,
	    just as for option RCTNGL (see \S\ref{sec:RCTGLPRSM}).
	    \ \\
	    Let us 
	    refer to a single rectangular brick as the Target Unit Cell (TUC).
	    The TUC 
	    is then repeated in the $y_\TF$- and $z_\TF$-directions, with 
	    periodicities {\tt PYAEFF}$\times \aeff$ and 
	    {\tt PZAEFF}$\times \aeff$, where 
	    $\aeff\equiv(3V_{\rm TUC}/4\pi)^{1/3}$,
	    where $V_{\rm TUC}$ 
	    is the total volume of solid material in one TUC.
	    This option requires 5 shape parameters:\newline
	    The pertinent line in {\tt ddscat.par} should read\\
	    \ \\
	${\tt SHPAR}_1~~{\tt SHPAR}_2~~{\tt SHPAR}_3~~{\tt SHPAR}_4~~{\tt SHPAR}_5$\\
	    \ \\
	where\\
	{\tt SHPAR$_1$} = (brick thickness)/$d$ in the $x_\TF$ direction\\
	{\tt SHPAR$_2$} = (brick thickness)/$d$ in the $y_\TF$ direction\\
	{\tt SHPAR}$_3$ = (brick thickness)/$d$ in the $z_\TF$ direction\\
	{\tt SHPAR}$_4$ = periodicity/$d$ in the $y_\TF$ direction\\
	{\tt SHPAR}$_5$ = periodicity/$d$ in the $z_\TF$ direction\\
	\ \\
	The overall size of the TUC (in terms of numbers of
	dipoles) is determined by parameters 
	{\tt SHPAR$_1$, SHPAR$_2$}, and {\tt SHPAR}$_3$.
	The periodicity in the $y_\TF$ and $z_\TF$ directions is determined
	by parameters {\tt SHPAR}$_4$ and {\tt SHPAR}$_5$.\\
	The physical size of the TUC is specified by the value of
	$\aeff$ (in physical units, e.g. cm), 
	specified in the file {\tt ddscat.par}.

	With target option {\tt RCTGL\_PBC}, the target
	becomes a periodic structure, of infinite extent in the target
	y- and z- directions.
	The target axes (in the TF) 
	are set to ${\hat{\bf a}}_1 = (1,0,0)_\TF$ --
	i.e., normal to the ``slab'' -- and
	${\hat{\bf a}}_2 = (0,1,0)_\TF$.
	The orientation of the incident radiation relative to the target
	is, as for all other targets, set by the usual orientation
	angles $\beta$, $\Theta$, and $\Phi$ 
	(see \S\ref{sec:target_orientation} above); for example,
	$\Theta=0$ would be for radiation incident normal to the slab.


\subsubsection{\bf Sample calculation in directory examples\_exp/RCTGL\_PBC}

Subdirectory {\tt examples\_exp/RCTGL\_PBC} contains {\tt ddscat.par}
for scattering by an infinite slab, constituted from $20\times1\times1$
dipole TUCs.
The wavelength $\lambda=0.5\micron$, and the slab thickness is $h=0.10\micron$.
The slab has refractive index $m=1.50+0.02i$.
The interdipole spacing $d=0.32\micron/20=0.005\micron$.
The TUC has dimension $V=0.10\micron\times0.005\micron\times0.005\micron$,
and hence 
$\aeff=(3V/4\pi)^{1/3}=(3\times0.10\times0.005\times0.005/4\pi)^{1/3}\micron =
0.0084195\micron$.
The incident radiation is at an angle $\theta_i=40^\circ$ relative
to the surface normal.

\subsubsection{\bf Sample calculation in directory 
               examples\_exp/RCTGL\_PBC\_NEARFIELD}

The directory {\tt examples\_exp/RCTGL\_PBC\_NEARFIELD} contains
{\tt ddscat.par} for calculation of scattering and absorption by
an infinite slab of material with refractive index $m=1.50+0.02i$ and
thickness $h=0.10\micron$ in
vacuo.  The incident radiation has wavelength $\lambda_{\rm vac}=0.5\micron$
and incidence angle $\theta_i=40^\circ$.
The interdipole spacing is set to $d=h/20 = 0.005\micron$.
This is the example problem shown in Fig.\ 7b of \citet{Draine+Flatau_2008a}.

The slab is treated as a periodic array of $1d\times1d\times20d$ structures
with periodicity $P_y=1d$ and $P_z=1d$.
The volume of the TUC is $V_{\rm TUC}=20d^3=2.50\times10^{-6}\micron^3$,
and the effective radius is
$\aeff=(3V_{\rm TUC}/4\pi)^{1/3}=0.0084195\micron$. 

\begin{figure}[h]
\begin{center}
\vspace*{-0.1cm}
%
\includegraphics[width=8.cm,angle=270]{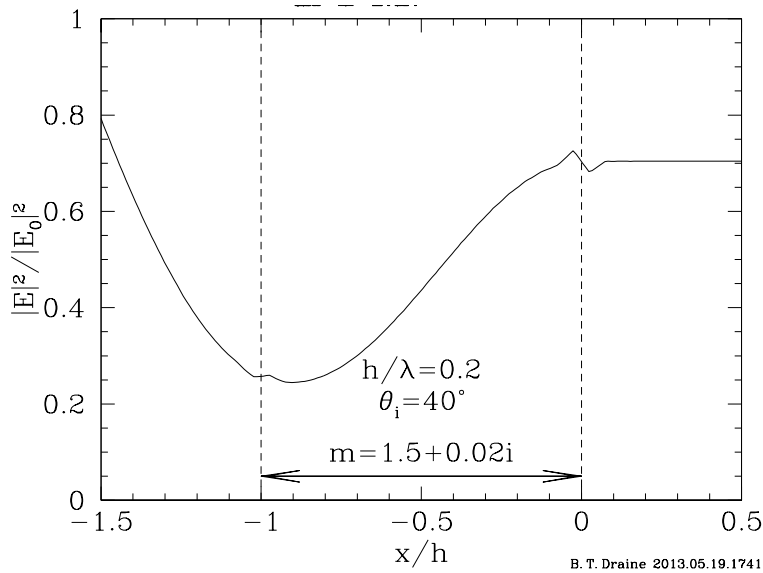}
\vspace*{-0.4cm}
\caption{\footnotesize
         Normalized macroscopic electric field intensity 
         $|\bE|^2/|\bE_0|^2$ along
         a line normal to a slab of thickness $h$, 
         refractive index $m=1.5+0.02i$.
         Radiation with $\lambda/h=5$ is incident with incidence
         angle $\theta_i=40^\circ$.
         The interdipole spacing is set to $d=h/20=0.01\lambda$.
         This is the same problem reported in Fig.\ 7b of 
         \citet{Draine+Flatau_2008a}.
         Eq.\ (\ref{eq:macro_vs_micro}) gives the relation between
         microscopic and macroscopic $\bE$ fields.
         }
\end{center}
\end{figure}
\subsubsection{\bf Sample calculation in directory 
               examples\_exp/RCTGL\_PBC\_NEARFLD\_B}

The directory {\tt examples\_exp/RCTGL\_PBC\_NEARFLD\_B} contains
{\tt ddscat.par} for calculation of scattering and absorption by
an infinite slab of material with refractive index $m=1.50+0.02i$ and
thickness $h=0.1\micron$ in
vacuo.  The incident radiation has wavelength $\lambda_{\rm vac}=0.5\micron$
and incidence angle $\theta_i=40^\circ$.
The interdipole spacing is set to $d=h/20 = 0.005\micron$.
This is the example problem shown in Fig. 7b of \citet{Draine+Flatau_2008a},
and is the same problem as in directory examples\_exp/RCTGL\_PBC\_NEARFIELD.

The slab is treated as a periodic array of $1d\times1d\times20d$ structures
with periodicity $P_y=1d$ and $P_z=1d$.
The volume of the TUC is $V_{\rm TUC}=20d^3=2.5\times10^{-6}\micron^3$,
and the effective radius is
$\aeff=(3V_{\rm TUC}/4\pi)^{1/3}=0.0084195\micron$. 

In this example, we set 
{\tt NRFLD}=2 so that, 
in addition to calculating $\bE$ in the nearfield volume,
\ddscatv\ will calculate $\bB$ throughout the nearfield volume.
With both $\bE$ and $\bB$ available, we can also compute the
Poynting vector $(c/4\pi)\bE\times\bB$.

\begin{figure}[h]
\begin{center}
\vspace*{-0.1cm}
\includegraphics[width=8.0cm,angle=0,
                 clip=true,trim=0.5cm 2.0cm 0.5cm 2.0cm]
                {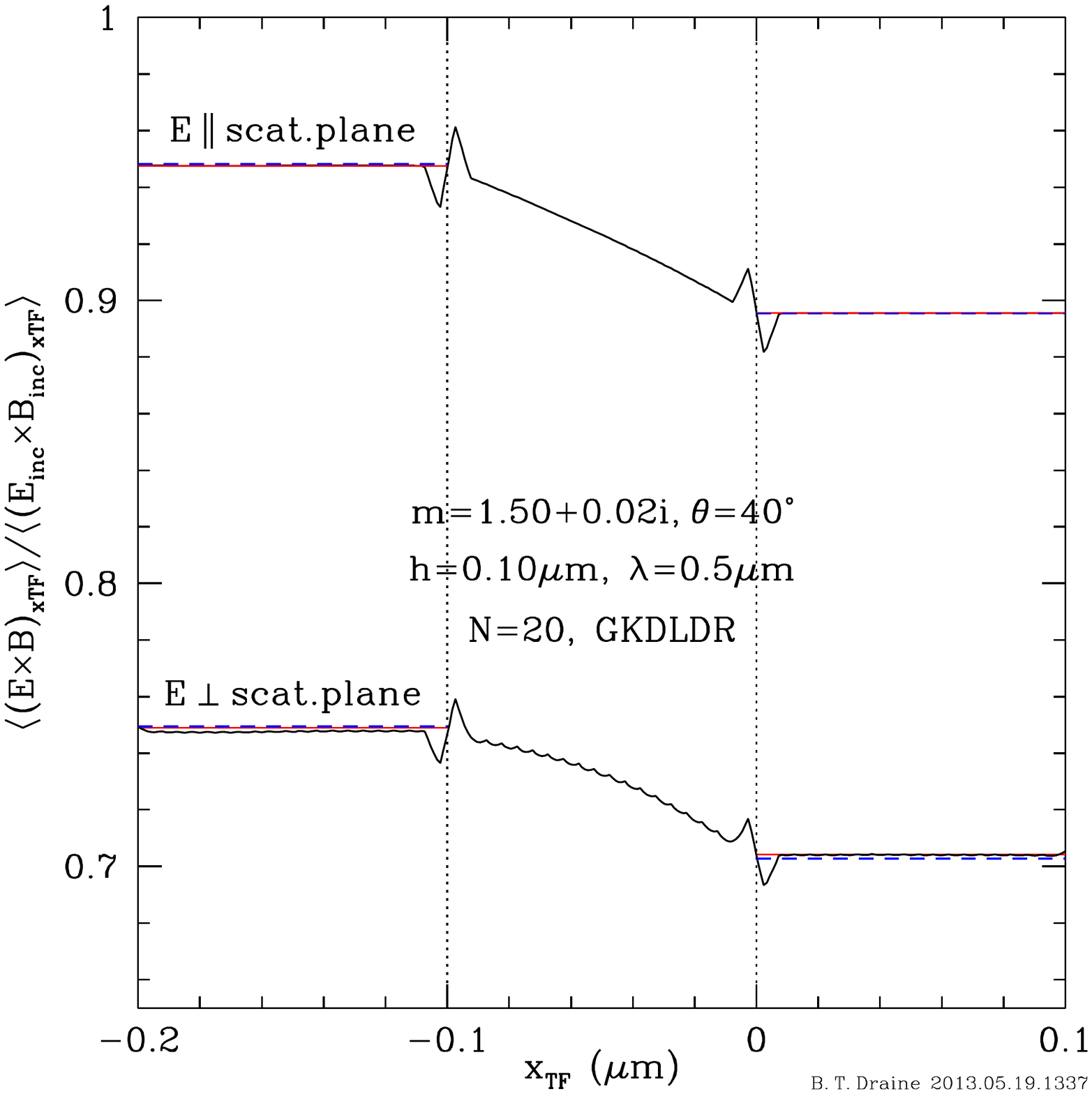}
\includegraphics[width=8.0cm,angle=0,
                 clip=true,trim=0.5cm 2.0cm 0.5cm 2.0cm]
                {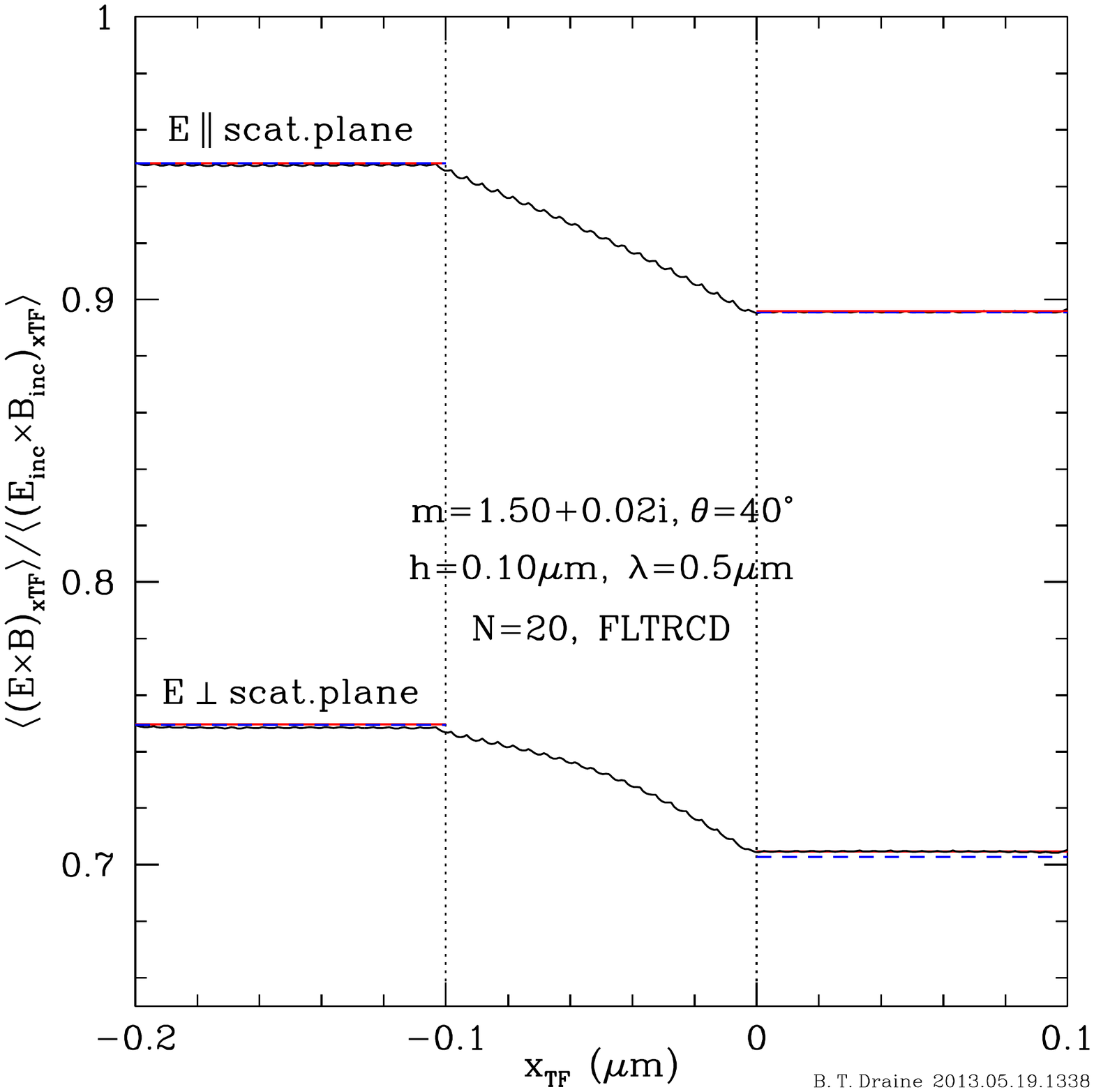}
\vspace*{-0.4cm}
\caption{\label{fig:example_RCTGL_PBC_NEARFLD_B}
         \footnotesize
         Component of Poynting flux normal to
         a slab of thickness $h=0.10\micron$, 
         refractive index $m=1.5+0.02i$.
         Radiation with $\lambda/h=5$ is incident with incidence
         angle $\theta_i=40^\circ$.
         The interdipole spacing is set to $d=h/20=0.005\micron$.
         This is the problem considered in Fig. 7b of 
         \citet{Draine+Flatau_2008a}.
         Left: obtained with DDA method GKDLDR.
         Right: obtained with DDA method FLTRCD.
         The red and blue lines show $(1-R)$ for $x_{\rm TF}<-0.1\micron$, 
         and $T$ for $x_{\rm TF}>0$,
         where $R$ and $T$ are the reflection and transmission coefficients.
         The blue (dashed) lines correspond to the
         exact solution for $T$ and $R$.  The red (solid) line uses the
         values of $T$ and $R$ computed by \ddscatv:
         $T_\parallel=S_{11}(\theta_T)+S_{12}(\theta_T)$,
         $R_\parallel=S_{11}(\theta_R)+S_{12}(\theta_R)$,
         $T_\perp=S_{11}(\theta_T)-S_{12}(\theta_T)$,
         $R_\perp=S_{11}(\theta_R)-S_{12}(\theta_R)$,
         with $S_{ij}$ from the output file 
         {\tt w000r000k000.sca}, $\theta_T=40^\circ$, and
         $\theta_R=140^\circ$.
         }
\end{center}
\end{figure}

In Figure \ref{fig:example_RCTGL_PBC_NEARFLD_B} we show
the component of the time-averaged Poynting flux normal to the slab,
$(c/4\pi)\langle(\bE\times\bB)\rangle\cdot \xtf$, divided by the time-averaged
value of the incident
Poynting flux, $(c/8\pi)E_0^2\cos\theta_i$.
Figure \ref{fig:example_RCTGL_PBC_NEARFLD_B}a shows results computed
using the standard DDA (method option {\tt GKDLDR}) and Figure 
\ref{fig:example_RCTGL_PBC_NEARFLD_B}b shows results computed using
the filtered couple dipole method (option {\tt FLTRCD}).
Both methods give accurate results.
The main difference is in the computed $\bE$ field near the surface of
the slab: the standard "point dipole" calculation (GKDLDR) has small-scale
structure in the computed $\bE$ near the surface, resulting in errors of
a few \% in the computed Poynting vector near the surface (see Figure
\ref{fig:example_RCTGL_PBC_NEARFLD_B}a).
The filtered coupled dipole method, on the other hand, explicitly filters
out the small-scale structure in the $\bE$ field, thereby suppressing
the "blips" in the Poynting flux near the two surfaces of the slab
(see Figure \ref{fig:example_RCTGL_PBC_NEARFLD_B}b).
Otherwise the two solutions are nearly identical, and both are smooth
near the middle of the slab.

The numerical accuracy can be assessed from Table \ref{tab:GKDLDR_vs_FLTRCD}.
For this problem, with $N=20$, 
both GKDLDR and FLTRCD give reflection and transmission coefficients
accurate to better than 0.5\%.
The absorption coefficients are somewhat less accurate, but the
fractional errors do not exceed 4.2\%.
Note that the filtered coupled dipole method (FLTRCD)
does not give greater accuracy than GKDLDR, at least for this problem.
If desired, the accuracy could be further improved by increasing $N$.

The CPU time required for the filtered coupled dipole calculations is
about twice that required for the GKDLDR method, because for this
calculation the dominant calculational task is calculation of the
elements of the dipole-dipole interaction matrix (or Green function).

\begin{table}[h]
\begin{center}
\caption{\label{tab:GKDLDR_vs_FLTRCD}Results for $m=1.5+0.02i$, $h/\lambda=0.2$, $\theta_i=40^\circ$}
{\footnotesize
\begin{tabular}{c c c c}
\hline
quantity                    & exact    & GKDLDR   & FLTRCD   \tabularnewline
\hline
$R_\parallel$               & $0.051875$ & $0.05179$ & $0.051740$\tabularnewline
$R_\perp$                   & $0.25046$  & $0.25047$ & $0.25042$ \tabularnewline
$T_\parallel$               & $0.89545$  & $0.89534$ & $0.89592$ \tabularnewline
$T_\perp$                   & $0.70280$  & $0.70524$ & $0.70478$ \tabularnewline
$A_\parallel$               & 0.052675 & $0.052482$ & $0.052259$ \tabularnewline
$A_\perp$                   & 0.046748 & $0.044970$ & $0.044801$ \tabularnewline
$R_\parallel+T_\parallel+A_\parallel$
                            & 1        & $0.99961$  & $0.99992$ \tabularnewline
$R_\perp+T_\perp+A_\perp$   & 1        & $1.00068$  & $1.00000$ \tabularnewline
\hline
$R_\parallel/{\rm exact}-1$ & 0         & $-0.0017$ & $-0.0026$ \tabularnewline
$R_\perp/{\rm exact}-1$     & 0        & $+0.0001$ & $-0.0002$ \tabularnewline
$T_\parallel/{\rm exact}-1$ & 0        & $-0.0001$ & $+0.0005$ \tabularnewline
$T_\perp/{\rm exact}-1$     & 0        & $+0.0035$ & $+0.0042$ \tabularnewline
$A_\parallel/{\rm exact}-1$ & 0        & $-0.0037$ & $-0.0079$ \tabularnewline
$A_\perp/{\rm exact}-1$     & 0        & $-0.0380$ & $-0.0416$ \tabularnewline
\hline
CPU time (s) (2.53GHz Intel, 1 core)           & --       & 1298.     & 2770. \tabularnewline
\hline
\end{tabular}
}
\end{center}
\end{table}

\index{target shape options!RECRECPBC}
\index{periodic boundary conditions}
\subsection{ RECRECPBC = Rectangular solid resting on top of another
                        rectangular solid, repeated periodically in
                        target y and z directions using
                        periodic boundary conditions}
            \label{sec:RECRECPBC}
	    The TUC consists of a single rectangular ``brick'', of material
	    1, resting on top of a second rectangular brick, of material 2.
	    The centroids of the two bricks along a line in the $\xtf$
	    direction.
	    The bricks are assumed to be homogeneous, and materials 1 and
	    2 are assumed to be isotropic.
	    The TUC 
	    is then repeated in the $y_\TF$- and $z_\TF$-directions, with 
	    periodicities {\tt SHPAR}$_4\times d$ and 
	    {\tt SHPAR}$_5\times d$. 
	    is the total volume of solid material in one TUC.
	    This option requires 8 shape parameters:\newline
	    The pertinent line in {\tt ddscat.par} should read\\
	    \ \\
	{\tt SHPAR$_1$ SHPAR$_2$ SHPAR$_3$ SHPAR$_4$ SHPAR$_5$
             SHPAR$_6$ SHPAR$_7$ SHPAR$_8$}\\
	    \ \\
	where\\
	{\tt SHPAR}$_1$ = (upper brick thickness)/$d$ in the 
	$\xtf$ direction\\
	{\tt SHPAR}$_2$ = (upper brick thickness)/$d$ in the 
	$\ytf$ direction\\
	{\tt SHPAR}$_3$ = (upper brick thickness)/$d$ in the 
	$\ztf$ direction\\
	{\tt SHPAR}$_4$ = (lower brick thickness)/$d$ in the 
	$\xtf$ direction\\
	{\tt SHPAR}$_5$ = (lower brick thickness)/$d$ in the 
	$\ytf$ direction\\
	{\tt SHPAR}$_6$ = (lower brick thickness)/$d$ in the 
	$\ztf$ direction\\
	{\tt SHPAR}$_7$ = periodicity/$d$ in the $\ytf$ direction\\
	{\tt SHPAR}$_8$ = periodicity/$d$ in the $\ztf$ direction\\
	\ \\
	The actual numbers of dipoles $N_{1x}$, $N_{1y}$, $N_{1z}$,
	along each dimension of the upper brick, and $N_{2x}$, $N_{2y}$,
	$N_{2z}$ along each dimension of the lower brick, must be integers.
	Usually, 
	$N_{1x}={\rm nint}({\tt SHPAR}_1)$,
	$N_{1y}={\rm nint}({\tt SHPAR}_2)$,
	$N_{1z}={\rm nint}({\tt SHPAR}_3)$,
	$N_{2x}={\rm nint}({\tt SHPAR}_4)$,
	$N_{2y}={\rm nint}({\tt SHPAR}_5)$,
	$N_{2z}={\rm nint}({\tt SHPAR}_6)$,
	where ${\rm nint}(x)$ is the integer nearest to $x$, but
	under some circumstances $N_{1x}$, $N_{1y}$, $N_{1z}$,
	$N_{2x}$, $N_{2y}$, $N_{2z}$ might be larger or smaller by 1 unit.
	
	The overall size of the TUC (in terms of numbers of
	dipoles) is determined by parameters
	{\tt SHPAR}$_1$ -- {\tt SHPAR}$_6$: 
\beq
N = \left(N_{1x}\times N_{1y}\times N_{1z}\right)
    + \left(N_{2x}\times N_{2y}\times N_{2z}\right)
\eeq
	The physical size of the TUC is specified by the value of
	$\aeff$ (in physical units, e.g. cm), 
	specified in the file {\tt ddscat.par}:
\beq
d = (4\pi/3N)^{1/3}\aeff
\eeq

	The periodicity in the $\ytf$ and $\ztf$ directions is determined
	by parameters {\tt SHPAR}$_7$ and {\tt SHPAR}$_8$.\\
        The periodicity should not be smaller than the extent of the
        target, so that one should have\\
        \ \\
        ${\tt SHPAR}_7 \geq {\rm max}({\tt SHPAR}_2,{\tt SHPAR}_5)$\\
        ${\tt SHPAR}_8 \geq {\rm max}({\tt SHPAR}_3,{\tt SHPAR}_6)$\\

	The target
	is a periodic structure, of infinite extent in the $\ytf$ and
	$\ztf$ directions.
	The target axes  
	are set to ${\hat{\bf a}}_1 = \xtf$ --
	i.e., normal to the ``slab'' -- and
	${\hat{\bf a}}_2 = \ytf$.
	The orientation of the incident radiation relative to the target
	is, as for all other targets, set by the usual orientation
	angles $\beta$, $\Theta$, and $\Phi$ 
	(see \S\ref{sec:target_orientation} above)
	specifying the orientation of the target axes ${\hat{\bf a}}_1$
	and  ${\hat{\bf a}}_2$ relative to the direction of incidence; 
	for example,
	$\Theta=0$ would be for radiation incident normal to the slab.

	The scattering directions are specified by specifying the
	diffraction order $(M,N)$; for each diffraction order one
	transmitted wave direction and one reflected wave direction
	will be calculated, with the dimensionless $4\times4$ scattering
	matrix $S^{(2d)}$ calculated for each scattering direction.
	At large distances from the infinite slab, the
	scattered Stokes vector in the (M,N) diffraction order is
\beq
        I_{sca,i}(M,N) = \sum_{j=1}^{4} S_{ij}^{(2d)} I_{in,j}
\eeq
where $I_{in,j}$ is the incident Stokes vector.  See
\citet{Draine+Flatau_2008a} for interpretation of the $S_{ij}$ as
transmission and reflection efficiencies.
\index{target shape options!SLBHOLPBC}
\index{periodic boundary conditions}
\subsection{ SLBHOLPBC = Target consisting of a periodic array of
             rectangular blocks, each containing a cylindrical hole
             \label{sec:SLBHOLPBC}}
        \label{sec:SLBHOLPBC}
        Individual blocks have extent $(a,b,c)$ in the $(\xtf,\ytf,\ztf)$
        directions,
        and the cylindrical hole has radius $r$.
        The period in the $\ytf$-direction is $P_y$, and
        the period in the $\ztf$-direction is $P_z$.\\
        The pertinent line in {\tt ddscat.par} should consist of\\
        {\tt SHPAR}$_1$ {\tt SHPAR}$_2$ {\tt SHPAR}$_3$ {\tt SHPAR}$_4$
        {\tt SHPAR}$_5$ {\tt SHPAR}$_6$\\
        where {\tt SHPAR}$_1 = a/d$ ($d$ is the interdipole spacing)\\
        {\tt SHPAR$_2$} = $b/a$ \\
        {\tt SHPAR$_3$} = $c/a$ \\
        {\tt SHPAR$_4$} = $r/a$ \\
        {\tt SHPAR$_5$} = $P_y/d$ \\
        {\tt SHPAR$_6$} = $P_z/d$.\\
        Ideally, $a/d=${\tt SHPAR}$_1$,
        $b/d=${\tt SHPAR}$_2\times${\tt SHPAR}$_1$, and
        $c/d=${\tt SHPAR}$_3\times${\tt SHPAR}$_1$ will be integers (so that the
        cubic lattice can accurately approximate the desired target).
        If $P_y=0$ and $P_z>0$ the target is periodic in the
        $\ztf$-direction only.\\
        If $P_y>0$ and $P_z=0$ the target is periodic in the 
        $\ytf$-direction only.\\
        If $P_y>0$ it is required that $P_y\geq b$, 
        and if $P_z>0$ it is required that $P_z\geq c$, so that
        the blocks do not overlap.\\
        With $P_y=b$ and $P_z=c$, the blocks are juxtaposed 
        to form a periodic array of cylindrical holes in a solid slab.\\
        Example: {\tt ddscat\_SLBHOLPBC.par} is a sample {\tt ddscat.par}
        for a periodic array of cylindrical holes in a slab of thickness
        $a$, with holes of radius $r$, and period $P_y$ and $P_d$
\index{target shape options!SPHRN\_PBC}
\index{periodic boundary conditions}
\subsection{ SPHRN\_PBC = Target consisting of group of N spheres, extended in
                     target y and z directions using
                     periodic boundary conditions}
            \label{sec:SPHRN_PBC}
            This option causes {{\bf DDSCAT}}\ to create a target consisting
	    of a periodic array of $N-$sphere structures, where one
	    $N-$sphere structure consists of $N$ spheres, just
	    as for target option {\tt NANSPH} (see \S\ref{sec:SPH_ANI_N}).  
	    Each sphere can be of arbitrary composition, and can be
	    anisotropic if desired.
	    Information 
	    for the description of one $N$-sphere structure is supplied via
	    an external file,
	    just as for target option {\tt NANSPH} -- 
	    see \S\ref{sec:SPH_ANI_N}).\\
	    \ \\
	    Let us refer to a single $N$-sphere structure as 
            the Target Unit Cell (TUC).
	    The TUC 
	    is then repeated in the $y_\TF$- and $z_\TF$-directions, with 
	    periodicities {\tt PYAEFF}$\times \aeff$ and 
	    {\tt PZAEFF}$\times \aeff$, where 
	    $\aeff\equiv(3V_{\rm TUC}/4\pi)^{1/3}$,
	    where $V_{\rm TUC}$ 
	    is the total volume of solid material in one TUC.
	    This option requires 3 shape parameters:\newline
	    {\tt DIAMX} = maximum extent of target in the target frame x
	                  direction/$d$\newline
            {\tt PYAEFF} = periodicity in target y direction/$\aeff$\newline
	    {\tt PZAEFF} = periodicity in target z direction/$\aeff$.

	    The pertinent line in {\tt ddscat.par} should read\\
	    \ \\
	{\tt SHPAR$_1$ SHPAR$_2$ SHPAR}$_3$ {\it `filename'} 
        (quotes must be used)\\
	    \ \\
	where\\
	{\tt SHPAR$_1$} = target diameter in $x$ direction 
	(in Target Frame) in units of $d$\\
	{\tt SHPAR$_2$}= {\tt PYAEFF}\\
	{\tt SHPAR}$_3$= {\tt PZAEFF}.\\
	{\it filename} is the name of the file specifying the locations and
	relative sizes of the spheres.\\
	\ \\
	The overall size of the TUC (in terms of numbers of
	dipoles) is determined by parameter {\tt SHPAR$_1$}, which is
	the extent of the multisphere target in the $x$-direction, in
	units of the lattice spacing $d$.
	The physical size of the TUC is specified by the value of
	$\aeff$ (in physical units, e.g. cm), 
	specified in the file {\tt ddscat.par}.

	The location of the spheres in the TUC, and their composition, is
	specified in file {\it `filename'}.
	{
	Please consult \S\ref{sec:SPH_ANI_N} above
	for detailed information concerning the information in this file,
	and its arrangement.}

	Note that while the spheres can be anisotropic and of differing
	composition, they can of course also be isotropic and of a single
	composition, in which case the relevant lines in the
	file {\it 'filename'} should read\\
	$x_1$ $y_1$ $z_1$ $r_1$ 1 1 1 0 0 0 \\
	$x_2$ $y_2$ $z_2$ $r_2$ 1 1 1 0 0 0 \\
	$x_3$ $y_3$ $z_3$ $r_3$ 1 1 1 0 0 0 \\
	...\\
	i.e., every sphere has isotropic composition {\tt ICOMP=1} and
	the three dielectric function orientation angles are set to zero.

	When the user uses target option {\tt SPHRN\_PBC}, the target now
	becomes a periodic structure, of infinite extent in the target
	y- and z- directions.
	The target axis ${\hat{\bf a}}_1 = \xlf = (1,0,0)_\TF$ --
	i.e., normal to the ``slab'' -- and target
	axis ${\hat{\bf a}}_2 = \ylf = (0,1,0)_\TF$.
	The orientation of the incident radiation relative to the target
	is, as for all other targets, set by the usual orientation
	angles $\beta$, $\Theta$, and $\Phi$ 
	(see \S\ref{sec:target_orientation} above).
	The scattering directions are, just as for other targets,
	determined by the scattering angles $\theta_s$, $\phi_s$ (see
	\S\ref{sec:scattering_directions} below).

	The scattering problem for this infinite structure, assumed to
	be illuminated by an incident monochromatic plane wave, is
	essentially solved ``exactly'', in the sense that the electric
	polarization of each of the constituent dipoles is due to the
	electric field produced by the incident plane wave plus {\it all}
	of the other dipoles in the infinite target.

	However, the assumed target will, of course, act as a perfect 
	diffraction
	grating if the scattered radiation is calculated as the coherent
	sum of all the oscillating dipoles in this periodic structure: the
	far-field 
	scattered intensity would be zero in all directions except those
	where the Bragg scattering condition is satisfied, and in those
	directions the far-field scattering intensity would be infinite.

	To suppress this singular behavior, we calculate the far-field
	scattered intensity as though the separate TUCs 
	scatter incoherently.
	The scattering efficiency $Q_\sca$ and the absorption efficiency
	$Q_\abs$ are defined to be the scattering and absorption cross
	section per TUC, divided by $\pi \aeff^2$, where 
	$\aeff\equiv(3V_{\rm TUC}/4\pi)^{1/3}$, where $V_{\rm TUC}$ is
	the volume of solid material per TUC.

	Note: the user is allowed to set the target periodicity in the
	target $y_\TF$ (or $z_\TF$) direction to values that could be smaller 
        than the total extent of one TUC in the target $y_\TF$ 
	(or $z_\TF$) direction.
	This is physically allowable, {\it provided that the spheres from
	one TUC do not overlap with the spheres from neighboring TUCs}.
	Note that \ddscat\ does {\it not} check for such overlap.
\subsubsection{\bf Sample calculation in directory examples\_exp/SPHRN\_PBC}
Subdirectory {\tt examples\_exp/SPHRN\_PBC} contains {\tt ddscat.par}
to calculate scattering by a doubly-periodic array with the target unit
cell consisting of a random cluster of 16 spheres.
The calculation is carried out with double precision arithmetic.
Because convergence is slow, the error tolerance is set to {\tt TOL = 5.e-5}
rather than the usual {\tt 1.e-5}, and the maximum number of iterations
allowed is increased to {\tt MXITER = 2000}.
\index{target shape options!TRILYRPBC}
\subsection{ TRILYRPBC = Three stacked rectangular blocks, 
            repeated periodically}
        \label{sec:TRILYRPBC}
	The target unit cell (TUC) consists of a stack of 3 rectangular blocks
	with centers on the $\xtf$ axis.
	The TUC is repeated in either the $\ytf$ direction, the $\ztf$
	direction, or both.
	A total of 11 shape parameters must be specified:\\
        {\tt SHPAR$_1$} = x-thickness of upper layer/$d$ [material 1]\\
        {\tt SHPAR$_2$} = y-width/$d$ of upper layer\\
        {\tt SHPAR}$_3$ = z-width/$d$ of upper layer\\
        {\tt SHPAR}$_4$ = x-thickness/$d$ of middle layer [material 2]\\
        {\tt SHPAR}$_5$ = y-width/$d$ of middle layer\\
        {\tt SHPAR}$_6$ = z-width/$d$ of middle layer\\
        {\tt SHPAR}$_7$ = x-width/$d$ of lower layer\\
        {\tt SHPAR}$_8$ = y-width/$d$ of lower layer\\
        {\tt SHPAR}$_9$ = z-width/$d$ of lower layer\\
        {\tt SHPAR}$_{10}$ = period/$d$ in y direction\\
        {\tt SHPAR}$_{11}$= period/$d$ in z direction


\section{Scattering Directions
\label{sec:scattering_directions}}

\subsection{ Isolated Finite Targets}

\index{scattering directions}
\index{$\theta_s$ -- scattering angle}
\index{$\phi_s$ -- scattering angle}
\index{CMDFRM -- specifying scattering directions}
\index{LFRAME}
\index{TFRAME}

{{\bf DDSCAT}}\ calculates scattering in selected directions, and elements of
the scattering matrix are reported in the output 
files {\tt w}{\it xxx}{\tt r}{\it yyy}{\tt k}{\it zzz}{\tt .sca} .
The scattering direction is specified through angles $\theta_s$ and $\phi_s$ 
(not to be confused with the angles $\Theta$ and $\Phi$ which specify the 
orientation of the target relative to the incident radiation!).

For isolated finite targets ({\it i.e.,} PBC {\it not} employed) 
there are two options for specifying the scattering direction, with the
option determined by the value of the string {\tt CMDFRM} read from
the input file {\tt ddscat.par}.

\begin{enumerate}
\item If the user specifies {\tt CMDFRM='LFRAME'}, 
      then the angles $\theta$, $\phi$
      input from {\tt ddscat.par} are understood to specify the 
      scattering directions relative to the Lab Frame 
      (the frame where the incident beam is in the $x-$direction).

      When {\tt CMDFRM='LFRAME'}, 
      the angle $\theta$ is simply the scattering angle $\theta_s$: the
      angle between 
      the incident beam (in direction $\xlf$) and the 
      scattered beam ($\theta_s=0$ for forward scattering, 
      $\theta_s=180^\circ$ for backscattering).

      The angle $\phi$ specifies the 
      orientation of the ``scattering 
      plane'' relative to the $\xlf - \ylf$ 
      plane.
      When $\phi=0$ the scattering plane is assumed to coincide with the
      $\xlf - \ylf$ plane.
      When $\phi=90^\circ$ the scattering plane is assumed to coincide with
      the ${\xlf-\zlf}$ plane.
      Within the scattering plane the scattering directions are specified by
      $0\leq\theta\leq180^\circ$.
      Thus:
\beq
\hat{\bf n}_s = 
\xlf\cos\theta + \ylf\sin\theta\cos\phi + \zlf\sin\theta\sin\phi
~~~,~~~
\eeq

\item If the user specifies {\tt CMDFRM='TFRAME'},
      then the angles $\theta$, $\phi$
      input from {\tt ddscat.par} are understood to specify the
      scattering directions $\bf\hat{n}_s$ relative to the Target Frame
      (the frame defined by target axes $\xtf$,
      $\ytf$, $\ztf$).
      $\theta$ is the angle between $\hat{\bf n}_s$ and $\xtf$,
      and $\phi$ is the angle between the $\hat{\bf n}_s-\xtf$ plane
      and the $\xtf-\ytf$ plane.
      Thus:
\beq
\hat{\bf n}_s = \xtf\cos\theta + \ytf\sin\theta\cos\phi + \ztf\sin\theta\sin\phi
~~~.~~~
\eeq

\end{enumerate}

Scattering directions for which the scattering properties are to
be calculated are set in the parameter file {\tt ddscat.par} by
specifying one or more scattering planes (determined by the value of $\phi_s$)
and for each scattering plane, the number and range of $\theta_s$ values.
The only limitation is that the number of scattering directions not exceed
the parameter {\tt MXSCA} in {\tt DDSCAT.f} (in the code as distributed it
is set to {\tt MXSCA=1000}).

\subsection{Scattering Directions for Targets that are Periodic in 1 Dimension
\label{sec:scattering_directions:1d}}

For targets that are periodic, scattering is only allowed in certain
directions \citep[see]{Draine+Flatau_2008a}.
If the user has chosen a PBC target (e.g,
{\tt CYLNDRPBC}, {\tt HEXGONPBC}, or {\tt RCTGL\_PBC}), {\tt SPHRN\_PBC}
but has set one of the periodicities to zero, then the target is
periodic in only one dimension -- e.g., {\tt CYLNDRPBC} could be used
to construct a single infinite cylinder.

In this case, the scattering directions are specified by giving an integral
diffraction order $M=0,\pm1,\pm2,...$ and one angle, the azimuthal angle
$\zeta$ around the target repetition axis.
The diffraction order $M$ determines the projection of ${\bf k}_{s}$
onto the repetition direction.
For a given order $M$, the scattering angles with $\zeta=0\rightarrow2\pi$
form a cone around the repetition direction.

For example, if ${\tt PYD}>0$ (target repeating in the $y_\TF$ direction),
then $M$ determines the value of $k_{sy}={\bf k}_s \cdot \hat{\bf y}_\TF$,
where $\hat{\bf y}_\TF$ is the unit vector in the Target Frame $y-$direction:
\beq
k_{sy} = k_{0y} + 2\pi M/L_y
\eeq
where $k_{0y}\equiv {\bf k}_0\cdot \hat{\bf y}_\TF$, where ${\bf k}_0$ 
is the incident
$k$ vector.
Note that  the
diffraction order $M$ {\it must} satisfy the condition 
\beq\label{eq:condition on M}
(k_{0y}-k_{0})(L_y/2\pi) < M < (k_0-k_{0y})(L_y/2\pi)
\eeq
where $L_y={\tt PYD}\times d$ is the periodicity along the $y$ axis in the
Target Frame.
$M=0$ is always an allowed diffraction order.
\index{PYD}

The azimuthal angle $\zeta$ defines a right-handed rotation of the
scattering direction around the target repetition axis.
Thus for a target repetition axis $\hat{\bf y}_\TF$,
\begin{eqnarray}
k_{sx} &=& k_\perp \cos\zeta ~~~,
\\
k_{sz} &=& k_\perp \sin\zeta ~~~,
\end{eqnarray}
where $k_\perp = (k_0^2-k_{sy}^2)^{1/2}$, with $k_{sy}=k_{0y}+2\pi M/L_y$.
For a target with repetition axis $z_\TF$,
\begin{eqnarray}
k_{sx} &=& k_\perp \cos\zeta ~~~,
\\
k_{sy} &=& k_\perp \sin\zeta ~~~,
\end{eqnarray}
where $k_\perp = (k_0^2-k_{sz}^2)^{1/2}$, $k_{sz}=k_{0z}+2\pi M/L_z$.

The user selects a diffraction order $M$ and the azimuthal angles $\zeta$
to be used for that $M$ 
via one line in {\tt ddscat.par}.  An example would be
to use {\tt CYLNDRPBC} to construct an infinite cylinder with the cylinder
direction in the $\hat{\bf y}_\TF$ direction:
e.g., {\tt examples\_exp/CYLNDRPBC/ddscat.par}~:
{\scriptsize
\begin{verbatim}
' ========== Parameter file for v7.3 ===================' 
'**** Preliminaries ****'
'NOTORQ' = CMTORQ*6 (DOTORQ, NOTORQ) -- either do or skip torque calculations
'PBCGS2' = CMDSOL*6 (PBCGS2, PBCGST, GPBICG, QMRCCG, PETRKP) -- CCG method
'GPFAFT' = CMETHD*6 (GPFAFT, FFTMKL) -- FFT method
'GKDLDR' = CALPHA*6 (GKDLDR, LATTDR, FLTRCD) -- DDA method
'NOTBIN' = CBINFLAG (NOTBIN, ORIGIN, ALLBIN)
'**** Initial Memory Allocation ****'
100 100 100 = dimensioning allowance for target generation
'**** Target Geometry and Composition ****'
'CYLNDRPBC' = CSHAPE*9 shape directive
1 64.499 2 1.0 0.0  = shape parameters 1 - 7
1         = NCOMP = number of dielectric materials
'../diel/m1.33_0.01' = file with refractive index 1
'**** Additional Nearfield calculation? ****'
0 = NRFLD (=0 to skip nearfield calc., =1 to calculate nearfield E)
0.0 0.0 0.0 0.0 0.0 0.0 (fract. extens. of calc. vol. in -x,+x,-y,+y,-z,+z)
'**** Error Tolerance ****'
1.00e-5 = TOL = MAX ALLOWED (NORM OF |G>=AC|E>-ACA|X>)/(NORM OF AC|E>)
'**** Maximum number of iterations ****'
200      = MXITER
'**** Integration limiter for PBC calculations ****'
1.00e-3 = GAMMA (1e-2 is normal, 3e-3 for greater accuracy)
'**** Angular resolution for calculation of <cos>, etc. ****'
0.5	= ETASCA (number of angles is proportional to [(3+x)/ETASCA]^2 )
'**** Vacuum wavelengths (micron) ****'
6.283185 6.283185 1 'LIN' = wavelengths (first,last,how many,how=LIN,INV,LOG)
'**** Refractive index of ambient medium'
1.0000 = NAMBIENT
'**** Effective Radii (micron) **** '
2.8555 2.8555 1 'LIN' = aeff (first,last,how many,how=LIN,INV,LOG)
'**** Define Incident Polarizations ****'
(0,0) (1.,0.) (0.,0.) = Polarization state e01 (k along x axis)
2 = IORTH  (=1 to do only pol. state e01; =2 to also do orth. pol. state)
'**** Specify which output files to write ****'
1 = IWRKSC (=0 to suppress, =1 to write ".sca" file for each target orient.
'**** Specify Target Rotations ****'
0.    0.   1  = BETAMI, BETAMX, NBETA  (beta=rotation around a1)
60.  60.   1  = THETMI, THETMX, NTHETA (theta=angle between a1 and k)
0.    0.   1  = PHIMIN, PHIMAX, NPHI (phi=rotation angle of a1 around k)
'**** Specify first IWAV, IRAD, IORI (normally 0 0 0) ****'
0   0   0    = first IWAV, first IRAD, first IORI (0 0 0 to begin fresh)
'**** Select Elements of S_ij Matrix to Print ****'
6	= NSMELTS = number of elements of S_ij to print (not more than 9)
11 12 21 22 31 41	= indices ij of elements to print
'**** Specify Scattered Directions ****'
'TFRAME' = CMDFRM (LFRAME, TFRAME for Lab Frame or Target Frame)
1 = NPLANES = number of scattering cones
0.  0. 180. 1  = OrderM zetamin zetamax dzeta for scattering cone 1
\end{verbatim}}

In this example, a single diffraction order $M=0$ is selected, and
$\zeta$ is to run from $\zeta_{\rm min}=0$ to $\zeta_{\rm max}=180^\circ$
in increments of $\delta\zeta = 0.05^\circ$.

There may be additional lines, one per diffraction order.  Remember, however,
that every diffraction order must satisfy eq.\ (\ref{eq:condition on M}).

\subsection{Scattering Directions for Targets for Doubly-Periodic Targets
\label{sec:scattering_directions:2d}}
\index{PBC = periodic boundary conditions}
\index{PYD}
\index{PZD}

If the user has specified nonzero periodicity in both the $y$ and $z$
directions, then the scattering directions are specified by two
integers -- the diffraction orders $M,N$ for the $\hat{\bf y},\hat{\bf z}$
directions.
The scattering directions are
\beq \label{eq:k_s(M,N)}
\bk_s = \pm\, \frac{k_n}{k_0} \left(\bk_0\cdot\xtf\right)\xtf + 
          \left(\bk_{0}\cdot\ytf+\frac{2\pi M}{L_y}\right)\ytf + 
          \left(\bk_{0}\cdot\ztf+\frac{2\pi N}{L_z}\right)\ztf
\eeq
\beq
k_n \equiv \left[k_0^2 - 
           \left(\bk_0\cdot\ytf+\frac{2\pi M}{L_y}\right)^2 +
           \left(\bk_0\cdot\ztf+\frac{2\pi N}{L_z}\right)^2
           \right]^{1/2}
\eeq
where the $+$ sign gives transmission, and the $-$ sign gives reflection.
The integers $M$ and $N$ {\it must} together satisfy the inequality
\beq\label{eq:condition on M and N}
(\bk_0\cdot\ytf + 2\pi M/L_y)^2 + (\bk_0\cdot\ztf + 2\pi N/L_z)^2 < k_{0}^2
\eeq
which, for small values of $L_y$ and $L_z$, may limit the scattering to only
$(M,N)=(0,0)$. [Of course, $(0,0)$ is {\it always} allowed].
For each $(M,N)$ specified in {\tt ddscat.par}, \ddscatv\ will calculate
the generalized Mueller matrix $S_{ij}^{(2d)}(M,N)$ --
see \S\ref{subsec:S^{(2d)}}.
At large distances from the infinite slab, the
scattered Stokes vector in the $(M,N)$ diffraction order is
\beq
        I_{sca,i}(M,N) = \sum_{j=1}^{4} S_{ij}^{(2d)}(M,N) I_{in,j}
\eeq
where $I_{in,j}$ is the incident Stokes vector, and $S_{ij}^{(2d)}(M,N)$
is the generalization of the $4\times4$ M\"uller scattering matrix to
targets that are periodic in 2-directions \citep{Draine+Flatau_2008a}.
There are distinct $S_{ij}^{(2d)}(M,N)$ for transmission and for
reflection, corresponding to the $\pm$ in eq.\ (\ref{eq:k_s(M,N)}).

\citet{Draine+Flatau_2008a} (eq. 69-71) show how 
the $S_{ij}^{(2d)}(M,N)$ are easily 
related to familiar "transmission coefficients"
and "reflection coefficients".

Here is {\tt examples\_exp/RCTGL\_PBC/ddscat.par} file as an example:
{\scriptsize
\begin{verbatim}
' =========== Parameter file for v7.3 ===================' 
'**** Preliminaries ****'
'NOTORQ' = CMTORQ*6 (DOTORQ, NOTORQ) -- either do or skip torque calculations
'PBCGS2' = CMDSOL*6 (PBCGS2,PBCGST, GPBICG, QMRCCG, PETRKP) -- CCG method
'GPFAFT' = CMDFFT*6 (GPFAFT, FFTMKL) -- FFT method
'GKDLDR' = CALPHA*6 (GKDLDR, LATTDR, FLTRCD) -- DDA method
'NOTBIN' = CBINFLAG (NOTBIN, ORIBIN, ALLBIN)
'**** Initial Memory Allocation ****'
100 100 100 = dimensioning allowance for target generation
'**** Target Geometry and Composition ****'
'RCTGL_PBC' = CSHAPE*9 shape directive
20 1 1 1 1 = shpar1 - shpar5 (see README.txt)
1         = NCOMP = number of dielectric materials
'../diel/m1.50_0.02'    = refractive index 1
'**** Additional Nearfield calculation? ****'
0 = NRFLD (=0 to skip nearfield calc., =1 to calculate nearfield E)
0.0 0.0 0.0 0.0 0.0 0.0 (fract. extens. of calc. vol. in -x,+x,-y,+y,-z,+z)
'**** Error Tolerance ****'
1.00e-5 = TOL = MAX ALLOWED (NORM OF |G>=AC|E>-ACA|X>)/(NORM OF AC|E>)
'**** Maximum number of iterations ****'
100      = MXITER
'**** Integration limiter for PBC calculations ****'
1.00e-2 = GAMMA (1e-2 is normal, 3e-3 for greater accuracy)
'**** Angular resolution for calculation of <cos>, etc. ****'
1.	= ETASCA (number of angles is proportional to [(3+x)/ETASCA]^2 )
'**** Vacuum wavelengths (micron) ****'
0.5 0.5 1 'LIN' = wavelengths (first,last,how many,how=LIN,INV,LOG)
'**** Refractive index of ambient medium'
1.000 = NAMBIENT
'**** Effective Radii (micron) **** '
0.0084195 0.0084195 1 'LIN' = aeff (first,last,how many,how=LIN,INV,LOG)
'**** Define Incident Polarizations ****'
(0,0) (1.,0.) (0.,0.) = Polarization state e01 (k along x axis)
2 = IORTH  (=1 to do only pol. state e01; =2 to also do orth. pol. state)
'**** Specify which output files to write ****'
1 = IWRKSC (=0 to suppress, =1 to write ".sca" file for each target orient.
'**** Prescribe Target Rotations ****'
0.   0.   1  = BETAMI, BETAMX, NBETA (beta=rotation around a1)
40. 40.   1  = THETMI, THETMX, NTHETA (theta=angle between a1 and k)
0.   0.   1  = PHIMIN, PHIMAX, NPHI (phi=rotation angle of a1 around k)
'**** Specify first IWAV, IRAD, IORI (normally 0 0 0) ****'
0   0   0    = first IWAV, first IRAD, first IORI (0 0 0 to begin fresh)
'**** Select Elements of S_ij Matrix to Print ****'
6       = NSMELTS = number of elements of S_ij to print (not more than 9)
11 12 21 22 31 41       = indices ij of elements to print
'**** Specify Scattered Directions ****'
'TFRAME' = CMDFRM (LFRAME, TFRAME for Lab Frame or Target Frame)
1 = NORDERS = number of diffraction orders for transmission
0. 0.
\end{verbatim}}

\section{Incident Polarization State\label{sec:incident_polarization}}

\index{polarization of incident radiation}
Recall that the ``Lab Frame'' is defined such that the incident radiation
is propagating along the $\xlf$ axis.
\ddscat\ allows the user to specify a 
general elliptical polarization
for the incident radiation, by specifying the (complex) polarization
vector ${\hat{\bf e}}_{01}$.
The orthonormal polarization state 
${\hat{\bf e}}_{02}=\xlf\times{\hat{\bf e}}_{01}^*$ is
generated automatically if {\tt ddscat.par} specifies {\tt IORTH=2}.

For incident linear polarization, one can simply set ${\hat{\bf
e}}_{01}={\hat{\bf y}}$ by specifying \\
{\tt (0,0) (1,0) (0,0)}\\
in {\tt ddscat.par}; then ${\hat{\bf e}}_{02}={\hat{\bf z}}$. 
For polarization
mode ${\hat{\bf e}}_{01}$ to correspond to right-handed circular
polarization, set ${\hat{\bf e}}_{01}=({\hat{\bf y}}+i{\hat{\bf
z}})/\surd2$ by specifying {\tt (0,0) (1,0) (0,1)} in {\tt ddscat.par}
(\ddscat\ automatically takes care of the normalization of
${\hat{\bf e}}_{01}$); then 
${\hat{\bf e}}_{02}=(i{\hat{\bf y}}+{\hat{\bf z}})/\surd2$, 
corresponding to left-handed circular polarization.

\section{Averaging over Scattering Directions: 
         $g(1)=\langle\cos\theta_s\rangle$, 
	etc.\label{sec:averaging_scattering}}
\subsection{Angular Averaging}
\index{averages over scattering angles}
\index{scattering -- angular averages}

\begin{figure}
\begin{center}
\vspace*{-0.9cm}
\includegraphics[width=8.3cm]{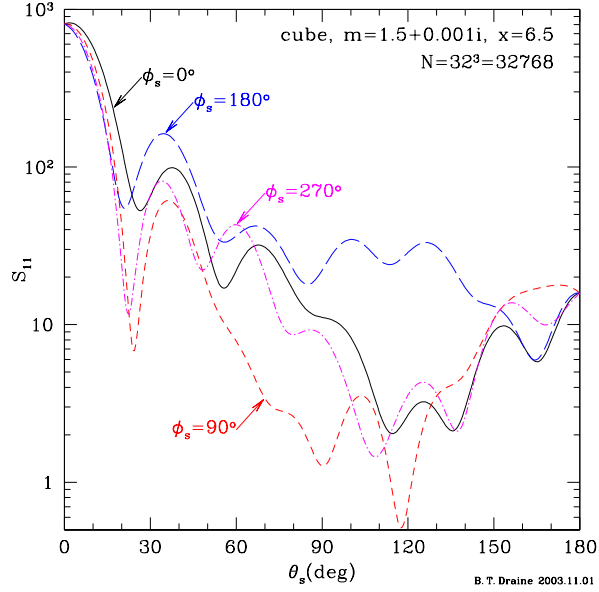}
\vspace*{-2.2cm}
\caption{\label{fig:cube_S11}
\footnotesize
Scattered intensity for incident unpolarized light for a cube
with $m=1.5+0.001i$ and $x=2\pi a_{\rm eff}/\lambda=6.5$ 
(i.e., $d/\lambda=1.6676$, where $d$ is the length of a side).
The cube is tilted with respect to the incident radiation, with
$\Theta=30^\circ$, and rotated by $\beta=15^\circ$ around its axis
to break reflection symmetry.
The Mueller matrix element $S_{11}$ is shown for 4 different scattering
planes.  
The strong forward scattering lobe is evident.
It is also seen that the scattered intensity is a strong function
of scattering angle $\phi_s$ as well as $\theta_s$ -- at a given
value of $\theta_s$ (e.g., $\theta_s=120^\circ$), 
the scattered intensity can vary by orders of magnitude
as $\phi_s$ changes by $90^\circ$.}
\end{center}
\end{figure}

\index{scattering by tilted cube}
An example of scattering by a nonspherical target is shown in Fig.
\ref{fig:cube_S11}, showing the scattering for a tilted cube.
Results are shown for
four different scattering planes.

{{\bf DDSCAT}}\ automatically carries out numerical integration of various
scattering properties,
including
\begin{itemize}
\item$\langle\cos\theta_s\rangle$;
\item$\langle\cos^2\theta_s\rangle$;
\item${\bf g}=\langle\cos\theta_s\rangle\xlf+
	\langle\sin\theta_s\cos\phi_s\rangle{\hat{\bf y}}+
	\langle\sin\theta_s\sin\phi_s\rangle{\hat{\bf z}}~$
	(see \S\ref{sec:force and torque calculation});
	
\item${\bf Q}_{\Gamma}$, provided
  option {\tt DOTORQ} is specified (see 
\S\ref{sec:force and torque calculation}).
\end{itemize}
The angular averages are 
accomplished by evaluating the scattered intensity for
selected scattering directions $(\theta_s,\phi_s)$, 
and taking the appropriately weighted
sum.
Suppose that we have $N_\theta$ different values of 
$\theta_s$,
\beq
\theta = \theta_j, ~~~j=1,..., N_\theta
~~~,
\eeq
and for each value of $\theta_j$, $N_\phi(j)$ different values of $\phi_s$:
\beq
\phi_s = \phi_{j,k}, ~~~k=1,...,N_\phi(j) ~~~.
\eeq
For a given $j$, the values of $\phi_{j,k}$ are assumed to be uniformly spaced:
$\phi_{j,k+1}-\phi_{j,k}=2\pi/N_\phi(j)$.
The angular average of a quantity $f(\theta_s,\phi_s)$ is
approximated by
\beq
\label{eq:average}
\langle f \rangle \equiv \frac{1}{4\pi}\int_{0}^{\pi}\sin\theta_s d\theta_s
\int_{0}^{2\pi}d\phi_s f(\theta_s,\phi_s)
~\approx~ \frac{1}{4\pi}\sum_{j=1}^{N_\theta}
                    \sum_{k=1}^{N_\phi(j)} f(\theta_j,\phi_{j,k}) \Omega_{j,k}
\eeq
\beq
\Omega_{j,k} = \frac{\pi}{N_\phi(j)}
\left[\cos(\theta_{j-1})-\cos(\theta_{j+1})\right]
~~~,
~~~j=2,...,N_\theta-1
~~~.
\eeq
\beq
\Omega_{1,k} = \frac{2\pi}{N_\phi(1)}
\left[1-\frac{\cos(\theta_1)+\cos(\theta_2)}{2}\right]
~~~,
\eeq
\beq
\Omega_{N_\theta,k} = 
\frac{2\pi}{N_\phi(N_\theta)}
\left[
\frac{\cos(\theta_{N_\theta-1})+\cos(\theta_{N_\theta})}{2}
+1\right]
~~~.
\eeq
For a sufficiently large number of scattering directions
\beq
N_\sca \equiv \sum_{j=1}^{N_\theta} N_\phi(j)
\eeq
the sum (\ref{eq:average}) approaches
the desired integral,
but the calculations can be a significant cpu-time
burden, so efficiency is an important consideration.

\subsection{Selection of Scattering Angles $\theta_s,\phi_s$}

\index{scattering angles $\theta_s$, $\phi_s$}
\index{$\theta_s$ -- scattering angle}
\index{$\phi_s$ -- scattering angle}
Beginning with {\bf DDSCAT 6.1}, an improved approach is taken to
evaluation of the angular average.
Since targets with large values of $x=2\pi a_{\rm eff}/\lambda$
in general have strong forward scattering, it is important to obtain
good sampling of the forward scattering direction.  
To implement a preferential
sampling of the forward scattering directions,
the scattering directions $(\theta,\phi)$ 
are chosen so that $\theta$ values correspond to equal intervals in
a monotonically increasing function $s(\theta)$.
The function $s(\theta)$ is chosen to have a 
negative second derivative $d^2s/d\theta^2 < 0$
so that the density of scattering directions will be higher for
small values of $\theta$.
\ddscat\ takes
\beq
s(\theta) = \theta + \frac{\theta}{\theta+\theta_0} ~~~.
\eeq
This provides increased resolution (i.e., increased $ds/d\theta$) for
$\theta < \theta_0$.
We want this for the forward scattering lobe, so we take
\beq
\theta_0 = \frac{2\pi}{1+x}
~~.
\eeq
The $\theta$ values run from $\theta=0$ to $\theta=\pi$,
corresponding to uniform increments 
\beq
\Delta s=\frac{1}{N_\theta-1}\left(\pi+\frac{\pi}{\pi+\theta_0}\right)
~~~.
\eeq
\index{$\eta$ -- parameter for selection of scattering angles}
If we now require that 
\beq
\max[\Delta\theta] \approx \frac{\Delta s}{(ds/d\theta)_{\theta=\pi} }= 
\eta \frac{\pi/2}{3+x}
~~~,
\eeq
we determine the number of values of $\theta$:
\beq
N_\theta = 1 + \frac{2(3+x)}{\eta} \frac{
\left[1+1/(\pi+\theta_0)\right]}{\left[1+\theta_0/(\pi+\theta_0)^2\right]}
~~~.
\eeq
Thus for small values of $x$, $\max[\Delta\theta]=30^\circ \eta$,
and for $x\gg 1$, $\max[\Delta\theta]\rightarrow 90^\circ\eta/x$.
For a sphere, minima in the scattering pattern are separated by
$\sim180^\circ/x$, so $\eta=1$ would be expected to marginally
resolve structure in the scattering function.
Smaller values of $\eta$ will obviously lead to improved sampling
of the scattering function, and more accurate angular averages.

\begin{figure}[h]
\begin{center}
\vspace*{-0.9cm}
\includegraphics[width=8.3cm]{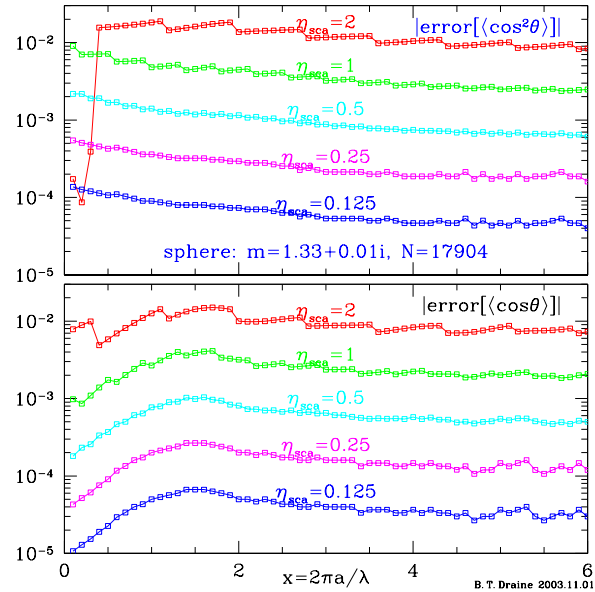}
\vspace*{-2.2cm}
\caption{\label{fig:err_eta_sphere}
\footnotesize
Errors in $\langle\cos\theta\rangle$ 
and $\langle\cos^2\theta\rangle$ calculated for a $N=17904$ dipole
pseudosphere with $m=1.33+0.01i$,
as functions of $x=2\pi a/\lambda$.
Results are shown 
for different values of the parameter $\eta$.
}
\end{center}
\end{figure}

\begin{figure}[h]
\begin{center}
\vspace*{-0.9cm}
\includegraphics[width=8.3cm]{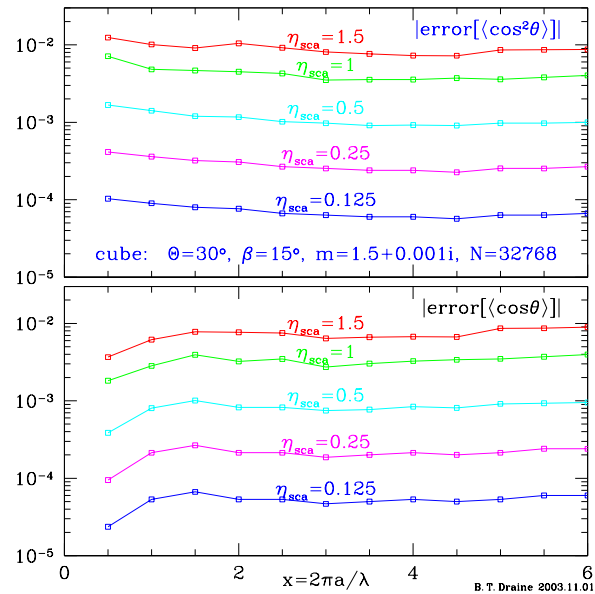}
\vspace*{-2.2cm}
\caption{\label{fig:err_eta_cube}
\footnotesize
Same as Fig.\ \ref{fig:err_eta_sphere}, but for a $N=32768$ dipole
cube with $m=1.5+0.001i$.
}
\end{center}
\end{figure}

The scattering angles $\theta_j$ used for the angular averaging are
then given by
\beq
\theta_j = \frac{(s_j-1-\theta_0)+
\left[(1+\theta_0-s_j)^2+4\theta_0s_j\right]^{1/2}}{2}
~~~,
\eeq
where
\beq
s_j = \frac{(j-1)\pi}{N_\theta-1}\left[1+\frac{1}{\pi+\theta_0}\right]
~~~j=1,...,N_\theta
~~~.
\eeq

For each $\theta_j$, we must choose values of $\phi$.
For $\theta_1=0$ and $\theta_{N_\theta}=\pi$ 
only a single value of $\phi$ is needed
(the scattering is independent of $\phi$ in these two directions).
For $0<\theta_j<\pi$ we use 
\beq
N_\phi=\max\left\{3,{\rm nint}\left[4\pi\sin(\theta_j)/
(\theta_{j+1}-\theta_{j-1})\right]\right\}
~~~,
\eeq
where ${\rm nint}=$ nearest integer.  This provides sampling in $\phi$
consistent with the sampling in $\theta$.

\subsection{Accuracy of Angular Averaging as a Function of $\eta$}

\index{$\eta$ -- parameter for choosing scattering angles}
Figure \ref{fig:err_eta_sphere} shows the absolute errors in
$\langle\cos\theta\rangle$ and $\langle\cos^2\theta\rangle$ calculated
for a sphere with refractive index $m=1.33+0.01i$
using the above prescription for choosing scattering angles.
The error is shown as a function of
scattering parameter $x$.
We see that accuracies of order 0.01 are attained with $\eta=1$, and
that the above prescription provides an accuracy which is approximately
independent of $x$.
We recommend using values of $\eta\leq 1$ unless accuracy in the
angular averages is not important.

\section{Scattering by Finite Targets: The Mueller Matrix
\label{sec:mueller_matrix}}
\index{Mueller matrix for scattering}
\index{$S_{ij}$ -- 4$\times$4 Mueller scattering matrix}

\subsection{\label{sec:IORTH=2}
            Two Orthogonal Incident Polarizations ({\tt IORTH=2})}
\index{IORTH parameter}
\index{ddscat.par!IORTH}

Subsection \ref{sec:IORTH=2} is intended for
those studying the internals of \ddscat\ to see how
it obtains the elements of the
scattering amplitude matrix
\citep[see][]{Bohren+Huffman_1983}
\beq
\left(
\begin{array}{c c}
S_2 & S_3 \\
S_4 & S_1
\end{array}
\right)
~~~.
\eeq
Unless you are interested in such computational details, 
you can skip this subsection and go forward to \S\ref{sec:Stokes}.

Throughout the following discussion, $\hat{\bf x}_\LF$, $\hat{\bf y}_\LF$,
$\hat{\bf z}_\LF$ are unit vectors defining the Lab Frame (thus
$\hat{\bf k}_0=\hat{\bf x}_\LF$).

\ddscat\ internally computes the scattering properties of the
dipole array in terms of a complex scattering matrix
$f_{ml}(\theta_s,\phi_s)$ \citep{Draine_1988}, where index $l=1,2$ denotes
the incident polarization state, $m=1,2$ denotes the scattered
polarization state, and $\theta_s$,$\phi_s$ specify the scattering
direction.  Normally {{\bf DDSCAT}}\ is used with {\tt IORTH=2} in
{\tt ddscat.par}, so that the scattering problem will be solved for
both incident polarization states ($l=1$ and 2); in this subsection it
will be assumed that this is the case.

Incident polarization states $l=1,2$ correspond to polarization states
${\hat{\bf e}}_{01}$, ${\hat{\bf e}}_{02}$; recall that polarization
state ${\hat{\bf e}}_{01}$ is user-specified, and 
${\hat{\bf e}}_{02}=
\hat{\bf k}_0\times{\hat{\bf e}}_{01}^*=
\hat{\bf x}_\LF\times{\hat{\bf e}}_{01}^*$.\footnote{
   \citet{Draine_1988} adopted the convention 
   $\hat{\bf e}_{02}=\hat{\bf x}\times\hat{\bf e}_{01}^*$.
   The customary definition of $\hat{\bf e}_{i\perp}$ is
   $\hat{\bf e}_{i\perp}\equiv\hat{\bf e}_{i\parallel}\times\hat{\bf k}_0$
   \citep{Bohren+Huffman_1983}.
   Thus, if $\hat{\bf e}_{01}=\hat{\bf e}_{i\parallel}$, then
   $\hat{\bf e}_{02}=-\hat{\bf e}_{i\perp}$.
   }$^,$\footnote{
   A frequent choice is $\hat{\bf e}_{01}=\hat{\bf y}_\LF$,
   with 
   $\hat{\bf e}_{02}=\hat{\bf x}_\LF\times\hat{\bf y}_\LF = \hat{\bf z}_\LF$.
   }
In the \ddscat\ code, which follows the convention in \citet{Draine_1988},
scattered polarization state $m=1$ corresponds to linear polarization of the
scattered wave parallel to the scattering plane 
(${\hat{\bf e}}_1={\hat{\bf e}}_{\parallel s}=\hat{\theta}_s$) and $m=2$
corresponds to linear polarization perpendicular to the scattering
plane (in the $+\hat{\phi}_s$ direction: $\hat{\bf e}_2=\hat{\phi}_s$).  
The scattering matrix $f_{ml}$ was defined \citep{Draine_1988} so that 
the scattered electric field ${\bf E}_s$ is related to the incident 
electric field ${\bf E}_i(0)$ at the origin 
(where the target is assumed to be located) by
\index{$f_{ij}$ -- amplitude scattering matrix}
\beq
\left(
\begin{array}{c}
	{\bf E}_s\cdot\hat{\theta}_s\\
	{\bf E}_s\cdot\hat{\phi}_s
\end{array}
\right)
=
{\exp(i{\bf k}_s\cdot{\bf r})\over kr}
\left(
\begin{array}{cc}
	f_{11}&f_{12}\\
	f_{21}&f_{22}
\end{array}
\right)
\left(
\begin{array}{c}
	{\bf E}_i(0)\cdot{\hat{\bf e}}_{01}^*\\
	{\bf E}_i(0)\cdot{\hat{\bf e}}_{02}^*
\end{array}
\right) ~~~.
\label{eq:f_ml_def}
\eeq
\index{$S_j$ -- scattering amplitude matrix}
The 2$\times$2 complex 
{\it scattering amplitude matrix} (with elements $S_1$, $S_2$,
$S_3$, and $S_4$) is defined so that \citep[see][]{Bohren+Huffman_1983}
\beq
\left(
\begin{array}{c}
	{\bf E}_s\cdot\hat{\theta}_s\\
	-{\bf E}_s\cdot\hat{\phi}_s
\end{array}
\right)
=
{\exp(i{\bf k}_s\cdot{\bf r})\over -ikr}
\left(
\begin{array}{cc}
	S_2&S_3\\
	S_4&S_1
\end{array}
\right)
\left(
\begin{array}{c}
	{\bf E}_i(0)\cdot{\hat{\bf e}}_{i\parallel}\\
	{\bf E}_i(0)\cdot{\hat{\bf e}}_{i\perp}
\end{array}
\right)~~~,
\label{eq:S_ampl_def}
\eeq
where
${\hat{\bf e}}_{i\parallel}$, ${\hat{\bf e}}_{i\perp}$ are (real) unit vectors 
for incident polarization
parallel and perpendicular to the scattering plane (with the
customary definition of 
${\hat{\bf e}}_{i\perp}={\hat{\bf e}}_{i\parallel}\times\hat{\bf k}_0=
{\hat{\bf e}}_{i\parallel}\times{\hat{\bf x}_\LF}$).

From (\ref{eq:f_ml_def},\ref{eq:S_ampl_def}) we may write
\beq
\left(
\begin{array}{cc}
	S_2&S_3\\
	S_4&S_1
\end{array}
\right)
\left(
\begin{array}{c}
	{\bf E}_i(0)\cdot{\hat{\bf e}}_{i\parallel}\\
	{\bf E}_i(0)\cdot{\hat{\bf e}}_{i\perp}
\end{array}
\right)
=
-i
\left(
\begin{array}{cc}
	f_{11}&f_{12}\\
	-f_{21}&-f_{22}
\end{array}
\right)
\left(
\begin{array}{c}
	{\bf E}_i(0)\cdot{\hat{\bf e}}_{01}^*\\
	{\bf E}_i(0)\cdot{\hat{\bf e}}_{02}^*
\end{array}
\right) ~~~.
\label{eq:fml_rel_S_v1}
\eeq

Let
\begin{eqnarray}
	a&\equiv& {\hat{\bf e}}_{01}^*\cdot\ylf~~~,\\
	b&\equiv& {\hat{\bf e}}_{01}^*\cdot\zlf~~~,\\
	c&\equiv& {\hat{\bf e}}_{02}^*\cdot\ylf~~~,\\
	d&\equiv& {\hat{\bf e}}_{02}^*\cdot\zlf~~~.
\end{eqnarray}
Note that since ${\hat{\bf e}}_{01}, {\hat{\bf e}}_{02}$ could be
complex (i.e., elliptical polarization),
the quantities $a,b,c,d$ are complex.
\index{polarization -- elliptical}
Then
\beq
\left(
\begin{array}{c}
	{\hat{\bf e}}_{01}^*\\
	{\hat{\bf e}}_{02}^*
\end{array}
\right)
=
\left(
\begin{array}{cc}
	a&b\\
	c&d
\end{array}
\right)
\left(
\begin{array}{c}
	\ylf\\
	\zlf
\end{array}
\right)
\eeq
and eq. (\ref{eq:fml_rel_S_v1}) can be written
\beq
\left(
\begin{array}{cc}
	S_2&S_3\\
	S_4&S_1
\end{array}
\right)
\left(
\begin{array}{c}
	{\bf E}_i(0)\cdot{\hat{\bf e}}_{i\parallel}\\
	{\bf E}_i(0)\cdot{\hat{\bf e}}_{i\perp}
\end{array}
\right)
=
i
\left(
\begin{array}{cc}
	-f_{11}&-f_{12}\\
	f_{21}&f_{22}
\end{array}
\right)
\left(
\begin{array}{cc}
	a&b\\
	c&d
\end{array}
\right)
\left(
\begin{array}{c}
	{\bf E}_i(0)\cdot\ylf \\
	{\bf E}_i(0)\cdot\zlf
\end{array}
\right) ~~~.
\label{eq:fml_rel_S_v2}
\eeq

The incident polarization states ${\hat{\bf e}}_{i\parallel}$ and
${\hat{\bf e}}_{i\perp}$ are related to $\ylf$, $\zlf$ by
\beq
\left(
\begin{array}{c}
        \hat{\bf{e}}_{i\parallel}\\
        \hat{\bf{e}}_{i\perp}
\end{array}
\right)
=
\left(
\begin{array}{cc}
	\cos\phi_s&	\sin\phi_s\\
	\sin\phi_s&	-\cos\phi_s
\end{array}
\right)
\left(
\begin{array}{c}
	\ylf\\ 
	\zlf
\end{array}
\right)
\eeq
\beq
\left(
\begin{array}{c}
	\ylf\\ 
	\zlf
\end{array}
\right)
=
\left(
\begin{array}{cc}
	\cos\phi_s&	\sin\phi_s\\
	\sin\phi_s&	-\cos\phi_s
\end{array}
\right)
\left(
\begin{array}{c}
	{\hat{\bf e}}_{i\parallel}\\
	{\hat{\bf e}}_{i\perp}
\end{array}
\right).
\label{eq:yz_vs_eis}
\eeq
The angle $\phi_s$ specifies the scattering plane, with
\beqa
\cos\phi_s &\!=\!& \hat{\phi}_s \cdot \hat{\bf z}_\LF \,=\, 
\hat{\bf e}_2 \cdot \hat{\bf z}_\LF ~~~,
\\
\sin\phi_s &\!=\!& -\hat{\phi}_s \cdot \hat{\bf y}_\LF \,=\, 
-\hat{\bf e}_2 \cdot \hat{\bf y}_\LF
~~~.
\eeqa
Substituting (\ref{eq:yz_vs_eis}) into (\ref{eq:fml_rel_S_v2}) we
obtain
\beq
\left(\!
\begin{array}{cc}
	S_2&S_3\\
	S_4&S_1
\end{array}
\!\right)
\left(\!
\begin{array}{c}
	{\bf E}_i(0)\cdot{\hat{\bf e}}_{i\parallel}\\
	{\bf E}_i(0)\cdot{\hat{\bf e}}_{i\perp}
\end{array}
\!\right)
=
i
\left(\!
\begin{array}{cc}
	-f_{11}&-f_{12}\\
	f_{21}&f_{22}
\end{array}
\!\right)
\left(\!
\begin{array}{cc}
	a&b\\
	c&d
\end{array}
\!\right)
\left(\!
\begin{array}{cc}
	\cos\phi_s&	\sin\phi_s\\
	\sin\phi_s&	-\cos\phi_s
\end{array}
\!\right)
\left(\!
\begin{array}{c}
	{\bf E}_i(0)\cdot{\hat{\bf e}}_{i\parallel}\\
	{\bf E}_i(0)\cdot{\hat{\bf e}}_{i\perp}
\end{array}
\!\right)
\label{eq:fml_rel_S_v3}
\eeq

Eq. (\ref{eq:fml_rel_S_v3}) must be true for all ${\bf E}_i(0)$; hence we
obtain an expression for the complex scattering amplitude matrix in terms
of the $f_{ml}$:
\index{$f_{ij}$ -- scattering amplitude matrix}
\index{$S_{j}$ -- scattering amplitude matrix}
\beq
\left(
\begin{array}{cc}
	S_2&S_3\\
	S_4&S_1
\end{array}
\right)
=
i
\left(
\begin{array}{cc}
	-f_{11}&-f_{12}\\
	f_{21}&f_{22}
\end{array}
\right)
\left(
\begin{array}{cc}
	a&	b\\
	c&	d
\end{array}
\right)
\left(
\begin{array}{cc}
	\cos\phi_s&\sin\phi_s\\
	\sin\phi_s&-\cos\phi_s
\end{array}
\right)~~~.
\label{eq:fml_rel_S_v4}
\eeq
This provides the 4 equations used in subroutine {\tt GETMUELLER} to
compute the scattering amplitude matrix elements:
\begin{eqnarray}
S_1 &=& -i\left[
	f_{21}(b\cos\phi_s-a\sin\phi_s)
	+f_{22}(d\cos\phi_s-c\sin\phi_s)
	\right]~~~,\label{eq:S_1_relto_f21andf22}\\
S_2 &=& -i\left[
	f_{11}(a\cos\phi_s+b\sin\phi_s)
	+f_{12}(c\cos\phi_s+d\sin\phi_s)
	\right]~~~,\\
S_3 &=& i\left[
	f_{11}(b\cos\phi_s-a\sin\phi_s)
	+f_{12}(d\cos\phi_s-c\sin\phi_s)
	\right]~~~,\\
S_4 &=& i\left[
	f_{21}(a\cos\phi_s+b\sin\phi_s)
	+f_{22}(c\cos\phi_s+d\sin\phi_s)
	\right] ~~~.
\end{eqnarray}

\subsection{\label{sec:Stokes}
            Stokes Parameters}
\index{Stokes vector $(I,Q,U,V)$}

It is both convenient and customary to characterize both incident and
scattered radiation by 4 ``Stokes parameters'' -- 
the elements of the ``Stokes vector''.
There are different conventions in the literature; we adhere to the
definitions of the Stokes vector ($I$,$Q$,$U$,$V$) adopted in the
excellent treatise by \citet{Bohren+Huffman_1983}, 
to which the reader is referred for further detail.
Here are some examples of Stokes vectors $(I,Q,U,V)=(1,Q/I,U/I,V/I)I$:
\begin{itemize}
\item $(1,0,0,0)I$ : unpolarized light (with intensity $I$);
\item $(1,1,0,0)I$ : 100\% linearly polarized with ${\bf E}$ parallel to the
        scattering plane;
\item $(1,-1,0,0)I$ : 100\% linearly polarized with ${\bf E}$ perpendicular
        to the scattering plane;
\item $(1,0,1,0)I$ : 100\% linearly polarized with ${\bf E}$ at +45$^\circ$
        relative to the scattering plane;
\item $(1,0,-1,0)I$ : 100\% linearly polarized with ${\bf E}$ at -45$^\circ$
        relative to the scattering plane;
\item $(1,0,0,1)I$ : 100\% right circular polarization ({\it i.e.,} negative
        helicity);
\item $(1,0,0,-1)I$ : 100\% left circular polarization ({\it i.e.,} positive
        helicity).
\end{itemize}

\subsection{Relation Between Stokes Parameters of Incident and Scattered
Radiation: The Mueller Matrix}
\index{Mueller matrix for scattering}
\index{$S_{ij}$ -- 4$\times$4 Mueller scattering matrix}

It is convenient to describe the scattering properties of a finite
target in terms of
the $4\times4$ Mueller matrix $S_{ij}$ relating the Stokes parameters 
$(I_i,Q_i,U_i,V_i)$ and
$(I_s,Q_s,U_s,V_s)$ of
the incident and scattered radiation:
\beq
\left(
\begin{array}{c}
	I_s\\
	Q_s\\
	U_s\\
	V_s
\end{array}
\right)
=
{1\over k^2r^2}
\left(
\begin{array}{cccc}
	S_{11}&S_{12}&S_{13}&S_{14}\\
	S_{21}&S_{22}&S_{23}&S_{24}\\
	S_{31}&S_{32}&S_{33}&S_{34}\\
	S_{41}&S_{42}&S_{43}&S_{44}
\end{array}
\right)
\left(
\begin{array}{c}
	I_i\\
	Q_i\\
	U_i\\
	V_i
\end{array}
\right)~~~.
\eeq
Once the amplitude scattering matrix elements are obtained, the Mueller
matrix elements can be computed \citep{Bohren+Huffman_1983}:
\begin{eqnarray}
S_{11}&=&\left(|S_1|^2+|S_2|^2+|S_3|^2+|S_4|^2\right)/2~~~,\nonumber\\
S_{12}&=&\left(|S_2|^2-|S_1|^2+|S_4|^2-|S_3|^2\right)/2~~~,\nonumber\\
S_{13}&=&{\rm Re}\left(S_2S_3^*+S_1S_4^*\right)~~~,\nonumber\\
S_{14}&=&{\rm Im}\left(S_2S_3^*-S_1S_4^*\right)~~~,\nonumber\\
S_{21}&=&\left(|S_2|^2-|S_1|^2+|S_3|^2-|S_4|^2\right)/2~~~,\nonumber\\
S_{22}&=&\left(|S_1|^2+|S_2|^2-|S_3|^2-|S_4|^2\right)/2~~~,\nonumber\\
S_{23}&=&{\rm Re}\left(S_2S_3^*-S_1S_4^*\right)~~~,\nonumber\\
S_{24}&=&{\rm Im}\left(S_2S_3^*+S_1S_4^*\right)~~~,\nonumber\\
S_{31}&=&{\rm Re}\left(S_2S_4^*+S_1S_3^*\right)~~~,\nonumber\\
S_{32}&=&{\rm Re}\left(S_2S_4^*-S_1S_3^*\right)~~~,\nonumber\\
S_{33}&=&{\rm Re}\left(S_1S_2^*+S_3S_4^*\right)~~~,\nonumber\\
S_{34}&=&{\rm Im}\left(S_2S_1^*+S_4S_3^*\right)~~~,\nonumber\\
S_{41}&=&{\rm Im}\left(S_4S_2^*+S_1S_3^*\right)~~~,\nonumber\\
S_{42}&=&{\rm Im}\left(S_4S_2^*-S_1S_3^*\right)~~~,\nonumber\\
S_{43}&=&{\rm Im}\left(S_1S_2^*-S_3S_4^*\right)~~~,\nonumber\\
S_{44}&=&{\rm Re}\left(S_1S_2^*-S_3S_4^*\right)~~~.
\end{eqnarray}
These matrix elements are computed in {\tt DDSCAT} and passed to subroutine
{\tt WRITESCA} which handles output of scattering properties.
Although the Muller matrix has 16 elements, only 9 are independent.

The user can select up to 9 distinct Muller matrix elements to be
printed out in the output files 
{\tt w}{\it xxx}{\tt r}{\it yyy}{\tt k}{\it zzz}{\tt .sca} and
{\tt w}{\it xxx}{\tt r}{\it yyy}{\tt ori.avg}; this choice is made by 
providing a list of indices in {\tt ddscat.par} 
(see Appendix \ref{app:ddscat.par}).

If the user does not provide a list of elements, 
{\tt WRITESCA} will provide a ``default'' set of 6 selected elements: 
$S_{11}$, $S_{21}$, $S_{31}$, $S_{41}$ 
(these 4 elements describe the intensity and polarization 
state for scattering of unpolarized incident radiation), 
$S_{12}$, and $S_{13}$.

In addition, {\tt WRITESCA} writes out the linear polarization $P$ of the
scattered light for incident unpolarized light 
\citep[see][]{Bohren+Huffman_1983}:
\beq
P = \frac{(S_{21}^2+S_{31}^2)^{1/2}}{S_{11}} ~~~.
\eeq

\subsection{Polarization Properties of the Scattered Radiation}
\index{polarization of scattered radiation}
The scattered radiation is fully characterized by its Stokes vector
$(I_s,Q_s,U_s,V_s)$.
As discussed in \citet{Bohren+Huffman_1983} (eq. 2.87), one can determine the
linear polarization of the Stokes vector by operating on it
by the Mueller matrix of an ideal linear polarizer:
\beq
{\bf S}_{\rm pol} = \frac{1}{2}
\left(\begin{array}{cccc}
  1&\cos2\xi&\sin2\xi&0\\
  \cos2\xi&\cos^22\xi&\cos2\xi\sin2\xi&0\\
  \sin2\xi&\sin2\xi\cos2\xi&\sin^22\xi&0\\
  0&0&0&0\\
\end{array}\right)
\eeq
where $\xi$ is the angle between the unit vector $\hat{\theta}_s$ parallel
to the scattering plane (``SP'') and the ``transmission'' axis of the linear
polarizer.  Therefore the intensity of light polarized parallel to
the scattering plane is obtained by taking $\xi=0$ and operating on
$(I_s,Q_s,U_s,V_s)$ to obtain
\begin{eqnarray}
I({\bf E}_s\parallel {\rm SP})&=&\frac{1}{2}(I_s + Q_s)\\
&=&\frac{1}{2k^2r^2}
\left[
(S_{11}+S_{21})I_i + 
(S_{12}+S_{22})Q_i + 
(S_{13}+S_{23})U_i + 
(S_{14}+S_{24})V_i
\right]~.~~~~~
\end{eqnarray}
Similarly, the intensity of light polarized perpendicular to the scattering
plane is obtained by taking $\xi=\pi/2$:
\begin{eqnarray}
I({\bf E}_s\perp {\rm SP}) &=& \frac{1}{2}(I_s - Q_s)\\
&=& \frac{1}{2k^2r^2}
\left[
(S_{11}-S_{21})I_i +
(S_{12}-S_{22})Q_i + 
(S_{13}-S_{23})U_i + 
(S_{14}-S_{24})V_i
\right]~.~~~~~
\end{eqnarray}

\subsection{Relation Between Mueller Matrix and Scattering Cross Sections}
\index{Mueller matrix for scattering}
\index{polarization of scattered radiation}
\index{$S_{ij}$ -- 4$\times$4 Mueller scattering matrix}
\index{differential scattering cross section $dC_\sca/d\Omega$}
Differential scattering cross sections can be obtained directly from the
Mueller matrix elements by noting that
\beq
I_s = \frac{1}{r^2} \left(\frac{dC_\sca}{d\Omega}\right)_{s,i} I_i
~~~.
\eeq
Let SP be the ``scattering plane'': the plane
containing the incident and scattered directions of propagation.
Here we consider some special cases:
\begin{itemize}
\item Incident light unpolarized: Stokes vector $s_i=I(1,0,0,0)$:
\begin{itemize}
  \item cross section for scattering with
    polarization ${\bf E}_s\parallel$ SP:
    \beq
      \frac{dC_\sca}{d\Omega} = 
      \frac{1}{2k^2}\left( |S_2|^2 + |S_3|^2 \right)
      =
      \frac{1}{2k^2}\left( S_{11}+S_{21}\right)
    \eeq
  \item cross section for scattering with
    polarization ${\bf E}_s\perp$ SP:
    \beq
      \frac{dC_\sca}{d\Omega} = 
      \frac{1}{2k^2}\left( |S_1|^2 + |S_4|^2 \right)
      =
      \frac{1}{2k^2}\left( S_{11}-S_{21}\right)
    \eeq
  \item total intensity of scattered light:
    \beq
      \frac{dC_\sca}{d\Omega} = 
      \frac{1}{2k^2}\left( |S_1|^2 + |S_2|^2 + |S_3|^2 + |S_4|^2 \right)
      =
      \frac{1}{k^2} S_{11}
    \eeq
   \end{itemize}
\item Incident light polarized with ${\bf E}_i\parallel$ SP:
  Stokes vector $s_i=I(1,1,0,0)$:
\begin{itemize}
  \item cross 
    section for scattering with polarization ${\bf E}_s\parallel$ to SP:
    \beq
      \frac{dC_\sca}{d\Omega} = \frac{1}{k^2} |S_2|^2 =
      \frac{1}{2k^2}\left(S_{11}+S_{12}+S_{21}+S_{22}\right)
    \eeq
  \item cross
    section for scattering with polarization ${\bf E}_s\perp$ to SP:
    \beq
      \frac{dC_\sca}{d\Omega} = \frac{1}{k^2} |S_4|^2 =
      \frac{1}{2k^2}\left(S_{11}+S_{12}-S_{21}-S_{22}\right)
    \eeq
  \item total scattering cross section:
    \beq
      \frac{dC_\sca}{d\Omega} = \frac{1}{k^2} \left(|S_2|^2+|S_4|^2\right) =
      \frac{1}{k^2} \left(S_{11}+S_{12}\right)
    \eeq
\end{itemize}
\item Incident light polarized with ${\bf E}_i\perp$ SP:
  Stokes vector $s_i=I(1,-1,0,0)$:
\begin{itemize}
  \item cross section
    for scattering with polarization ${\bf E}_s\parallel$ to SP:
    \beq
      \frac{dC_\sca}{d\Omega} = \frac{1}{k^2} |S_3|^2 =
      \frac{1}{2k^2}\left(S_{11}-S_{12}+S_{21}-S_{22}\right)
    \eeq
  \item cross section
    for scattering with polarization ${\bf E}_s\perp$ SP:
    \beq
      \frac{dC_\sca}{d\Omega} = \frac{1}{k^2} |S_1|^2 = 
      \frac{1}{2k^2}\left(S_{11}-S_{12}-S_{21}+S_{22}\right)
    \eeq
  \item total
    scattering cross section:
    \beq
      \frac{dC_\sca}{d\Omega} = \frac{1}{k^2} \left(|S_1|^2+|S_3|^2\right) =
      \frac{1}{k^2}\left(S_{11}-S_{12}\right)
    \eeq
\end{itemize}
\item Incident light linearly polarized at angle $\gamma$ to SP:
  Stokes vector $s_i = I(1,\cos2\gamma,\sin2\gamma,0)$:
\begin{itemize}
  \item cross section for scattering with polarization 
    ${\bf E}_s\parallel$ to SP:
    \beq
      \frac{dC_\sca}{d\Omega} = \frac{1}{2k^2}\left[
	(S_{11}+S_{21}) + 
	(S_{12}+S_{22})\cos2\gamma + 
	(S_{13}+S_{23})\sin2\gamma\right]
      \eeq
    \item cross section for scattering with polarization
      ${\bf E}_s\perp$ SP:
      \beq
        \frac{dC_\sca}{d\Omega} = \frac{1}{2k^2}
        \left[
	(S_{11}-S_{21}) + 
	(S_{12}-S_{22})\cos2\gamma + 
	(S_{13}-S_{23})\sin2\gamma
	\right]
      \eeq
    \item total scattering cross section:
      \beq
	\frac{dC_\sca}{d\Omega} = \frac{1}{k^2}\left[
	  S_{11} + S_{12}\cos2\gamma + S_{13}\sin2\gamma\right]
	\eeq
      \end{itemize}
\end{itemize}

\subsection{One Incident Polarization State Only ({\tt IORTH=1})}
\index{IORTH}
\index{ddscat.par!IORTH}

In some cases it may be desirable to limit the calculations to a
single incident polarization state -- for example, when each solution
is very time-consuming, and the target is known to have some symmetry
so that solving for a single incident polarization state may be
sufficient for the required purpose.  In this case, set {\tt IORTH=1}
in {\tt ddscat.par}.

When {\tt IORTH=1}, only $f_{11}$ and $f_{21}$ are available;
hence, {{\bf DDSCAT}}\ cannot
automatically generate the Mueller matrix elements.
In this case, the output routine {\tt WRITESCA} writes out the quantities
$|f_{11}|^2$, $|f_{21}|^2$, ${\rm Re}(f_{11}f_{21}^*)$, and
${\rm Im}(f_{11}f_{21}^*)$ for each of the scattering directions.

The differential scattering cross section for scattering with polarization
${\bf E}_s \parallel$ and $\perp$ to the scattering plane are
\begin{eqnarray}
\left(\frac{dC_\sca}{d\Omega}\right)_{s,\parallel} &=& \frac{1}{k^2}|f_{11}|^2
\\
\left(\frac{dC_\sca}{d\Omega}\right)_{s,\perp} &=& \frac{1}{k^2}|f_{21}|^2
\end{eqnarray}

Note, however, that if {\tt IPHI} is greater than 1, {{\bf DDSCAT}}\
will automatically set {\tt IORTH=2} even if {\tt ddscat.par}
specified {\tt IORTH=1}: this is because when more than one value of
the target orientation angle $\Phi$ is required, there is no
additional ``cost'' to solve the scattering problem for the second
incident polarization state, since when solutions are available for
two orthogonal states for some particular target orientation, the
solution may be obtained for another target orientation differing only
in the value of $\Phi$ by appropriate linear combinations of these
solutions.  Hence we may as well solve the ``complete'' scattering
problem so that we can compute the complete Mueller matrix.

\section{Scattering by Periodic Targets: Generalized Mueller Matrix
\label{sec:generalized mueller matrix}}
\index{Mueller matrix for infinite targets, periodic in 1-d}
\index{Mueller matrix for infinite targets, periodic in 2-d}

The Mueller scattering matrix $S_{ij}(\theta)$
described above
was originally defined \citep[see, e.g.,]{Bohren+Huffman_1983} 
to describe scattering of incident plane waves
by finite targets, such as aerosol particles or dust grains.

\citet{Draine+Flatau_2008a} have extended the Mueller scattering matrix
formalism to also apply to targets that are periodic and infinite in 
one or two dimensions (e.g., an infinite chain of particles, or a
two-dimensional array of particles).

\subsection{Mueller Matrix $S_{ij}^{(1d)}$ for Targets 
Periodic in One Direction}

The dimensionless
$4\times4$ matrix $S_{ij}^{(1d)}(M,\zeta)$ describes the scattering properties
of targets that are periodic in one dimension.
In the radiation zone,
for incident Stokes vector $I_{\rm in}$, the scattered Stokes vector in
direction $(M,\zeta)$ (see \S\ref{sec:scattering_directions:1d})
is \citep[see][eq.\ 63]{Draine+Flatau_2008a}
\beq
I_{{\rm sca},i}(M,\zeta) =
\frac{1}{k_0R}\sum_{j=1}^4 S_{ij}^{(1d)}(M,\zeta) \, I_{{\rm in},j}
~~~,~~~
\eeq
where $R$ is the distance from the target repetition axis.

\ddscatv\ reports the 
scattering matrix elements $S_{ij}^{(nd)}$ in the output files
{\tt w}{\it aaa}{\tt r}{\it bbb}{\tt k}{\it ccc}{\tt .sca}. 

\subsection{\label{subsec:S^{(2d)}}
   Mueller Matrix $S_{ij}^{(2d)}(M,N)$ for Targets Periodic in Two Directions}

For targets that are periodic in two directions, the scattering intensities
are described by 
the dimensionless
$4\times4$ matrix $S_{ij}^{(2d)}(M,N)$.
In the radiation zone,
for incident Stokes vector $I_{{\rm in}}$, the scattered Stokes
vector in scattering order $(M,N)$ is
\citep[see][eq.\ 64]{Draine+Flatau_2008a}
\beq
I_{{\rm sca},i}(M,N) =
\sum_{j=1}^4 S_{ij}^{(2d)}(M,N) \, I_{{\rm in},j}
~~~,~~~
\eeq
where
integers $(M,N)$ define the scattering order 
(see \S\ref{sec:scattering_directions:2d}).  For given $(M,N)$ there
are actually two possible scattering directions -- corresponding to
transmission and reflection (see \S\ref{sec:scattering_directions:2d}),
and for each there is a value of $S_{ij}^{(2d)}(M,N)$, which we
will denote $S_{ij}^{(2d)}(M,N,{\rm tran})$ and
$S_{ij}^{(2d)}(M,N,{\rm refl})$.

For targets with 2-d periodicity, the scattering matrix elements
$S_{ij}^{(2d)}(M,N)$ are directly related to the usual
transmission coefficient and reflection coefficient
\citep[see][eq.\ 69-71]{Draine+Flatau_2008a}.
For unpolarized incident radiation, the reflection coefficient $R$,
transmission coefficient $T$, and absorption coefficient $A$ are
just
\beqa
T &=& \sum_{M,N} S_{11}^{(2d)}(M,N,{\rm tran})
\\
R &=& \sum_{M,N} S_{11}^{(2d)}(M,N,{\rm refl})
\\
A &=& 1 - T - R
~~~.
\eeqa



\index{anisotropic materials!BETADF!THETADF!PHIDF!Dielectric Frame (DF)}
\section{Composite Targets with Anisotropic Constituents
         \label{sec:composite anisotropic targets}}

Section \ref{sec:target_generation} includes
targets composed of anisotropic materials
(ANIELLIPS, ANIRCTNGL, ANI\_ELL\_2, ANI\_ELL\_3).
However, in each of these cases it is assumed that the dielectric tensor
of the target material is diagonal in the 
\index{Target Frame}
``Target Frame''.  For targets
consisting of a single material (in a single domain), it is obviously possible
to choose the ``Target Frame'' to coincide with a frame in which the
dielectric tensor is diagonal.

However, for inhomogeneous targets containing anisotropic materials as,
e.g., inclusions, the optical axes of the constituent material may be
oriented differently in different parts of the target.  In this case,
it is obviously not possible to choose a single reference frame such that
the dielectric tensor is diagonalized for all of the target material.

To extend DDSCAT to cover this case, we must allow for off-diagonal elements
of the dielectric tensor, and therefore of the dipole polarizabilities.
It will be assumed that the dielectric tensor $\epsilon$ is symmetric
(this excludes magnetooptical materials -- check this).

Let the material at a given location in the grain have a dielectric
tensor 
\index{dielectric tensor}
with complex eigenvalues $\epsilon^{(j)}$, $j=1-3$.  These are the
diagonal elements of the dielectric tensor in a frame where it is diagonalized
(i.e., the frame coinciding with the principal axes of the dielectric tensor).
Let this frame in which the dielectric tensor is diagonalized have
unit vectors $\hat{\bf e}_j$ corresponding to the principal axes of
the dielectric tensor.
We need to describe the orientation of the 
\index{Dielectric Frame (DF)}
``Dielectric Frame'' (DF)
-- defined by the ``dielectric axes'' $\hat{\bf e}_j$ --
relative to the Target Frame.
It is convenient to do so with rotation angles analogous to the rotation
angles $\Theta$, $\Phi$, and $\beta$ used
to describe the orientation of the Target Frame in the Lab Frame:
Suppose that we start with unit vectors $\hat{\bf e}_j$ aligned with the
target frame $\hat{\bf x}_j$.
\begin{enumerate}
\index{$\Theta_\DF$}
\item Rotate the DF through an angle $\theta_\DF$ 
around axis $\ytf$,
so that $\theta_\DF$ is now the angle between $\hat{\bf e}_1$ and
$\xtf$.
\index{$\Phi_\DF$}
\item Now rotate the DF through an angle $\phi_\DF$ 
around axis $\xtf$,
in such a way that $\hat{\bf e}_2$ remains in the $\xtf-\hat{\bf e}_1$
plane.
\index{$\beta_\DF$}
\item Finally, rotate the DF through an angle $\beta_\DF$ around axis
$\hat{\bf e}_1$.
\end{enumerate}
The unit vectors $\hat{\bf e}_i$ are related to the TF basis vectors
$\xtf$, $\ytf$, $\ztf$ by:
\begin{eqnarray}
{\hat{\bf e}}_1 &=&   \xtf \cos\theta_\DF
+ \ytf \sin\theta_\DF \cos\phi_\DF
+ \ztf \sin\theta_\DF \sin\phi_\DF
	\\
{\hat{\bf e}}_2 &=& - \xtf \sin\theta_\DF \cos\beta_\DF
+ \ytf [\cos\theta_\DF \cos\beta_\DF \cos\phi_\DF-
\sin\beta_\DF \sin\phi_\DF] \nonumber\\
&&+ \ztf [\cos\theta_\DF \cos\beta_\DF \sin\phi_\DF+
\sin\beta_\DF \cos\phi_\DF]
	\\
{\hat{\bf e}}_3 &=&   \xtf \sin\theta_\DF \sin\beta_\DF 
- \ytf [\cos\theta_\DF \sin\beta_\DF \cos\phi_\DF+
\cos\beta_\DF \sin\phi_\DF] \nonumber\\
  &&         - \ztf [\cos\theta_\DF \sin\beta_\DF 
\sin\phi_\DF-\cos\beta_\DF \cos\phi_\DF]
\end{eqnarray}
or, equivalently:
\begin{eqnarray}
\xtf &=&   {\hat{\bf e}}_1 \cos\theta_\DF
           - {\hat{\bf e}}_2 \sin\theta_\DF \cos\beta_\DF
           + {\hat{\bf e}}_3 \sin\theta_\DF \sin\beta_\DF \\
\ytf &=&   {\hat{\bf e}}_1 \sin\theta_\DF \cos\phi_\DF
           + {\hat{\bf e}}_2 [\cos\theta_\DF \cos\beta_\DF \cos\phi_\DF-\sin\beta_\DF \sin\phi_\DF]
\nonumber\\
&&           - {\hat{\bf e}}_3 [\cos\theta_\DF \sin\beta_\DF \cos\phi_\DF+\cos\beta_\DF \sin\phi_\DF]
\\
\ztf &=&   {\hat{\bf e}}_1 \sin\theta_\DF \sin\phi_\DF
           + {\hat{\bf e}}_2 [\cos\theta_\DF \cos\beta_\DF \sin\phi_\DF+\sin\beta_\DF \cos\phi_\DF]
\nonumber\\
&&           - {\hat{\bf e}}_3 [\cos\theta_\DF \sin\beta_\DF \sin\phi_\DF-\cos\beta_\DF \cos\phi_\DF]
\end{eqnarray}
Define the rotation matrix 
$R_{ij}\equiv \hat{\bf x}_i\cdot\hat{\bf e}_j$:
{\footnotesize
\beq 
R_{ij}=
\eeq
\beq
\left(
\begin{array}{ccc}
\cos\theta_\DF &
\sin\theta_\DF\cos\phi_\DF & 
\sin\theta_\DF\sin\phi_\DF \\
\!\!\!\!\!\!-\!\sin\theta_\DF\cos\beta_\DF\! &
\cos\theta_\DF\cos\beta_\DF\cos\phi_\DF\!-\!
\sin\beta_\DF\sin\phi_\DF &
\cos\theta_\DF\cos\beta_\DF\sin\phi_\DF\!+\!
\sin\beta_\DF\cos\phi_\DF\!\!\!\!
\\
\sin\theta_\DF\sin\beta_\DF &
\!\!\!-\!\cos\theta_\DF\sin\beta_\DF\cos\phi_\DF\!-\!
\cos\beta_\DF\sin\phi_\DF\!\! &
\!-\!\!\cos\theta_\DF\sin\beta_\DF\sin\phi_\DF\!+\!
\cos\beta_\DF\cos\phi_\DF\!\!\!\!\!
\end{array}
\right)
\eeq
and its inverse
\beq
(R^{-1})_{ij} =
\eeq
\beq
\left(
\begin{array}{ccc}
\cos\theta_\DF &
-\sin\theta_\DF\cos\beta_\DF &
\sin\theta_\DF\sin\beta_\DF 
\\
\!\!\!\sin\theta_\DF\cos\phi_\DF\!\! &
\!\!\cos\theta_\DF\cos\beta_\DF\cos\phi_\DF\!-\!
\sin\beta_\DF\sin\phi_\DF &
-\cos\theta_\DF\sin\beta_\DF\cos\phi_\DF\!-\!
\cos\beta_\DF\sin\phi_\DF\!\!\!\!
\\
\!\!\!\sin\theta_\DF\sin\phi_\DF &
\!\!\cos\theta_\DF\cos\beta_\DF\sin\phi_\DF\!+\!
\sin\beta_\DF\cos\phi_\DF\!\! &
-\cos\theta_\DF\sin\beta_\DF\sin\phi_\DF\!+\!
\cos\beta_\DF\cos\phi_\DF\!\!\!\!
\end{array}
\right)
\eeq
}
The dielectric tensor $\balpha$ is diagonal in the DF.
If we calculate the dipole polarizabilility tensor in the DF it will
also be diagonal, with elements
\beq
\balpha^\DF =
\left(
\begin{array}{ccc}
\alpha_{11}^\DF & 0 & 0 \\
0 & \alpha_{22}^\DF & 0 \\
0 & 0 & \alpha_{33}^\DF
\end{array}
\right)
\eeq
The polarizability tensor in the TF is given by
\beq
\label{eq:alpha^TF from alpha^DF}
(\alpha^{\rm TF})_{im} =
R_{ij} (\alpha^\DF)_{jk} (R^{-1})_{km}
\eeq

Thus, we can describe a general anisotropic material if we provide,
for each lattice site, the three diagonal elements $\epsilon_{jj}^\DF$
of the dielectric
tensor in the DF, and the three rotation angles $\theta_\DF$,
$\beta_\DF$, and $\phi_\DF$.
We first use the LDR prescription and the elements to obtain the
three diagonal elements $\alpha_{jj}^\DF$ of the polarizability
tensor in the DF.
We then calculate the polarizability tensor $\alpha^{\rm TF}$ using
eq.\ (\ref{eq:alpha^TF from alpha^DF}).

\section{Near-Field Calculations: $\bE$ and $\bB$ 
         Within or Near the Target
            \label{sec:nearfield}
            \label{sec:micro_vs_macro}
            }
\index{Electric field within or near the target}
\index{Magnetic field within or near the target}

\ddscatseventhree\ includes options for calculating the electric field
$\bE$ and magnetic field $\bB$ within or near the target.

\begin{itemize}
\item {\tt NRFLD}=0 : no near-field calculations.
\item {\tt NRFLD}=1 : calculate $\bE$ in and near the target.
\item {\tt NRFLD}=2 : calculate both $\bE$ and $\bB$ in and near the
target.
\end{itemize}

The near-field calculation of $\bE$ by \ddscatseventhree\ 
returns the {\it macroscopic}
electric field $\bE_{\rm macro}$, as opposed to the microscopic electric
field $\bE_{\rm micro}$.
The distinction between $\bE_{\rm micro}$ and $\bE_{\rm macro}$ is
discussed in textbooks on electromagnetism \citep[e.g.,][]{Jackson_1975}.

\index{macroscopic $\bE_{\rm macro}$ vs. microscopic $\bE_{\rm micro}$}
\index{microscopic $\bE_{\rm micro}$ vs. macroscopic $\bE_{\rm macro}$}

In brief,
$\bE_{\rm micro}$ is the electric field seen by an "atom" in a solid,
or a "point dipole" in the DDA.  This includes the contributions to the
electric field by nearby atoms in the solid, or nearby point dipoles in the
DDA.
The polarization of the material is $\bP=n\alpha\bE_{\rm micro}$,
where $\alpha$ is the molecular polarizability, and $n$ is the atomic
density of molecules.

The macroscopic field is the field such that 
${\bf D} = \epsilon \bE_{\rm macro}$.
Since ${\bf D}=\bE_{\rm macro} + 4\pi\bP$, this implies
$\bP=(1/4\pi)(\epsilon-1)\bE_{\rm macro}$.

\index{Clausius-Mossotti relation}

The Clausium-Mossotti relation gives 
\beq
\frac{4\pi n\alpha}{3} = \frac{\epsilon-1}{\epsilon+2}
~~~.
\eeq
Thus
\beq \label{eq:macro_vs_micro}
\bE_{\rm macro} = \left(\frac{3}{\epsilon+2}\right) \bE_{\rm micro} ~~~.
\eeq
In vacuum, $\bE_{\rm macro}=\bE_{\rm micro}$.

Because \ddscatseventhree\ is for nonmagnetic materials
(i.e., magnetic permeability $\mu = 1$), the macroscopic and microscopic
magnetic fields are the same: $\bB=\bB_{\rm macro}=\bB_{\rm micro}$.

\subsection{Running \ddscatseventhree\ with NRFLD = 1}

For some scientific applications (e.g., surface-enhanced Raman
scattering) one wishes to calculate the electric field $\bE$ near the
the target surface.  It may also be of interest to calculate the
electromagnetic field at positions within the target volume.
\ddscatv\ includes the capability for fast calculations of $\bE$ in
and near the target, using methods described by
\citet{Flatau+Draine_2012}.

When \ddscat\ is run with {\tt NRFLD}=1\index{NRFLD}, \ddscat\
will be run twice.
The first run, using a ``minimal'' computational volume just enclosing the
physical target (or target unit cell when used for periodic targets), 
creates stored files\\
\indent\indent{\tt w}{\it xxx}{\tt r}{\it yyy}{\tt k}{\it zzz}{\tt .pol}{\it n}\\
for {\it n}=1 and (if {\tt IORTH=2}) {\it n}=2.
These files contain the stored solution for the polarization field
$\bP_j$ for lattice points
within the minimal computational volume.

\ddscatseventhree\ then takes the stored solution $\bP_j$ and
proceeds to calculate $\bE$ at a lattice of points in an ``extended'' 
rectangular volume
that can be specified to be larger than the original ``minimal'' computational
volume.  The calculation is done rapidly using FFT methods, as described
by \citet{Flatau+Draine_2012}.
The result is then stored in output files\\
\indent\indent{\tt w}{\it xxx}{\tt r}{\it yyy}{\tt k}{\it zzz}{\tt .E}{\it n}\\
for {\it n}=1 and (if {\tt IORTH=2}) {\it n}=2.

\subsection{Running \ddscatseventhree\ with NRFLD = 2}

When {\tt NRFLD}=2, \ddscatseventhree\ will calculate both $\bE_{\rm macro}$
and $\bB$
within the user-specified volume containing the
target or target unit cell.
The results will be stored in binary output files\\
\indent\indent{\tt w}{\it xxx}{\tt r}{\it yyy}{\tt k}{\it zzz}{\tt .E}{\it n}\\
\indent\indent{\tt w}{\it xxx}{\tt r}{\it yyy}{\tt k}{\it zzz}{\tt .B}{\it n}\\
for {\it n}=1 and (if {\tt IORTH=2}) {\it n}=2.

\subsection{The Binary Files {\tt w}{\it xxx}{\tt r}{\it yyy}{\tt k}{\it zzz}{\tt .E}{\it n}}

For each of the points $j=1,...,N_{xyz}$ in the extended volume where
$\bE$ was to be calculated, the binary file 
{\tt w}{\it xxx}{\tt r}{\it yyy}{\tt k}{\it zzz}{\tt .E}{\it n}
contains:
\begin{itemize}
\item The composition identifer $I_{{\rm comp},j}$ at each lattice site.
      {\tt ICOMP(K,IX,IY,IZ)}= composition identifier for directions 
      {\tt K=1-3} at locations {\tt IX,IY,IZ}.
      Vacuum sites have {\tt ICOMP=0}.
\item The polarization $\bP_j$.  $\bP_j$ will be zero at all points
      outside the target (the ambient medium, taken to be vacuum).
\item The (macroscopic) field $\bE_{\sca,j}$ at each point $j$ 
      produced by the polarization of
      the target (not including the dipole at $j$).
      (The total field at $j$ is just 
      $\bE_j=\bE_{\inc,j}+\bE_{\sca,j}$).
\item The (macroscopic) incident field $\bE_{\inc,j}$ at each point $j$.
\item The diagonal elements of the polarizability tensor $\balpha_j$
      at each point.  An exact solution would satisfy
      $\bP_j=\balpha_j(\bE_{\inc,j}+\bE_{\sca,j})$.
      Thus the stored $\bP$, $\balpha$, $\bE_\inc$, and $\bE_\sca$ allow
      the accuracy of the numerical solution to be verified.
\item The composition identifer $I_{{\rm comp},j}$ at each lattice site.
      (The vacuum sites have $I_{{\rm comp},j}=0$).
\item The complex dielectric function $\epsilon_j$ for compositions
      $j=1$, ... ,{\tt NCOMP}.
\end{itemize}
The {\tt w}{\it xxx}{\tt r}{\it yyy}{\tt k}{\it zzz}{\tt .E}{\it n}
files can be quite large.
For example, the files {\tt w000r000k000.E1} and {\tt w000r000k000.E2}
created in the sample calculation in {\tt examples\_exp/ELLIPSOID\_NEARFIELD}
(see \S\ref{sec:ELLIPSOID_NEARFIELD}) are each 90 MBbytes.
Some of the stored data is easily recomputed (e.g., $\bE_\inc$ and
$\balpha$) or
in principle redundant ($\bP_j$ could in principle be obtained from
$\balpha$, $\bE_\inc$, and $\bE_\sca$) but it is convenient to have
them at hand.

\subsection{The Binary Files {\tt w}{\it xxx}{\tt r}{\it yyy}{\tt k}{\it zzz}{\tt .EB}{\it n}}

For each of the points $j=1,...,N_{xyz}$ in the extended volume where
$\bE$ and $\bB$ were to be calculated, the binary file 
{\tt w}{\it xxx}{\tt r}{\it yyy}{\tt k}{\it zzz}{\tt .EB}{\it n}
contains:
\begin{itemize}
\item The composition identifer $I_{{\rm comp},j}$ at each lattice site.
      {\tt ICOMP(K,IX,IY,IZ)}= composition identifier for directions 
      {\tt K=1-3} at locations {\tt IX,IY,IZ}.
      Vacuum sites have {\tt ICOMP=0}.
\item The polarization $\bP_j$.  $\bP_j$ will be zero at all points
      outside the target (the ambient medium, taken to be vacuum).
\item The (macroscopic) field $\bE_{\sca,j}$ at each point $j$ 
      produced by the polarization of
      the target (not including the dipole at $j$).
      (The total field at $j$ is just 
      $\bE_j=\bE_{\inc,j}+\bE_{\sca,j}$).
\item The (macroscopic) incident field $\bE_{\inc,j}$ at each point $j$.
\item The diagonal elements of the polarizability tensor $\balpha_j$
      at each point.  An exact solution would satisfy
      $\bP_j=\balpha_j(\bE_{\inc,j}+\bE_{\sca,j})$.
      Thus the stored $\bP$, $\balpha$, $\bE_\inc$, and $\bE_\sca$ allow
      the accuracy of the numerical solution to be verified.
\item The magnetic field $\bB_{\sca,j}$ at each point $j$
      produced by the oscillating polarizations of th target (not including
      the dipole at $j$).
      (The total magnetic field at $j$ is just
      $\bB_j=\bB_{\inc,j}+\bB_{\sca,j}$).
\item The magnetic field $\bB_{\inc,j}$ of the incident wave at each point $j$.
\end{itemize}
The {\tt w}{\it xxx}{\tt r}{\it yyy}{\tt k}{\it zzz}{\tt .E}{\it n}
files can be quite large.
For example, the files {\tt w000r000k000.EB1} and {\tt w000r000k000.EB2}
created in the sample calculation in {\tt examples\_exp/ELLIPSOID\_NEARFLD\_B}
(see \S\ref{sec:ELLIPSOID_NEARFIELD}) are each 133 MBbytes.
Some of the stored data is easily recomputed (e.g., $\bE_\inc$ and
$\balpha$) or
in principle redundant ($\bP_j$ could in principle be obtained from
$\balpha$, $\bE_\inc$, and $\bE_\sca$) but it is convenient to have
them at hand.

\section{Post-Processing of Near-Field Calculations
         \label{sec:postprocessing}}

\ddscatv\ is designed to obtain solutions to Maxwell's equations for
arbitrary targets illuminated by an external source of monochromatic
radiation.  Selected information (cross sections for absorption and scattering,
and far-field scattering properties) are automatically calculated by
\ddscatv.  However, if near-field calculations have been requested
(by specifying {\tt NRFLD=1} or {\tt 2} in {\tt ddscat.par} -- 
see \S\ref{sec:nearfield}),
the complete nearfield solutions will be stored on disk to allow
subsequent post-processing for visualization, etc.

To facilitate such post-processing, we provide a Fortran-90 code
{\tt DDPOSTPROCESS.f90} that can be used to extract $\bE$ (and $\bB$ if
magnetic field calculations were requested by specifying {\tt NRFLD=2}).
{\tt DDPOSTPROCESS.f90} also outputs some data suitable for
visualization by VTK (see \S\ref{sec:VTK}).
More importantly, {\tt DDPOSTPROCESS.f90} is easily modifiable by
the user, e.g., to output data in formats compatible with other tools that the
user may be accustomed to using, such as \Matlab.\footnote{%
   \Matlab\ users may wish to 
   check {\tt http://www.google.com/p/ddscat/} to
   see if a \Matlab\ version of ddpostprocess is available.}

\subsection{The Program ddpostprocess
            \label{sec:ddpostprocess}}
\index{DDPOSTPROCESS.f90}
\index{postprocessing}
\index{visualization}

A separate program, {\tt DDPOSTPROCESS.f90}, is provided to conveniently
read the stored\\ 
\indent\indent{\tt w}{\it xxx}{\tt r}{\it yyy}{\tt k}{\it zzz}{\tt .E}{\it n}\\
files.  To create the {\tt ddpostprocess} executable, position yourself in the
{\tt /src} directory and type\\
\indent\indent {\tt make ddpostprocess}\\
which will compile {\tt DDPOSTPROCESS.f90} and
create an executable {\tt ddpostprocess}.

\medskip
The program {\tt ddpostprocess} is constructed to:
\begin{enumerate}

\item Read a user-specified
binary file (of the form 
{\tt w}{\it xxx}{\tt r}{\it yyy}{\tt k}{\it zzz}{\tt .E}{\it n}, or
{\tt w}{\it xxx}{\tt r}{\it yyy}{\tt k}{\it zzz}{\tt .EB}{\it n}) .
The filename is specified in a parameter file
{\tt ddpostprocess.par} . \index{ddpostprocess.par}
The {\tt ddpostprocess} executable 
will automatically determine whether the input file
contains only $\bP$ and $\bE$ (i.e., was produced by \ddscatv\
with {\tt NRFLD}=1) or if it also contains $\bB$ (i.e., was produced
by \ddscatv\ with {\tt NRFLD}=2).

\item Read in control parameter {\tt ILINE} (0 or 1) determining whether
to calculate and write out $\bE$ (and $\bB$, if available)
at points along a user-specified line.

\item Read in control parameter {\tt IVTR} (0 or 1) determining whether to
create files for subsequent visualization using VTK-based software.

\item If {\tt ILINE=1}
   \begin{itemize}
   \item Read in seven parameters defining points along a line:
   $x_A,y_A,z_A,x_B,y_B,z_B$ and $N_{AB}$.
   $x_A,y_A,z_A,x_B,y_B,z_B$ are coordinates in the target frame (TF),
   given in physical units.

   \item Obtain the complex $\bE$ field (and complex $\bB$ if available)
   at $N_{AB}$ uniformly-spaced points
   along the line connecting points $(x_A,y_A,z_A)$  and $(x_B,y_B,z_B)$.

   \item repeat for as many lines specifying 
   $x_A,y_A,z_A,x_B,y_B,z_B$ and $N_{AB}$
   as are present in the parameter file {\tt ddpostprocess.par}.
   \end{itemize}
\end{enumerate}

The program {\tt ddpostprocess} writes output to an ascii output file
{\tt ddpostprocess.out}.  The first 17 lines include information about the
calculation (e.g., $\aeff$, $d$, number of dipoles in target,
$\lambda$, $\bE_\inc$, $\bB_\inc$, incident Poynting vector) 
and column headings.
Beginning with line 18, {\tt ddpostprocess.out} gives 
physical coordinates $x_\TF$, $y_\TF$, $z_\TF$ 
(in the ``Target Frame''), and the real and imaginary
parts of the $x$, $y$, and $z$ components of the electric field
$\bE=\bE_\inc+\bE_\sca$ at $N_{AB}$ points running from
$(x_A,y_A,z_A)$ to $(x_B,y_B,z_B)$.

We have carried out nearfield calculations for the problem of
an Au sphere with
radius $a=0.39789\micron$ ($D=0.7958\micron$)
illuminated by a plane wave with $\lambda=0.5\micron$
The Au
target has refractive index $m=0.96+1.01i$.
The nearfield calculation of $\bE$ is for a volume extending $0.5D$ beyond
the spherical target in the
+x,-x,+y,-y,+z,-z directions.
This is the example problem in {\bf examples\_exp/ELLIPSOID\_NEARFIELD},
where you can find the {\tt ddscat.par} file.

The {\tt ddpostprocess.par} file in 
{\bf examples\_exp/ELLIPSOID\_NEARFIELD} consists
of: \index{ddpostprocess.par}
{\footnotesize\begin{verbatim}
'w000r000k000.E1'            = name of file with E stored
'VTRoutput'                  = prefix for name of VTR output files
1   = IVTR (set to 1 to create VTR output)
1   = ILINE (set to 1 to evaluate E along a line)
-0.59684 0.0 0.0 0.59684 0.0 0.0 501  = XA,YA,ZA, XB,YB,ZB (phys units), NAB
\end{verbatim}}
This calls for 501 equally-spaced points along a line passing through the
center of the sphere, running from
$(x_\TF,y_\TF,z_\TF)=(-.59684,0,0)$ to $(+0.59684,0,0)$.

For the $(x,0,0)$ track running through the center of the sphere
the output {\tt ddpostprocess.out} file will look like (for brevity,
here we show only every 10th line of output along the track...):
{\scriptsize
\begin{verbatim}
3.9789E-01 = a_eff (radius of equal-volume sphere)
1.6408E-02 = d = dipole spacing
     59728 = N = number of dipoles in target or TUC
      96      96      96 = NX,NY,NZ = extent of computational volume
5.0000E-01 = wavelength in vacuo
   1.00000 = refractive index of ambient medium
   0.20619   0.00000   0.00000 = k_{inc,TF} * d
   0.00000   0.00000 = Re(E_inc,x) Im(E_inc,x) at x_TF=0,y_TF=0,z_TF=0
   1.00000   0.00000 = Re(E_inc,y) Im(E_inc,y) "
   0.00000   0.00000 = Re(E_inc,z) Im(E_inc,z) "
   0.00000   0.00000 = Re(B_inc,x) Im(B_inc,x) "
   0.00000   0.00000 = Re(B_inc,y) Im(B_inc,y) "
   1.00000   0.00000 = Re(B_inc,z) Im(B_inc,z) "
   1.00000   0.00000   0.00000 = 2*(4pi/c)*<S_inc> where <S_inc>=time-averaged incident Poynting vector
[Poynting vector S = (Sx,Sy,Sz) =  (c/4pi)*Re(E)xRe(B) ]
   1.00000 = 2*(4pi/c)*|<S_inc>|
   x_TF       y_TF       z_TF      Re(E_x)   Im(E_x)   Re(E_y)   Im(E_y)   Re(E_z)   Im(E_z)
-5.968E-01  0.000E+00  0.000E+00  -0.00000  -0.00000   0.30023  -0.70908  -0.00000   0.00000
-5.734E-01  0.000E+00  0.000E+00  -0.00000  -0.00000   0.62277  -0.54872   0.00000   0.00000
-5.500E-01  0.000E+00  0.000E+00   0.00000  -0.00000   0.90488  -0.34299  -0.00000   0.00000
-5.266E-01  0.000E+00  0.000E+00  -0.00000  -0.00000   1.11247  -0.10490   0.00000   0.00000
-5.032E-01  0.000E+00  0.000E+00  -0.00000  -0.00000   1.24200   0.13835   0.00000   0.00000
-4.798E-01  0.000E+00  0.000E+00  -0.00000   0.00000   1.27287   0.36285  -0.00000   0.00000
-4.564E-01  0.000E+00  0.000E+00  -0.00000  -0.00000   1.20527   0.54763  -0.00000   0.00000
-4.330E-01  0.000E+00  0.000E+00  -0.00000  -0.00000   1.04366   0.67745  -0.00000   0.00000
-4.096E-01  0.000E+00  0.000E+00   0.00000  -0.00000   0.79740   0.73725   0.00000   0.00000
-3.862E-01  0.000E+00  0.000E+00  -0.00000  -0.00000  -0.16350   0.73080  -0.00000   0.00000
-3.628E-01  0.000E+00  0.000E+00  -0.00000  -0.00000  -0.27552   0.50952  -0.00000  -0.00000
-3.394E-01  0.000E+00  0.000E+00  -0.00000  -0.00000  -0.30770   0.30251   0.00000  -0.00000
-3.160E-01  0.000E+00  0.000E+00   0.00000  -0.00000  -0.29075   0.14597  -0.00000  -0.00000
-2.926E-01  0.000E+00  0.000E+00   0.00000  -0.00000  -0.24434   0.03789   0.00000  -0.00000
-2.692E-01  0.000E+00  0.000E+00   0.00000  -0.00000  -0.18603  -0.03123   0.00000  -0.00000
-2.458E-01  0.000E+00  0.000E+00   0.00000  -0.00000  -0.12775  -0.06575   0.00000  -0.00000
-2.224E-01  0.000E+00  0.000E+00   0.00000   0.00000  -0.07632  -0.07883   0.00000  -0.00000
-1.989E-01  0.000E+00  0.000E+00   0.00000   0.00000  -0.03688  -0.07503   0.00000   0.00000
-1.755E-01  0.000E+00  0.000E+00   0.00000   0.00000  -0.00836  -0.06318   0.00000  -0.00000
-1.521E-01  0.000E+00  0.000E+00   0.00000   0.00000   0.00933  -0.04767  -0.00000   0.00000
-1.287E-01  0.000E+00  0.000E+00   0.00000   0.00000   0.01830  -0.03206  -0.00000  -0.00000
-1.053E-01  0.000E+00  0.000E+00   0.00000   0.00000   0.02123  -0.01840  -0.00000  -0.00000
-8.192E-02  0.000E+00  0.000E+00   0.00000   0.00000   0.01968  -0.00804   0.00000   0.00000
-5.851E-02  0.000E+00  0.000E+00  -0.00000   0.00000   0.01607  -0.00067  -0.00000   0.00000
-3.511E-02  0.000E+00  0.000E+00  -0.00000  -0.00000   0.01157   0.00347  -0.00000   0.00000
-1.170E-02  0.000E+00  0.000E+00  -0.00000  -0.00000   0.00725   0.00537  -0.00000   0.00000
 1.170E-02  0.000E+00  0.000E+00  -0.00000  -0.00000   0.00370   0.00556   0.00000  -0.00000
 3.511E-02  0.000E+00  0.000E+00   0.00000  -0.00000   0.00118   0.00473  -0.00000  -0.00000
 5.851E-02  0.000E+00  0.000E+00  -0.00000   0.00000  -0.00037   0.00350  -0.00000   0.00000
 8.192E-02  0.000E+00  0.000E+00  -0.00000   0.00000  -0.00099   0.00233  -0.00000  -0.00000
 1.053E-01  0.000E+00  0.000E+00  -0.00000  -0.00000  -0.00109   0.00151   0.00000  -0.00000
 1.287E-01  0.000E+00  0.000E+00  -0.00000  -0.00000  -0.00092   0.00137   0.00000   0.00000
 1.521E-01  0.000E+00  0.000E+00  -0.00000   0.00000  -0.00083   0.00198   0.00000  -0.00000
 1.755E-01  0.000E+00  0.000E+00  -0.00000   0.00000  -0.00119   0.00355   0.00000  -0.00000
 1.989E-01  0.000E+00  0.000E+00  -0.00000  -0.00000  -0.00231   0.00627   0.00000  -0.00000
 2.224E-01  0.000E+00  0.000E+00  -0.00000  -0.00000  -0.00435   0.01031   0.00000  -0.00000
 2.458E-01  0.000E+00  0.000E+00  -0.00000  -0.00000  -0.00764   0.01644   0.00000   0.00000
 2.692E-01  0.000E+00  0.000E+00   0.00000  -0.00000  -0.01199   0.02493   0.00000   0.00000
 2.926E-01  0.000E+00  0.000E+00  -0.00000  -0.00000  -0.01729   0.03739   0.00000   0.00000
 3.160E-01  0.000E+00  0.000E+00   0.00000  -0.00000  -0.02279   0.05492  -0.00000   0.00000
 3.394E-01  0.000E+00  0.000E+00   0.00000   0.00000  -0.02689   0.07987   0.00000   0.00000
 3.628E-01  0.000E+00  0.000E+00   0.00000  -0.00000  -0.02611   0.11547   0.00000   0.00000
 3.862E-01  0.000E+00  0.000E+00   0.00000  -0.00000  -0.01723   0.15689   0.00000   0.00000
 4.096E-01  0.000E+00  0.000E+00   0.00000  -0.00000   0.08404   0.15656   0.00000   0.00000
 4.330E-01  0.000E+00  0.000E+00   0.00000   0.00000   0.00918   0.16548   0.00000   0.00000
 4.564E-01  0.000E+00  0.000E+00   0.00000   0.00000  -0.06139   0.16754  -0.00000   0.00000
 4.798E-01  0.000E+00  0.000E+00   0.00000   0.00000  -0.12922   0.15953  -0.00000  -0.00000
 5.032E-01  0.000E+00  0.000E+00   0.00000   0.00000  -0.19302   0.13802  -0.00000   0.00000
 5.266E-01  0.000E+00  0.000E+00   0.00000  -0.00000  -0.25016   0.10058  -0.00000   0.00000
 5.500E-01  0.000E+00  0.000E+00   0.00000  -0.00000  -0.29827   0.04870  -0.00000  -0.00000
 5.734E-01  0.000E+00  0.000E+00   0.00000  -0.00000  -0.33092  -0.01905  -0.00000   0.00000
 5.968E-01  0.000E+00  0.000E+00   0.00000   0.00000  -0.34716  -0.09712  -0.00000   0.00000
\end{verbatim}}

In the example given, the incident wave is propagating in the $+x$ direction.
The first layer of dipoles in the rectangular target is at $x/d=-31.5$, and the
last layer is at $x/d=-0.5$.
The ``surface'' of the target is at $x/d=-32$ ($x=-0.50\micron$) and
$x/d=0$ ($x=0$).

Figure \ref{fig:E_ellipsoid_nf} shows how $|E|^2$ varies along a line passing
through the center of the Au sphere.
As expected, the wave is strongly suppressed in the interior of the Au sphere.

\begin{figure}[h]
\begin{center}
\vspace*{-0.1cm}
\includegraphics[width=8.3cm,angle=270]{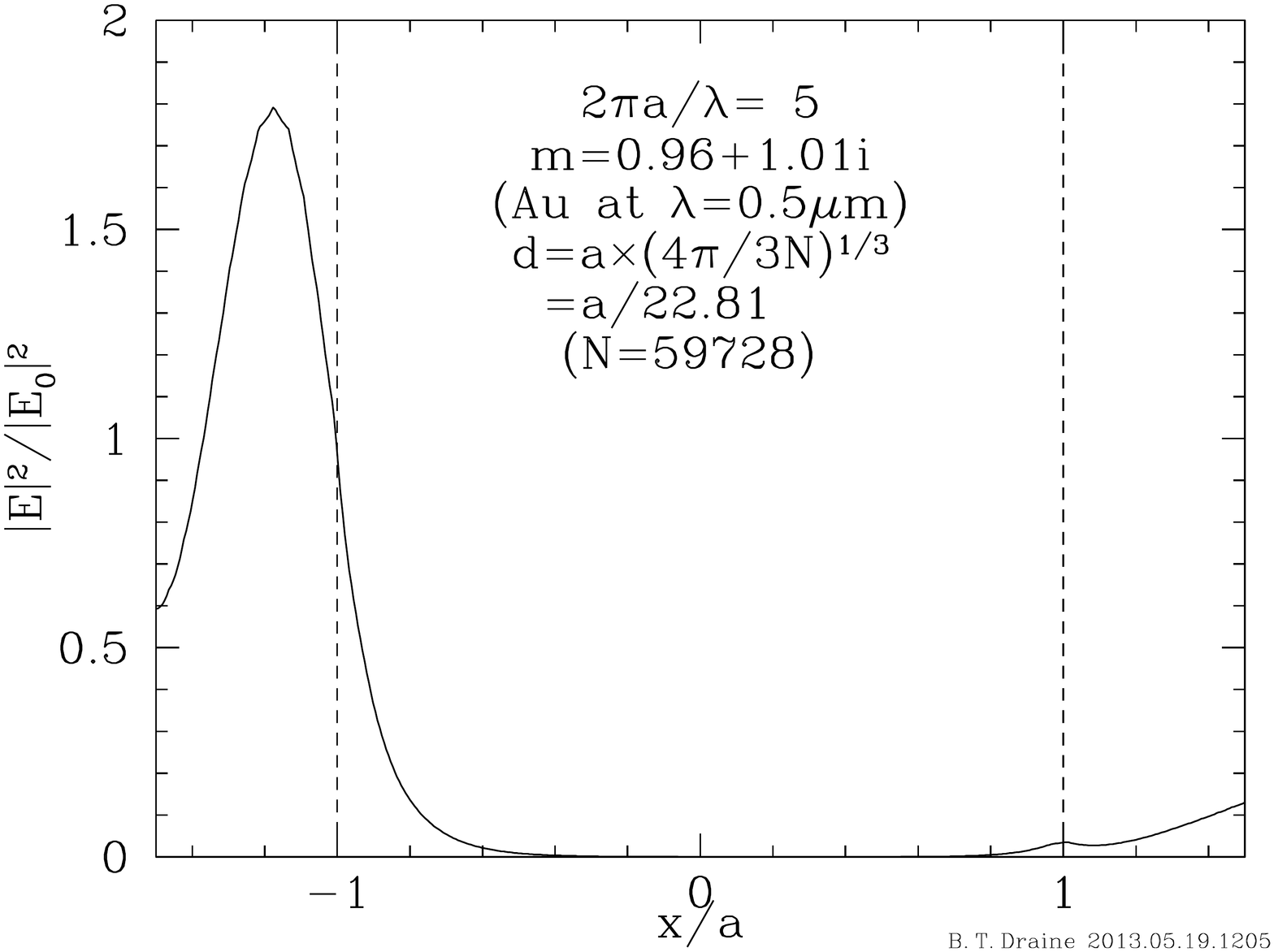}
\vspace*{-0.4cm}
\caption{\label{fig:E_ellipsoid_nf}
         \footnotesize
         Solid line: $|E|^2$ on track passing through center of Au sphere for
         the sample problem in examples\_exp/ELLIPSOID\_NEARFIELD,
         calculated using target option {\tt ELLIPSOID}.
         Note how weak the $\bE$ field is inside the Au.
         Dashed line: $|E|^2$ on track passing near the corner of
         the slab.
         Incident radiation is propagating in the $+x$ direction.
}
\end{center}
\end{figure}

\subsection{Modifying DDPOSTPROCESS.f90}

{\tt DDPOSTPROCESS.f90} is written to call subroutine {\tt readnf.f90}, which
reads $\bP_j$, $\bE_{\inc,j}$, 
$\bE_{\sca,j}$
from the stored file written by \ddscatv\ subroutine NEARFIELD
(see {\tt nearfield.f90}), as well as $\bB_{\inc,j}$ and $\bB_{\sca,j}$ if the
magnetic field has also been computed.

{\tt DDPOSTPROCESS.f90} includes some calls to VTK routines to prepare data
for VTK visualization.  It is written in standard Fortran 90, and
can be readily modified to do additional calculations with $\bE$ and
$\bB$ returned by  output data in other formats.
 
\section{Displaying Target Shapes}

\subsection{VTRCONVERT}
\index{visualization}
\index{VTRCONVERT}

It is often desirable to be able to display the target shape. We are
providing a program VTRCONVERT.f90 which allows to convert DDSCAT
shape format to VTK format. VTK is used world-wide in many advanced
visualization applications such as: ParaView, VisIt, 3DSlicer, or
MayaVi2. 

Every time \ddscatseventhree\ is run, it will create a "target.out" file.
To obtain a target.out file without running \ddscatseventhree\, the
user can run {\bf calltarget} to create a "target.out" file.
For example

\indent\indent calltarget < sphere40x40x40.shp

\noindent where file {}``sphere40x40x40.shp'' is

\indent\indent ELLIPSOID \\
\indent\indent 40 40 40 \\
\indent\indent 0 0 0

\noindent will create an ASCII file \textquotedbl{}target.out\textquotedbl{}
which is a list of the occupied lattice sites. The format of \textquotedbl{}target.out\textquotedbl{}
is the same as the format of the {}``shape.dat'' files read by \ddscat\ 
if option
FROM\_FILE is used in ddscat.par (when you run the \ddscat\ code). The DDSCAT
format is very simple but is not compatible with 
modern graphics programs such as "ParaView" or
"Mayavi2".  Therefore we have created a program {\bf VTRCONVERT} to convert
between DDSCAT format and VTK format.

The VTRCONVERT.f90 reads target shape data 
\textquotedblleft{}target.out\textquotedblright{}
and converts it to VTR format. The calling sequence is 

\indent\indent VTRCONVERT target.out output

where target.out is the name of the DDSCAT shape file created by 
``calltarget''.
The code writes two output files: 
{}``output\_1.vtr\textquotedblright{} and {}``output.pvd''.
These files can be directly read in by {}``ParaView'' or {}``Mayavi2''. 

If you are familar with PERL you can execute a script ``shapes.pl''.
It will create several shape files.

\subsection{\label{sec:VTK}
            What is VTK?}
\index{visualization!VTK: Visualization ToolKit}
\index{VTK: Visualization ToolKit}

The Visualization Toolkit (VTK) is an open-source, freely available
software system for 3D computer graphics and visualization. VTK is
used world-wide in many advanced visualization applications such as:
ParaView, VisIt, 3DSlicer, or MayaVi2. See for example

\indent\indent http://en.wikipedia.org/wiki/ParaView 
\index{Paraview}
\index{visualization!Paraview}

\indent\indent http://en.wikipedia.org/wiki/MayaVi
\index{MayaVi}
\index{visualization!MayaVi}

Our converter code VTRCONVERT.f90 is written in FORTRAN90 
and relies on the public
domain software module vtr.f90 written by Jalel Chergui -- see

\indent\indent http://www.limsi.fr/Individu/chergui/pv/PVD.htm

vtr.f90 defines 5 subroutines which enable writing ASCII data
in XML VTK format. The code allows the user to write a 3D rectilinear mesh
defined by x, y and z components.


\subsection{How to plot shapes once you have VTR/PVD files.}
\index{Paraview}
\index{visualization!ParaView}
\index{visualization!VTK: Visualization ToolKit}

Go to http://paraview.org/paraview/resources/software.html
and download the ParaView executable to your system 
(ParaView is available for Windows, Linux, and MacOS)

To plot a contour of a surface:

\begin{enumerate}
\item In the toolbar, go to File/Open 
      and select the file, e.g., \textquotedbl{}shapes/cylinder80x40.pvd
      \textquotedbl{} (note that in ParaView one needs to open
{*}.pvd file) 
\item Under Object Inspector/Properties
 click \textquotedbl{}apply\textquotedbl{}
\end{enumerate}

To add spheres indicating dipole positions

\begin{enumerate}
\item In the toolbar, go to Filters/Common/Contour 
\item under Object Inspector/Properties click 
      \textquotedbl{}apply\textquotedbl{} 
\item In the toolbar, go to Filters/Common/Glyph
\item Under Object Inspector/Properties one needs to change
several settings 
\item Glyph Type - change to \textquotedbl{}sphere\textquotedbl{} 
\item Then

  \begin{enumerate}
  \item change \textquotedbl{}Radius\textquotedbl{} to 0.1 (make radius
  of a dipole smaller) as an initial choice.
  \item change \textquotedbl{}Maximum Number of Points\textquotedbl{} to 5000 
      (or perhaps more depending on shape, but
      the points you have the longer it takes to plot)
  \item toggle \textquotedbl{}Mask Points\textquotedbl{} to off 
  \item toggle \textquotedbl{}Random Mode\textquotedbl{} to off 
      (plot dipoles in their positions) 
  \end{enumerate} 

\item Click \textquotedbl{}Apply\textquotedbl{} in \textquotedbl{}object
inspector\textquotedbl{}\\
The initial choice of 0.1 for the radius in step 6 may
not produce "touching" spheres -- you can return to step 6
and
iterate until you like the appearanace
\item To rotate the figure, left-click on it and "drag" left-right
and/or up-down to rotate around the vertical and/or horizontal axes. 
\item To add arrow showing target axes ${\bf a}_1$
  \begin{enumerate}
  \item In the toolbar, go to File/Open/a1a2$\_$1.pvd \\
  This file has just one point (at position (0,0,0)) and
  two vectors a1(3) and a2(3) which are "target axes".
  \item Under Object Inspector/Properties click "Apply".
  \item In the Toolbar, go to Filters/Common/Glyph
  \item Under Object Inspector/Properties 

    \begin{enumerate}
    \item "Vectors" will show "a1"
    \item change Glyph Type to "arrow"
    \item change "Glyph Type" to "arrow"
    \item change "Tip Radius" to 0.03 (improves appearance; suit yourself)
    \item change "Tip Length" to 0.1  (improves appearance...)
    \item change "Shaft Radius" to 0.01 (improves appearance...)
    \item click "edit" next to "Set Scale Factor"
    \item change "Scale Factor" to some value like "40"
    \item toggle "Mask Points" to off
    \item toggle "Random Mode" to off
    \item click "Apply"
    \end{enumerate}
  \end{enumerate}

\item To add arrow showing target axis ${\bf a}_2$
  \begin{enumerate}
  \item In the toolbar, go to File/Open/a1a2$\_$1.pvd
  \item Under Object Inspector/Properties click "Apply".
  \item In the Toolbar, go to Filters/Common/Glyph
  \item Under Object Inspector/Properties 
    \begin{enumerate}
    \item "Vectors" will show "a1" -- change this to "a2"
    \item change "Glyph Type" to "arrow"
    \item change "Tip Radius" to 0.03 (improves appearance...)
    \item change "Tip Length" to 0.1  (improves appearance...)
    \item change "Shaft Radius" to 0.01 (improves appearance...)
    \item click "edit" next to "Set Scale Factor"
    \item change "Scale Factor" to some value like "40"
    \item toggle "Mask Points" to off
    \item toggle "Random Mode" to off
    \item click "Apply"
   \end{enumerate}
\end{enumerate}

\item To save the image: in the toolbar, go to File/Save Screenshot
\begin{enumerate}
\item select desired resolution (default may be acceptable) and click \textquotedbl{}Ok\textquotedbl{}.
\item enter the desired filename
\item select the file type 
(options are .jpg, .png, .pdf, .tif, .ppm, .bmp)
\item click \textquotedbl{}OK\textquotedbl{}
\end{enumerate}
\end{enumerate}

\noindent If you wish to add additional features to images,
please consult the ParaView documentation.

\begin{figure}[h]
\begin{center}
\includegraphics[width=8.0cm,angle=0]{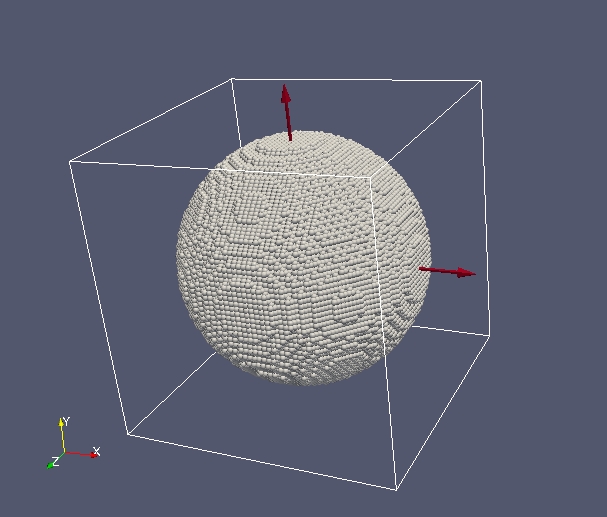}
\caption{\footnotesize
         Visualization with ParaView of the dipole realization of a sphere,
         produced following above instructions, using files in directory
examples\_exp/ELLIPSOID.}
\end{center}
\end{figure} 

\section{Visualization of the Electric Field
         \label{sec:visualization of E}
         }

The following instructions assume the user has the MayaVi2 graphics package installed.
If you don't have it yet, go to\\
\indent\indent http://code.enthought.com/projects/mayavi/mayavi/installation.html\\
for installation options.

\begin{enumerate}
\item First run example ELLIPSOID\_NEARFIELD. This will produce
      output\_1.vtr file. 
\item Start Mayavi2
\item File/load data/open file output
\item Point to examples\_exp/ELLIPSID\_NEARFIELD/VTRoutput\_1.vtr
\item colors and legends/add module/outline
\item outline/right click/add module/contour grid plane
\item In Mayavi2 object editor you can position slider between (0,95).
      Choose value in the middle. Set "filled contours" to on. 
      Increase number of contours to 64.
\item One can add more "contour grid planes" to illustrate the results in
different cross sections.
\end{enumerate}

\begin{figure}
\begin{center}
\includegraphics[width=12.0cm,angle=0]{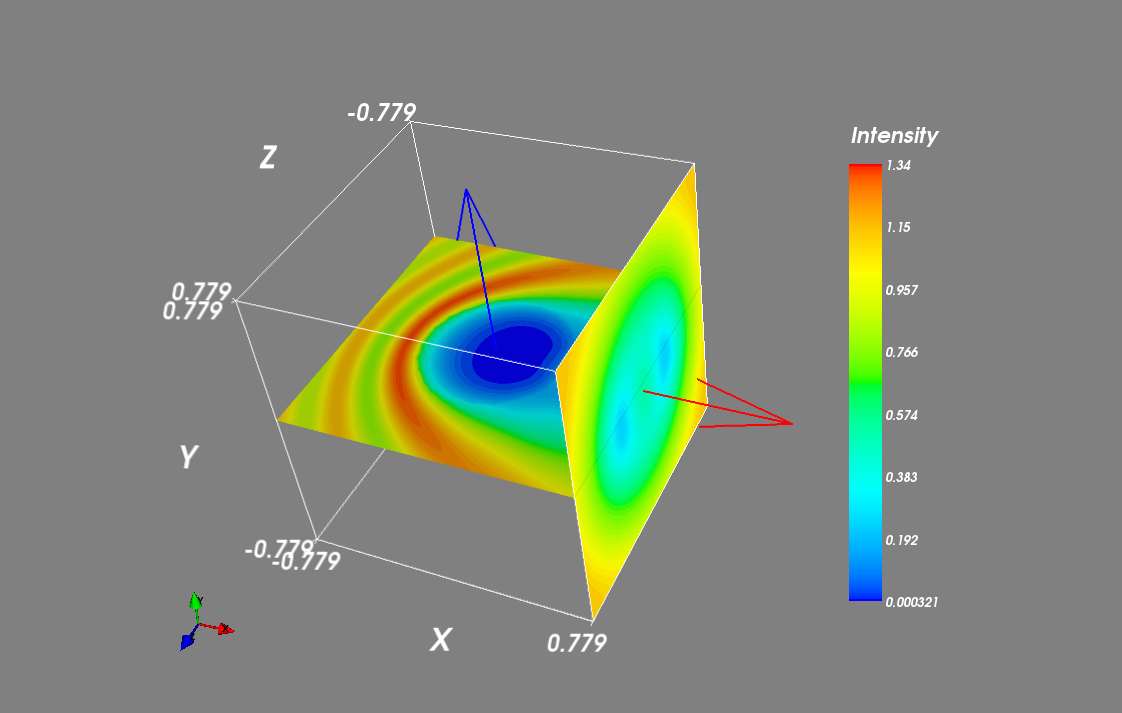}
\caption{\footnotesize
         $|\bE|/|\bE_0|$ on two planes, one passing through the center
         and one passing near the $a=0.398\micron$ Au sphere
         calculated in examples\_exp/ELLIPSOID\_NEARFIELD.
         The incident wave, with $\lambda=0.5\micron$, 
         is propagating with ${\bf k}_0\parallel\xtf$.
         and $\bE_0\parallel\ytf$.
         The axes are labelled in $\micron$.
         This is the same problem as the results shown in Figure
         \ref{fig:E_ellipsoid_nf}.
         This figure was generated by MayaVi2.}
\end{center}
\end{figure}

\section{Finale}

This User Guide is somewhat inelegant, but we hope that it will prove
useful.  The structure of the {\tt ddscat.par} file is intended to be
simple and suggestive so that, after reading the above notes once, the
user may not have to refer to them again.

Known bugs in
\ddscat\ will be posted 
at the \ddscat\ web site,\\
\hspace*{3em}{\tt http://code.google.com/p/ddscat/wiki/ReleaseNotes}\\
and the latest version of \ddscat\ can be found at\\
\hspace*{3em}{\tt http://code.google.com/p/ddscat/}

\medskip
Users are encouraged to provide B. T. Draine 
({\tt draine@astro.princeton.edu}) with their
email address; email notification of bug fixes,
and any new releases of \ddscat, 
will be made known to those who do.

\index{SCATTERLIB}
P. J. Flatau maintains the  
``SCATTERLIB - Light Scattering Codes Library"
at\\
\hspace*{3em}{\tt http://code.google.com/p/scatterlib} .\\
Emphasis is on providing source
codes (mostly FORTRAN). However, other information related to scattering 
on spherical and
non-spherical particles is collected: an extensive list of references 
to light scattering methods, refractive index, etc. 
This URL page contains section on the discrete dipole approximation.

\medskip
Concrete suggestions for improving {{\bf DDSCAT}}\ (and this User Guide) are
welcomed.

\medskip
Users of \ddscat\ should cite appropriate papers describing DDA theory and its
implementation.  The following papers may be relevant
(pdfs are included in the /doc directory)
\begin{itemize}
\item \citet{Draine_1988}: 
basic theory of the DDA, including radiative reaction, and application to
anisotropic materials;
\item \citet{Goodman+Draine+Flatau_1990}: introduction of FFT methods to greatly
accelerate DDA calcualations
\item \citet{Draine+Flatau_1994}: review of the DDA, including 
demonstrations of accuracy and convergence;
\item \citet{Draine+Flatau_2008a}: extension of the DDA to periodic
structures, and generalization of the Mueller scattering matrix to
describe scattering by 1-d and 2-d periodic structures.
\item \citet{Flatau+Draine_2012}: implementation of efficient
near-field calculations using FFTs.
\item This UserGuide.
\end{itemize}

\section{Acknowledgments}

\begin{itemize}
\index{ESELF}
\item The routine {\tt ESELF} making use of the FFT was originally written by 
Jeremy Goodman, Princeton University Observatory.  
\item The FFT routine {\tt FOURX}, used in the comparison of different
FFT routines,
is based on a FFT routine written by Norman
Brenner \citet{Brenner_1969}. 
\index{GPFAPACK}
\item The {\tt GPFAPACK} package was written by Clive Temperton 
\citep{Temperton_1992}, and
generously made available by him for use with {\tt DDSCAT}.
\item Much of the work involved in modifying {\tt DDSCAT} to use {\tt MPI} 
was done by Matthew Collinge, Princeton University.
\item The conjugate gradient routine {\tt ZBCG2} was written by
M.A. Botchev \\
({\tt http://www.math.utwente.nl/$\sim$botchev/}), based on
earlier work by D.R. Fokkema (Dept. of Mathematics, Utrecht University).
\item Art Lazanoff (NASA Ames Research Center) did most of the coding necessary
to use OpenMP and to use the {\tt DFTI} library routine from the
\Intel\ Math Kernel Library, as well as considerable testing of the new
code.  
\item The conjugate gradient routine {\tt qpbicg.f90} was written
by P.C. Chaumet and A. Rahmani \citep{Chaumet+Rahmani_2009}.
\item The conjugate gradient routine {\tt qmrpim2.f90} is based on f77
code written by P.C. Chaumet and A. Rahmani.
\item {\tt vtr.f90} was written by Jalel Chergui (LIMSI-CNRS).
\item Ian Wong helped write subroutine BSELF and helped implement
the filtered couple dipole option.
\end{itemize}
We are deeply 
indebted to all of these authors for making their work and code available.  

We wish also to acknowledge bug reports and suggestions from {{\bf DDSCAT}}
users, including
Rodrigo Alcaraz de la Osa, 
V. Choliy,
Michel Devel,
Souraya Goumri-Said,
Bo Hu,
Bala Krishna Juluri,
Stefan Kniefl,
Henrietta Lemke, 
Georges Levi, 
Shuzhou Li,
Wang Lin,
Paul Mulvaney, 
Timo Nousianen, 
Stuart Prescott,
Honoh Suzuki, 
Sanaz Vahidinia,
Bernhard Wasserman,
Mike Wolff,
and Hui Zhang.

Development of {{\bf DDSCAT}}\ was supported in part by 
National Science Foundation
grants AST-8341412, AST-8612013, AST-9017082, AST-9319283, 
AST-9616429, AST-9988126, AST-0406883, and AST-1008570 to BTD,
in part by support from the Office of Naval Research 
Young Investigator Program to PJF, in part by DuPont Corporate Educational
Assistance to PJF, and in part by the United Kingdom Defence Research Agency.

\bibliography{/u/draine/work/bib/btdrefs}
\newpage

\begin{appendix}
\section{Understanding and Modifying {\tt ddscat.par}\label{app:ddscat.par}}

In order to use DDSCAT to perform the specific calculations of interest to
you, it will be necessary to modify the {\tt ddscat.par} file.  
Here we list the sample {\tt ddscat.par} file for the example problem in
{\tt examples\_exp/RCTGLPRSM}, followed by a discussion of
how to modify this file as needed.
Note that all numerical input data in DDSCAT is read with free-format 
{\tt READ(IDEV,*)...} statements.  
Therefore you do not need to worry about the precise format in which integer 
or floating point numbers are entered on a line.
The crucial thing is that lines in {\tt ddscat.par} containing numerical 
data have the correct number of data entries, with any informational 
comments appearing {\it after} the numerical data on a given line.
\index{CMDFFT -- specifying FFT method}
\index{CMDFRM -- specifying scattering directions}

{\scriptsize
\begin{verbatim}
' ========= Parameter file for v7.3 ===================' 
'**** Preliminaries ****'
'NOTORQ' = CMDTRQ*6 (NOTORQ, DOTORQ) -- either do or skip torque calculations
'PBCGS2' = CMDSOL*6 (PBCGS2, PBCGST, GPBICG, PETRKP, QMRCCG) -- CCG method
'GPFAFT' = CMDFFT*6 (GPFAFT, FFTMKL) -- FFT method
'GKDLDR' = CALPHA*6 (GKDLDR, LATTDR, FLTRCD) -- DDA method
'NOTBIN' = CBINFLAG (NOTBIN, ORIBIN, ALLBIN) -- specify binary output
'**** Initial Memory Allocation ****'
100 100 100 = dimensioning allowance for target generation
'**** Target Geometry and Composition ****'
'RCTGLPRSM' = CSHAPE*9 shape directive
16 32 32  = shape parameters 1 - 3
1         = NCOMP = number of dielectric materials
'../diel/Au_evap' = file with refractive index 1
'**** Additional Nearfield calculation? ****'
0 = NRFLD (=0 to skip nearfield calc., =1 to calculate nearfield E)
0.0 0.0 0.0 0.0 0.0 0.0 (fract. extens. of calc. vol. in -x,+x,-y,+y,-z,+z)
'**** Error Tolerance ****'
1.00e-5 = TOL = MAX ALLOWED (NORM OF |G>=AC|E>-ACA|X>)/(NORM OF AC|E>)
'**** maximum number of iterations allowed ****'
300     = MXITER
'**** Interaction cutoff parameter for PBC calculations ****'
1.00e-2 = GAMMA (1e-2 is normal, 3e-3 for greater accuracy)
'**** Angular resolution for calculation of <cos>, etc. ****'
0.5	= ETASCA (number of angles is proportional to [(3+x)/ETASCA]^2 )
'**** Vacuum wavelengths (micron) ****'
0.5000 0.5000 1 'LIN' = wavelengths (first,last,how many,how=LIN,INV,LOG)
'**** Refractive index of ambient medium'
1.000 = NAMBIENT
'**** Effective Radii (micron) **** '
0.246186 0.246186 1 'LIN' = aeff (first,last,how many,how=LIN,INV,LOG)
'**** Define Incident Polarizations ****'
(0,0) (1.,0.) (0.,0.) = Polarization state e01 (k along x axis)
2 = IORTH  (=1 to do only pol. state e01; =2 to also do orth. pol. state)
'**** Specify which output files to write ****'
1 = IWRKSC (=0 to suppress, =1 to write ".sca" file for each target orient.
'**** Prescribe Target Rotations ****'
0.    0.   1  = BETAMI, BETAMX, NBETA  (beta=rotation around a1)
0.    0.   1  = THETMI, THETMX, NTHETA (theta=angle between a1 and k)
0.    0.   1  = PHIMIN, PHIMAX, NPHI (phi=rotation angle of a1 around k)
'**** Specify first IWAV, IRAD, IORI (normally 0 0 0) ****'
0   0   0    = first IWAV, first IRAD, first IORI (0 0 0 to begin fresh)
'**** Select Elements of S_ij Matrix to Print ****'
6	= NSMELTS = number of elements of S_ij to print (not more than 9)
11 12 21 22 31 41	= indices ij of elements to print
'**** Specify Scattered Directions ****'
'LFRAME' = CMDFRM (LFRAME, TFRAME for Lab Frame or Target Frame)
2 = NPLANES = number of scattering planes
0.   0. 180.  5 = phi, thetan_min, thetan_max, dtheta (in deg) for plane 1
90.  0. 180.  5 = phi, thetan_min, thetan_max, dtheta (in deg) for plane 2
\end{verbatim}
}
{\footnotesize
\begin{tabular}{l l}
Lines	&Comments\\
1-2	&comment lines \\
3	&{\tt NOTORQ} if torque calculation is not required; \\
	&{\tt DOTORQ} if torque calculation is required. \\
4	&{\tt PBCGS2} is recommended; 
	other options are {\tt PBCGST}, {\tt GPBICG}, {\tt PETRKP}, and 
        {\tt QMRCCG} 
        (see \S\ref{sec:choice_of_algorithm}).\\
5	&{\tt GPFAFT} is supplied as default, but {\tt FFTMKL} is recommended 
	if {\tt DDSCAT} has been compiled with\\
	& the \Intel\ Math Kernel Library (see \S\S\ref{sec:MKL},
	\protect{\ref{sec:choice_of_fft}}).\\
6	&{\tt GKDLDR} is recommended as the DDA method if the refractive index
         $m$ is not too large, but\\ 
        & {\tt FLTRCD} may be better if $|m|$ is large.
         (see \S\protect{\ref{sec:DDA_method}})\\
7	&{\tt NOTBIN} for no unformatted binary output.\\
	&{\tt ORIBIN} for unformatted binary dump of orientational averages only; \\
        &{\tt ALLBIN} for full unformatted binary dump (\S\ref{subsec:binary}); \\
8       &comment line \\
9	&initial memory allocation NX,NY,NZ.  These must be large enough to
        accomodate the target that\\
        &will be generated.\\
10	&comment line \\
11	&specify choice of target shape (see \S\ref{sec:target_generation} for
	description of options {\tt RCTGLPRSM}, {\tt ELLIPSOID}, \\
        &{\tt TETRAHDRN}, ...)\\
12	&shape parameters {SHPAR1}, {SHPAR2}, {SHPAR3}, ... 
	(see \S\ref{sec:target_generation}).\\
13	&number of different dielectric constant tables (see \S\ref{sec:dielectric_func}).\\
14	&name(s) of dielectric constant table(s) (one per line).\\
15	&comment line\\
16      &{\tt NRFLD} = 0, 1, 2 to skip, do nearfield calculation of $\bE$, do
        nearfield calculation of both $\bE$ and $\bB$\\
17      &6 non-negative numbers $r_1,...,r_6$ specifying fractional extension of
         comptutational volume\\
        & (in $-\xtf,+\xtf,-\ytf,+\ytf,-\ztf,+\ztf$
         direction (see \S\ref{sec:nearfield calc})\\
18      &commment line\\
20	&{\tt TOL} = error tolerance $h$: maximum allowed value of 
	$|A^\dagger E-A^\dagger AP|/|A^\dagger E|$ [see eq.(\ref{eq:err_tol})].\\
21	&comment line\\
22      &{\tt MXITER}$=$ maximum number of conjugate-gradient iterations 
         allowed\\
23      &comment line\\
24	&{\tt GAMMA}$=$ interaction cutoff parameter $\gamma$ 
         (see Draine \& Flatau 2009)\\
        &the value of $\gamma$ does not affect calculations for 
         isolated targets\\
25      &comment line\\
26	&{\tt ETASCA} -- parameter $\eta$ controlling angular averages (\S\ref{sec:averaging_scattering}).\\
27	&comment line\\
28	&$\lambda$ -- vacuum wavlenghts: first, last, how many, how chosen.\\
29	&comment line\\
30      &{\tt NAMBIENT}$=$ (real) refractive index of ambient medium.\\
31      & comment line\\
32	&$a_{\rm eff}$ -- first, last, how many, how chosen.\\
33	&comment line\\
34	&specify x,y,z components of (complex) incident polarization ${\hat{\bf e}}_{01}$ (\S\ref{sec:incident_polarization})\\
35	&{\tt IORTH} = 2 to do both polarization states (normal);\\
	&{\tt IORTH} = 1 to do only one incident polarization.\\
36      &comment line\\
37	&{\tt IWRKSC} = 0 to suppress writing of ``.sca'' files;\\
	&{\tt IWRKSC} = 1 to enable writing of ``.sca'' files.\\
38      &comment line\\
39	&$\beta$ (see \S\ref{sec:target_orientation}) -- first, last, how many .\\
40	&$\Theta$ (see \S\ref{sec:target_orientation}) -- first, last, how many.\\
41	&$\Phi$ (see \S\ref{sec:target_orientation}) -- first, last, how many.\\
42	&comment line\\
43	&{\tt IWAV0 IRAD0 IORI0} -- starting values of integers
	{\tt IWAV IRAD IORI} (normally {\tt 0 0 0}).\\
44	&comment line\\
45	&$N_{S}$ = number of scattering matrix elements (must be $\leq9$)\\
46	&indices $ij$ of $N_S$ elements of the scattering matrix $S_{ij}$\\
47	&comment line\\
48	&specify whether scattered directions are to be
         specified by {\tt CMDFRM='LFRAME'} or {\tt 'TFRAME'}.\\
49      &{\tt NPLANES}$=$ number of scattering planes to follow\\
50	&$\phi_s$ for first scattering plane, $\theta_{s,min}$, 
        $\theta_{s,max}$, how many $\theta_s$ values;\\
51,...	&$\phi_s$ for 2nd,... scattering plane, ...
\end{tabular}
}
\newpage
\section{{\tt w{\it xxx}r{\it yyy}.avg} Files\label{app:w000r000ori.avg}}

The file {\tt w000r000ori.avg} contains the results for the first wavelength
({\tt w000}) and first target radius ({\tt r000})
averaged over orientations ({\tt .avg}).
The {\tt w000r000ori.avg} file generated by the sample calculation 
in {\tt examples\_exp/RCTGLPRSM}
should look like the following:
\vspace*{-0.4em}
{\scriptsize
\begin{verbatim}
 DDSCAT --- DDSCAT 7.3.0 [12.12.29]   
 TARGET --- Rectangular prism; NX,NY,NZ=  16  32  32                          
 GKDLDR --- DDA method
 PBCGS2 --- CCG method
 RCTGLPRSM --- shape 
   16384     = NAT0 = number of dipoles
  0.06346821 = d/aeff for this target [d=dipole spacing]
    0.015625 = d (physical units)
  AEFF=      0.246186 = effective radius (physical units)
  WAVE=      0.500000 = wavelength (in vacuo, physical units)
K*AEFF=      3.093665 = 2*pi*aeff/lambda
NAMBIENT=    1.000000 = refractive index of ambient medium
n= ( 0.9656 ,  1.8628),  eps.= ( -2.5374 ,  3.5975)  |m|kd=  0.4120 for subs. 1
   TOL= 1.000E-05 = error tolerance for CCG method
( 1.00000  0.00000  0.00000 ) = target axis A1 in Target Frame
( 0.00000  1.00000  0.00000 ) = target axis A2 in Target Frame
  NAVG=   962 = (theta,phi) values used in comp. of Qsca,g
( 0.19635  0.00000  0.00000 ) = k vector (latt. units) in Lab Frame
( 0.00000, 0.00000 )( 1.00000, 0.00000 )( 0.00000, 0.00000 )=inc.pol.vec. 1 in LF
( 0.00000, 0.00000 )( 0.00000, 0.00000 )( 1.00000, 0.00000 )=inc.pol.vec. 2 in LF
   0.000   0.000 = beta_min, beta_max ;  NBETA = 1
   0.000   0.000 = theta_min, theta_max; NTHETA= 1
   0.000   0.000 = phi_min, phi_max   ;   NPHI = 1

 0.5000 = ETASCA = param. controlling # of scatt. dirs used to calculate <cos> etc.
 Results averaged over    1 target orientations
                   and    2 incident polarizations
          Qext       Qabs       Qsca      g(1)=<cos>  <cos^2>     Qbk       Qpha
 JO=1:  3.6134E+00  1.4308E+00  2.1827E+00  4.1463E-01 7.5169E-01 6.1831E-01 -1.2614E-01
 JO=2:  3.6135E+00 1.4308E+00 2.1827E+00  4.1463E-01 7.5169E-01 6.1831E-01 -1.2614E-01
 mean:  3.6134E+00  1.4308E+00  2.1827E+00  4.1463E-01 7.5169E-01 6.1831E-01 -1.2614E-01
 Qpol= -6.4373E-06                                                  dQpha= -4.9174E-07
         Qsca*g(1)   Qsca*g(2)   Qsca*g(3)   iter  mxiter  Nsca
 JO=1:  9.0502E-01 -3.2556E-09 -1.8964E-06     18    300    962
 JO=2:  9.0502E-01 -7.7139E-08  1.4648E-07     18    300    962
 mean:  9.0502E-01 -4.0197E-08 -8.7497E-07
            Mueller matrix elements for selected scattering directions in Lab Frame   
 theta    phi    Pol.    S_11        S_12        S_21       S_22       S_31       S_41
  0.00   0.00  0.00000  7.5115E+01 -1.2970E-04 -1.297E-04  7.512E+01  6.785E-05 -2.220E-06
  5.00   0.00  0.00768  7.2167E+01 -5.5400E-01 -5.540E-01  7.217E+01  6.960E-05  6.113E-06
 10.00   0.00  0.03116  6.3968E+01 -1.9932E+00 -1.993E+00  6.397E+01  6.634E-05  1.229E-05
 15.00   0.00  0.07180  5.2244E+01 -3.7512E+00 -3.751E+00  5.224E+01  5.818E-05  1.351E-05
 20.00   0.00  0.13160  3.9245E+01 -5.1645E+00 -5.165E+00  3.924E+01  4.661E-05  1.559E-05
 25.00   0.00  0.21215  2.7088E+01 -5.7468E+00 -5.747E+00  2.709E+01  3.576E-05  1.251E-05
 30.00   0.00  0.31112  1.7234E+01 -5.3617E+00 -5.362E+00  1.723E+01  2.802E-05  9.135E-06
 35.00   0.00  0.41156  1.0271E+01 -4.2270E+00 -4.227E+00  1.027E+01  1.873E-05  6.317E-06
 40.00   0.00  0.46174  6.0117E+00 -2.7759E+00 -2.776E+00  6.012E+00  1.343E-05  1.892E-06
 45.00   0.00  0.38394  3.7970E+00 -1.4578E+00 -1.458E+00  3.797E+00  9.607E-06 -1.182E-06
 50.00   0.00  0.20310  2.8383E+00 -5.7645E-01 -5.765E-01  2.838E+00  7.253E-06 -1.685E-06
 55.00   0.00  0.08917  2.4759E+00 -2.2077E-01 -2.208E-01  2.476E+00  6.675E-06 -2.270E-06
 60.00   0.00  0.12808  2.2936E+00 -2.9377E-01 -2.938E-01  2.294E+00  6.400E-06 -1.932E-06
 65.00   0.00  0.28428  2.1074E+00 -5.9909E-01 -5.991E-01  2.107E+00  5.960E-06 -1.838E-06
 70.00   0.00  0.49621  1.8862E+00 -9.3596E-01 -9.360E-01  1.886E+00  5.212E-06 -1.174E-06
 75.00   0.00  0.70066  1.6630E+00 -1.1652E+00 -1.165E+00  1.663E+00  4.518E-06 -7.433E-07
 80.00   0.00  0.83815  1.4710E+00 -1.2329E+00 -1.233E+00  1.471E+00  3.404E-06 -2.239E-07
 85.00   0.00  0.87905  1.3178E+00 -1.1584E+00 -1.158E+00  1.318E+00  2.064E-06 -6.282E-07
 90.00   0.00  0.84390  1.1878E+00 -1.0024E+00 -1.002E+00  1.188E+00  1.096E-06 -6.092E-07
 95.00   0.00  0.78583  1.0596E+00 -8.3269E-01 -8.327E-01  1.060E+00  6.710E-07 -4.426E-07
100.00   0.00  0.75672  9.2363E-01 -6.9893E-01 -6.989E-01  9.236E-01  1.610E-07 -4.866E-08
105.00   0.00  0.78540  7.9128E-01 -6.2148E-01 -6.215E-01  7.913E-01  2.381E-07  2.968E-07
110.00   0.00  0.85741  6.9404E-01 -5.9507E-01 -5.951E-01  6.940E-01  8.433E-07  8.885E-07
115.00   0.00  0.89355  6.7418E-01 -6.0241E-01 -6.024E-01  6.742E-01  1.297E-06  1.225E-06
120.00   0.00  0.81395  7.7422E-01 -6.3018E-01 -6.302E-01  7.742E-01  1.748E-06  1.515E-06
125.00   0.00  0.65813  1.0320E+00 -6.7920E-01 -6.792E-01  1.032E+00  2.553E-06  1.802E-06
130.00   0.00  0.51335  1.4856E+00 -7.6264E-01 -7.626E-01  1.486E+00  3.536E-06  2.497E-06
135.00   0.00  0.40829  2.1850E+00 -8.9213E-01 -8.921E-01  2.185E+00  4.473E-06  3.066E-06
140.00   0.00  0.33090  3.2004E+00 -1.0590E+00 -1.059E+00  3.200E+00  6.583E-06  4.014E-06
145.00   0.00  0.26514  4.6139E+00 -1.2233E+00 -1.223E+00  4.614E+00  9.927E-06  4.933E-06
150.00   0.00  0.20350  6.4859E+00 -1.3199E+00 -1.320E+00  6.486E+00  1.326E-05  6.138E-06
155.00   0.00  0.14593  8.8035E+00 -1.2847E+00 -1.285E+00  8.804E+00  1.727E-05  7.720E-06
160.00   0.00  0.09522  1.1432E+01 -1.0886E+00 -1.089E+00  1.143E+01  1.989E-05  7.600E-06
165.00   0.00  0.05409  1.4098E+01 -7.6253E-01 -7.625E-01  1.410E+01  2.215E-05  7.833E-06
170.00   0.00  0.02413  1.6425E+01 -3.9641E-01 -3.964E-01  1.642E+01  2.194E-05  7.723E-06
175.00   0.00  0.00604  1.8022E+01 -1.0884E-01 -1.088E-01  1.802E+01  2.147E-05  6.818E-06
180.00   0.00  0.00000  1.8591E+01 -1.5259E-05 -1.526E-05  1.859E+01  1.746E-05  5.261E-06
  0.00  90.00  0.00000  7.5115E+01  1.2970E-04  1.297E-04  7.512E+01 -6.792E-05 -2.304E-06
  5.00  90.00  0.00767  7.2167E+01 -5.5382E-01 -5.538E-01  7.217E+01 -7.026E-05  1.632E-06
 10.00  90.00  0.03116  6.3968E+01 -1.9931E+00 -1.993E+00  6.397E+01 -6.760E-05  6.943E-06
 15.00  90.00  0.07180  5.2244E+01 -3.7513E+00 -3.751E+00  5.224E+01 -6.176E-05  1.252E-05
 20.00  90.00  0.13160  3.9245E+01 -5.1645E+00 -5.164E+00  3.924E+01 -5.102E-05  1.746E-05
 25.00  90.00  0.21216  2.7088E+01 -5.7468E+00 -5.747E+00  2.709E+01 -3.823E-05  2.014E-05
 30.00  90.00  0.31112  1.7234E+01 -5.3617E+00 -5.362E+00  1.723E+01 -2.451E-05  2.102E-05
 35.00  90.00  0.41156  1.0271E+01 -4.2270E+00 -4.227E+00  1.027E+01 -1.188E-05  1.911E-05
 40.00  90.00  0.46174  6.0117E+00 -2.7758E+00 -2.776E+00  6.012E+00 -1.197E-06  1.526E-05
 45.00  90.00  0.38394  3.7970E+00 -1.4578E+00 -1.458E+00  3.797E+00  6.698E-06  1.039E-05
 50.00  90.00  0.20309  2.8383E+00 -5.7643E-01 -5.764E-01  2.838E+00  1.136E-05  5.124E-06
 55.00  90.00  0.08916  2.4759E+00 -2.2076E-01 -2.208E-01  2.476E+00  1.306E-05  6.010E-07
 60.00  90.00  0.12807  2.2936E+00 -2.9375E-01 -2.937E-01  2.294E+00  1.260E-05 -2.887E-06
 65.00  90.00  0.28427  2.1074E+00 -5.9906E-01 -5.991E-01  2.107E+00  1.042E-05 -4.946E-06
 70.00  90.00  0.49620  1.8862E+00 -9.3594E-01 -9.359E-01  1.886E+00  7.425E-06 -5.874E-06
 75.00  90.00  0.70066  1.6630E+00 -1.1652E+00 -1.165E+00  1.663E+00  4.307E-06 -5.601E-06
 80.00  90.00  0.83814  1.4710E+00 -1.2329E+00 -1.233E+00  1.471E+00  1.319E-06 -4.684E-06
 85.00  90.00  0.87905  1.3178E+00 -1.1584E+00 -1.158E+00  1.318E+00 -1.179E-06 -3.851E-06
 90.00  90.00  0.84389  1.1878E+00 -1.0023E+00 -1.002E+00  1.188E+00 -2.932E-06 -2.944E-06
 95.00  90.00  0.78583  1.0596E+00 -8.3267E-01 -8.327E-01  1.060E+00 -3.904E-06 -2.553E-06
100.00  90.00  0.75672  9.2362E-01 -6.9892E-01 -6.989E-01  9.236E-01 -3.952E-06 -2.432E-06
105.00  90.00  0.78540  7.9127E-01 -6.2146E-01 -6.215E-01  7.913E-01 -3.187E-06 -2.802E-06
110.00  90.00  0.85741  6.9402E-01 -5.9506E-01 -5.951E-01  6.940E-01 -1.639E-06 -3.211E-06
115.00  90.00  0.89354  6.7417E-01 -6.0240E-01 -6.024E-01  6.742E-01  4.964E-07 -3.506E-06
120.00  90.00  0.81394  7.7422E-01 -6.3017E-01 -6.302E-01  7.742E-01  3.042E-06 -3.363E-06
125.00  90.00  0.65811  1.0320E+00 -6.7919E-01 -6.792E-01  1.032E+00  5.488E-06 -2.384E-06
130.00  90.00  0.51334  1.4856E+00 -7.6263E-01 -7.626E-01  1.486E+00  7.604E-06 -5.760E-07
135.00  90.00  0.40828  2.1850E+00 -8.9210E-01 -8.921E-01  2.185E+00  8.701E-06  1.924E-06
140.00  90.00  0.33089  3.2005E+00 -1.0590E+00 -1.059E+00  3.200E+00  8.507E-06  5.022E-06
145.00  90.00  0.26513  4.6139E+00 -1.2233E+00 -1.223E+00  4.614E+00  6.576E-06  8.175E-06
150.00  90.00  0.20349  6.4859E+00 -1.3198E+00 -1.320E+00  6.486E+00  3.323E-06  1.081E-05
155.00  90.00  0.14593  8.8035E+00 -1.2847E+00 -1.285E+00  8.804E+00 -1.197E-06  1.261E-05
160.00  90.00  0.09522  1.1432E+01 -1.0885E+00 -1.089E+00  1.143E+01 -6.047E-06  1.320E-05
165.00  90.00  0.05409  1.4098E+01 -7.6252E-01 -7.625E-01  1.410E+01 -1.070E-05  1.235E-05
170.00  90.00  0.02413  1.6425E+01 -3.9637E-01 -3.964E-01  1.642E+01 -1.452E-05  1.062E-05
175.00  90.00  0.00604  1.8022E+01 -1.0885E-01 -1.089E-01  1.802E+01 -1.681E-05  8.054E-06
180.00  90.00  0.00000  1.8591E+01  1.5259E-05  1.526E-05  1.859E+01 -1.742E-05  5.249E-06
\end{verbatim}
}
\newpage\section{ w{\it xxx}r{\it yyy}k{\it zzz}.sca Files
	\label{app:w000r000k000.sca}}

The {\tt w000r000k000.sca} file contains the results for the first
wavelength ({\tt w000}), first target radius ({\tt r000}),
and first orientation ({\tt k000}).
The {\tt w000r000k000.sca} file created by the sample calculation 
in {\tt examples\_exp/RCTGLPRSM} should look
like the following:
{\scriptsize
\begin{verbatim}
 DDSCAT --- DDSCAT 7.3.0 [12.12.29]   
 TARGET --- Rectangular prism; NX,NY,NZ=  16  32  32                          
 GKDLDR --- DDA method
 PBCGS2 --- CCG method
 RCTGLPRSM --- shape 
   16384     = NAT0 = number of dipoles
  0.06346821 = d/aeff for this target [d=dipole spacing]
    0.015625 = d (physical units)
----- physical extent of target volume in Target Frame ------
     -0.250000      0.000000 = xmin,xmax (physical units)
     -0.250000      0.250000 = ymin,ymax (physical units)
     -0.250000      0.250000 = zmin,zmax (physical units)
  AEFF=      0.246186 = effective radius (physical units)
  WAVE=      0.500000 = wavelength (in vacuo, physical units)
K*AEFF=      3.093665 = 2*pi*aeff/lambda
NAMBIENT=    1.000000 = refractive index of ambient medium
n= ( 0.9656 ,  1.8628),  eps.= ( -2.5374 ,  3.5975)  |m|kd=  0.4120 for subs. 1
   TOL= 1.000E-05 = error tolerance for CCG method
( 1.00000  0.00000  0.00000 ) = target axis A1 in Target Frame
( 0.00000  1.00000  0.00000 ) = target axis A2 in Target Frame
  NAVG=   962 = (theta,phi) values used in comp. of Qsca,g
( 0.19635  0.00000  0.00000 ) = k vector (latt. units) in TF
( 0.00000, 0.00000 )( 1.00000, 0.00000 )( 0.00000, 0.00000 )=inc.pol.vec. 1 in TF
( 0.00000, 0.00000 )( 0.00000, 0.00000 )( 1.00000, 0.00000 )=inc.pol.vec. 2 in TF
( 1.00000  0.00000  0.00000 ) = target axis A1 in Lab Frame
( 0.00000  1.00000  0.00000 ) = target axis A2 in Lab Frame
( 0.19635  0.00000  0.00000 ) = k vector (latt. units) in Lab Frame
( 0.00000, 0.00000 )( 1.00000, 0.00000 )( 0.00000, 0.00000 )=inc.pol.vec. 1 in LF
( 0.00000, 0.00000 )( 0.00000, 0.00000 )( 1.00000, 0.00000 )=inc.pol.vec. 2 in LF
 BETA =  0.000 = rotation of target around A1
 THETA=  0.000 = angle between A1 and k
  PHI =  0.000 = rotation of A1 around k
 0.5000 = ETASCA = param. controlling # of scatt. dirs used to calculate <cos> etc.
          Qext       Qabs       Qsca      g(1)=<cos>  <cos^2>     Qbk       Qpha
 JO=1:  3.6134E+00  1.4308E+00  2.1827E+00  4.1463E-01 7.5169E-01 6.1831E-01 -1.2614E-01
 JO=2:  3.6135E+00  1.4308E+00  2.1827E+00  4.1463E-01 7.5169E-01 6.1831E-01 -1.2614E-01
 mean:  3.6134E+00  1.4308E+00  2.1827E+00  4.1463E-01 7.5169E-01 6.1831E-01 -1.2614E-01
 Qpol= -6.4373E-06                                                  dQpha= -4.9174E-07
         Qsca*g(1)   Qsca*g(2)   Qsca*g(3)   iter  mxiter  Nsca
 JO=1:  9.0502E-01 -3.2556E-09 -1.8964E-06     18    300    962
 JO=2:  9.0502E-01 -7.7139E-08  1.4648E-07     18    300    962
 mean:  9.0502E-01 -4.0197E-08 -8.7497E-07
            Mueller matrix elements for selected scattering directions in Lab Frame   
 theta    phi    Pol.    S_11        S_12        S_21       S_22       S_31       S_41
  0.00   0.00  0.00000  7.5115E+01 -1.2970E-04 -1.297E-04  7.512E+01  6.785E-05 -2.220E-06
  5.00   0.00  0.00768  7.2167E+01 -5.5400E-01 -5.540E-01  7.217E+01  6.960E-05  6.113E-06
 10.00   0.00  0.03116  6.3968E+01 -1.9932E+00 -1.993E+00  6.397E+01  6.634E-05  1.229E-05
 15.00   0.00  0.07180  5.2244E+01 -3.7512E+00 -3.751E+00  5.224E+01  5.818E-05  1.351E-05
 20.00   0.00  0.13160  3.9245E+01 -5.1645E+00 -5.165E+00  3.924E+01  4.661E-05  1.559E-05
 25.00   0.00  0.21215  2.7088E+01 -5.7468E+00 -5.747E+00  2.709E+01  3.576E-05  1.251E-05
 30.00   0.00  0.31112  1.7234E+01 -5.3617E+00 -5.362E+00  1.723E+01  2.802E-05  9.135E-06
 35.00   0.00  0.41156  1.0271E+01 -4.2270E+00 -4.227E+00  1.027E+01  1.873E-05  6.317E-06
 40.00   0.00  0.46174  6.0117E+00 -2.7759E+00 -2.776E+00  6.012E+00  1.343E-05  1.892E-06
 45.00   0.00  0.38394  3.7970E+00 -1.4578E+00 -1.458E+00  3.797E+00  9.607E-06 -1.182E-06
 50.00   0.00  0.20310  2.8383E+00 -5.7645E-01 -5.765E-01  2.838E+00  7.253E-06 -1.685E-06
 55.00   0.00  0.08917  2.4759E+00 -2.2077E-01 -2.208E-01  2.476E+00  6.675E-06 -2.270E-06
 60.00   0.00  0.12808  2.2936E+00 -2.9377E-01 -2.938E-01  2.294E+00  6.400E-06 -1.932E-06
 65.00   0.00  0.28428  2.1074E+00 -5.9909E-01 -5.991E-01  2.107E+00  5.960E-06 -1.838E-06
 70.00   0.00  0.49621  1.8862E+00 -9.3596E-01 -9.360E-01  1.886E+00  5.212E-06 -1.174E-06
 75.00   0.00  0.70066  1.6630E+00 -1.1652E+00 -1.165E+00  1.663E+00  4.518E-06 -7.433E-07
 80.00   0.00  0.83815  1.4710E+00 -1.2329E+00 -1.233E+00  1.471E+00  3.404E-06 -2.239E-07
 85.00   0.00  0.87905  1.3178E+00 -1.1584E+00 -1.158E+00  1.318E+00  2.064E-06 -6.282E-07
 90.00   0.00  0.84390  1.1878E+00 -1.0024E+00 -1.002E+00  1.188E+00  1.096E-06 -6.092E-07
 95.00   0.00  0.78583  1.0596E+00 -8.3269E-01 -8.327E-01  1.060E+00  6.710E-07 -4.426E-07
100.00   0.00  0.75672  9.2363E-01 -6.9893E-01 -6.989E-01  9.236E-01  1.610E-07 -4.866E-08
105.00   0.00  0.78540  7.9128E-01 -6.2148E-01 -6.215E-01  7.913E-01  2.381E-07  2.968E-07
110.00   0.00  0.85741  6.9404E-01 -5.9507E-01 -5.951E-01  6.940E-01  8.433E-07  8.885E-07
115.00   0.00  0.89355  6.7418E-01 -6.0241E-01 -6.024E-01  6.742E-01  1.297E-06  1.225E-06
120.00   0.00  0.81395  7.7422E-01 -6.3018E-01 -6.302E-01  7.742E-01  1.748E-06  1.515E-06
125.00   0.00  0.65813  1.0320E+00 -6.7920E-01 -6.792E-01  1.032E+00  2.553E-06  1.802E-06
130.00   0.00  0.51335  1.4856E+00 -7.6264E-01 -7.626E-01  1.486E+00  3.536E-06  2.497E-06
135.00   0.00  0.40829  2.1850E+00 -8.9213E-01 -8.921E-01  2.185E+00  4.473E-06  3.066E-06
140.00   0.00  0.33090  3.2004E+00 -1.0590E+00 -1.059E+00  3.200E+00  6.583E-06  4.014E-06
145.00   0.00  0.26514  4.6139E+00 -1.2233E+00 -1.223E+00  4.614E+00  9.927E-06  4.933E-06
150.00   0.00  0.20350  6.4859E+00 -1.3199E+00 -1.320E+00  6.486E+00  1.326E-05  6.138E-06
155.00   0.00  0.14593  8.8035E+00 -1.2847E+00 -1.285E+00  8.804E+00  1.727E-05  7.720E-06
160.00   0.00  0.09522  1.1432E+01 -1.0886E+00 -1.089E+00  1.143E+01  1.989E-05  7.600E-06
165.00   0.00  0.05409  1.4098E+01 -7.6253E-01 -7.625E-01  1.410E+01  2.215E-05  7.833E-06
170.00   0.00  0.02413  1.6425E+01 -3.9641E-01 -3.964E-01  1.642E+01  2.194E-05  7.723E-06
175.00   0.00  0.00604  1.8022E+01 -1.0884E-01 -1.088E-01  1.802E+01  2.147E-05  6.818E-06
180.00   0.00  0.00000  1.8591E+01 -1.5259E-05 -1.526E-05  1.859E+01  1.746E-05  5.261E-06
  0.00  90.00  0.00000  7.5115E+01  1.2970E-04  1.297E-04  7.512E+01 -6.792E-05 -2.304E-06
  5.00  90.00  0.00767  7.2167E+01 -5.5382E-01 -5.538E-01  7.217E+01 -7.026E-05  1.632E-06
 10.00  90.00  0.03116  6.3968E+01 -1.9931E+00 -1.993E+00  6.397E+01 -6.760E-05  6.943E-06
 15.00  90.00  0.07180  5.2244E+01 -3.7513E+00 -3.751E+00  5.224E+01 -6.176E-05  1.252E-05
 20.00  90.00  0.13160  3.9245E+01 -5.1645E+00 -5.164E+00  3.924E+01 -5.102E-05  1.746E-05
 25.00  90.00  0.21216  2.7088E+01 -5.7468E+00 -5.747E+00  2.709E+01 -3.823E-05  2.014E-05
 30.00  90.00  0.31112  1.7234E+01 -5.3617E+00 -5.362E+00  1.723E+01 -2.451E-05  2.102E-05
 35.00  90.00  0.41156  1.0271E+01 -4.2270E+00 -4.227E+00  1.027E+01 -1.188E-05  1.911E-05
 40.00  90.00  0.46174  6.0117E+00 -2.7758E+00 -2.776E+00  6.012E+00 -1.197E-06  1.526E-05
 45.00  90.00  0.38394  3.7970E+00 -1.4578E+00 -1.458E+00  3.797E+00  6.698E-06  1.039E-05
 50.00  90.00  0.20309  2.8383E+00 -5.7643E-01 -5.764E-01  2.838E+00  1.136E-05  5.124E-06
 55.00  90.00  0.08916  2.4759E+00 -2.2076E-01 -2.208E-01  2.476E+00  1.306E-05  6.010E-07
 60.00  90.00  0.12807  2.2936E+00 -2.9375E-01 -2.937E-01  2.294E+00  1.260E-05 -2.887E-06
 65.00  90.00  0.28427  2.1074E+00 -5.9906E-01 -5.991E-01  2.107E+00  1.042E-05 -4.946E-06
 70.00  90.00  0.49620  1.8862E+00 -9.3594E-01 -9.359E-01  1.886E+00  7.425E-06 -5.874E-06
 75.00  90.00  0.70066  1.6630E+00 -1.1652E+00 -1.165E+00  1.663E+00  4.307E-06 -5.601E-06
 80.00  90.00  0.83814  1.4710E+00 -1.2329E+00 -1.233E+00  1.471E+00  1.319E-06 -4.684E-06
 85.00  90.00  0.87905  1.3178E+00 -1.1584E+00 -1.158E+00  1.318E+00 -1.179E-06 -3.851E-06
 90.00  90.00  0.84389  1.1878E+00 -1.0023E+00 -1.002E+00  1.188E+00 -2.932E-06 -2.944E-06
 95.00  90.00  0.78583  1.0596E+00 -8.3267E-01 -8.327E-01  1.060E+00 -3.904E-06 -2.553E-06
100.00  90.00  0.75672  9.2362E-01 -6.9892E-01 -6.989E-01  9.236E-01 -3.952E-06 -2.432E-06
105.00  90.00  0.78540  7.9127E-01 -6.2146E-01 -6.215E-01  7.913E-01 -3.187E-06 -2.802E-06
110.00  90.00  0.85741  6.9402E-01 -5.9506E-01 -5.951E-01  6.940E-01 -1.639E-06 -3.211E-06
115.00  90.00  0.89354  6.7417E-01 -6.0240E-01 -6.024E-01  6.742E-01  4.964E-07 -3.506E-06
120.00  90.00  0.81394  7.7422E-01 -6.3017E-01 -6.302E-01  7.742E-01  3.042E-06 -3.363E-06
125.00  90.00  0.65811  1.0320E+00 -6.7919E-01 -6.792E-01  1.032E+00  5.488E-06 -2.384E-06
130.00  90.00  0.51334  1.4856E+00 -7.6263E-01 -7.626E-01  1.486E+00  7.604E-06 -5.760E-07
135.00  90.00  0.40828  2.1850E+00 -8.9210E-01 -8.921E-01  2.185E+00  8.701E-06  1.924E-06
140.00  90.00  0.33089  3.2005E+00 -1.0590E+00 -1.059E+00  3.200E+00  8.507E-06  5.022E-06
145.00  90.00  0.26513  4.6139E+00 -1.2233E+00 -1.223E+00  4.614E+00  6.576E-06  8.175E-06
150.00  90.00  0.20349  6.4859E+00 -1.3198E+00 -1.320E+00  6.486E+00  3.323E-06  1.081E-05
155.00  90.00  0.14593  8.8035E+00 -1.2847E+00 -1.285E+00  8.804E+00 -1.197E-06  1.261E-05
160.00  90.00  0.09522  1.1432E+01 -1.0885E+00 -1.089E+00  1.143E+01 -6.047E-06  1.320E-05
165.00  90.00  0.05409  1.4098E+01 -7.6252E-01 -7.625E-01  1.410E+01 -1.070E-05  1.235E-05
170.00  90.00  0.02413  1.6425E+01 -3.9637E-01 -3.964E-01  1.642E+01 -1.452E-05  1.062E-05
175.00  90.00  0.00604  1.8022E+01 -1.0885E-01 -1.089E-01  1.802E+01 -1.681E-05  8.054E-06
180.00  90.00  0.00000  1.8591E+01  1.5259E-05  1.526E-05  1.859E+01 -1.742E-05  5.249E-06
\end{verbatim}
}
\newpage\section{{\tt w{\it xxx}r{\it yyy}k{\it zzz}.pol{\it n}} Files
	\label{app:w000r000k000.poln}}

Binary files {\tt w{\it xxx}r{\it yyy}k{\it zzz}.pol{\it n}} are written to disk
only when nearfield calculations are done (parameter {\tt NRFLD}=1).
The {\tt w000r000k000.pol1} file contains the polarization solution for the
first wavelength ({\tt w000}), first target radius ({\tt r000}),
first orientation ({\tt k000}), and first incident polarization ({\tt pol1}).
In order to limit the size of this file, it has been written as an
{\it unformatted} or "binary" file.  
This preserves full machine precision for the data, is quite compact,
and can be read efficiently, but unfortunately the file is {\bf not}
fully portable because 
different computer architectures (e.g., Linux vs. MS Windows)
have adopted different standards for storage of ``unformatted'' data.
However, anticipating that many users will be computing within a single
architecture, the distribution version of \ddscat\ uses this format.

Additional warning: even on a single architecture, users should be alert to the
possibility that different compilers may follow
different conventions for reading/writing unformatted files.
  
The file contains the following information:
\begin{itemize}
\item The location of each dipole in the target frame.
\item $(k_x,k_y,k_z)d$, where $\bf k$ is the incident $k$ vector.
\item $(E_{0x},E_{0y},E_{0z})$, the complex polarization vector of the
incident wave.
\item $\alpha^{-1}d^3$, the inverse of the symmmetric complex
      polarizability tensor for each of the dipoles in the target.
\item $(P_x,P_y,P_z)$, the complex polarization vector for each of the
      dipoles.
\end{itemize}
The interested user should consult the routine {\tt writepol.f} to
see how this information has been organized in the unformatted file.

\section{{w{\it xxx}r{\it yyy}k{\it zzz}.E{\it n}} Files
         \label{app:w000r000k000.En}}

Binary files {\tt w{\it xxx}r{\it yyy}k{\it zzz}.E{\it n}} are written to disk
only when nearfield calculations of $\bE$ (but not $\bB$) 
are done (parameter {\tt NRFLD}=1).
The {\tt w000r000k000.E1} file contains information describing the problem and
the solution at "grid points" throughout the extended "computational volume" specified
for the nearfield calculation (see \S\ref{sec:nearfield}).
These binary files are very large, but have been written in a way to simplify
subsequent use for visualization using, e.g., 
the program {\tt DDPOSTPROCESS.f90}
(see \S\ref{sec:visualization of E}).
The interested user can examine subroutine {\tt nearfield.f90} that writes
the file, or
program {\tt DDPOSTPROCESS.f90} that reads the file, to see how the
information is organized.

A user who finds the file size to be a serious problem may wish to modify the
code in {\tt nearfield.f90} to suppress writing of some of the arrays
(e.g., the diagonal elements of the {\bf A} matrix), retaining only
the data of specific interest (e.g., the $\bE$ field).  Of course,
and modifications to {\tt WRITE} statements in {\tt nearfield.f90} will require corresponding
changes to {\tt READ} statements in subroutine {\tt readnf.f90} that 
{\tt DDPOSTPROCESS.f90} calls to read from the stored data files.

\section{{w{\it xxx}r{\it yyy}k{\it zzz}.EB{\it n}} Files
         \label{app:w000r000k000.En}}

Binary files {\tt w{\it xxx}r{\it yyy}k{\it zzz}.EB{\it n}} are written to disk
only when nearfield calculations are done for both $\bE$ and $\bB$
(parameter {\tt NRFLD}=2).
The {\tt w000r000k000.EB1} file contains information describing the problem and
the solution at "grid points" throughout the extended "computational volume" specified
for the nearfield calculation (see \S\ref{sec:nearfield}).
These binary files are very large, but have been written in a way to simplify
subsequent use for visualization using, e.g., 
the program {\tt DDPOSTPROCESS.f90}
(see \S\ref{sec:visualization of E}).
The interested user can examine subroutine {\tt nearfield.f90} that writes
the file, or
program {\tt DDPOSTPROCESS.f90} that reads the file, to see how the
information is organized.

A user who finds the file size to be a serious problem may wish to modify the
code in {\tt nearfield.f90} to suppress writing of some of the arrays
(e.g., the diagonal elements of the {\bf A} matrix), retaining only
the data of specific interest (e.g., $\bB_{\sca}$).  Of course,
and modifications to {\tt WRITE} statements in {\tt nearfield.f90} will require corresponding
changes to {\tt READ} statements in {\tt DDPOSTPROCESS.f90}.
\end{appendix}
\newpage

\addcontentsline{toc}{section}{Index}
\input UserGuide.ind
\end{document}